UNIVERSITÀ DEGLI STUDI DI TRIESTE

FACOLTÀ DI INGEGNERIA
DIPARTIMENTO DI ELETTROTECNICA, ELETTRONICA ED INFORMATICA

TESI DI LAUREA
IN
DISPOSITIVI ELETTRONICI

# Progetto di un detector a camera di ionizzazione per esperimenti SAXS

Laureando:
Francesco VOLTOLINA

Relatore:
Chiar.mo Prof. Sergio CARRATO

Correlatore:
Dr. Ralf Hendrik MENK

# Contents

# Progetto di un detector a camera di ionizzazione per esperimenti SAXS

Il lavoro presentato in questa Tesi di Laurea è nato dalla collaborazione tra l'IPL (Image Processing Laboratory) del *Dipartimento di Elettrotecnica, Elettronica ed Informatica* dell'Università degli Studi di Trieste ed il *Laboratorio di Strumentazione e Detectors* presso il sincrotrone ELETTRA di Trieste. In quel periodo era attivo presso Elettra un Progetto Europeo finalizzato al miglioramento delle strutture dedicate agli *esperimenti di X-ray scattering* (SAXS) nei principali sincrotroni europei.

Facevano parte di questo progetto molte strutture di ricerca ed in particolare i sincrotroni HASYLAB (Hamburger Synchrotron, Germany), il CCLRC (Daresbury Laboratory, UK), l'ESRF (European Synchrotron Radiation Facility in Grenoble, France) e l'Università di Siegen (Germany): una parte del tempo dedicato a questa tesi è stato speso proprio presso il *Gruppo di Fisica dei Detector* dell'Università di Siegen.

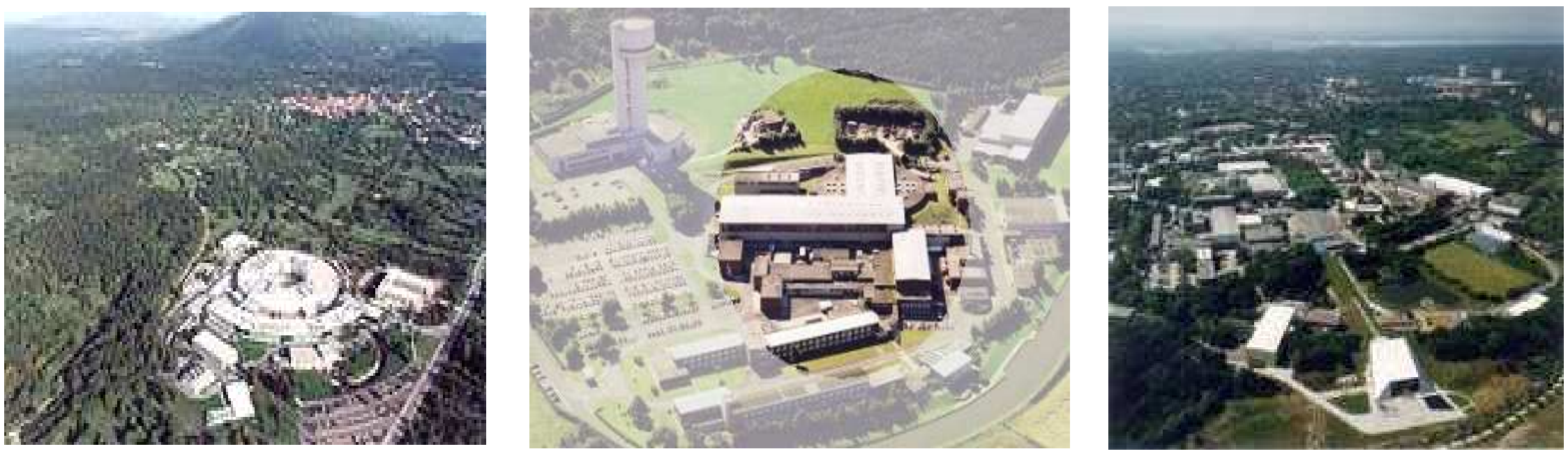

Figure I: Fotografie relative ad alcune installazioni per la produzione di luce di sincrotrone: Elettra (Trieste) -sinistra-, CCLRC (Daresbury) -centro-, HASYLAB (Hamburg) -destra-

Il principale problema degli esperimenti realizzati con le attuali sorgenti di luce di sincrotrone di terza generazione, quali Elettra, ma anche delle precedenti (meno brillanti) è che la rapida evoluzione tecnologica ed il relativo *incremento in brillanza* non è stato seguito da un paragonabile *incremento delle prestazioni dei detector* utilizzati negli esperimenti. Il miglioramento di questi dispositivi era un punto prioritario nell'ambito di questo progetto che si proponeva di consentire un utilizzo più efficace del potenziale insito nelle attuali macchine di luce di sincrotrone. Specificatamente, questo Progetto Europeo era finalizzato allo sviluppo di due detector di avanzate caratteristiche da utilizzarsi negli esperimenti di tipo SAXS: questo termine è l'acronimo inglese di *scattering di Raggi-X ai piccoli angoli* con cui si indica una tecnica sperimentale, basata sul fenomeno fisico dello scattering elastico, usata per sondare la struttura di oggetti con dimensioni caratteristiche comprese tra 1 e 1000 nm quali polimeri, micro-emulsioni o tessuti organici.

L'origine della teoria utilizzata in questi esperimenti risale alla *Legge di Bragg*, datata 1913, che spiegava il fenomeno per cui i cristalli riflettano i Raggi-X solamente sotto specifici angoli di incidenza. L'imponente lavoro di *Debye*, *Fournet and Guinier* sulla teoria del SAXS, negli anni '30, segnò una svolta fondamentale nell'utilizzo delle odierne metodologie. In accordo con la Legge di Bragg, l'informazione relativa a strutture con scala molecolare si trova per angoli di diffrazione inferiori ai 5 gradi: particolarizzando questa informazione alle caratteristiche dimensionali e alle specifiche energie dei raggi X utilizzati nella *stazione sperimentale della linea di luce SAXS presso ELETTRA*, otteniamo i requisiti richiesti ad un generico detector adatto a questi esperimenti.

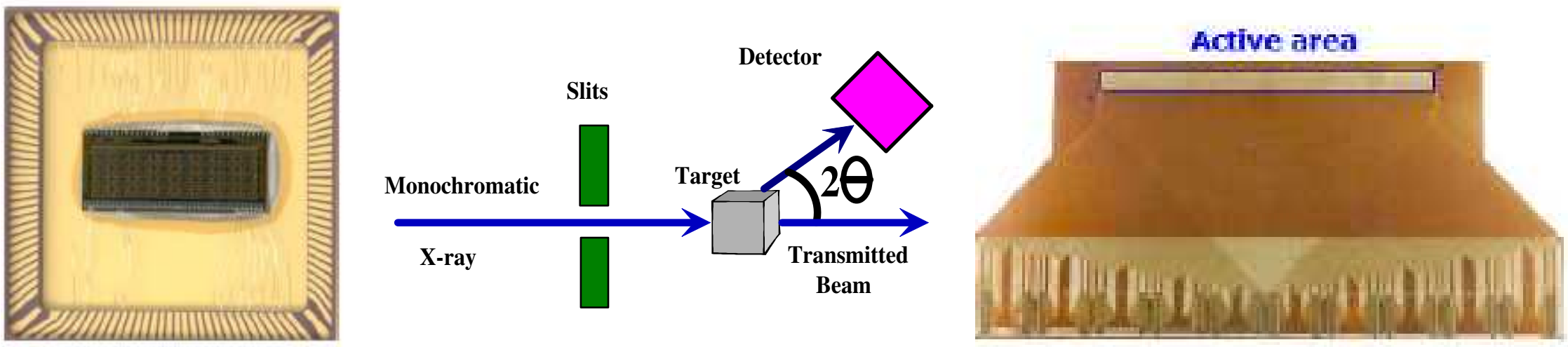

Figure II: Il circuito integrato JAMEX -sinistra-, un tipico esperimento di diffrazione eseguito con Raggi-X -centro-, lo speciale anodo costruito per questo detector -destra-

L'oggetto di questa tesi di laurea concerne proprio la progettazione e sviluppo di un simile dispositivo, attuata nell'ambito del sopra citato Progetto Europeo: esso avrà la forma di una *camera di ionizzazione multi-segmentata* contraddistinta da un'altissima dinamica e velocità di conteggio, caratteristiche che la rendono adatta agli esperimenti più critici. Benchè all'inizio del lavoro di tesi questo detector si trovasse in uno *stadio iniziale di sviluppo* e molte possibilità erano ancora disponibili per la sua effettiva realizzazione, lo speciale anodo era ormai pronto ed i circuiti integrati sviluppati appositamente per la sua lettura, chiamati JAMEX, erano disponibili a livello di prototipi.

Era perciò necessario progettare e realizzare, in tempi ridotti, una scheda elettronica capace di interfacciare questi ultimi dispositivi con un *sistema di acquisizione dati* di comune commercializzazione, seppure molto avanzato, di cui bisognava decidere la definitiva configurazione. Bisognava anche trovare la soluzione più efficace per realizzare effettivamente il detector: nello specifico c'era ancora libertà di scelta su come integrare i vari componenti, sulla forma del contenitore e la relativa struttura meccanica, ma anche relativamente all'interfaccia utente e all'eventuale protocollo di comunicazione. In tabella I.1 sono riassunte le caratteristiche richieste al nuovo detector per la linea SAXS.

| | |
|---|---|
| lunghezza della regione attiva | 192 mm |
| dimensioni esterne | $\leq$ 500 mm nella direzione del fascio |
| numero di pixels | 1280 |
| risoluzione temporale | $\cong$ 100 $\mu s$ |
| gamma dinamica a guadagno fissato | 10.000:1 minimo |
| regolazione di guadagno in passi | 16 = 4bit |

Table I.1: Prestazioni richieste al nuovo detector per la linea di luce SAXS

## I.1 Valutazione di differenti ipotesi costruttive

La prima soluzione costruttiva che si è voluta considerare per questo detector è quella che si può definire lo standard per le *apparecchiature di elevata complessità*, costruite in esemplare unico, utilizzate in generale per fini di ricerca.

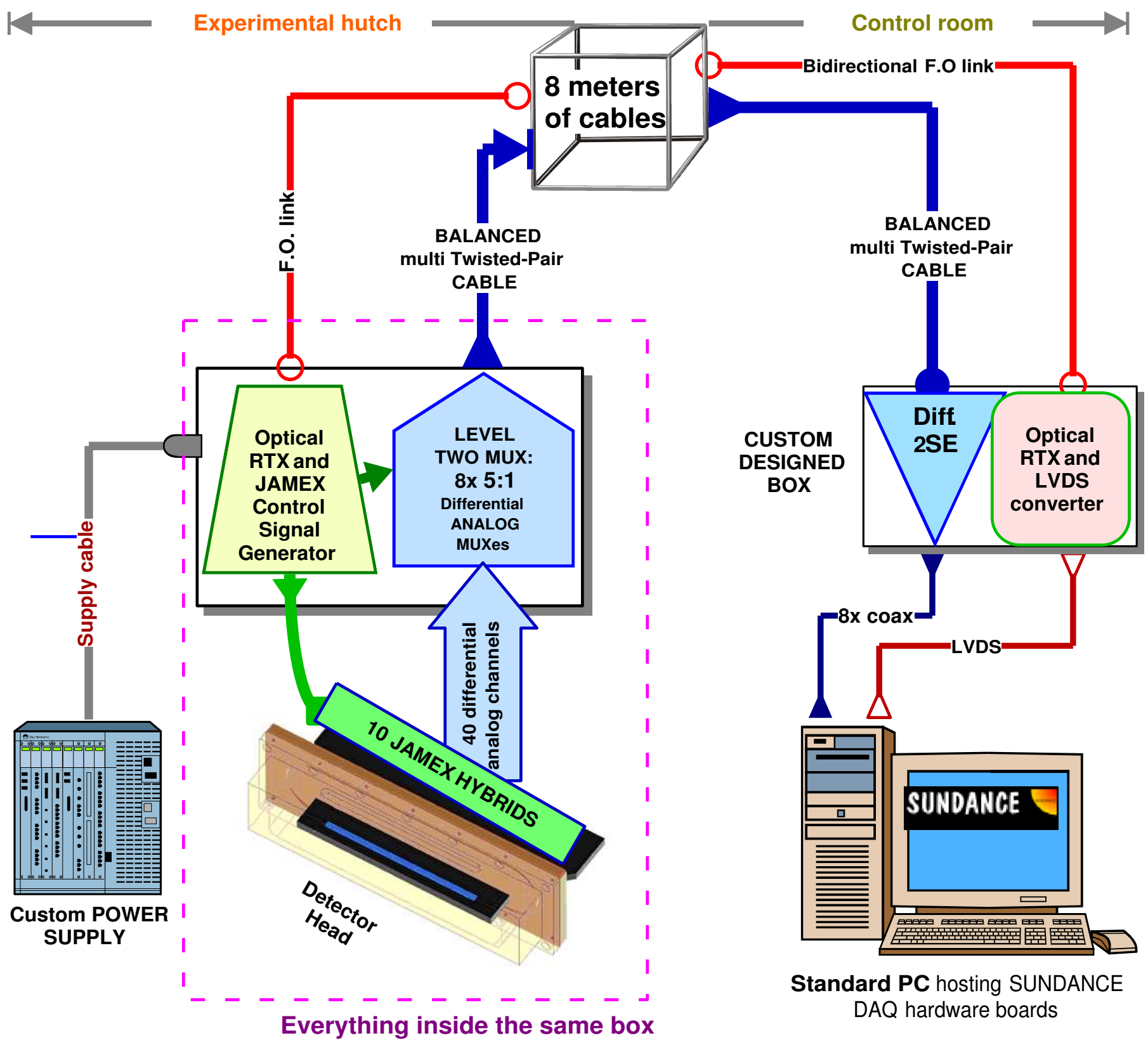

Figure III: Rappresentazione schematica dei componenti impiegati in un'eventuale soluzione costruttiva con due stadi di multiplazione analogica in cascata

Si era pensato di creare un sistema diviso in due parti principali: il detector, ospitato nella *camera sperimentale* e il sistema di acquisizione dati, integrato in un normale PC situato nella *sala di controllo*.

Il `detector` è costituito dall'anodo, i circuiti integrati `JAMEX` dedicati alla lettura delle cariche (che integrano un primo stadio di multiplazione analogica di tipo 32:1), un secondo stadio di multiplazione, necessario per ridurre il numero di canali analogici fisici da trasferire al sistema di acquisizione ad un numero accettabile, e la relativa elettronica di supporto. Il `sistema di acquisizione`, prodotto dalla ditta *Sundance*, è collegato al detector tramite un numero di cavi compreso tra 4 e 40, a seconda delle caratteristiche del secondo stadio di multiplazione.

La prima parte del lavoro di tesi ha avuto come oggetto lo studio e realizzazione di un adatto circuito di multiplazione, poichè questo aspetto era considerato il più critico di tutto il progetto, specialmente a causa della ricaduta sulle prestazioni finali, in termini di rumore e dinamica, dell'intero detector.

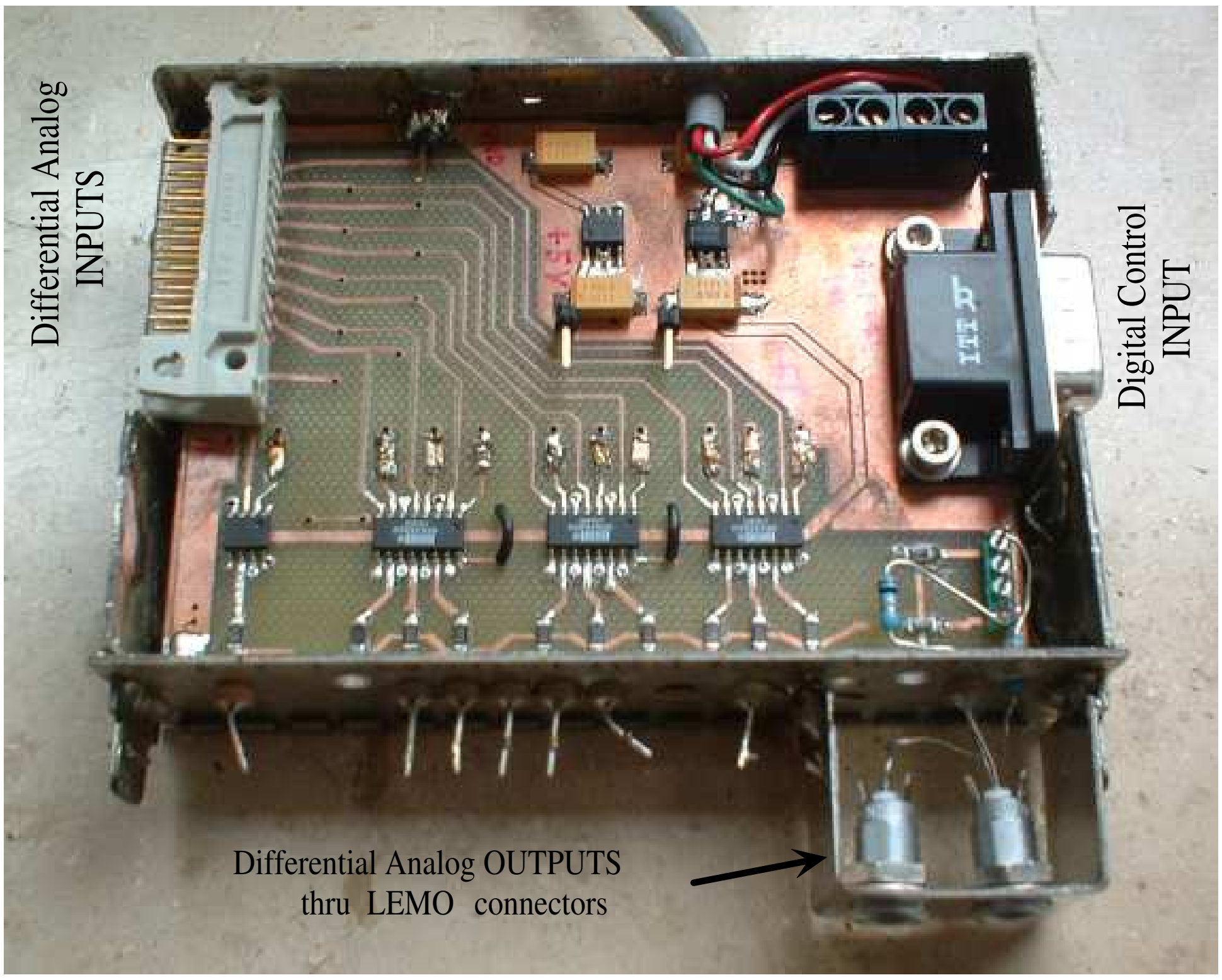

Figure IV: Immagine relativa al circuito di multiplazione analogica chiamato MIDA

Alcuni integrati specializzati dedicati alla multiplazione analogica sono stati valutati e testati durante questo periodo iniziale; alla fine si è deciso di realizzare un prototipo (caratterizzato da una multiplazione 5:1) adatto a valutare le prestazioni di un simile circuito nel contesto dell'intero progetto. Tale prototipo, soprannominato MIDA, è risultato idoneo a *garantire la necessaria dinamica*, di circa 14 bit, ai canali analogici multiplati che saranno collegati agli ingressi del sistema di acquisizione.

Si era ormai giunti a questo stadio del progetto, quando i nuovi risultati derivanti dalle misure sulla versione finale del circuito integrato JAMEX, e in particolare alcune limitazioni sulla massima frequenza operativa, hanno cambiato completamente il corso dell'intero progetto.

## I.2 Definizione di un modello per la soluzione prescelta

La nuova configurazione, scelta per la versione definitiva di questo detector, prevede di configurare gli JAMEX con un fattore di multiplazione 64:1 invece di 32:1, modificando la maschera usata nella produzione dei circuiti integrati di preserie.

In questo modo si può fare a meno di utilizzare il critico stadio di multiplazione secondario, in quanto il numero di canali richiesti al sistema di acquisizione assume un valore realizzabile e pari a 20.

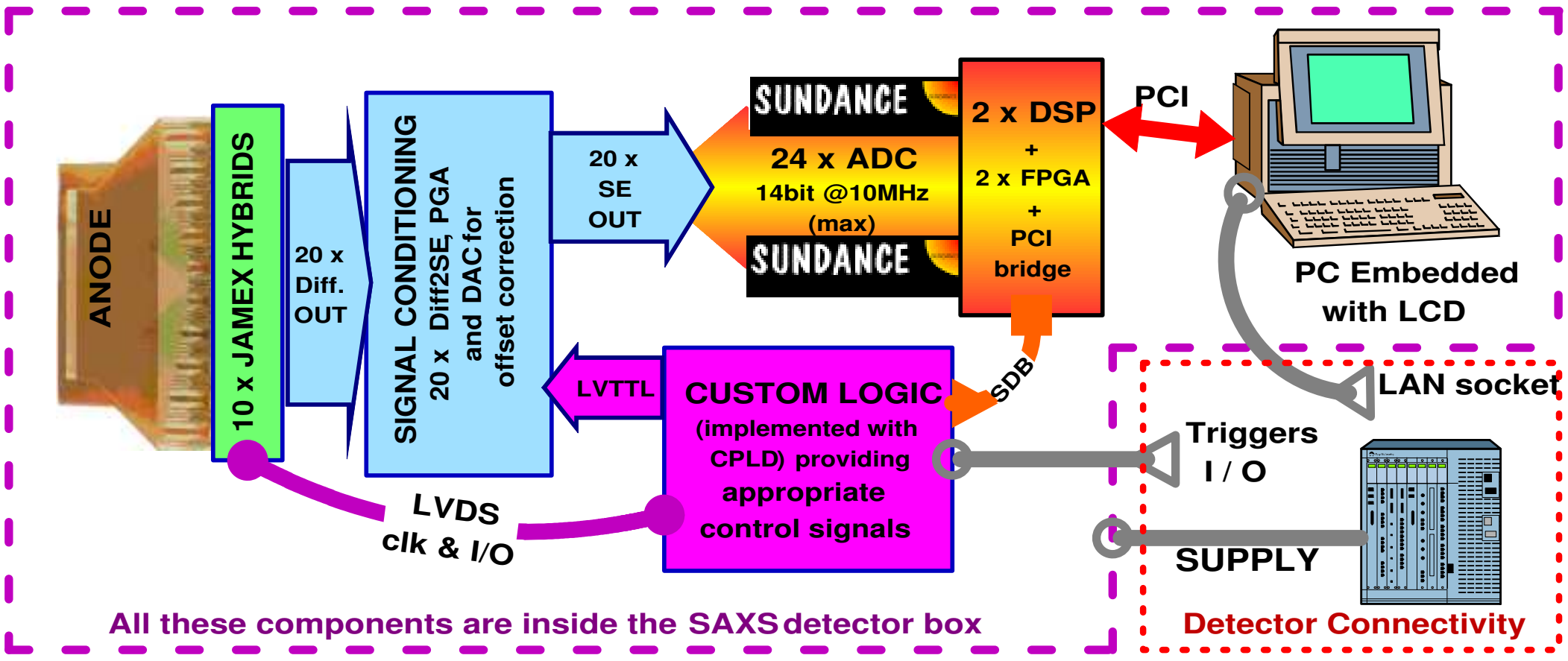

Figure V: Rappresentazione schematica della nuova configurazione scelta per il detector

L'idea di scegliere un Personal Computer compatto per applicazioni industriali, invece di un normale PC, ha permesso di continuare lo sviluppo del detector in termini di sistema monolitico.

Con questo termine si vuole sottolineare il fatto che il nuovo detector assumerà la forma di una scatola di dimensioni compatte, contenente tutta la necessaria elettronica, e collegata con il mondo esterno tramite un semplice cavo Ethernet.

Risulta conveniente che il nuovo detector si comporti, dal punto di vista di un qualsiasi utente della linea di luce SAXS, come un generico computer al quale ci si possa collegare via rete locale. L'accesso alle risorse del detector sarà garantito ai soli utenti autorizzati e in questo caso lo scambio di dati sarà generalmente limitato al trasferimento dei parametri richiesti per una misura e alla lettura dei relativi risultati.

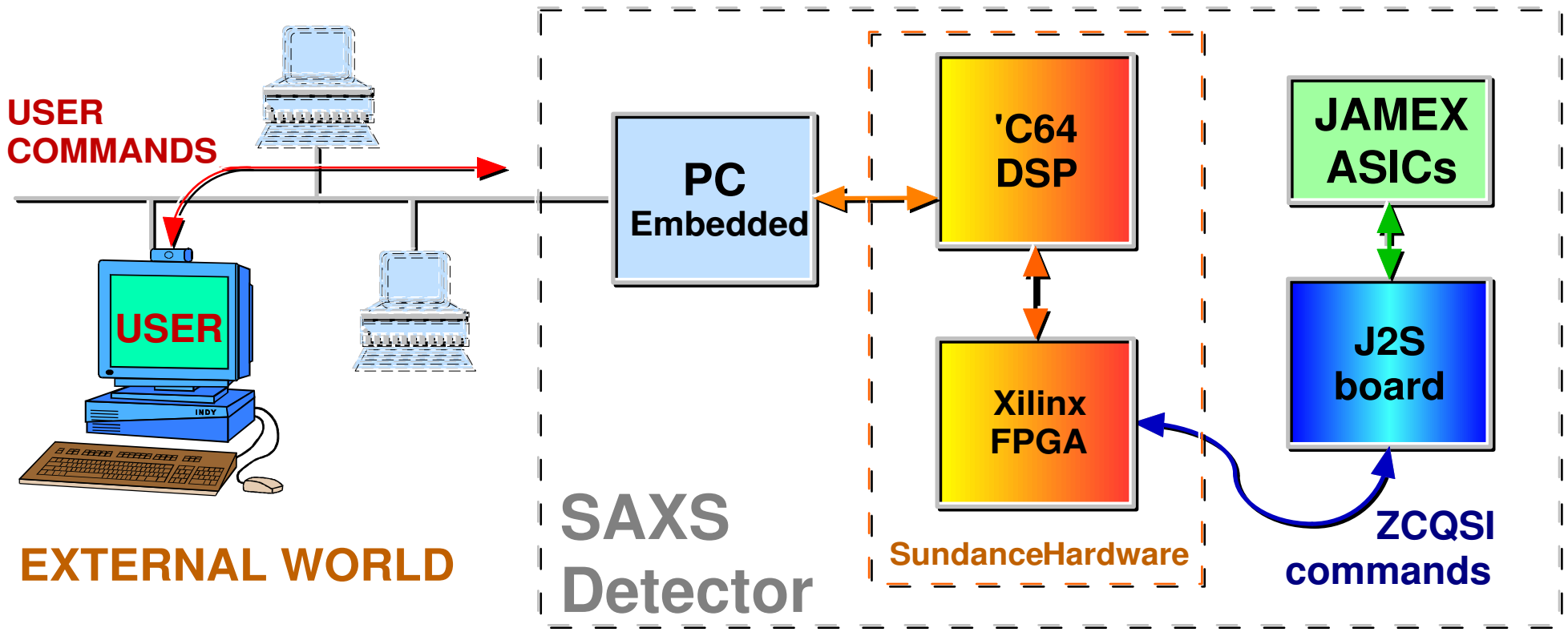

Figure VI: I principali percorsi di scambio dati del nuovo SAXS detector

Per definire il comportamento del sistema si è creata una rappresentazione del tipo macchina a stati (vedi Figura VII) che ha permesso di validare un gruppo di comandi che è stato creato appositamente per questo detector. Questi comandi possono essere differenziati in quattro principali categorie:

1. **ACTIVITY COMMANDS** includono i due comandi START e STOP che controllano a livello macroscopico l'attività del detector
2. **SET COMMANDS** comprende tutti i comandi della `INIT phase`, essi prevedono una trasmissione di dati dall'UTENTE al DETECTOR
3. **GET COMMANDS** sono comandi che determinano una trasmissione di dati dal DETECTOR all'UTENTE
4. **WRITE COMMANDS** sono comandi dedicati al tracciamento di possibili malfunzionamenti

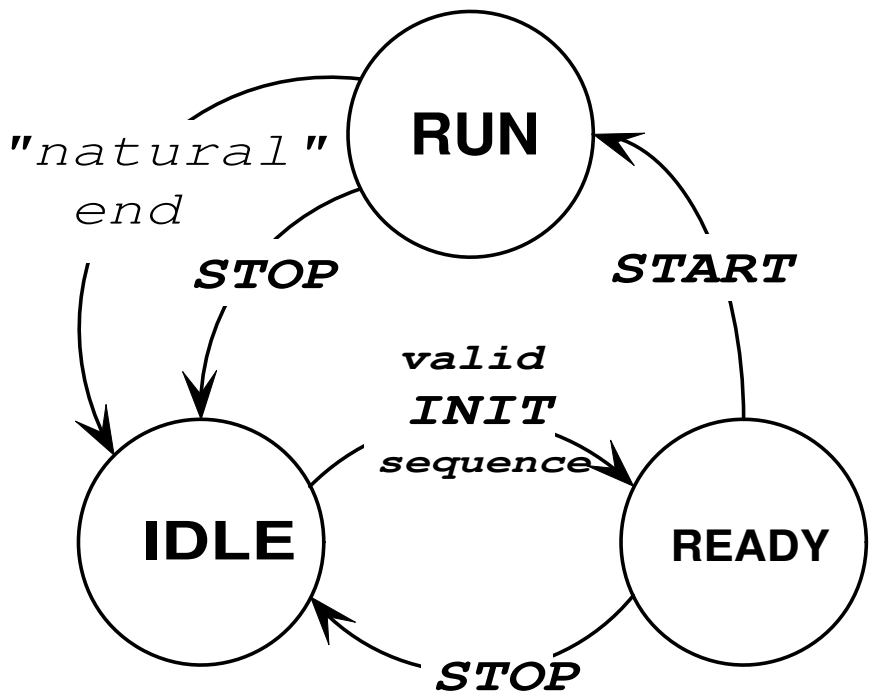

Figure VII: La macchina a stati

Si vuole ora introdurre il complesso circuito che è stato realizzato nell'ambito di questa tesi per interfacciare il sistema di acquisizione Sundance agli speciali circuiti integrati `JAMEX`. Questo circuito è stato chiamato `J2S board` (vedi Figura VI) e verrà diffusamente trattato nel Capitolo 4. Esso comunica con il sistema di acquisizione tramite una speciale interfaccia seriale chiamata `ZCQSI` che verrà implementata a livello hardware grazie ad un CPLD (*Sistema Logico Complesso Programmabile*).

L'importanza dell'interfaccia ZCQSI è dovuta al fatto che essa rappresenta il canale dedicato alla trasmissione dei comandi necessari alla *configurazione e controllo dei circuiti* `JAMEX` oltre a quelli usati per la selezione di alcune funzionalità proprie della scheda J2S.

Tramite un opportuna configurazione dei segnali dell'interfaccia ZCQSI è possibile selezionare uno dei seguenti comandi:

- `Write_PGA_GAIN` imposta il guadagno dei canali analogici
- `Write_CONTROL` seleziona la modalità di test ed il DAC da programmare
- `Write_SPI` questo comando permette di programmare il DAC selezionato
- `Write_JAMEX` scrive il bitstream di configurazione negli `JAMEX`
- `Read_JAMEX` legge il bitstream di configurazione degli `JAMEX`

Questi comandi sono eseguiti fornendo adatte sequenze di segnali a livello di interfaccia ZCQSI: un esempio, relativo ad una simulazione, viene fornito qui di seguito.

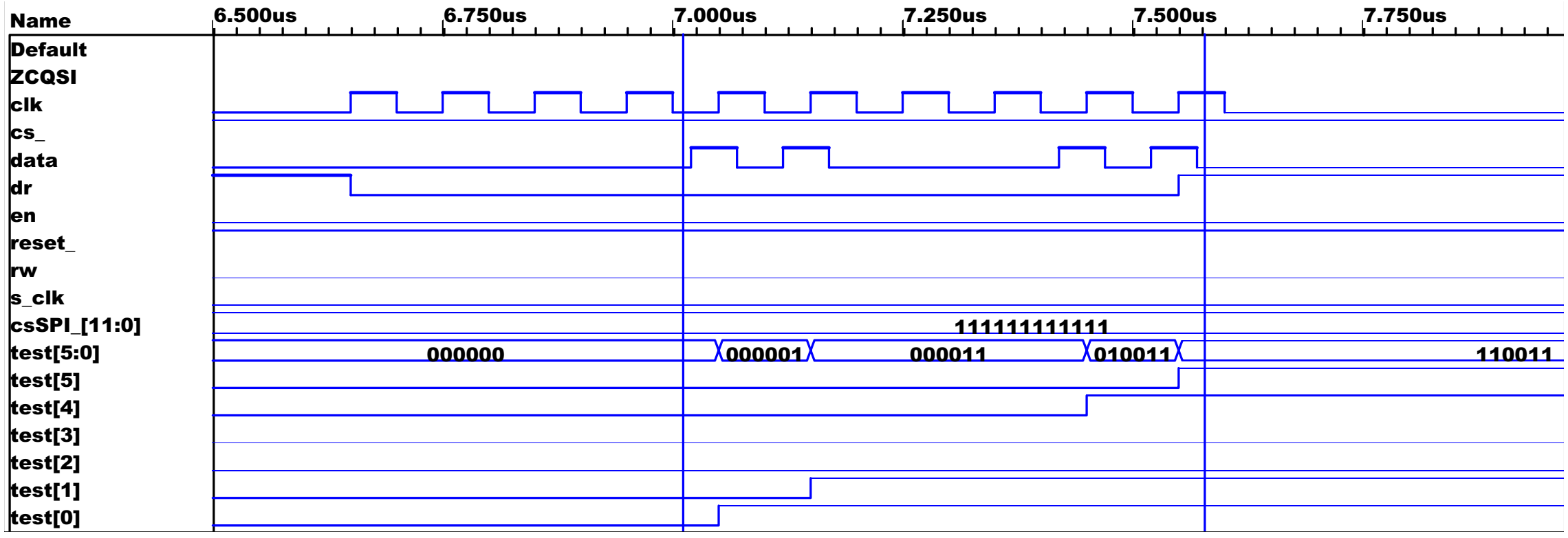

Figure VIII: La sequenza necessaria alla corretta esecuzione del comando `Write_CONTROL`

## I.3 Sistema di acquisizione Sundance

La decisione di utilizzare un sistema di acquisizione *commercialmente disponibile* caratterizza il progetto di questo detector e nasce dal desiderio di ridurre il tempo necessario a sviluppare l'intero sistema.

La ditta Sundance produce un sistema modulare e liberamente configurabile che fornisce una soluzione ad alta tecnologia facilmente scalabile per accomodare eventuali incrementi di prestazioni o aumento del numero di canali trattati. La configurazione prescelta per questo detector consiste di due schede madri, capaci di ospitare un massimo di 4 moduli in formato `TIM-40` ciascuna, dotate di un connettore a standard PCI che permette di utilizzarle anche in un normale Personal Computer.

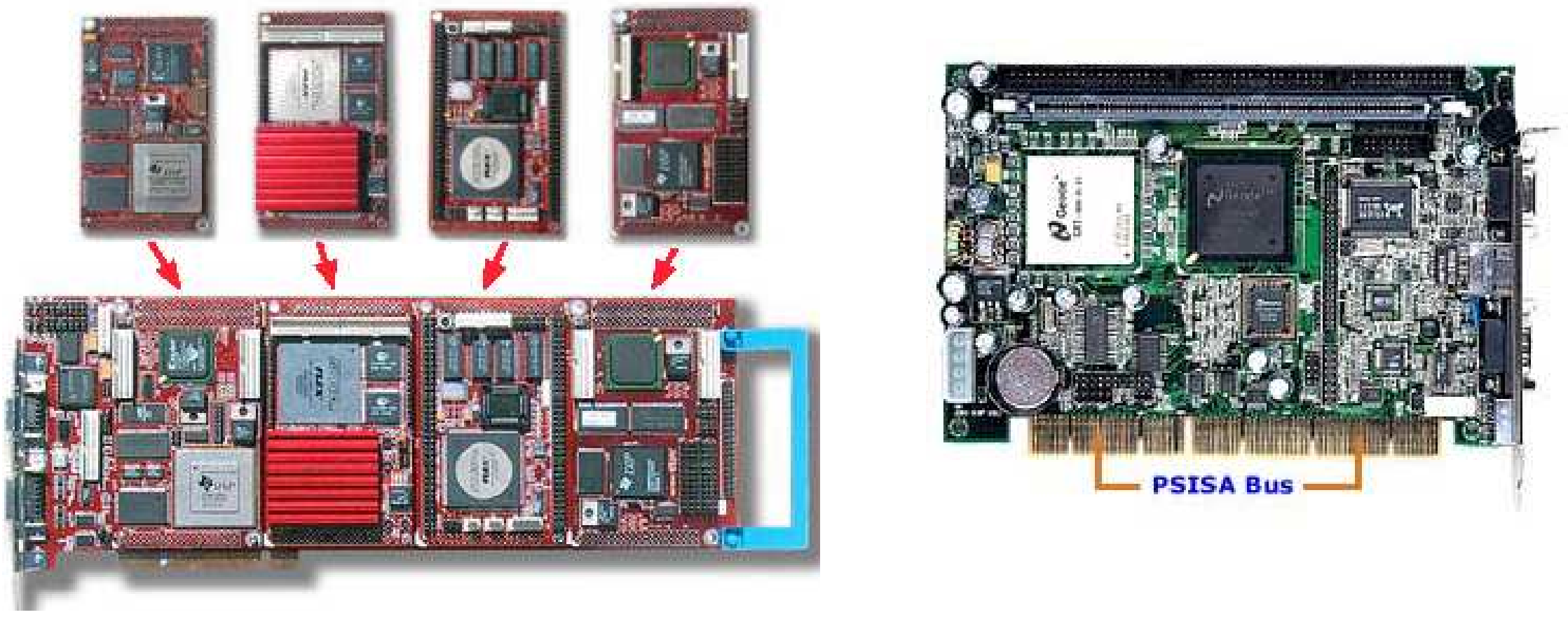

Figure IX: Il sistema modulare Sundance -sinistra- e il compatto PC industriale -destra-

Nel nostro caso, la prima scheda madre ospiterà tre moduli ADC ad 8 canali, mentre la seconda due potenti moduli dedicati al processamento digitale di segnali (DSP) con un'ampia capacità di memoria ed elevata reconfigurabilità garantita da una capace FPGA.

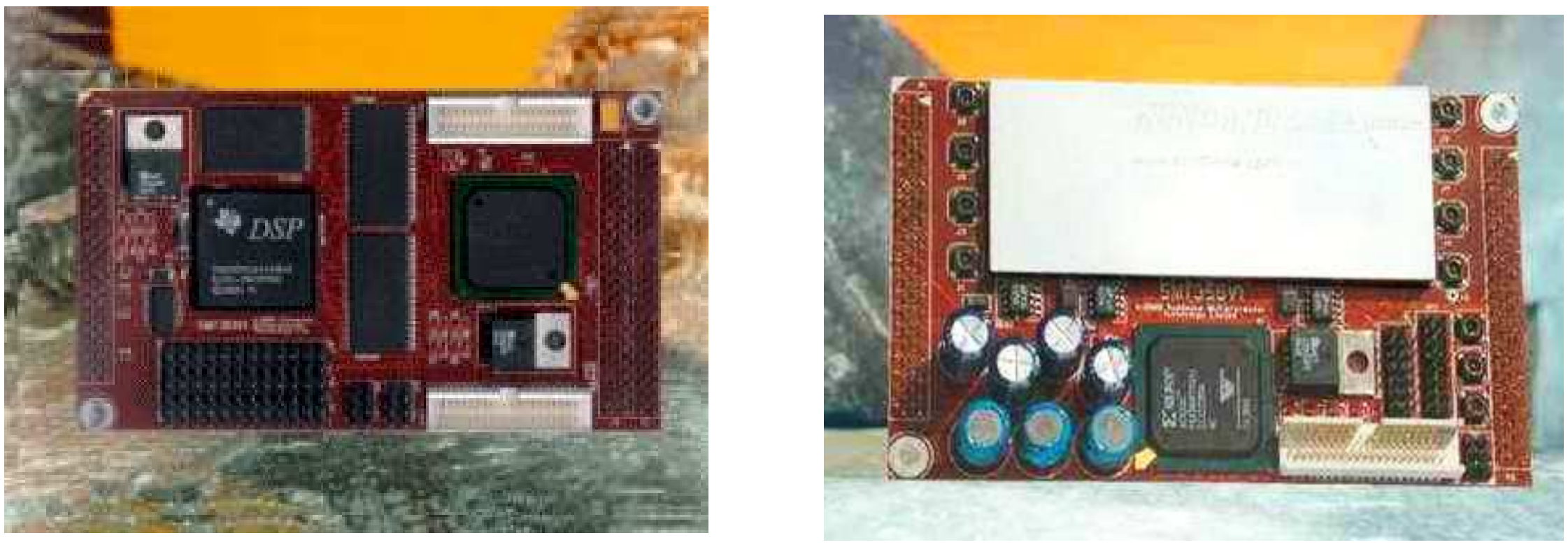

Figure X: Il modulo DSP-FPGA Sundance -sinistra- e il modulo ADC ad 8 canali -destra-

L'intero sistema esegue un'applicazione chiamata "`JAMEX` application" che, a causa della complessità dell'intero sistema, è stata sviluppata inizialmente con il coinvolgimento diretto di Sundance Italia e in seguito autonomamente.

Essa è composta da un *programma di controllo* eseguito sul Personal Computer, alcuni *algoritmi* che vengono eseguiti sui 2 DSP del sistema e una parte di *firmware* che comprende delle funzionalità addizionali implementate con una programmazione specifica della FPGA.

Il sistema di sviluppo comprende l'ambiente integrato *Code Composer Studio* di Texas Instruments e il sistema operativo in tempo reale di *3L Diamond*, oltre ai numerosi tool di supporto proprietari di Sundance e al compilatore Visual C++ di Microsoft.

## I.4 Caratteristiche delle schede J2S

Benchè il sistema di acquisizione Sundance fosse stato progettato in modo da essere il più versatile possibile, chiaramente non era pensabile che esso potesse interfacciarsi direttamente ai circuiti integrati `JAMEX`.

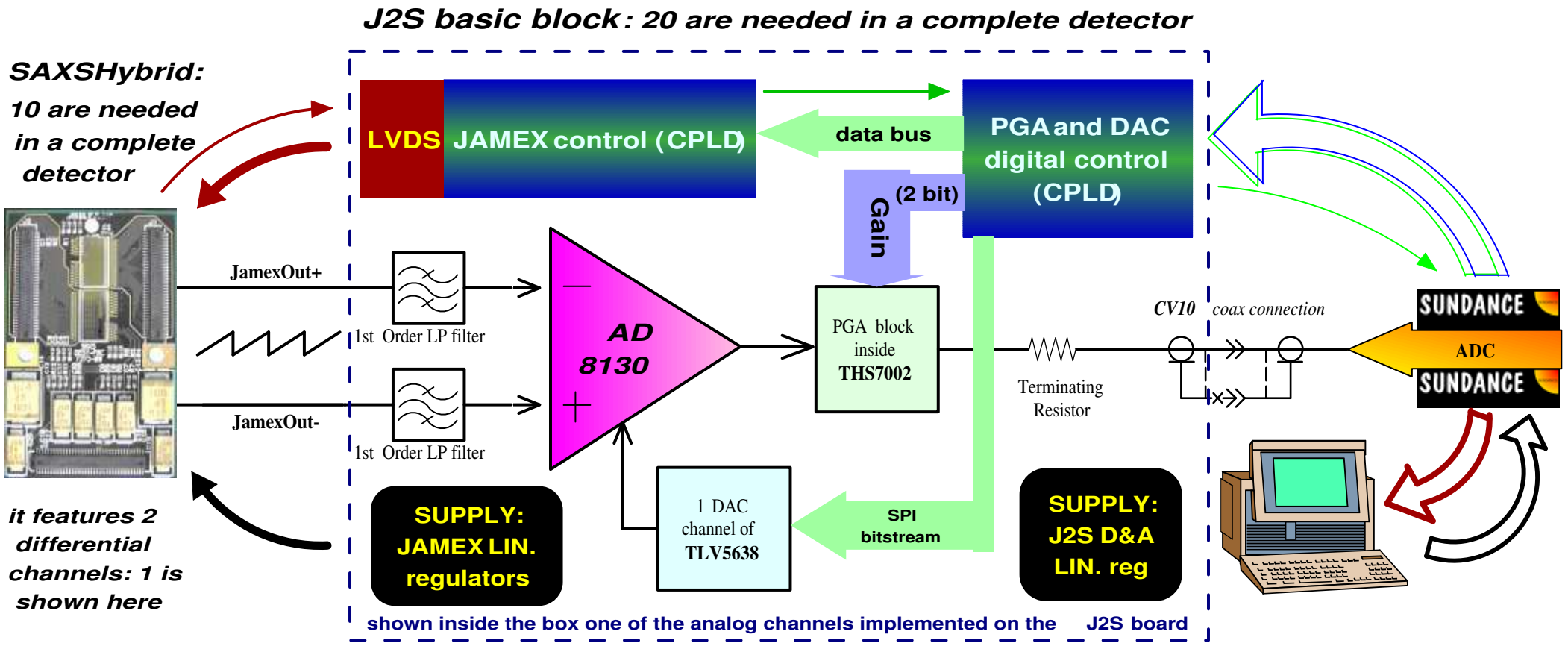

Figure XI: Esemplificazione delle funzionalità realizzate dalle schede J2S

Come si può vedere dalla figura soprastante, le schede J2S che sono state progettate e realizzate nell'ambito di questa tesi di laurea rappresentano il mezzo che consente l'interfacciamento tra le due realtà sopra citate.

Per ragioni di pura convenienza progettuale, quella che originariamente era stata pensata come una scheda unica, chiamata appunto J2S, è stata realizzata in maniera distribuita: queste componenti sono chiamate schede J2S.PIG, J2S.MINI e J2S.SR.

La scheda **J2S.PIG** è stata prodotta in 60 esemplari: esse permettono una connessione diretta, per mezzo di un connettore specifico miniaturizzato, alle schede `SAXS Hybrid`, prodotte dal gruppo FEC presso il Sincrotrone di Amburgo, che ospitano 2 integrati `JAMEX` ciascuna e si connettono direttamente all'anodo del detector.

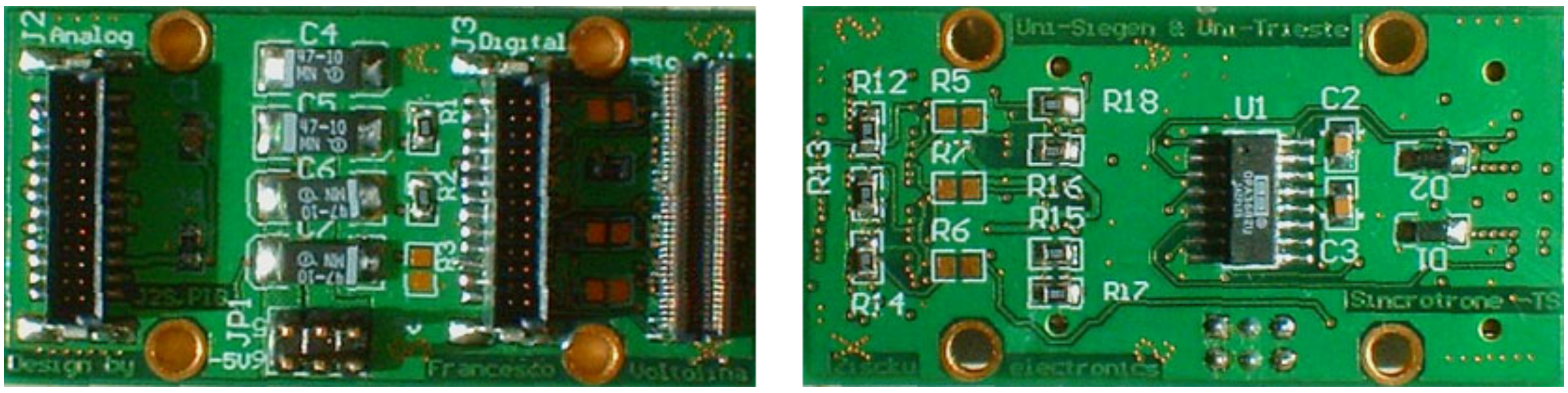

Figure XII: La scheda `J2S.PIG` in versione definitiva, con i componenti montati, è qui riprodotta in *dimensioni reali*: vista da sopra -sinistra- e vista da sotto -destra-

Questa scheda fornisce supporto meccanico ai SAXS Hybrid, separa convenientemente le connessioni analogiche da quelle digitali e implementa lo stadio di potenza del generatore di segnali di test. Il detector finale utilizzerà un numero di schede J2S.PIG pari a 10, una per ogni SAXS Hybrid, che verranno collegate alla scheda J2S.FULL da appropriati cavi piatti multipolari.

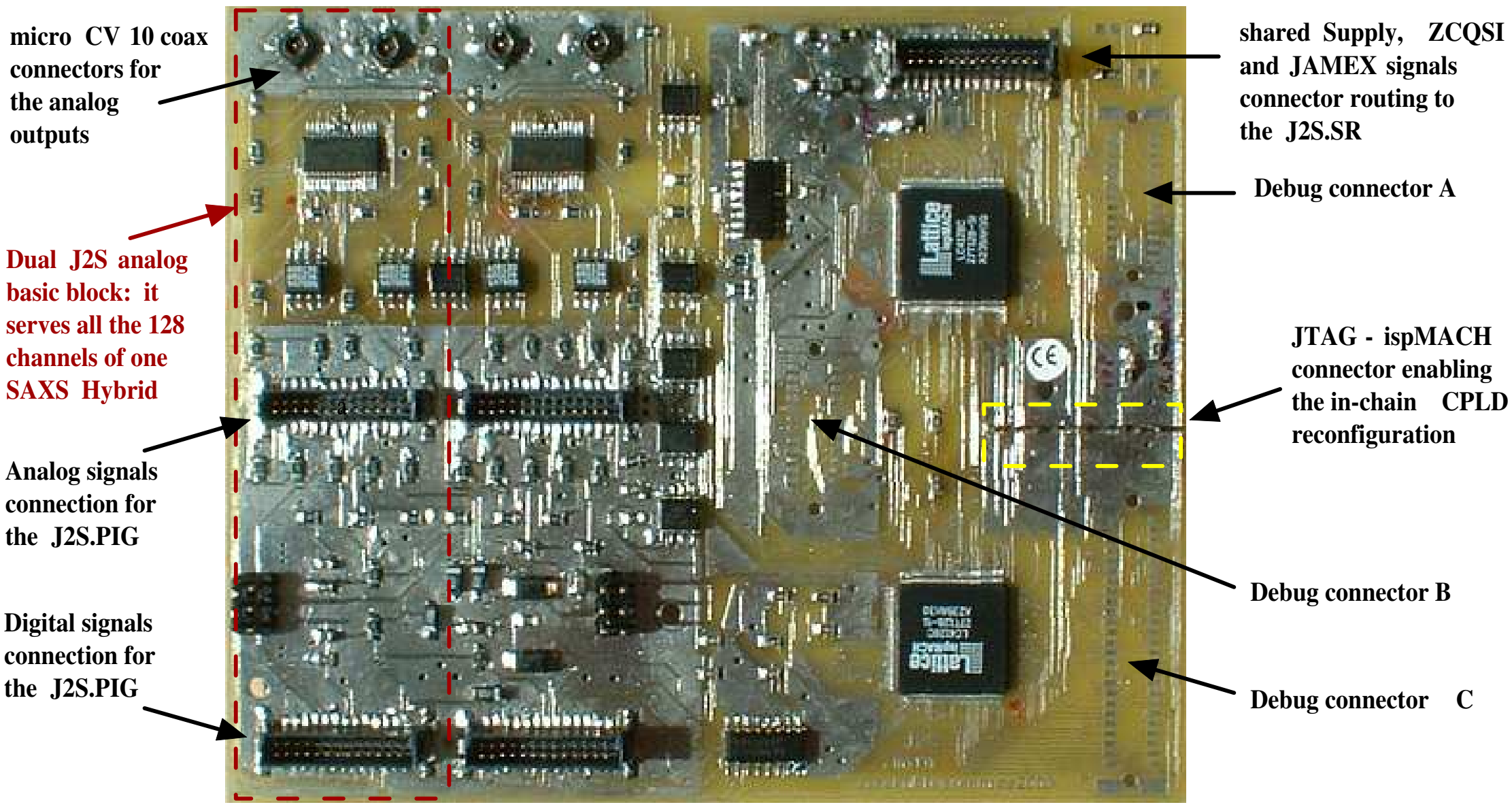

Figure XIII: Foto della scheda J2S.MINI montata

Una versione ridotta di questa seconda scheda, chiamata **J2S.MINI**, è stata prodotta allo scopo di testare la validità delle scelte progettuali destinate ad essere implementate nella versione completa.

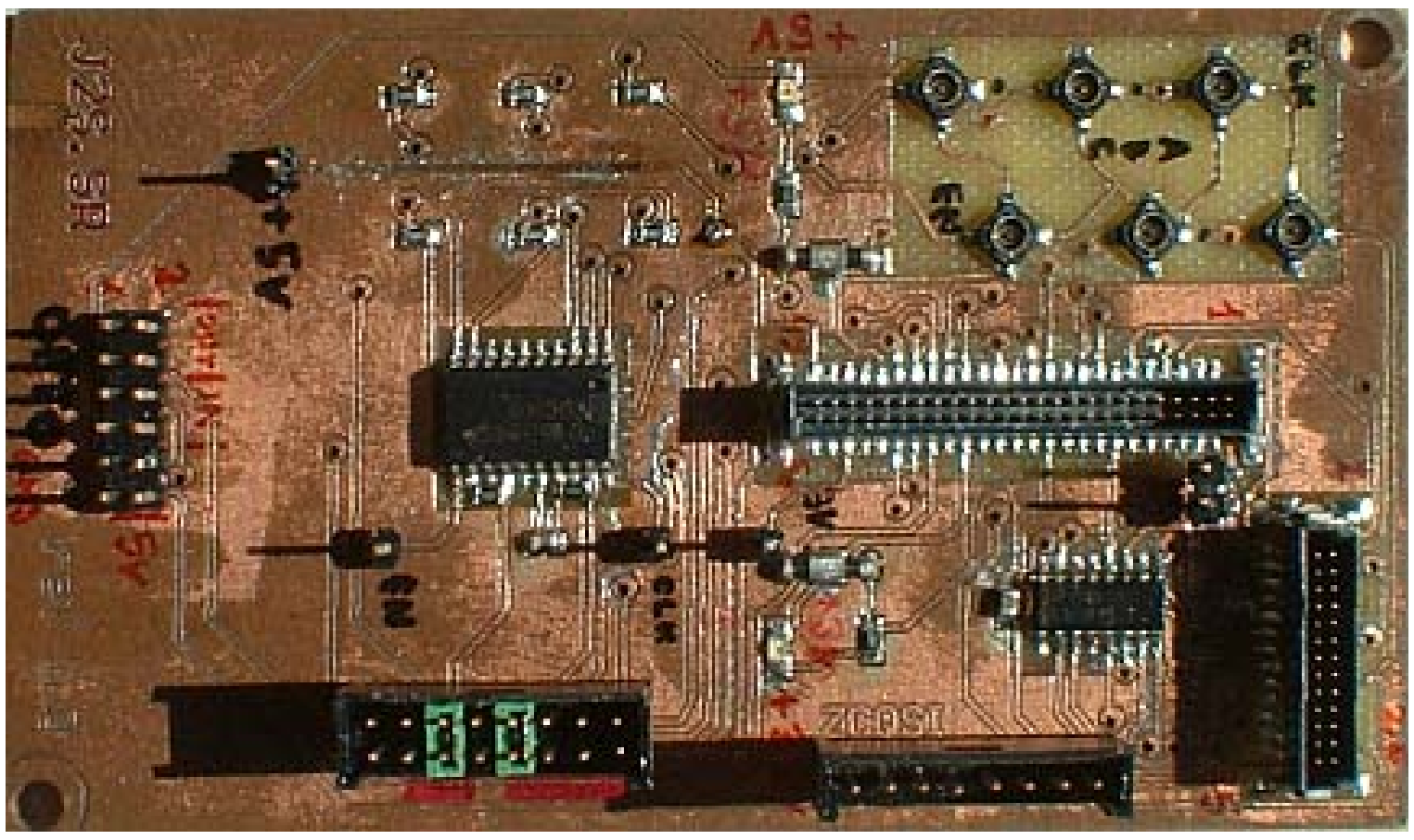

Figure XIV: Foto della scheda J2S.SR in *dimensioni reali*

Il numero di canali analogici supportati dalla scheda J2S.MINI è limitato a 4, ma la sua parte digitale è analoga a quella destinata alla versione finale. Questa scheda adatta le uscite analogiche differenziali degli `JAMEX` agli ingressi single-ended delle schede Sundance e allo stesso tempo attua un filtraggio passa-basso e un'amplificazione programmabile. Inoltre, questa scheda implementa lo stadio di controllo relativo al generatore di segnali di test e consente anche il controllo individuale dell'offset di uscita dei canali analogici grazie ad un DAC associato ad ogni canale: questa soluzione permette di sfruttare al massimo la gamma di ingresso, e quindi la risoluzione, offerta dal sistema di acquisizione Sundance.

Anche i segnali di controllo per gli `JAMEX` sono generati sulla scheda J2S.MINI: essa si interfaccia via ZCQSI con il sistema Sundance attraverso una scheda di conversione dei segnali chiamata J2S.SR. Il clock principale destinato agli `JAMEX` viene convertito in LVDS (acronimo inglese di sistema di Segnalazione Differenziale a Basso Voltaggio) da questa ultima scheda: in questo modo l'unico segnale potenzialmente dannoso, in quanto attivo durante ogni misura effettuata con il detector, vede ridotta drasticamente la sua capacità interferente.

## I.5 Integrazione del prototipo e relative misure

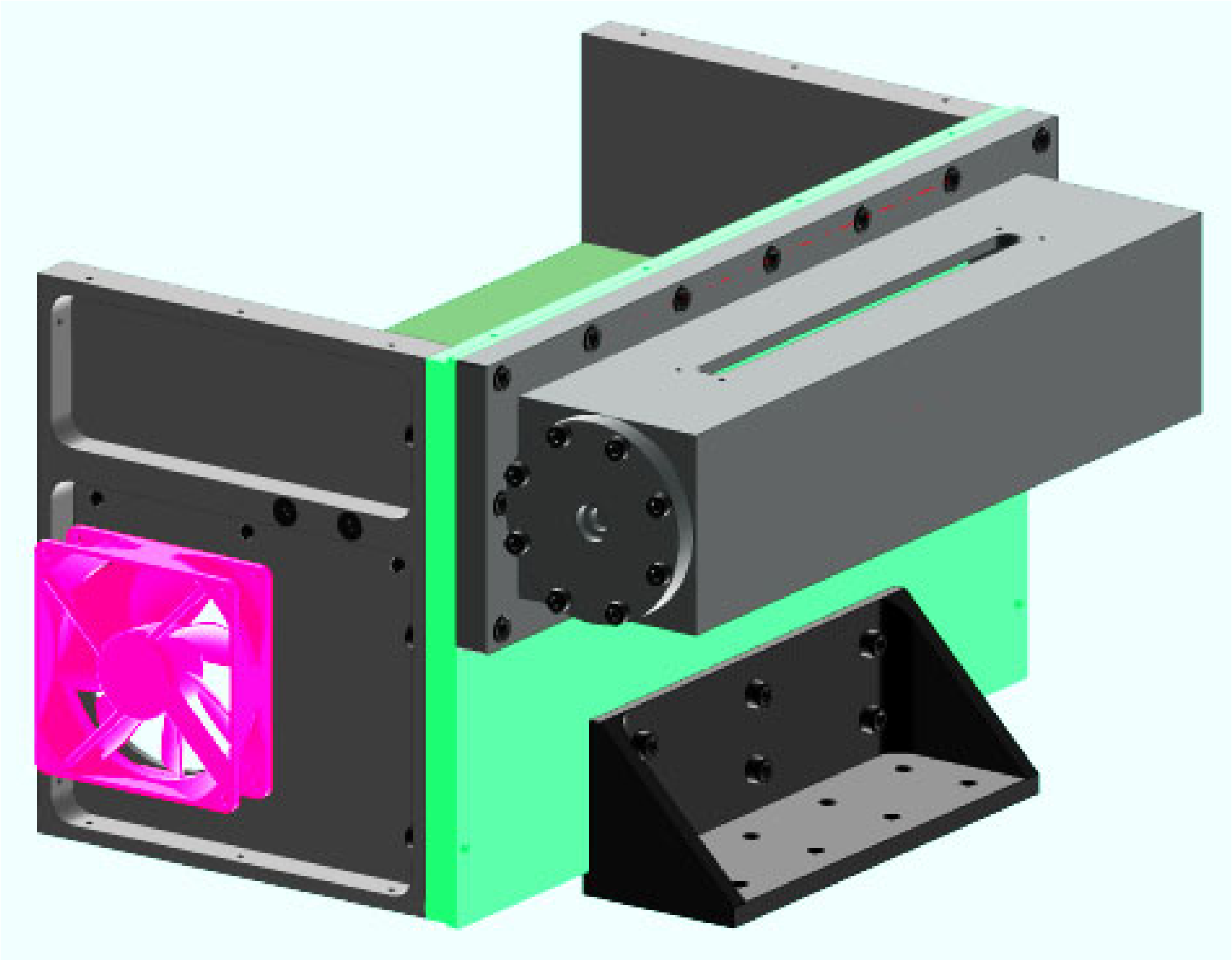

Figure XV: Simulazione fotorealistica del contenitore del detector

Grazie al supporto fornito dall'*ufficio di progettazione meccanica* ad Elettra e dal *laboratorio meccanico* dell'Università di Siegen, è stato possibile ottenere un contenitore, creato su misura, adatto ad ospitare il detector. Questo componente si è dimostrato estremamente utile per verificare che tutte le ipotesi di montaggio fossero valide in pratica e che tutti i componenti vi alloggiassero come previsto.

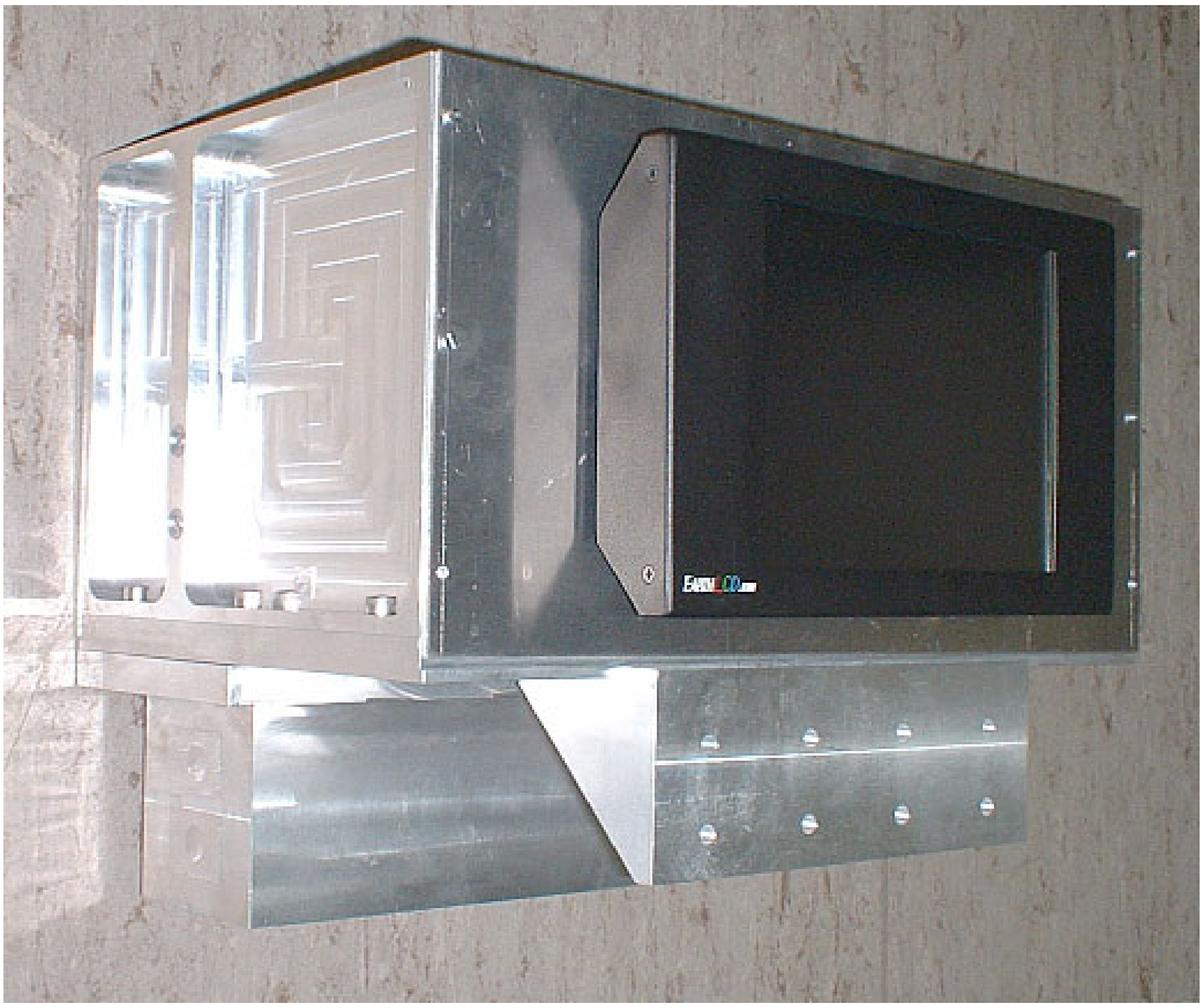

Figure XVI: Fotografia relativa al detector SAXS corredato di schermo LCD

Uno schermo LCD industriale di dimensioni compatte fornisce informazioni sullo stato del detector e può essere disattivato per evitare interferenze durante la misura. La maggior parte dei test sono stati effettuati con il detector separato nelle sue componenti, per ragioni di convenienza ed accessibilità.

Il sistema di acquisizione Sundance è stato montato all'interno di un contenitore standard per Personal Computer opportunamente modificato per accogliere il compatto PC industriale. Si sono comunque ricreate le condizioni più simili possibile a quelle in cui il detector si troverà ad operare, limitando il volume dedicato al sistema di acquisizione e alloggiando le schede J2S negli spazi previsti all'interno del detector. Diverse sessioni di prove, mirate alla valutazione del funzionamento delle differenti componenti del detector, hanno fornito risultati incoraggianti.

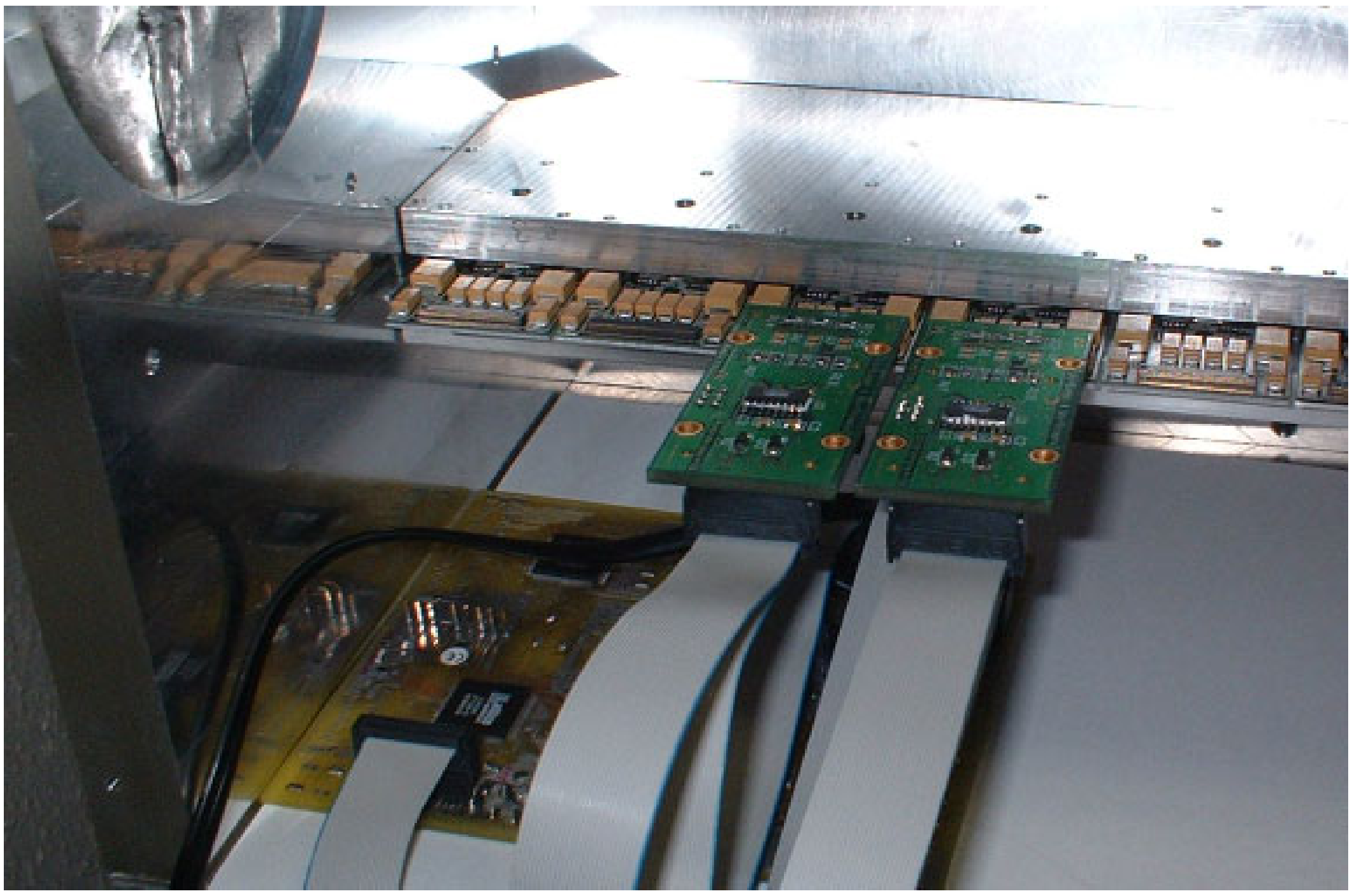

Figure XVII: Una vista parziale dell'interno del detector con i 10 `SAXS Hybrids` montati sull'anodo: due di questi sono collegati tramite schede J2S.PIG alla scheda J2S.MINI

In particolare, i test effettuati sui canali analogici hanno confermato la bontà del progetto J2S, il quale introduce un rumore contenuto, anche al massimo guadagno permesso dal PGA, e quindi permette lo sfruttamento completo della massima risoluzione e dinamica consentite dal sistema di acquisizione Sundance.

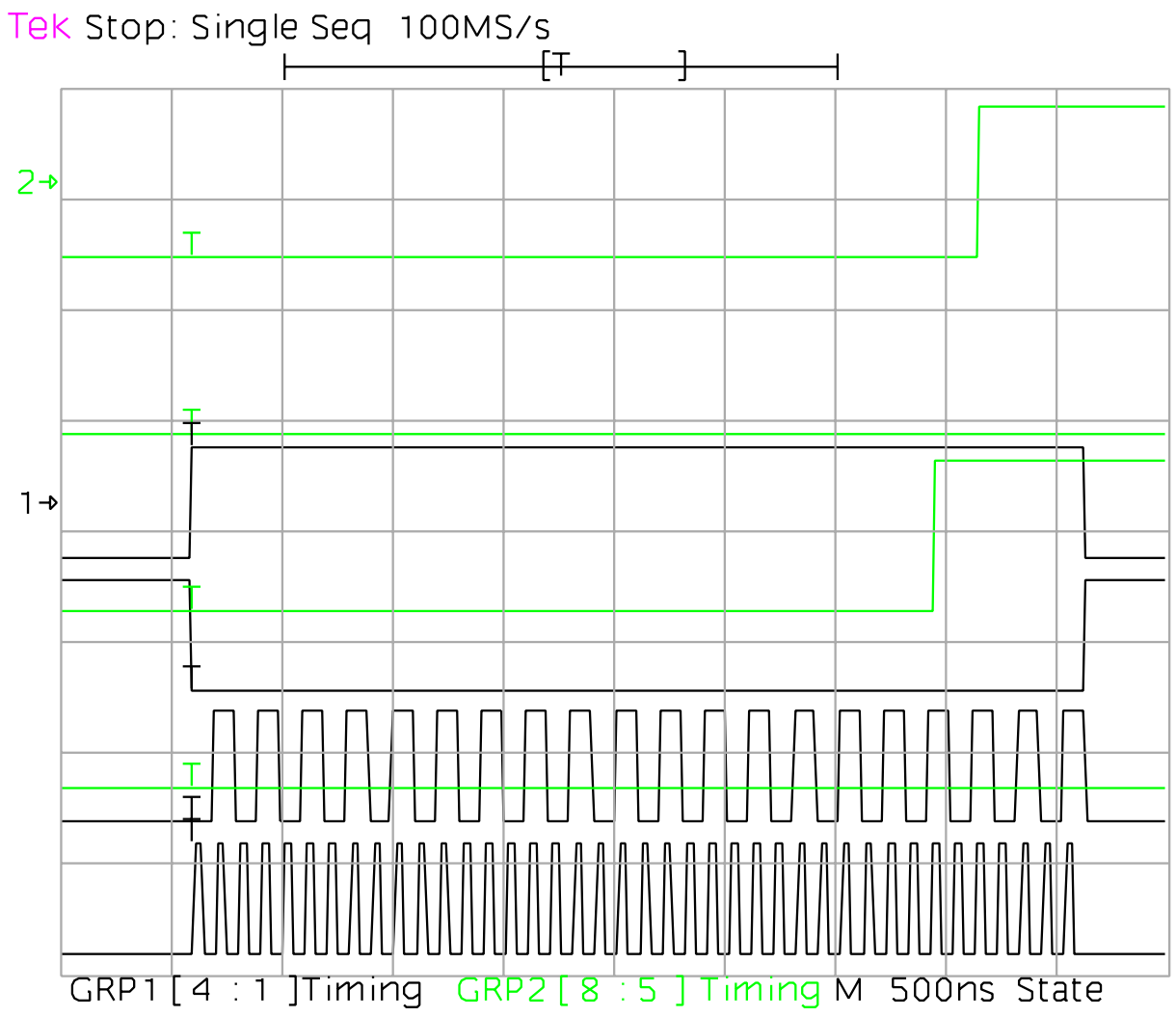

Figure XVIII: Timings misurati per l'esecuzione del comando `Write_PGA_GAIN`

Anche la logica implementata sulla scheda J2S.MINI, testata con l'utilizzo di un oscilloscopio digitale a 16 canali (Tektronix TLS216), ha mostrato un comportamento coerente con le simulazioni e quindi la possibilità di interfacciarsi convenientemente con il sistema di acquisizione Sundance via interfaccia ZCQSI.

Il lavoro svolto in questa tesi di laurea ha portato alla realizzazione di un "prototipo in stadio avanzato" di detector per l'acquisizione di dati in esperimenti di tipo SAXS. Molti aspetti di questa realizzazione sono stati sviluppati e portati a termine, ma il sistema necessita ancora di vari perfezionamenti prima di poter essere provato sul campo, ovvero in un esperimento alla linea di luce SAXS di Elettra.

L'attuale incognita del sistema è legata alla non correttezza di alcuni segnali forniti dalla parte di *firmware* proprietario sviluppato per le schede Sundance. Nell'immediato futuro verrà modificato e corretto il VHDL relativo a queste funzionalità in modo da garantire una corretta programmazione degli `JAMEX`. Alcune modifiche al software della "JAMEX Application" consentiranno un corretto funzionamento in *tempo reale* del detector, anche utilizzando tempi di integrazione minori di 2.7 ms (che rappresentano l'attuale limite fornito dalla struttura software del sistema di acquisizione) fino al limite di 128 $\mu$s consentito dagli `JAMEX`.

UNIVERSITÄT SIEGEN

Theorie und Praxis für Karrieren von morgen

UNIVERSITÀ DEGLI STUDI DI TRIESTE

UNIVERSITÄT SIEGEN FACHBEREICH 7 -PHYSIK-

ELETTRA -SINCROTRONE TRIESTE S.C.P.A.-

TRIESTE *DEEI* -FACULTY OF ENGINEERING-
SIEGEN -DETEKTORPHYSIK UND ELEKTRONIK-

TESI DI LAUREA IN ELECTRONIC DEVICES

# Design of a MicroStrips Ionization Chamber Detector Aimed to SAXS Experiments

Thesis student:
Francesco VOLTOLINA

Supervisor:
Chiar.mo Prof. Sergio CARRATO

Co-supervisor:
Dr. Ralf Hendrik MENK

# Introduction

The work presented in this Tesi di Laurea arises from a collaboration between the IPL (Image Processing Laboratory) of the *Dipartimento di Elettrotecnica, Elettronica ed Informatica* (DEEI) at the University of Trieste and the *Instrumentation and Detector Laboratory* belonging to the ELETTRA Synchrotron Light Source of Trieste. Under the European Contract ERBFMGECT 980104 a project was active at ELETTRA with the objective to improve facilities for time resolved small angle X-ray scattering (SAXS) experiments within Europe.

Partners of the project were, together with ELETTRA, other large scale facilities like the HASYLAB (Hamburger Synchrotron, Germany), the CCLRC (Daresbury Laboratory, UK), the ESRF (European Synchrotron Radiation Facility in Grenoble, France) and the University of Siegen (Germany). In particular the latest half of the time spent working on this project was based in the *Arbeitsgruppe Detektorphysik und Elektronik* belonging to the Faculty of Physics at the University of Siegen.

Several third generation rings dedicated to the production of synchrotron radiation are now operating and are designed for very low emittance but extreme brilliance. In the last decade the improvements achieved at modern synchrotron light sources and the associated optics result in a tremendous increase in brilliance and subsequently photon flux, which was however not accompanied by an equivalent development of the x-ray detection devices. There now exists a situation where the detectors are generally the weakest component of a beamline system. Analysis shows that the potential of many beamlines can be degraded by several orders of magnitude by the lack of appropriate detectors and their associated electronics and data acquisition systems. This observation is valid not only for third generation sources that are currently available (like ELETTRA in Trieste) but also for less brilliant second-generation sources like those in Daresbury and Hamburg (see Figure 1).

*Improving detectors is therefore a strategic issue to exploit more efficiently the potential of all synchrotron radiation sources*: in the framework of this European project originated the idea to develop and produce two detection devices capable to cover as wide range of experiments as possible. They will feature exceptionally high performances in extracting the highest amount of information from the experiments in which they are going to be used. One detector was a *high rate one-dimensional microstrip ionization chamber* which will have an enormous count rate capability, while the other is

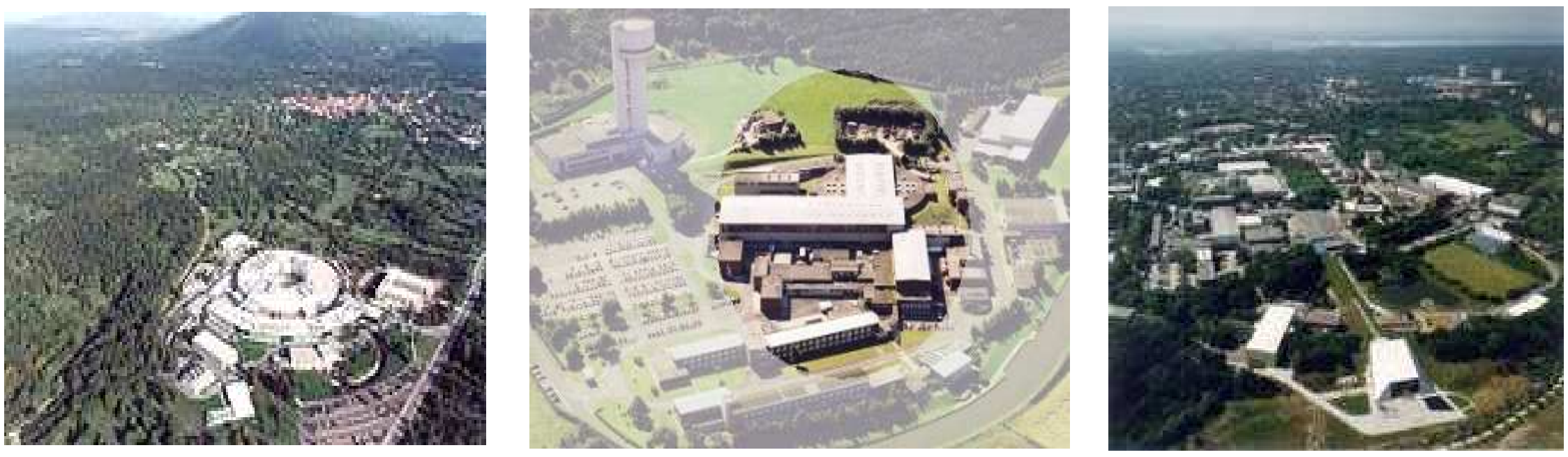

Figure 1: Elettra(Trieste) -left-, CCLRC(Daresbury) -center-, HASYLAB(Hamburg) -right-

a bidimensional single photon counter which has an excellent dynamic range but at lower rates. In order to realize the final version of the one dimensional detector, which is also the object of this Tesi di Laurea, primarily it was necessary to design a *custom frontend electronics aimed to interface specialized custom Application Specific Integrated Circuits (ASICs) chips to a commercially available hardware* consisting in modular ADC and DSP cards. Since the final configuration of the detector was not yet finalized some degrees of freedom were still available in order to choose the best feasible solution. In particular it was still possible to modify the final mask of the ASIC chip in order to have an output stage with single 64 to 1 analog multiplexing or to have 2 separate outputs driven from a dual 32 channels multiplexer for each ASIC. Since the required total number of microstrips to be sensed was 1280, the number of analog channels for the required acquisition system was needed to be in the range of *20 to 40 independent and simultaneously sampling channels* at speeds in excess of 1 MHz.

Considering the complexity of such a system and the availability of fast ADCs with suitable resolution but maximum speed well in excess for this application it was conceived the possibility of designing an *additional analog multiplexing stage* in order to *reduce the number of analog to digital converters* used in the system and thus to simplify the whole electronic required. Nevertheless, it was clear that such a solution was to be taken into account only if it was not going to degrade the overall specification of the entire system that was needed to be close to state of art in order to fulfil the initial requirements.

An initial preliminary feasibility study, in which different simplified configurations were designed, built and evaluated, has been followed with the final version detector overall design that included not only the frontend electronics design but also the interface protocol specifications, the overall mechanical structure devoted to shield and contain the final system, and careful software debugging and developing to characterize the overall functionality of the complex system.

In order to *clarify the appropriate design constraint for this X-ray detector*, the next section will give a short introduction about the type of experiments in which this new device is going to be used and the specific environment for which it has to be tailored.

# Small Angle X-ray Scattering

Small Angle X-ray Scattering (SAXS) is essentially an experimental technique, based on the physical phenomenon of elastic scattering, used to derive the structure of various objects in the spatial range from 1 nm to 1000 nm. This is the size range of many relevant structures in polymers, liquid crystals, nanoporous materials, microemulsions, etc.

Roots for this theory can be found in the `Bragg's law` derived by the English physicists Sir W.H. Bragg and his son Sir W.L. Bragg in 1913 to explain why the cleavage faces of crystals appear to reflect X-ray beams at specific angles of incidence $\theta$.

X-rays primarily interact with electrons in atoms: when X-ray photons collide with these electrons, some photons from the incident beam will be deflected away from the direction where they originally travel. If the wavelength of these scattered X-rays does not change the process is called elastic scattering (also Thompson Scattering) because only momentum has been transferred in the scattering process. These are the X-rays that we measure and we are interested in any generic diffraction experiment because it results that the scattered X-rays carry information about the electron distribution in materials.

Bragg's law states that the dimensions *d* of an object are reciprocal to the angles to which the X-rays are scattered (large structures scatter to low angles and vice versa). Bragg's Law refers to the simple equation: $n\lambda = 2d \cdot \sin\theta$ . The variable *d* is the distance between atomic layers in a crystal, the variable $\lambda$ is the wavelength of the incident X-ray beam, *n* is an integer giving the reflection order and $2\theta$ is the scattering angle.

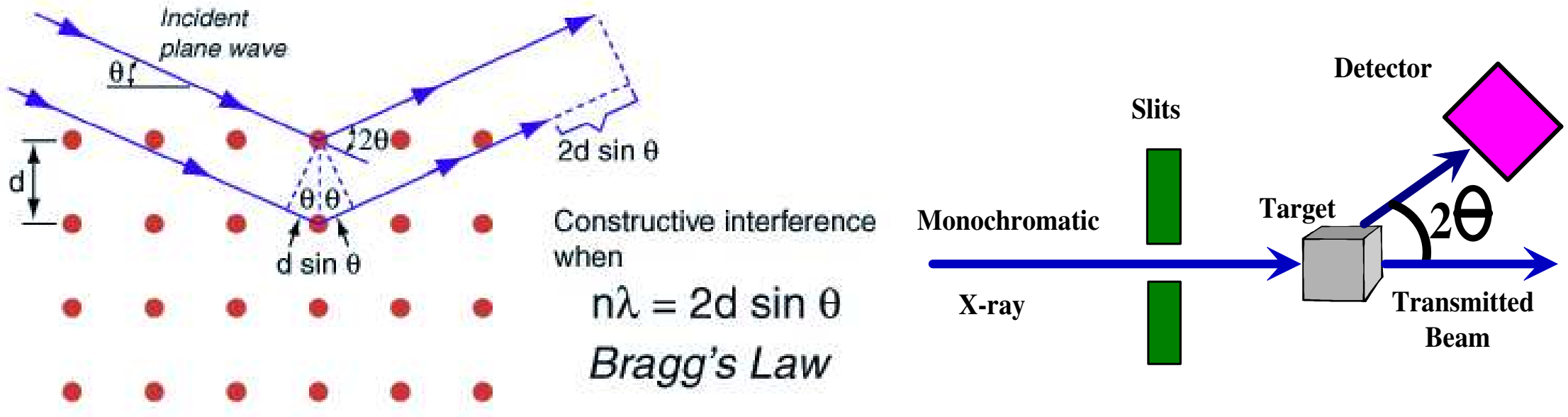

Figure 2: Bragg's Law example -left- and Simple X-ray diffraction setup -right-

When atoms are arranged in a periodic fashion, as in crystals, the diffracted waves from different atoms will interfere with each other originating constructive interference peaks with the same symmetry that characterizes the distribution of atoms. These diffracted waves, and the resultant intensity distribution sensed by an appropriate detector, are strongly modulated by this interaction. Measuring the diffraction pattern therefore allows us to deduce the distribution of atoms in a material. The Braggs were awarded the Nobel Prize in physics in 1915 for their work in determining crystal structures and although Bragg's law was originally used to explain the interference pattern of X-rays scattered by crystals, many diffraction methods has been developed since then to study the structure of all states of matter with any kind of beam.

In the example in Figure 2 we assume the scattering points to be atoms, but Bragg's Law applies to scattering centers consisting of any periodic distribution of electron density. In other words, the law holds true if the atoms are replaced by molecules or collections of molecules, such as colloids, polymers, proteins and virus particles.

Focusing back to the SAXS technique, we predict that scattering will occur whenever a beam of X-rays (whose wavelength is around 0.1 nm) hits upon a material, liquid or solid, that contains inhomogeneities larger than the wavelength of the X-rays. The scattering pattern measured by a suitable detector contains information about the structure of these inhomogeneities. Usually the scattering experiment sees a scattering density that in the case of X-rays is the electron density. Important is not the absolute density but the difference between the particle and the surrounding medium: this gives what is usually called "contrast". The result of the experiment is the *Fourier transform* of the contrast distribution. By comparing the experimental data to theoretical calculated intensities or performing an inverse Fourier transform it is possible to retrieve the information about the contrast distribution that relates to the mass density distribution of particles inside the sample.

The fundamentals of the SAXS theory were developed by `Debye, Fournet and Guinier` in the 1930's. Considering again the Bragg's Law we can obtain a minimum size d = $\lambda/2$ for the measurement of the dimension of the features in the sample in a diffraction experiment but it does not predict a maximum size. In fact the origin point in the inverse space means infinite size of the feature that we are measuring.

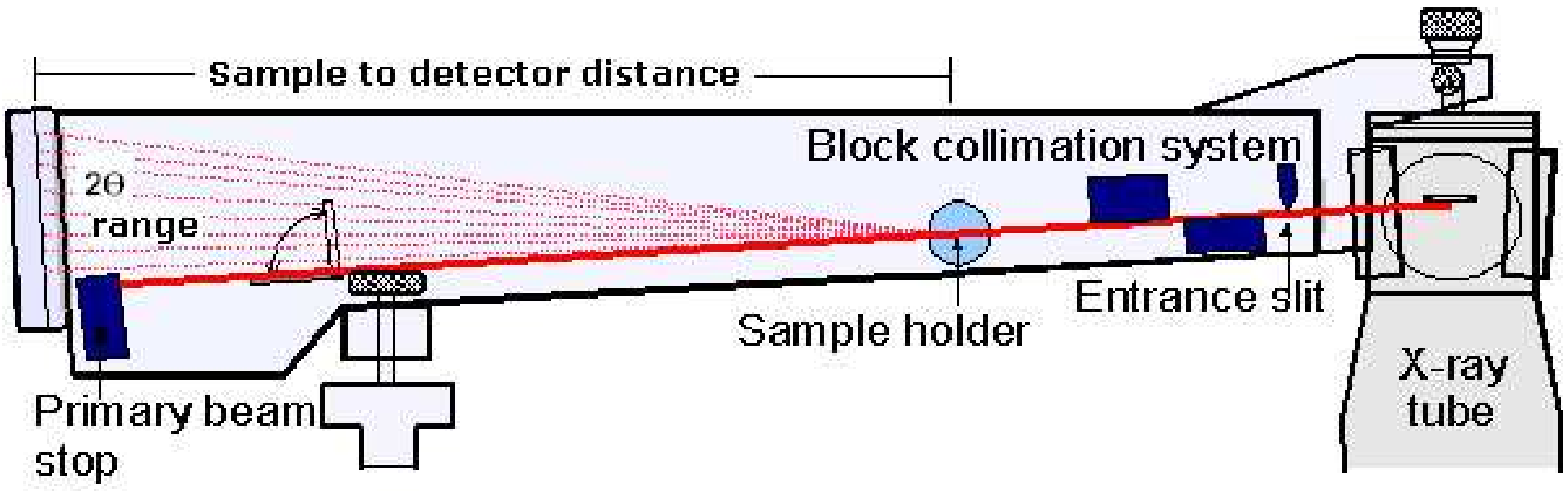

Figure 3: Typical Kratky camera setup used in SAXS experiments

Theoretically, affording infinite resolution near the origin, it would be possible to measure arbitrary big structures. This field was pioneered again by Guinier in his research on metals, where many important morphologies have characteristic sizes much larger than the atomic scale, during his development of the theory of Guinier-Preston zones. Guinier became one of the fathers of the *massive interest in diffraction techniques aimed* *at large-scale structures* in the 1950's (ref. [1]). Bragg's Law nevertheless predicts that information pertaining to molecular scale structures would be seen at diffraction angles $2\theta \leq 5°$. Coming back to the real word experiments we can say it is possible to design specialized instruments that have enough resolution near the origin to measure down to less than 1/1000 of a degree enabling the measurement of structure sizes up to 1000 nm using X-rays: this is indeed the goal of our design. Large specimen to detector distances, between 0.5 m and 10 m, and high quality collimating optics are used to achieve good signal to noise ratio in SAXS measurements.

The advent of `Synchrotron Radiation` (SR) in the 1970's and the development of efficient and fast electronic detectors made feasible a range of experiments on a number of different structures that would not otherwise be studied with conventional X-ray equipment like the widely used and well known `Kratky camera` in Figure 3 (ref. [2]). In recent years synchrotron facilities have become widely used as preferred sources for X-ray diffraction measurements. Synchrotron radiation is emitted by electrons traveling at near light speed in a circular storage ring.

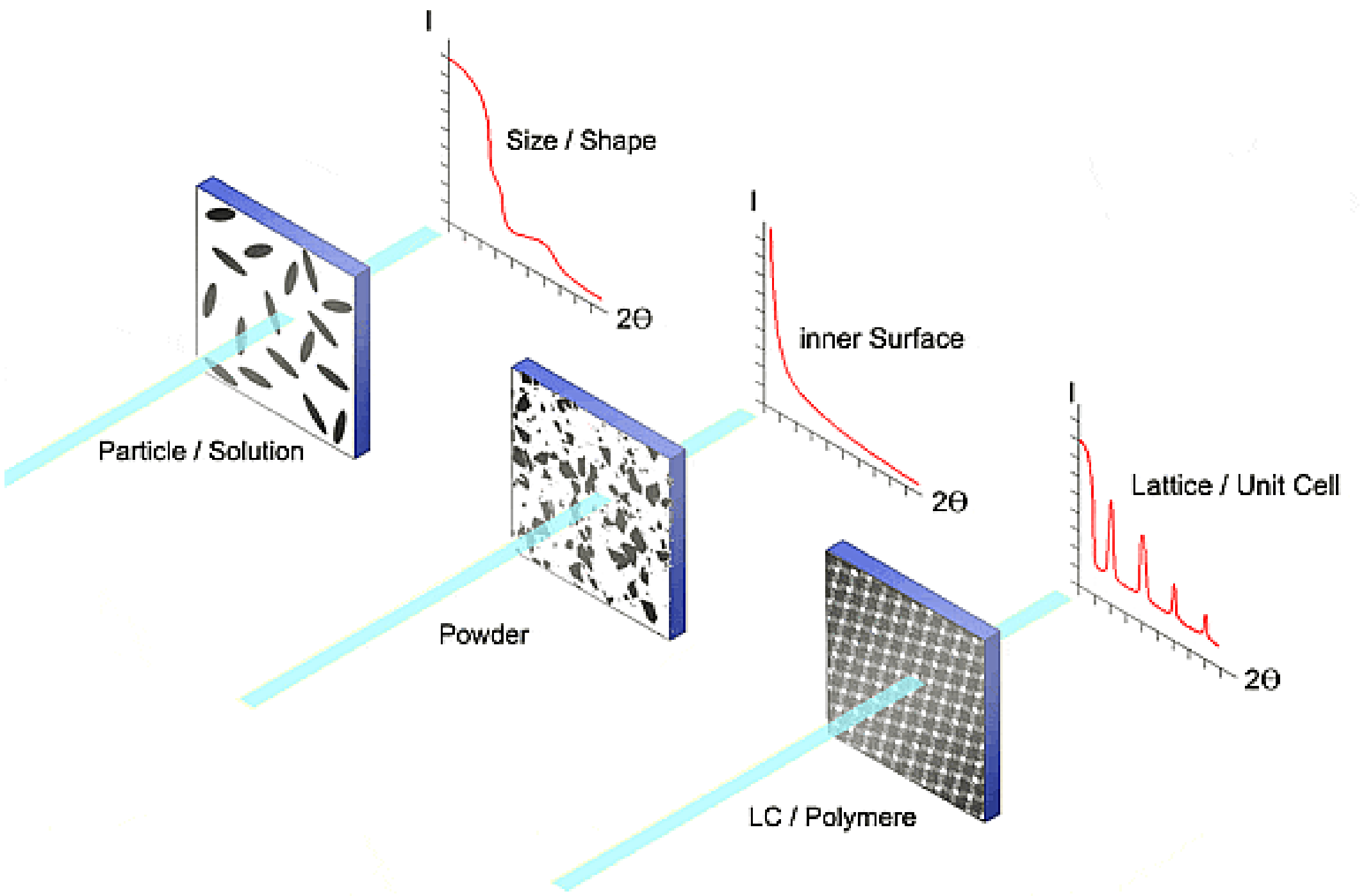

Figure 4: Example of some typical SAXS patterns for isotropic systems

These powerful sources, which are thousands to millions of times more intense than laboratory X-ray tubes, have become invaluable tools for a wide range of structural investigations and brought advances in numerous fields of science and technology. SAXS measurements with synchrotron radiation are technically challenging not just because of the small angular separation of the direct beam to the scattered beam. Indeed, as already mentioned, in the case of synchrotron radiation the intensity of the direct beam is several orders of magnitude more intense than a conventional X-ray tube SAXS setup. In addition, the measured SAXS intensity can show a large variation in magnitude, demanding a high dynamic range to the detector, across the $2\theta$-range (range of scattering angles, like in Figure 4) of interest, particularly in fractal systems.

Due to the lack of detectors which can both detect single photons and cope with the high counting rates in proximity of the direct beam (nevertheless a beam stop mask is used) it is often necessary to run the same experiment two or more times (using different absorption filters) in order to get acceptable statistics for the whole $2\theta$-range needed.

Absorbers of different thickness are used to reduce counting rates to a level that actual detector systems are capable to digest. This is not only an *ineffective way of using the powerful SR beams available nowadays* but it makes also impossible any enhanced analysis like for example to study the time dependence in the (sub)-millisecond range in non-repetitive experiments (ref. [3]).

# SAXS beamline at ELETTRA: detector's requirements and constraints

The high-flux SAXS beamline at ELETTRA (also codenamed BL 5.2 L, see Figures 5 and [4]) is mainly intended for time-resolved studies on fast structural transitions in the sub-millisecond time region in solutions and partly ordered systems with a SAXS-resolution of 1 to 140 nm in the real-space (ref. [5]).

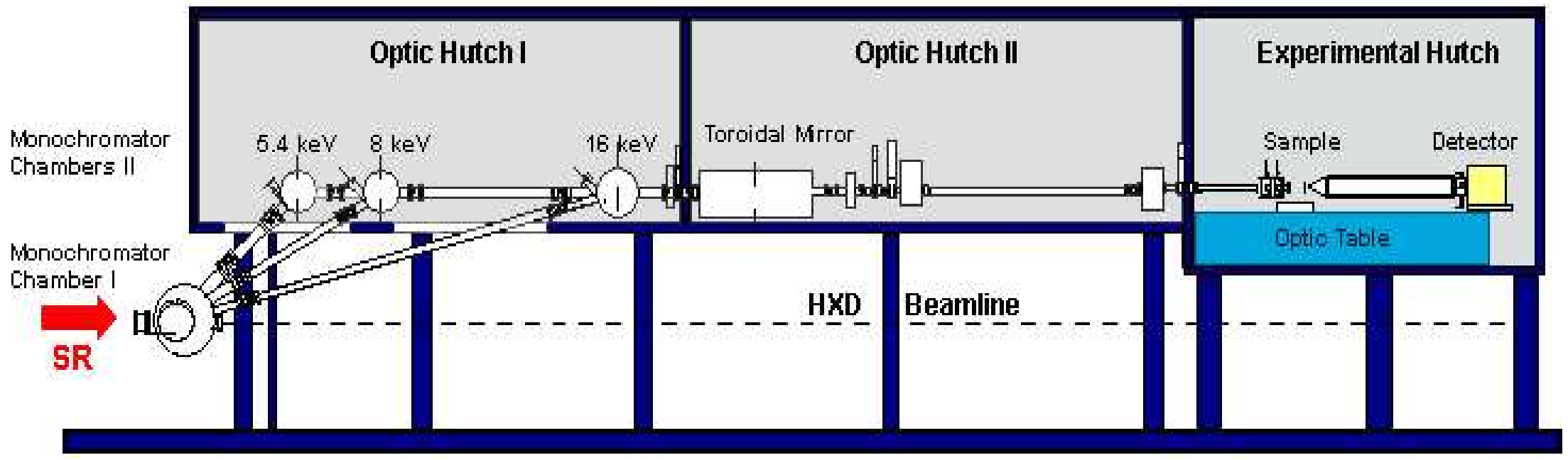

Figure 5: Overall structure of the BL 5.2 L beamline

The photon source is the 57-pole wiggler whose beam is shared and used simultaneously with a Macromolecular Crystallography beamline. The wiggler delivers a very intense radiation between 4 and 25 keV of which the SAXS beamline accepts 3 discrete energies: the standard 8 keV as well as 5.4 and 16 keV (whose respective wavelengths are 0.154, 0.23 and 0.077 nm). These additional energies make the beamline applicable also to experiments concerning very thin (e.g single muscle fibers) and optically thick (high Z) specimen like that often used in material science and solid state physics. The beamline optics consists of a flat, asymmetric-cut double crystal monochromator and a double focusing toroidal mirror. A versatile SAXS experimental station has been setup with the option to use an additional Wide Angle X-ray Scattering (WAXS) detector for taking simultaneously diffraction patterns in the range of 0.1 - 0.9 nm. The main area of research and relative scientific applications for the SAXS beamline at Elettra are:

- Time-Resolved Studies $\geq 11\mu$s
- Low Contrast Solution Scattering
- Grazing Incidence Surface Diffraction
- Micro-Spot Scanning
- X-ray Fluorescence Analysis
- Simultaneously Performed Small- and Wide-Angle Measurements (SWAXS) on: Gels, Liquid Crystals, (Bio)-Polymers, Amorphous Materials and Muscles

Depending on the desired resolution in time and in reciprocal space either on isotropic or anisotropic scattering of the sample, a one-dimensional position sensitive (delay line type) or a two-dimensional area detectors (CCD-type) are employed. The sample station is mounted movable onto an optical table that allows optimization of the sample detector distance with respect to the required SAXS resolution and sample size.

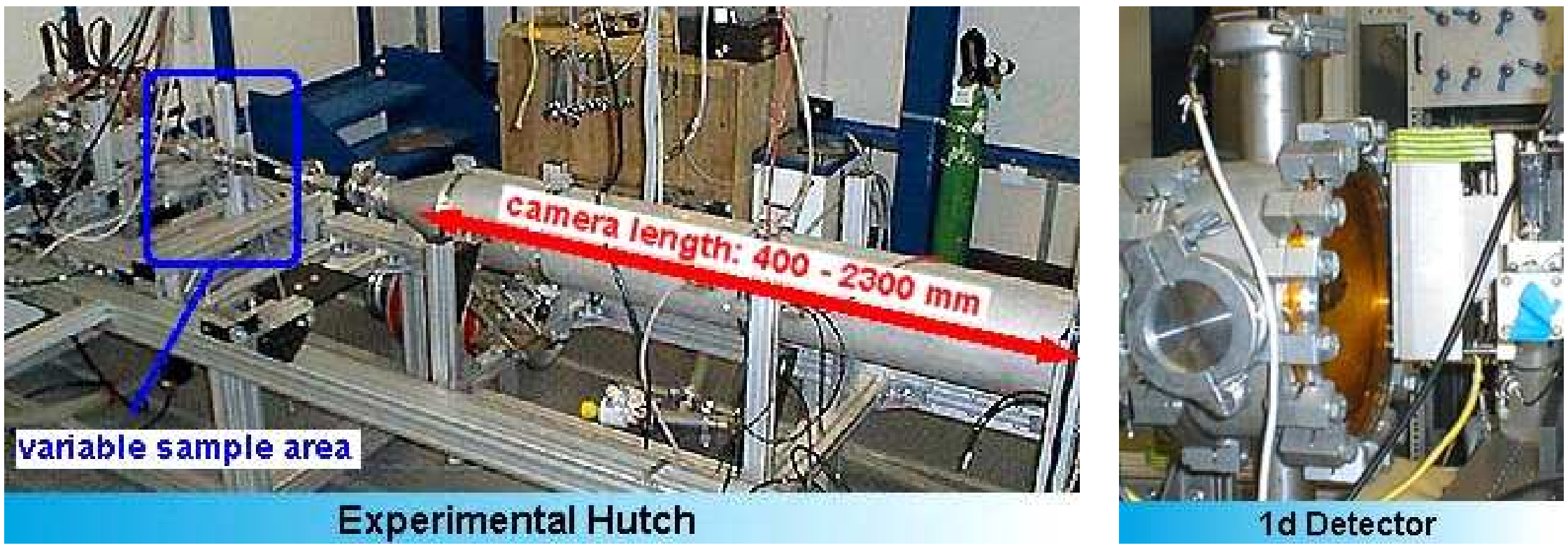

Figure 6: A few particulars of the SAXS beamline experimental station

In the generic SAXS setup used at the beamline a high intensity beam of monochromatic and collimated X-rays is available in the experimental hutch (see Figure 6) and eventually scattered by a sample. Detectors are usually mounted at the end of an evacuated camera, because if the X-rays pass through air they would be partly absorbed and partly scattered off the gas molecules creating errors in the data. Two detectors can be simultaneously used: one on the long variable-length vacuum camera, up to 2.75 meters in length, to measure scattering angles $2\theta$ up to 9 degrees (SAXS-range), and a second detector mounted at higher angles for the WAXS range. Usually a lead beam-stop is used to block the direct beam after the sample in order to prevent the full intensity beam from reaching directly the SAXS-detector. It is also good practice if the X-ray beam intensity is recorded just before the sample because the intensity of the X-ray beamline fluctuates with time and in this way it is possible to normalize the data. Scattering images from the detectors are recorded on a computer for real time monitoring and offline processing.

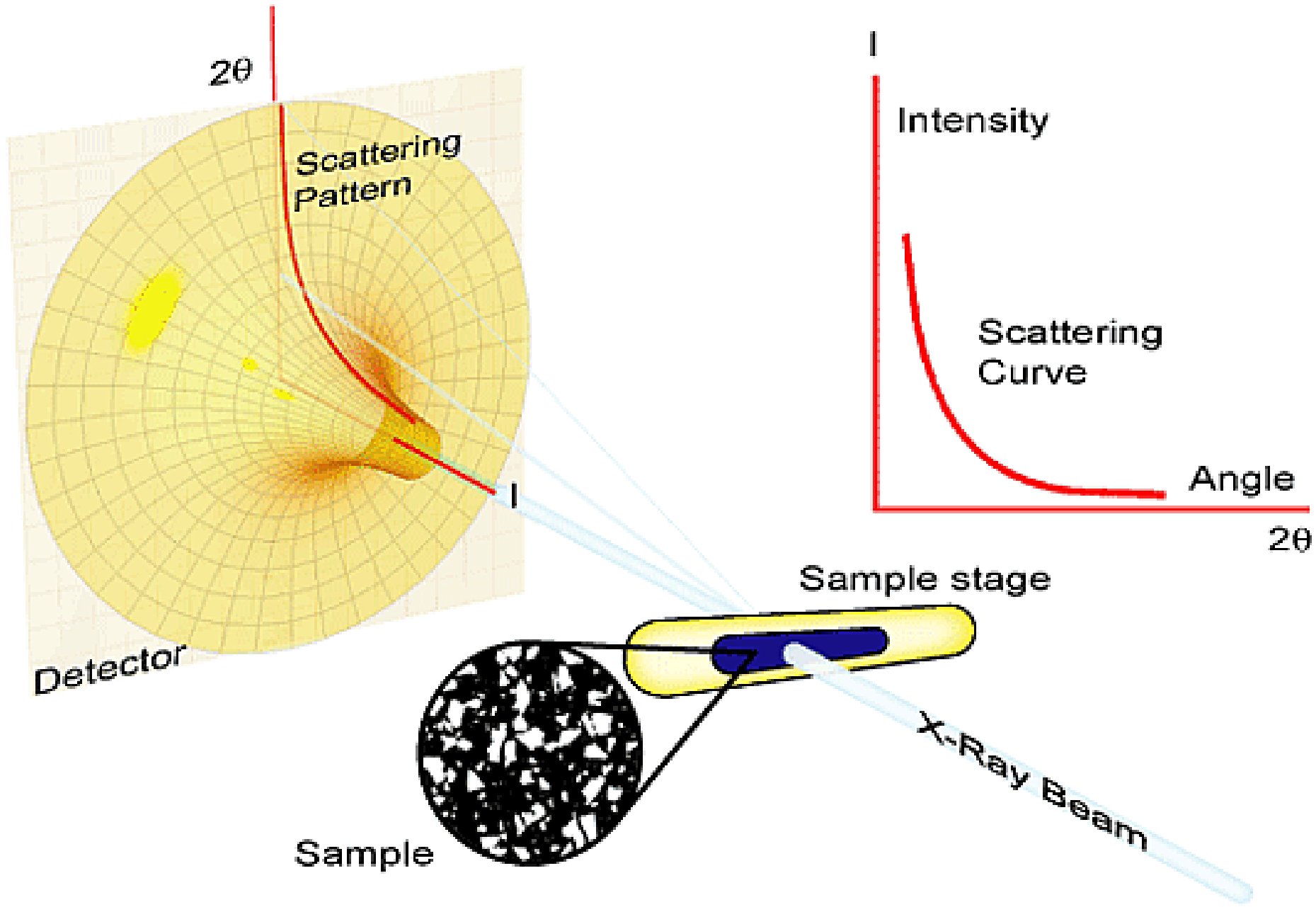

Figure 7: Circular symmetry manifested by the SAXS pattern of an isotropic systems

The current interest and aim of the project is to find a *better substitute for the 1-dimensional position sensitive X-ray detector* that is used for isotropic scattering experiments (see Figure 7): it is a Gabriel type from the name of its inventor `A.Gabriel` (ref. [6]). It is a position sensitive proportional counter that was offering, at the time it was invented, some new interesting features in the method of position encoding compared with previous systems. The fact that the encoding is achieved using an inductance-capacity (L-C) line connected to cathode strips, rather than by resistance-capacity encoding (high resistance) using the anode wire, was giving definite advantages to this new detector over the preesisting devices. Near its anode an intense electric field is applied and the incoming X-ray photons ionize the gas in the detector. Many free electrons are immediately accelerated and they have sufficient energy to ionize other atoms in the gas. A very fast multiplication phenomenon appears that is called "avalanche". On the cathode readout strips an inducted charge arrives perpendicular to the impact point of the avalanche.

The position of this charge is determined by the *delay-line method* that is based on measuring the time arrival difference of 2 signals that are those relative to the charge traveling to the left and to the right along the delay line and reaching the 2 readout nodes at the opposite ends of the delay-line itself: the difference in arrival times of the charge at each end correlates to the absolute avalanche position (see Figure 8). Since the read-out electronics is entirely based on timing circuits it inherently provides timing and position information: high position resolution require high speed and high timing precision. While the absolute position resolution is limited by the timing accuracy of the `time-to-digital converters (TDC)`, the relative position resolution is only limited by the digital range of the TDC and not by the signal-to-noise ratio. Theoretically such a delay-line detector operates properly at particle rates up to 10 MHz for the entire detector (Global Rate) and only the data acquisition capability of the hardware or software will set the limits.

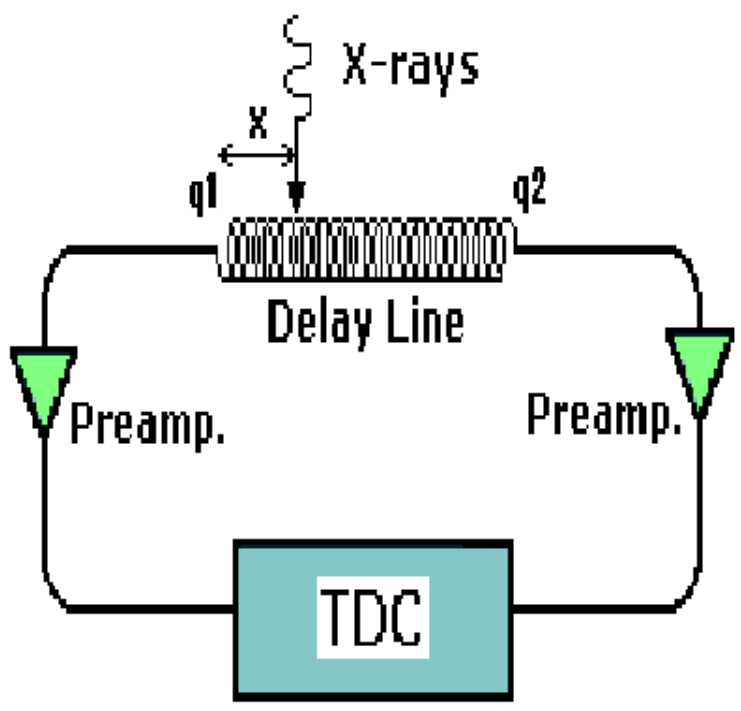

Figure 8: Delay-line readout

The delay-line method will only be able to cope with multi-hits events (group of particles coming within a few microseconds) if combined with multiple stop/start TDC for read-out and hence increased complexity. The favorable properties of the delay-line technique can only be fully exploited if the digital read-out chain following the front-end electronic circuits can keep up with the signal flow and if the time digitization is also as fast. More realistically all proportional counters suitable for this application suffer from low count rate capability and rates are generally limited to around 35 kHz for the whole detector at high-intensity SR beamlines due to slow recovery from the high space charges.

The specific *Gabriel detector in use at the SAXS beamline achieves a time resolution in the range of 10 $\mu$s (rates of about 100 kHz) and position resolution of 100 $\mu$m.* The final version of the one-dimensional gaseous detector proposed to replace the Gabriel one in the SAXS beamline is based upon a recent innovative design (see Figure 9) and allows exposure times in the microsecond range (ref. [7]). It is a high rate multi-element ionization chamber (integrating detector) which meets the high requirements for advanced time resolved SAXS measurements and represents a significant step forward to a detector that better suits experiments with high-intensity synchrotron radiation. Gas filled detectors, with delay line detection, suffer from *two main limitations* that affect their speed: both the local count rate and the global count rate are limited.

When a large number of photons hits the detector in a small area many electron-ion pairs are created. Due to the relatively slow drift velocities of the ions a local cloud of ions will form and create a field screening effect called “space charge”. This space charge significantly decreases the detector efficiency. The solution to this problem is to reduce the drift length of the ions and to neutralize them as quickly as possible.

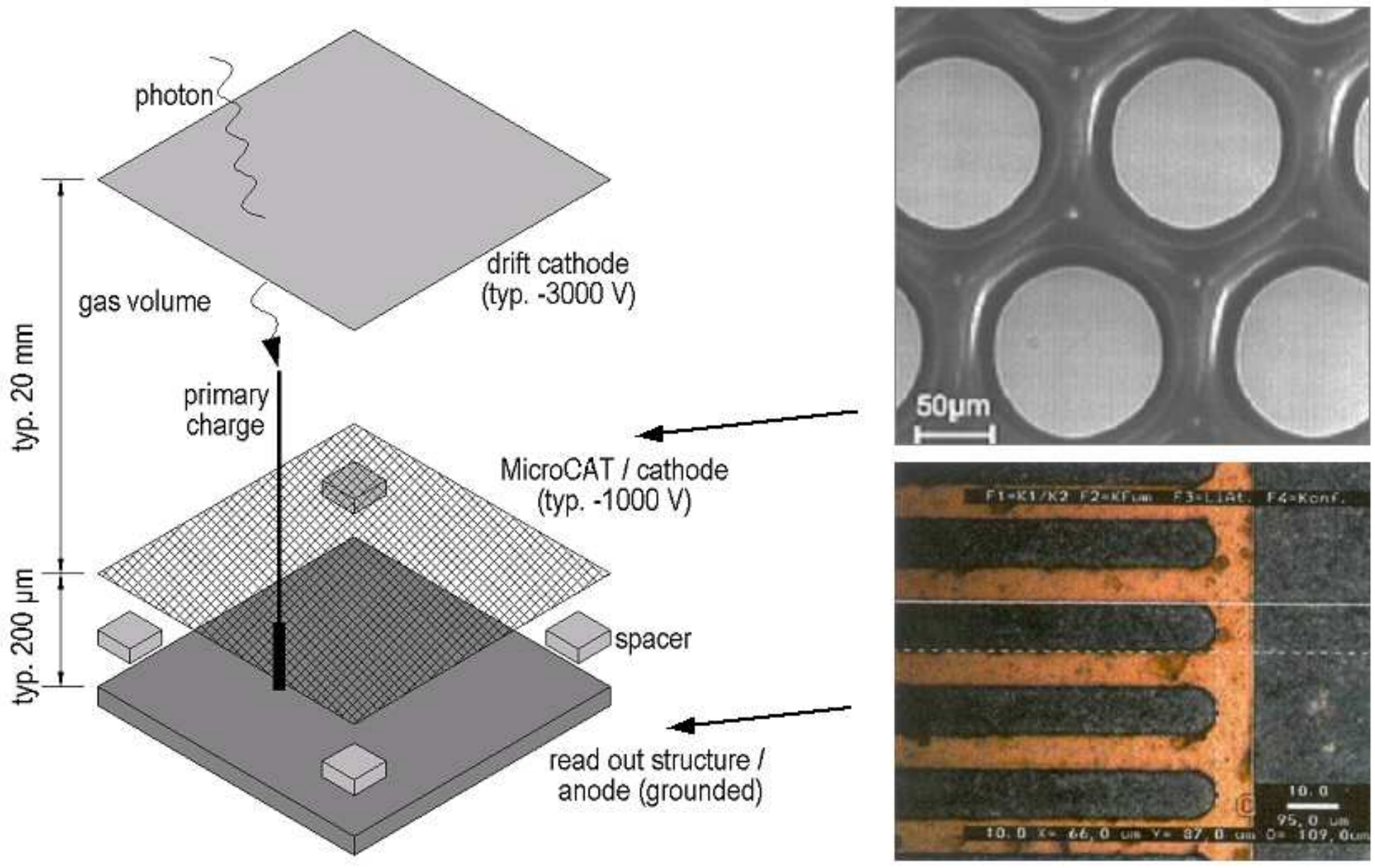

Figure 9: New SAXS detector working principle

The high-count rate capability of the new SAXS detector is based on the use of recently developed *MicroCAT* structures (Arbeitsgruppe Detektorphysik und Elektronik, Siegen, Germany [8, 9]). It was around 1995 that `M.Lemonnier` (LURE, Orsay, France) invented the *CAT* (*compteur à trou*) detector (ref. [10]). This device is made of a thin metal sheet perforated with small holes. When supplied with negative high voltage, and placed close above a grounded anode, gas gain arises because of the strong increase of electrical field in the vicinity of the holes. Beside the CAT, other similar detector types have been developed by other groups, like the *MICROMEGAS* by `Y.Giomataris` (Saclay, France) and the *GEM* (*Gaseous Electron Multiplication technology* [11]) by `F.Sauli` (CERN, Switzerland). The MicroCAT structure gets its name since the working principle is related to that of the original CAT structure, while having 2 main differences: reduced hole diameters and larger open areas allowed by the hexagonal arrangement of the holes. With this technology it is possible to obtain an increase in local count-rate of several orders of magnitude while allowing excellent spatial resolution. The global count rate limitation for delay line based gas filled detectors originates from the fact that all strips are read through the *single readout port* called the “delay line”.

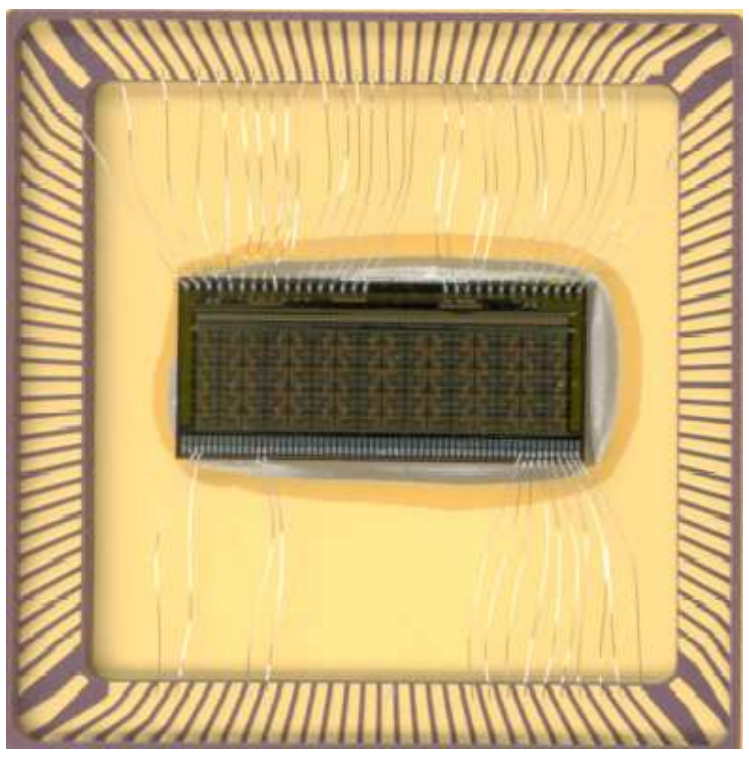

Figure 10: The JAMEX ASIC

While considerable progress has been made to speed up this process the real solution to the problem is to *read all strips in parallel* but such a parallelization is only possible by using Application Specific Integrated Circuits (ASIC) like the JAMEX in Figure 10. Proportional counters are already used successfully for the recording of low photon fluxes, but in order to be able to record larger fluxes, even at a single strip position, it was proposed to build for this specific application a *multi-cell position-sensitive ionization chamber* similar to that used in the project of the *non invasive coronary angiography* (ref. [12]) at **HASYLAB** (DESY, Hamburg). In order to apply the concept to SAXS measurements *four major improvements* had to be developed:

- smaller overall dimensions and lower complexity of the entire system
- position resolution of around $100\ \mu$m instead of $300\ \mu$m
- higher sensitivity (requiring lower noise of the read out electronics)
- higher dynamic range

The *first requirement* is accomplished having the acquisition electronics built around a compact, and commercially available, multi channel ADC system developed by the company Sundance (Reno, USA). However, the Sundance system does not match in the first place all the requirements of the proprietary ASIC chips used. Thus both systems have to be connected via an interface card whose design is also the object of this Tesi di Laurea and will be discussed extensively in a later chapter.

Regarding the *second improvement* some research was already carried out using a finer structured readout anode: a different prototype was built and a position resolution of $180\ \mu$m was achieved which has improved further to the actual $150\ \mu$m final design. This readout anode (see Figure 11) was built using an ultra thin multilayer process on a Kapton foil, which was subsequently glued on a very precise aluminum board of 10 mm thickness. The aluminum plate on which the kapton foil is mounted is cast rather than pressed. This gives a $10\ \mu$m standard deviation flatness specification over the full active area of 192 x 10 mm$^2$: this is crucial for obtaining homogeneous gas gain operation of the detector. The active area is composed of 1280 anode strips with a pitch of $150\ \mu$m and a very high typical resistance between two adjacent strips of around $10^{13}\ \Omega$.

The *third requirement* is addressed in several ways: the acquisition electronics that is performing the readout of the microstrips is based on a new version of the analog sampling, storage and multiplexing ASIC device (called JAMEX, see next Chapter and Appendix A) derived from that already used in many particle physics experiments (such as the *CLEO-II Silicon Vertex Detector* [13]) that has been specifically improved for this specific application. Moreover, a monolithic aluminum structure contains all the electronics of the detector, in different volumes, separated by appropriate EMI shielding. This configuration allows the shortest available paths for the analog signals thus avoiding their corruption. Appropriate screening of any digital signal, extensive use of differential signaling and careful grounding design will help to preserve the inherent high resolution of the ADC system used.

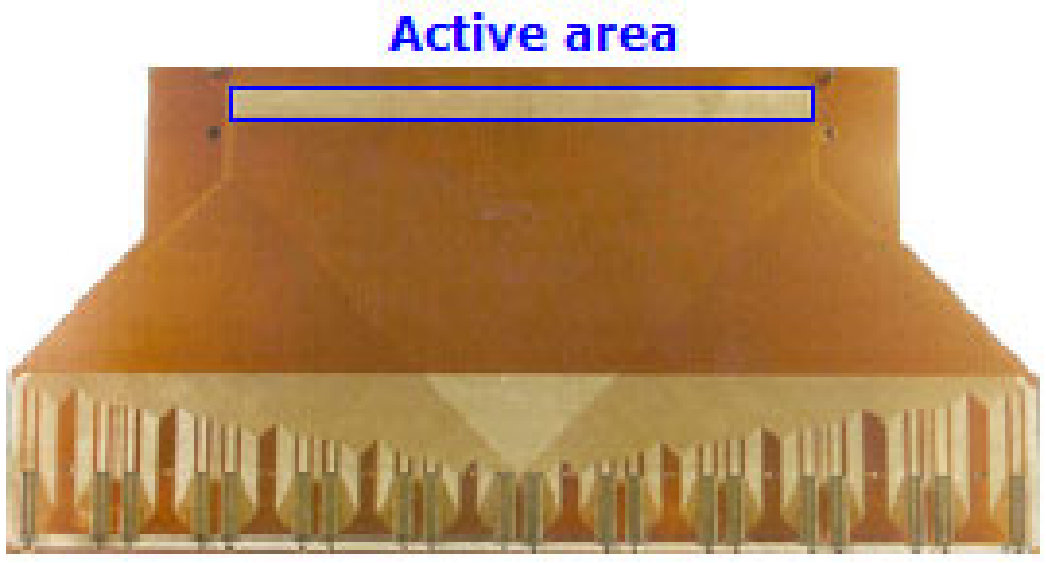

Figure 11: Photograph of the anode

The *fourth point*, to design a detector with high dynamic range, is very important and means that, in the extreme case, the detector can record also the direct beam: this fact drastically improves the measurements of the small angle region by avoiding an absorber, usually large, for the central part of the pattern and the associated scattering. Calculations for the saturation behavior of this kind of ionization chamber based on measurements obtained at ELETTRA and DESY and on results from a simulation program show that the direct beam with fluxes up to $10^{12}$ photons/mm$^2$ can be measured (ref. [14]). Experiments utilizing an electron amplification structure (such as CAT or GEM structures) have shown that, even with this integrating system, measurements on a sub-photon noise level are possible but with a limited precision of around 20% which reflects the inherent single photon energy resolution of a generic gaseous detector (ref. [11, 15]).

## Structure of this Tesi di Laurea

The work presented in this Tesi di Laurea started with a training period at the *Laboratorio di Strumentazione e Detector* belonging to the ELETTRA Synchrotron Light Source in Trieste. The initial goal was to achieve in short time a basic level of knowledge in the field of X-ray detectors. The dynamic environment at ELETTRA with its tendency to encourage the sharing of knowledge between researcher from all around the World resulted in a very important step forward the concretization of the general configuration for the SAXS detector.

Few travels characterized the one year period spent at ELETTRA in Trieste: a longer period, of approximately one month, accompanying the co-supervisor at the University of Siegen and two shorter trips to Manchester and Hamburg to deal with some technical aspects of the detector. It was decided to move back to the University of Siegen after a period with base in Trieste that was dedicated to the developing of the main electronic boards and to the ordering of most of the hardware and software components needed for the final version of the detector. In the year passed at the University of Siegen all the knowledge shared with the persons working in the *Arbeitsgruppe Detektorphysik und Elektronik* and the various facilities available there proved to be invaluable in obtaining a beta version of the final detector. Though the beta frontend electronics was built with a reduced number of analog channels it is organized in order to allow full testing, while the reconfiguration capability is useful to determine the best final design.

In order to explain all the aspects of this work in a simple way it was decided to structure this Tesi di Laurea in five chapters that closely reflect the main stages in the detector hardware development:

- *The first chapter* is a review of the different possible configurations for this detector, the JAMEX ASIC is presented and some earlier detector configurations are tested till the disclosure of the final JAMEX characteristics. Its final specifications brought to a total redesign of the final detector configuration that will be explained in the successive chapters.

- *The second chapter* gives a theoretical description of the final detector organization with all the specifications and interface-related issues: protocols, user commands and file structures.

- *The third chapter* shortly reviews the Sundance hardware and focus on the specific design issues of this implementation.

- *The fourth chapter* presents the Printed Circuit Boards designed to be the frontend electronics of this detector and called the J2S boards family.

- *The fifth chapter* discuss the final prototype implementation, gives an overview of some secondary aspects in the development of this detector, presents the measurements and the relative upgrading strategy to fit the final detector.

# 1

# Evaluation of several project hypotheses

This detector is supposed to convey, as we have already mentioned in the introduction, all the experience obtained from the many people involved in the developing of the four versions of the *NIKOS detectors* (ref. [16]) to produce a far smaller device featuring improved spatial resolution and comparable dynamic range. The NIKOS system was developed for experimental *intravenous coronary angiography* at the Hamburger Synchrotronstrahlungslabor HASYLAB at DESY and its specially designed fast low-noise, two-line ionization chamber detector was built in a no compromise way using all the latest available electronics components, to achieve an impressive dynamic range of up to 300.000:1, with 336 pixel per line and a minimum readout time of 200 $\mu$s (max. rates of about 5 kHz but used at 1.25 kHz).

The key idea of this detector is that the *ionization current of each strip is integrated and digitized individually* with 20-bit resolution while 4 different gains can be selected, in order to optimally adjust the available high dynamic range to the specific measurement being performed.

This approach leads naturally to a really complicated electronic device which necessary has to be made modular and nevertheless requires a lot of space and is exclusively suited for a research activity, with continuous support from skilled technical people devoted to keep the entire system working with top specification. Many hundreds top quality ADCs, hybrid technology preamplifiers and shapers, fiber optical data communication and several racks full of custom made electronics were the key to such exceptional performances: all this was required to ensure that the image quality of the entire diagnostic method would not have been impaired by the detector.

All this considered and having said that the detector we are going to build is expecting to show comparable dynamic, higher spatial resolution and small sizes can evidently rise perplexity: it is true that *few years has passed since the previous design* and many electronic devices get important performance improvement over just few months, because of the strong research originated from the dynamic consumer market demands, but the risk is to be too much optimistic.

Hence we need to focus our attention on the *main components* involved in such a design and the *relative constraints*:

- **preamplifier stage** (space restricts the choice to SMD Operational Amplifiers or specialized ASIC)
- **Analog to Digital Converter** (strict space and power dissipation requirements)
- **Signal Processing Electronics** (more freedom is here available)

It is needed as well to consider the *expected performances* that are specified in Table 1.1: we can immediately infer from the number of channels that this detector will need 1280 integrating preamplifiers. In comparison the NIKOS detector was having half this number of channels realized on several tenths of modular cards built in ceramic hybrid technology with laser trimming for optimal component matching.

| | |
|---|---|
| detector active size | 192 mm |
| ’box’ physical size | $\leq$ 500 mm in beam direction |
| number of pixels | 1280 |
| time resolution | $\cong$ 100 $\mu s$ |
| dynamic range @ single gain | 10.000:1 minimum |
| number of gain steps | 16 = 4bit |

Table 1.1: Specifications for the SAXS detector.

Moreover, the ionization chamber and preamplifiers box of that detector, alone, was taking a volume roughly 10 times bigger than the entire detector we want to build. In this case for the integrating preamplifier stage there is no alternative to the use of a specially designed ASIC.

The choice related to the Analog to Digital Converters (ADCs) configuration is wider, hence a careful evaluation of different strategy was needed to be considered together with their positive and negative points.

## 1.1 Overview of two main acquisition system topologies

Recalling some history of the ADC we can note that high performances ADCs were relatively expensive components with a high power consumption until 10 years ago, when the technology received a strong evolutionary boost due to the growing interest in DSP processing. Hence in the past decade the tendency was to avoid, if viable, using more than one ADC for each system, meaning that was imperative to use some kind of “Analog Multiplexing” if more than one channel was to be measured. The need for sampling all the channels at exactly the same time forced designers to include many “Sample and Hold” (S/H) circuits, one for each channel, before the multiplexer inputs. To summarize the situation: the amplified signals are stored in many S/H circuits, an analog multiplexer is used to scan the S/H outputs and a single ADC is used to convert sequentially the analog signals to produce a serial digital output signal.

Nowadays the tendency is exactly the opposite: it is preferred, whenever it is possible, to dedicate a singular ADC with integrated S/H circuit to each channel which has to be measured. This happened because massive research led to the modern VLSI process for producing ADCs which allows very fine features to be realized on silicon permitting a production with very high yield of small components with low dissipation: high performance ADCs with high number of bits can be now realized using the relatively inexpensive CMOS processes.

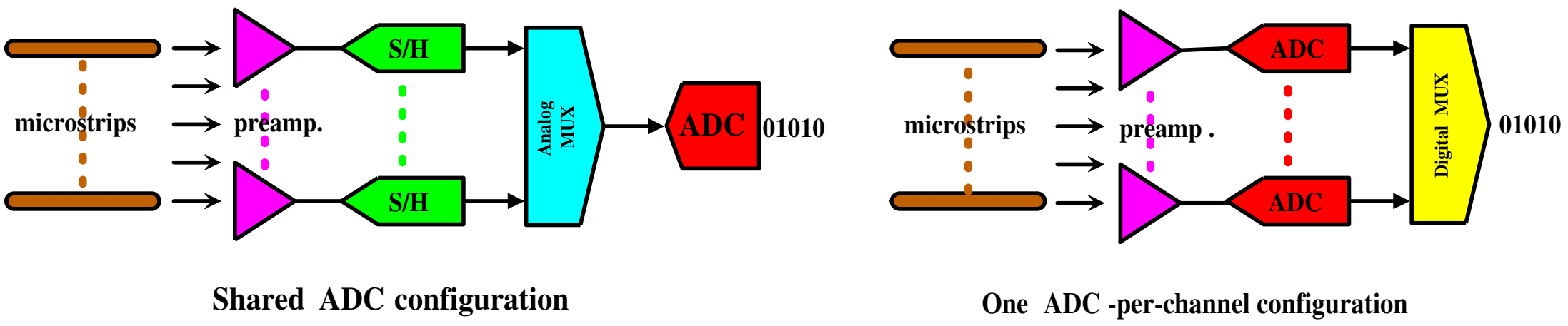

Figure 1.1: The two main acquisition system topologies

This two different approaches to the design of an acquisition system (see Fig. 1.1) can be synthetically referred as:

**Shared-ADC** configuration versus **ADC-per-Channel** configuration. The main pro of an ADC-per-Channel configuration is that each channel has its own ADC and the final conversion to a serial data stream is a *completely digital process*: it is enough to scan the digital outputs of all the ADCs used in the device (“Digital Multiplexing”) avoiding all the problems connected to analog multiplexing: this operation is performed entirely in the digital domain and so any deterioration of the signal is eliminated, at least theoretically. Also the current advancing in the development of low-power Programmable Logic Devices (PLD), allowing a quick development of complex digital logic, allowed actual designers of modern multichannel acquisition systems to switch without regret to the ADC-per-channel topology.

Actually, a careful ADC-per-channel design can reach *performances practically limited only from the quality of the selected ADC device*. On the other side, a practical limit is given because of the power dissipation and physical dimension of these devices which makes *at least impractical* to build a system using thousands ADCs, even if using the incredibly small micropower devices available today.

With its hundreds ADCs and preamplifiers the NIKOS detector is thus to be considered an extreme application of the ADC-per-Channel approach, having the primary objective to overcome the limitations imposed by analog multiplexers and in general the many problems connected with a Shared ADC configuration.

We can summarize the *problems arising with a Shared ADC configuration* in few points:

1. The analog multiplexer provide *limited channel separation*: this has the nasty effect of high crosstalk to neighboring channels and is worse when many channels are being measured.

2. The *dynamic range* of a viable S/H and analog multiplexer circuitry is limited to approximately 80 dB (roughly 14 bit). This means that relatively high preamplifier gain is needed to raise the input noise above the noise level of the S/H and MUX if top performance are needed. Regrettably, being limited the maximum amplitude of the signals passing through the multiplexer, a high amplifier gain decreases the input range and creates problems, related to the offset amplification, that eliminate the possibility of DC amplifiers.

3. *Charge leakage currents* in the S/H circuits causes gain differences between channels.

4. *Charge injection* caused by the switching of the analog multiplexer imposes glitches on the S/H output.

5. In *systems with many channels* (more than 100), feeding one single multiplexer, proper routing of all the sensitive analog signals to the single Analog MUX becomes a formidable task since interference problems become huge.

6. The output of the Analog MUX is a *staircase like signal* with the sequential S/H voltages. This means that, for every consecutive channel, the ADC has to quickly settle on a new signal level. This imposes extreme demands to the ADC analog input stage (and on any intermediate buffer amplifier stages which may be needed to prevent the loading of the MUX output) regarding its bandwidth, frequency response and settling time. Effects like overshoot and ringing lead to further deterioration of the sampling precision.

On the other hand, since we must be fair, we need to raise the question if the ADC-per-Channel approach does not introduce additional errors as a result of the differences among the ADCs. It is in fact reasonable to think that using a Shared-ADC topology at least ensures that any ADC error is the same, since we are using one ADC for all channels, and thus it is easier to correct with an easy postprocessing. Although this assumption is not false, it is worth remind that the differences between modern ADCs, in a properly designed circuit topology, are much smaller than the errors introduced by any analog multiplexing circuit.

Coming back to our original problem we realize that the dimension constraints and the specific geometry of the SAXS detector require an analog front-end amplification system with very high density of electronics. Therefore it was *impossible to continue to think in terms of discrete components* to realize the 1280 charge sensitive preamplifiers and relative digitization circuits: it was necessary to rethink the readout system in a modular way based on an ASIC that will be presented in the next section.

## 1.2 Introducing the JAMEX64I4 ASIC

The use of CMOS technology is very common to realize special ASICs implementing a complete readout system tailored to an X-ray detector, but this type of transistor technology has the disadvantage of the high noise components at low-frequency. The selected ASIC device has derived, as we already said in the introduction, from that already used in many particle physics experiments and has been specifically improved for this specific application. The `Ingenieurbuero Werner Buttler`, an independent design house for design, developing and testing of monolithic integrated circuits (connected through collaboration with the Fraunhofer Institute of Microelectronics Circuits and Systems, Duisburg, GERMANY) developed an entire family of integrated readout ASIC devices: `CAMEX64A` (a CMOS readout system for stripline detector), `CAMEX64B` (a CMOS aimed to readout of X-ray CCD), `JAMEX64IR Type B and C` (monolithic readout system for IR-Diodes) and `JAMEX64A`, a JFET/CMOS readout system for stripline detectors.

The ASIC that will be used in this SAXS detector has derived from this experience and in particular from the last device mentioned. The JAMEX64I4 was produced using a 1 $\mu$m JFET/CMOS process, which is a special microelectronic CMOS process that allows a junction field effect transistor (JFET) to be implemented together with the standard NMOS and PMOS transistors. The JFET has some properties that well suit this kind of analog readout systems: it exhibits very high impedance input, drastically lower noise at low-frequency, and finally the gate of the transistor is less sensitive to ESD damage. Using this process we have the *ideal combination* of a CMOS transistors for analog multiplexing and digital switching parts and a JFET for the low noise analog preamplifiers, all on the same chip.

Two versions of the JAMEX64I4 ASIC, apart from the prototypes, were produced especially for this project: the first version (and the only available when this thesis work started) was featuring 64 channels with JFET input stage, each consisting of a resettable analogue integrator with digital selectable gain, a sample and hold stage featuring an 8 times correlated double-sampler and one *64:1 analogue multiplexer* with 1 single ended analogue output channel.

The *correlated double-sampler* is a shaper belonging to the category of the time-variant filters and can be analyzed exactly only in the time domain. A short explanation, in 3 steps, of its operation is given here for the simple double-correlated sampler (see Figure 1.2) and assumes that the signal is superimposed to a slowly fluctuating baseline:

1. the baseline V$n_i$ is sampled prior to the arrival of the signal
2. the sum of the signal and the baseline V$s_{i+1}$+V$n_{i+1}$ is sampled
3. the previous baseline sampled is subtracted (V$s_{i+1}$+V$n_{i+1}$-V$n_i$) to obtain the original signal

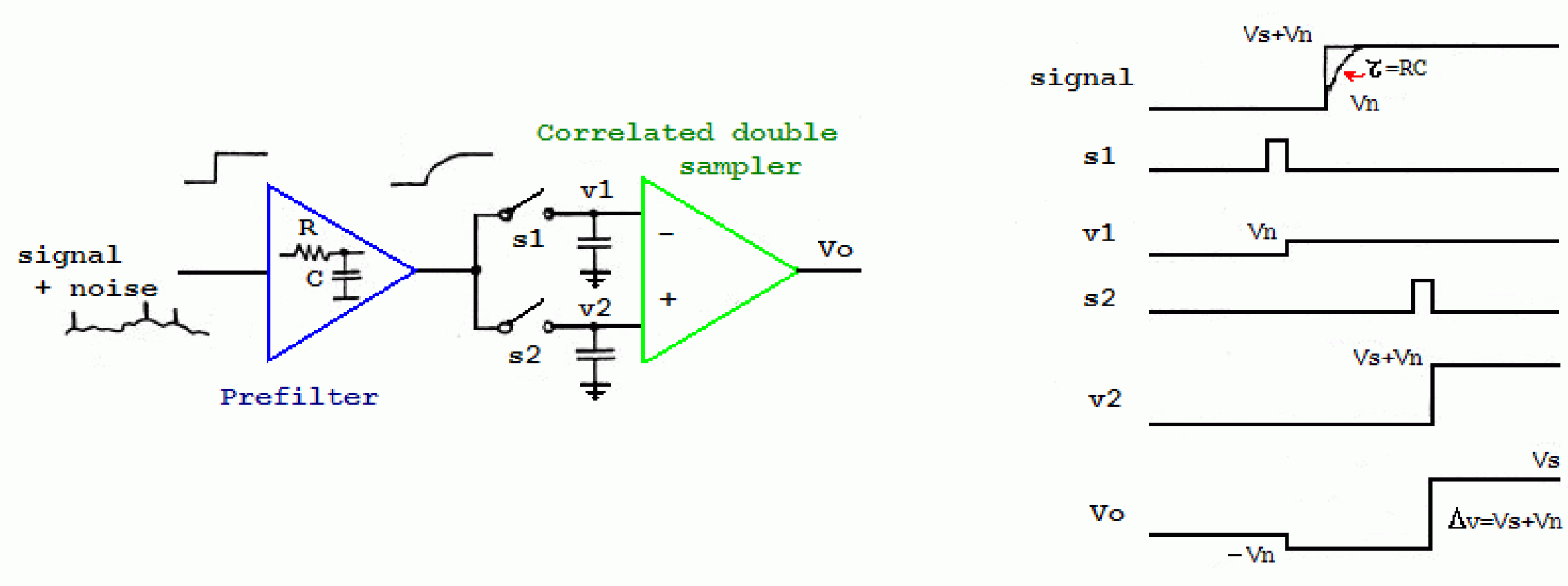

Figure 1.2: Simple double-correlated sampler example and relative waveforms

From the specifications of the ASIC, revealing the presence of an analog multiplexer, we acknowledge the actual impossibility of pursuing the pure one ADC-per-channel approach for this design. To handle the 1280 channels of the SAXS detector it is necessary to use a front-end readout chip with integrated multiplexing to reduce the space taken from the high number of analog channels and at the same time to lower the number of required ADCs. Nevertheless, having everything implemented on a small ASIC represents a better starting point than a generic Shared ADC configuration implemented with discrete components and consequently we can expect to obtain performances laying somewhere in between the two configurations. We already gave some figures regarding the maximum dynamic range that is expected to be less than 14 bits for a Shared ADC approach and less than 22 bits for an adc-per-channel design: reasonable performances for our solution, being somewhere in between, are fitting the specifications given in Table 1.1.

We cannot exclude that in a near future, considering the recent progress of CMOS micropower ADC technology, either the 64 analog channels or the 64 relative ADCs with reference and digital multiplexer could be implemented together on an ASIC similar in size to the JAMEX. A very compact and effective solution would result but, to face the reality, the major modification to layout, the need to shift to some kind of sub-micron 'exotic' process and the very likely problems of noise and reciprocal digital-analog crosstalk suggest this cannot be a short-term goal.

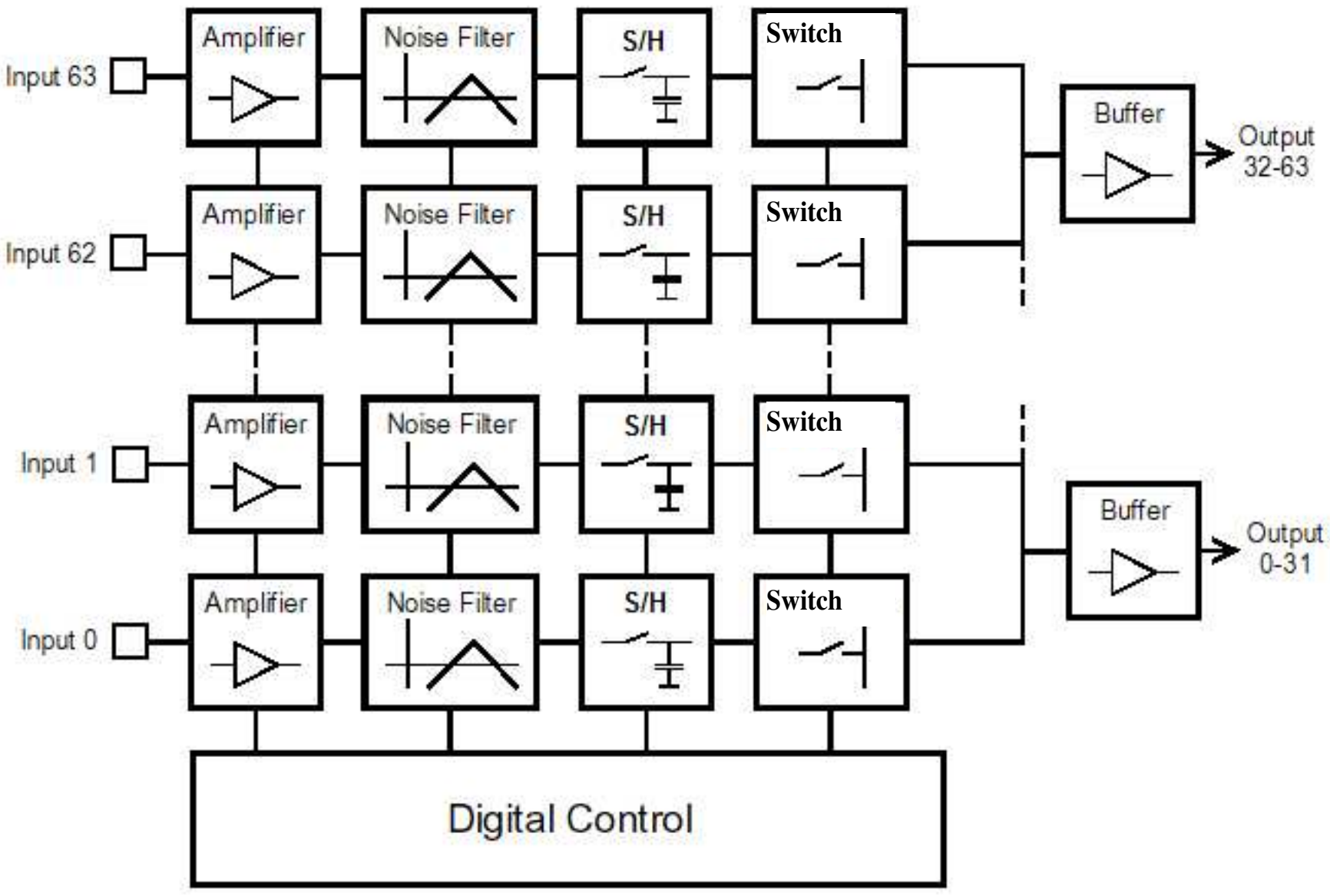

Figure 1.3: The original JAMEX configured like two *32:1* multiplexers

Coming back to the JAMEX64I4 it was found in the preliminary tests that the analog part of the readout chip fits the requirements of the SAXS detector (ref. [7]), but the chip was not easy to handle because a large number of digital signals was necessary to operate the chip. These were single ended digital TTL signals, thus with large amplitude, producing huge amount of noise. Another weak point to be improved was the analog output of the ASIC: the buffer was not able to operate with long cables, hence requiring an external high-current buffer. It was close after the start of this thesis work that a noise problem, related to the first amplification stage, was identified during testing of a chip similar in design to the planned JAMEX64I4 v2 (datasheet in Appendix A). It was decided to delay production of the JAMEX64I4 v2 until this problem was rectified thus allowing some additional time to decide about few minor changes for the ASIC, such as input load capacitance tuning and whether to configure it like *two 32:1 output multiplexers* (see Figure 1.3) or a *single 64:1* multiplexer (see Figure 1.4) with single output buffer.

The last possibility was particularly important since it was determining the number of *secondary analog channels* conveying the signals to the ADCs in the final system:

**20** if opting for the *64:1* configuration, **40** when selecting the two *32:1* option, with the foreseen option of reducing them to **8** with a *secondary analog multiplexing system* on which we will focus in the next section.

The second version of the JAMEX64I4 is the final version that will be used to build the hybrids for the SAXS gas detector: two of these chips should be assembled together on one hybrid that is essentially a small PCB with bonding sites for the ASICs bare die, plus few additional passive components and the required fine pitch connectors.

These 10 hybrids, mounted on the anode structure in the detector housing, will fit the requirements for the initial X-ray detection stage. New components will be present in the second generation of the JAMEX64I4 v2: for the timing of the switches an externally programmable timing register is provided which keeps the flexibility of changing the timing by serial digital programming. In this way only one external signal, the master clock, is necessary to run the programmed switching sequence during the measurement. Also the preamplifier gains, programmable in 16 steps with values in common to groups of 8 channels, are serially programmable and stored in an internal register.

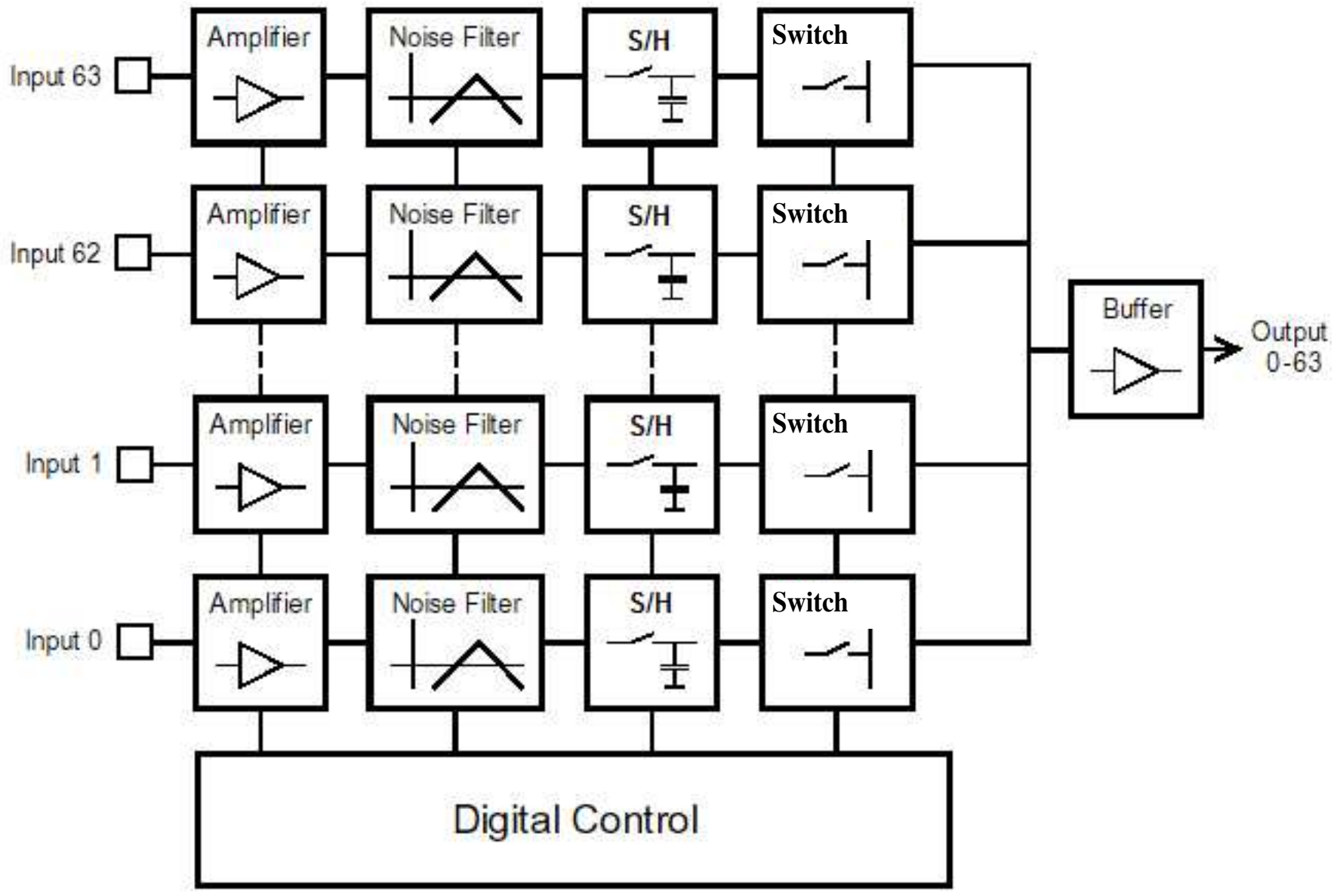

Figure 1.4: The JAMEX used in the detector is configured like a single *64:1* multiplexer

The interface for the external digital signals has changed from single ended TTL signals to LVDS (Low Voltage Differential Signaling). All these changes help to keep the peak-up from the digital signals during operation at a very low level. Finally, the output driver for the output analog signals is upgraded in the second version to a differential current line driver. This buffer allows a longer cable connection from the ASIC to an ADC without an external driver. Even in this case the use of differential signals reduces the sensitivity to external peak-up sources.

In short, due to the many improvements in this second version of the ASIC, the readout system is far easier to operate and less sensitive to its electrical environment.

## 1.3 *Dual-Multiplexing* JAMEX64I4 solution

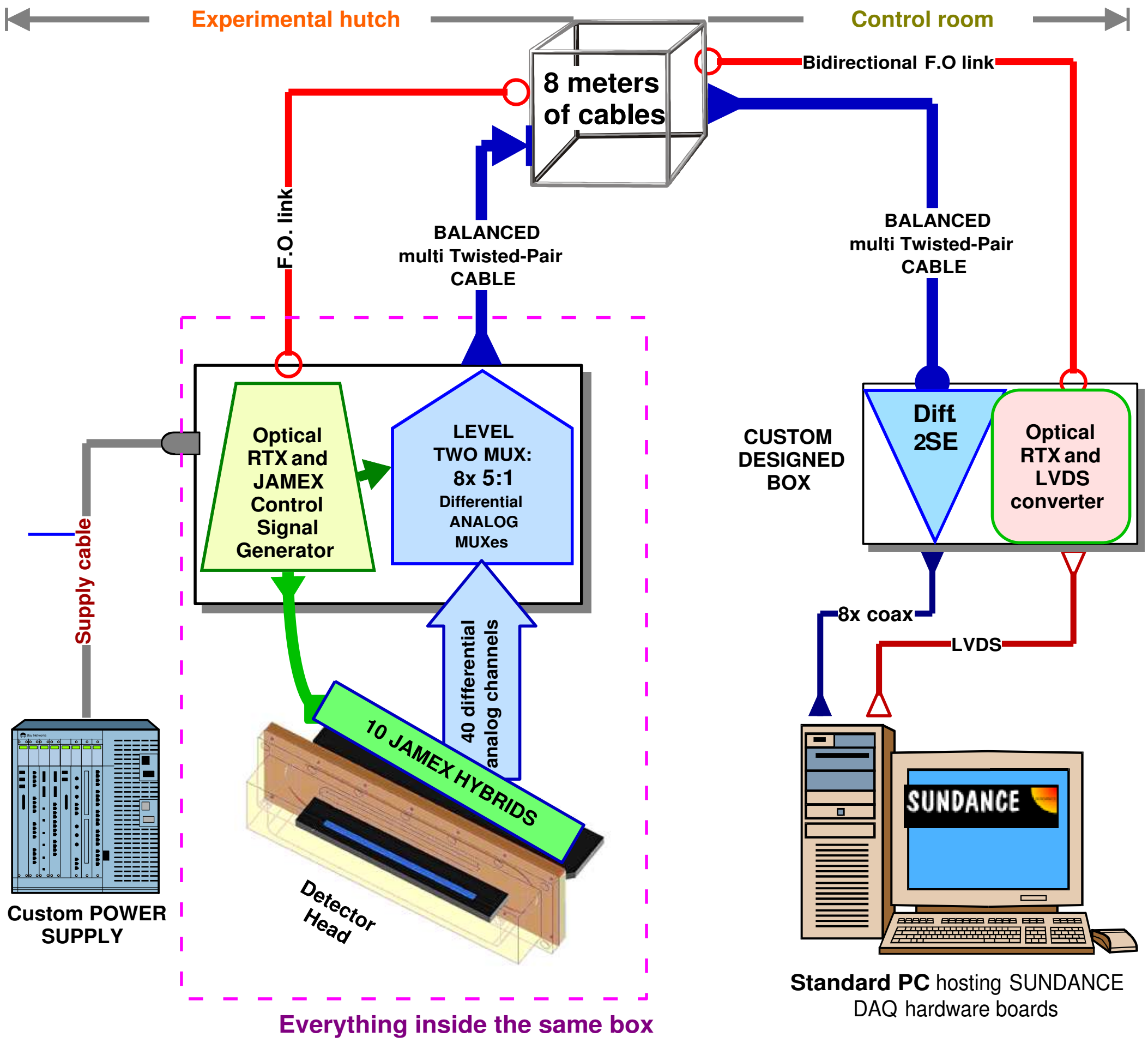

Figure 1.5: Schematic view of the components involved in a SAXS detector configured with two levels of analog multiplexers

The first design solution to be investigated for this SAXS detector was easy to put in a schematic way, as we can see from the upper figure, but not so accordingly simple to realize especially because of the strict requirements on the dynamic range of the analog channels.

The X-ray detectors used for research purposes in Synchrotron Radiation Sources are traditionally built without paying much attention to aspects like *ergonomy* or *user friendly* setup. Under this assumption if such a detector is composed of multiple boxes, interconnected with custom made cables of considerable section and length, it will not be blamed for this and the objective result is that many specialized detectors are realized exactly in this fashion. Looking to the positive side of this configuration, it allows easy upgrade and accessibility of the device in case of faults: in these circumstances, since the detector is usually a very expensive unique-made part which cannot be simply swapped with a new one in case of malfunction, this is a very useful characteristic. A second added value of this configuration, not less important, is that different parts can be replaced or improved with time, allowing to have a detector ready for the measurements in a short time and with the idea to achieve the expected performances in a more or less near future.

Even if this SAXS detector was needed to be built in *four* exemplars, not just a single one, the first idea was to realize it in the way we just presented, since it was considered too much effort to begin a deep "engineerization" of this device because such an effort would suit better to a higher volume production. Having decided to proceed in this way, the first issue to solve was to minimize the complexity of the *Digital Acquisition system* (DAQ) and, more importantly, to reduce the number of long signal cables interconnecting the *detector Box*, containing the JAMEX Hybrids and situated in the *Experimental hutch*, with the DAQ itself hosted in a Standard PC in the *Control Room*.

This was not a trivial problem: using the JAMEX64I4 v2 with two 32:1 MUXes would have required the impressive number of 40 twisted pair wires just for the analog signals connections. After a survey at the Elettra SAXS Beamline the distance to be covered appeared not less than 8 meters. A close collaboration with the personnel of the beamline -since the best choice was to ask hints from the future users of this detector- suggested that such a number of wires, although perfectly acceptable in a High Energy Physics experiment, is not suiting the environment of a generic beamline for a series of reasons:

- the configuration of the beamline is changing very often in relation to the specific experiment performed: detectors need often to be replaced or repositioned, this is very difficult with such a huge number of cables
- experience with similar systems showed that, if they work fine in the laboratory, when moved to the electrically noisy beamline environment they will suffer an evident decrease in performance and quite luckily became unacceptably unreliable because of either the electrical noise pick-up and electrostatic discharges
- space is limited in the beamline and usually very crowded, a small detector would be really welcome

The straightforward way to reduce the number of cables is of course to multiplex again the 40 analog signals available. A quick consideration of the generic hardware specifications available will bring directly to the required design: the Sundance DAQ system is modular and based on cards supporting DSP-FPGA processing, the number of channels being an integer multiple of 8, and specified as having 14 bits resolution at 10 MHz, the JAMEX ASIC will be operated at 1 MHz primary clock, giving an integration time slot of 128 $\mu$s which fixes the Multiplexer Clock (**M_CLK**) for the ASIC to be run at no less than 250 kHz. Using a level two **5 inputs to 1 output** analog multiplexer (hence the name *Dual-Multiplexing* solution) will limit the number of differential signals being transferred at the reasonable number of **8**, while requiring the multiplexer of the JAMEX to be run at exactly 250 kHz, a quarter of the rate of the primary clock. The level two multiplexer will run at 1.25 MHz clock (**MUX_CLK**), meaning that each of the inputs will be selected sequentially at this rate and in a circular manner, allowing the sampling rate (**ADC_CLK**) of the DAQ system to be a convenient multiple: either 2.5 MHz or 5 MHz can be a reasonable choice that satisfy Shannon's Theorem.

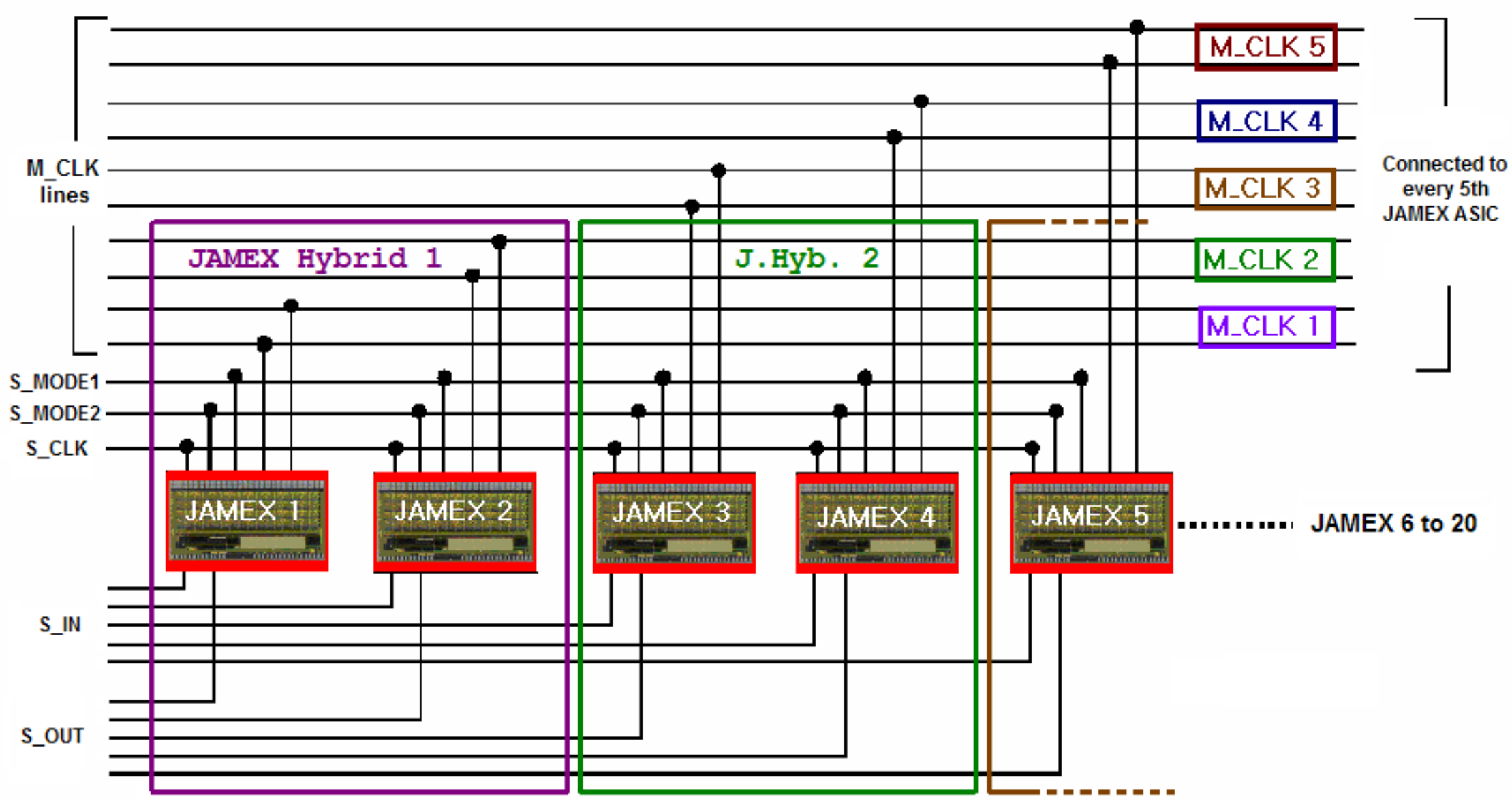

Figure 1.6: Required connections for the 20 JAMEX ASICs

In this way each analog channel of the DAQ system will be demanded to digitize the signals relative to a group of 160 microstrips belonging to the detector anode. In Figure 1.6 it is presented a convenient way to connect the digital control signals for all the 20 JAMEX ASICs used in a complete detector using this Dual-Multiplexing solution.

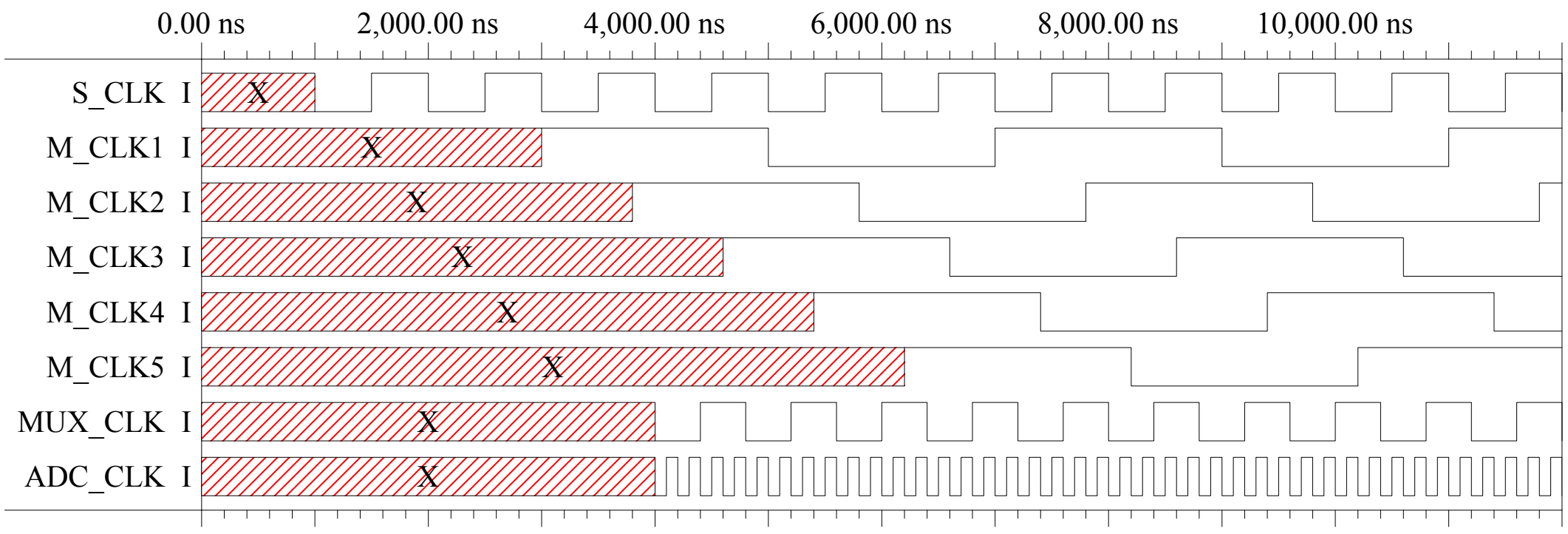

Figure 1.7: Proposed timings for the *dual-multiplexing* configuration

We note that it is convenient to route the primary clock for all the JAMEX (S_CLK) and the mode selection pins (S_MODE1 and 2) in a fully bussed way, since they are common to all the chips. The JAMEX multiplexer clocks (M_CLK1 to 5) can be implemented like interleaved and partially bussed signals, while all the other digital control signals would need to be individually routed (S_IN and S_OUT in particular). The timings for the main signals used in this system are given in Fig. 1.7; since they repeat identically when the system is operating at regime condition, they are presented for a slice of just 12 $\mu$s. The important thing to highlight is the constant delay of an increasing multiple of 800 ns for the consecutive signals M_CLK1 to 5. This trick allows for proper settling time of the analog signal at the output of the JAMEX ASIC and assures that the sampling of the output waveform is always performed at a constant position, provided all the clocks of the system are derived from a unique source. The choice of *fiber optical communication* (indicated as `F.O. link` in Figure 1.5) to transfer the digital signals and synchronization data, between the detector head and the DAQ equipment in the control room, is justified by immunity to noise and the overall reliability.

## 1.4 Evaluation of different analog multiplexers

Since the *level two analog multiplexer* was considered the most critical part of the previously discussed SAXS detector configuration, it was decided to setup some test circuit solutions. This was done in order to get some measured data from this specific component, thus having the possibility to make some comparison not just based on the given specification, and eventually to find weak points not apparent from the datasheets.

Summarizing our configuration, the level two analog multiplexer will run at a commutation rate of 1.25 MHz and the JAMEX multiplexer at 250 kHz or 500 kHz, depending on the configuration, restricting the choice of the candidate component drastically. The good point is that the differential current line driver at the JAMEX output guarantees that the level two multiplexer is driven in low-impedance mode so that the charge transmission from one channel to another can be expected to be very low. It is worth to remind that the output of the JAMEX MUX is a *staircase like signal* with sequential S/H voltages of different input channels (see Fig. 1.8) which are likely to generate fairly hopping signals instead of smooth transitions: the settling time (to 0.01% of the value) for the JAMEX output is close to 100 ns. We require the sum of settling and access time of the level two multiplexer output (to 0.01% of the value) to be less than one tenth of the commutation period. This is the main critical parameter: here most of the standard multiplexer fails, but this is required in order to avoid sampling of wrong non-settled values. We also need low noise, low crosstalk and low, or at least stable, offset at the output. Also the ability to drive an external cable without additional buffer would be a good plus for such a component since the output signal will need to travel for around 8 meters. All this characteristics brought to our attention two devices primarily aimed to Hi-Speed Video Multiplexing: the *Analog Devices AD9300*, already used in some projects of the laboratory with good success, and a newer device from Maxim, the *MAX440*.

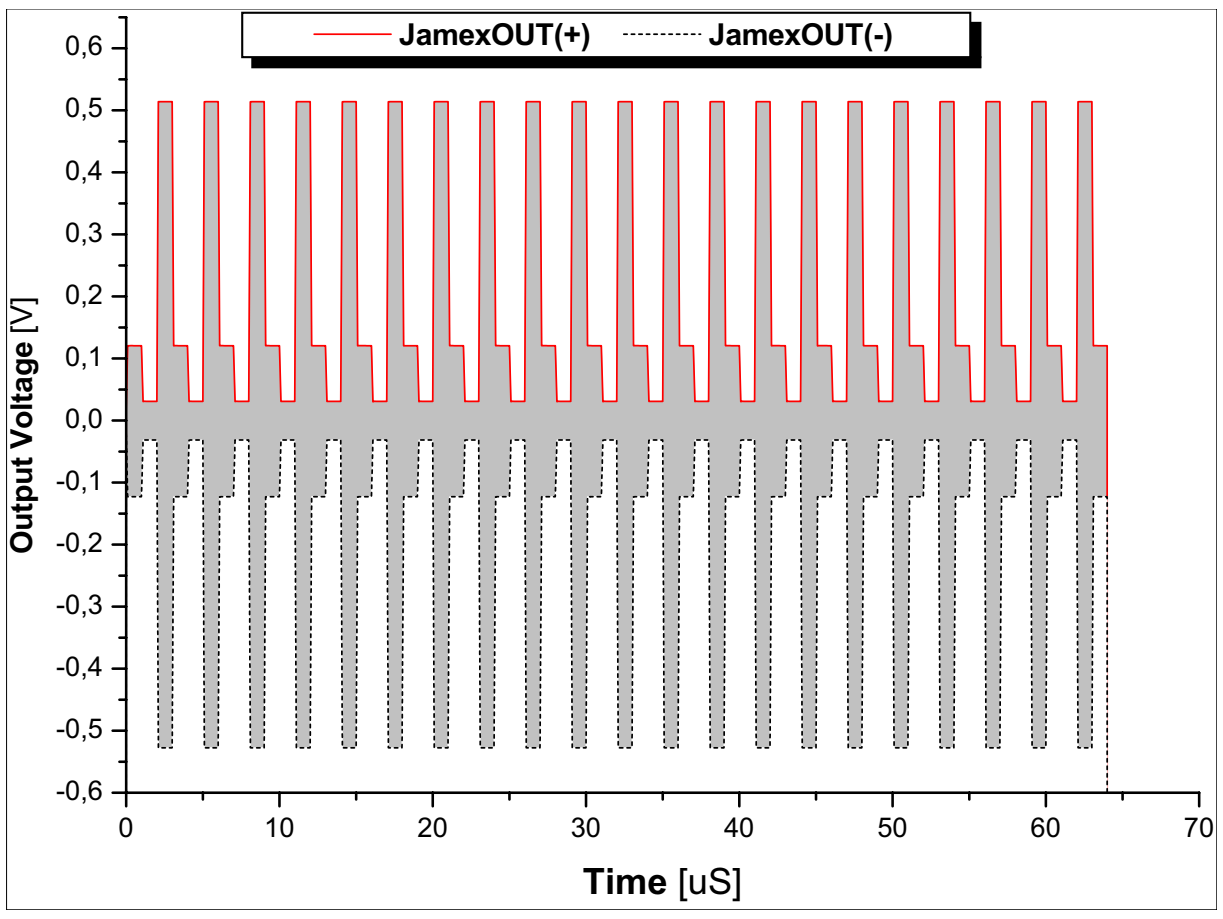

Figure 1.8: Simulated JAMEX64 differential outputs for M_CLK= 1 MHz and 3 different TST_IN1-3 values

### 1.4.1 Testing the Analog Devices and Maxim multiplexers

Although the Analog Devices **AD9300** (ref. [17]) intended application is the routing of video signals, the harmonic and dynamic attributes of the device make it appropriate for other applications requiring greater dynamic range and precision than those in video: Medical Imaging, Radar Systems and generic Data Acquisition. Its four channels of input signals can be randomly switched at megahertz rates to the single output. In addition, and crucial for our design which requires at least 5 input channels, multiple devices can be configured in parallel arrangements to increase the number of inputs: this is possible because the output of the device is in a high-impedance state when the chip is not enabled. When the chip is enabled, the unit acts as a buffer with high input impedance and low output impedance. It is manufactured in an advanced bipolar process that provides fast wideband switching capabilities while maintaining crosstalk rejection of 72 dB at 10 MHz. The device can be operated from 10 V to 15 V dual power supplies and is available in a 16-pin ceramic DIP that we used for our tests.

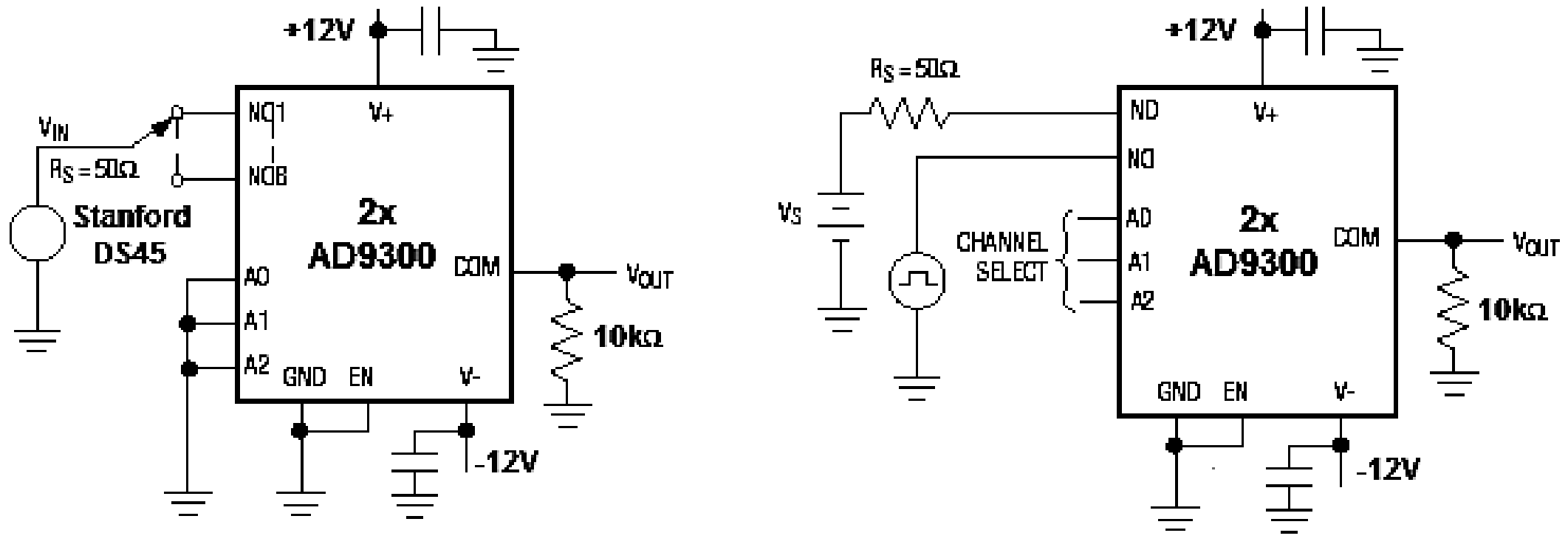

Figure 1.9: Measurements on the AD9300: off isolation -left- and crosstalk, access time or offset in dynamic conditions (depending on the specific configuration) -right-

Due to the simplicity of these test circuits (see Fig. 1.9), they were built directly soldering the two chips (8 channels in total) on a *multi-holes double sided pcb* with integrated ground plane, keeping all the connections as short as possible and providing suitable decoupling capacitors and terminating resistors. We used for these tests a Digital Storage Oscilloscope (Lecroy WavePro 960), a Signal Generator (Stanford DS45) and an Integrated Logic Analyzer and Pattern Generator (HP16702A) useful to generate the commutation patterns that control the multiplexer. We can summarize in few points the results from the measurements of the AD9300 analog multiplexer circuit:

- good settling+access time and crosstalk: $\approx$ 80 ns and close to 80dB at 1 MHz
- noise performance: $\approx\ 400\ \mu V_{rms}$ wideband (0-10 MHz) at the output
- high offset at the output: often above 3 mV and complex in behaviour

Other bad points were the high power dissipation (a complete multiplexer system for a full detector would dissipate an estimated 300mW $\times\ 32\ \ \cong$ 10W) and the lack of an integrated buffer to properly drive a long cable connection. These facts, together with the variable offset at the output, that would introduce relevant errors in the measurements, and the poor noise performances, which would require additional amplification stages and other tricks in order to obtain anything better than 12 bits resolution from our design (since the maximum amplitude of the JAMEX output signals is $\approx$1.2 Volt), suggested to continue experimenting with the second candidate.

The **MAX440** (ref. [18]) is a low-voltage 8-channel monolithic multiplexer-amplifier that combines low glitch commutations with high speed of operation and integrates an output buffer optimized for a fixed gain of +2 and ideally suited for driving back-terminated cables to provide unity-gain at the cable outputs. Its use provides some advantages: a complete multiplexer system for a full detector would require *just 16 chips* and hence dissipate an estimated 350mW $\times\ 16\ \cong$ 5W (this value already includes the dissipation of the integrated cable drivers) and the device can be operated from a lower $\pm$5 V dual power supply, thus simplifying the overall design. Some circuits were built similar to that used for the AD9300 multiplexer and the same instrumentation and methodology was used for the tests. An additional test was added to check the output buffer performance in driving a sample of 120$\Omega$ (balanced, consisting in 8 twisted pair couples) cable. Again we obtained not really encouraging results: slightly worst settling+access time and crosstalk if compared to the AD9300, higher noise justified by the presence of an integrated output buffer and an extremely high offset voltage at the output (above 10 mV). Additional tests using the Pattern Generator to drive the analog inputs of the chip, in order to characterize the complex offset behaviour, proofed that the offset at the output is strongly influenced from the voltage present at the not-enabled inputs. Considering the characteristics of the outputs voltage of the JAMEX ASIC it was considered *not viable* a circuit solution using such a kind of components since the errors introduced from the level two multiplexer would have been far higher than the signals we want to measure.

### 1.4.2 A different timing concept tested with the MPC507 MUX

Since the option to use a fast level two multiplexer appeared not to be compatible with the minimum specification required for this SAXS detector, it was considered worth to try a slower multiplexer. The approach of using a fast level two multiplexer had been chosen in order to allow the slower possible frequency (250 kHz) for the JAMEX multiplexer: this was considered convenient in order to guarantee the best quality of the output signal. But, since this multiplexer can run up to a speed of 2 MHz, it is possible to conceive another timing scheme that fulfil our needs: one integration time slot of the JAMEX ASIC has a duration of 128 $\mu$s, enough long to run 5 times the following sequence:

`for n = 1 to 5`

1. Enable the `n-input` of the multiplexer and allow 4 $\mu$s for settling
2. Read all the sequential 32 output values of the `n-JAMEX` ($32 \times 500$ ns $= 16$ $\mu$s)

while having still 28 $\mu$s free of activity in each cycle that is useful to avoid reading the JAMEX multiplexer in the initial part of its integration time, where it is more sensitive to noise. With this scheme, it is possible to run the level two multiplexer at a mere 50 kHz commutation rate, requiring an access time affordable to many devices.

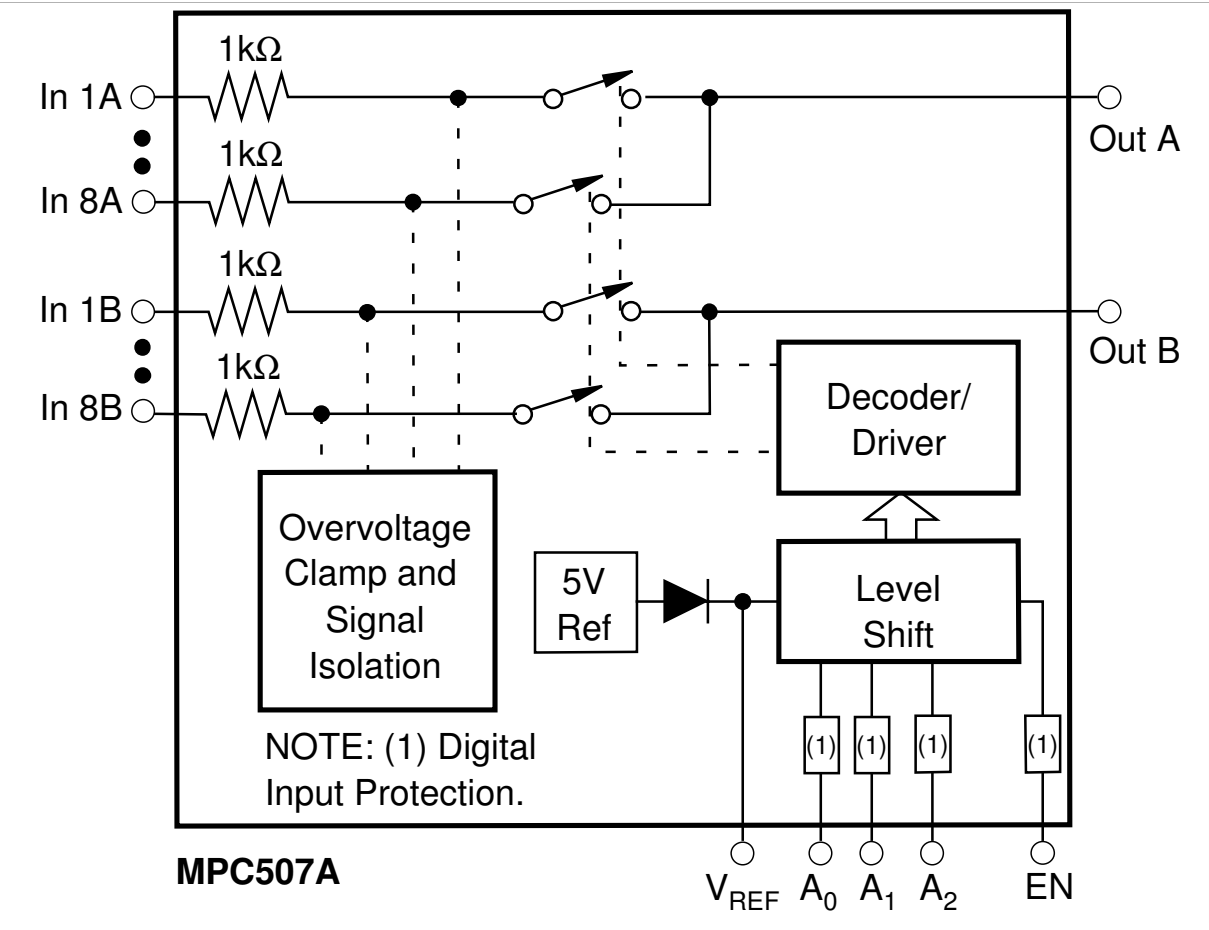

Figure 1.10: Functional diagram of the MPC507A

Having some samples of the Burr-Brown **MPC507A** (ref. [19]), a low power Differential 8-Channel CMOS Analog Multiplexer, already available in the laboratory raised the optimism to use such a device. A quick look at the Functional Diagram (see Fig. 1.10) shows that 8 of these devices (also available in a convenient SOIC package) would be enough for a full system, giving the advantage of a suitable access time and very low dissipation. The reality came out very soon considering the specifications about crosstalk and settling time, also confirmed from a quick circuit setup and relative measures, showing that it is impossible to use the MPC507A like a proper switch with a signal frequency on the analog channel higher than 1 kHz. To give an idea, feeding a 2 MHz square wave (simulating the output of the JAMEX ASIC) to a channel in disable state was producing at the output the same waveform attenuated of just a factor of 10. Also the evident distortion for signal frequencies higher than 100 kHz and the absence of an integrated buffer to drive the external transmission cable made it less attractive.

## 1.5 The OPA682 buffer amplifier and the MIDA design

The quest for the right component to implement the level two multiplexer reached an end with the Texas Instruments **OPA682** *Wideband, Fixed Gain BUFFER AMPLIFIER With Disable* (ref. [20]). The OPA682 provides an easy to use, broadband fixed gain buffer amplifier block. Depending on the external connections, the internal -laser trimmed-resistor network may be used to realize a voltage buffer with either a fixed gain of +2 or a gain of +1 or -1. Operating on a very low 6 mA supply current, this device offers a slew rate and output power normally associated with a much higher supply current. With a dual ±5V supply, the OPA682 can deliver an 8 $V_{pp}$ output swing with over 100 mA drive current and 200 MHz bandwidth. It is worth also to note the fast *settling time* of 12 ns (to 0.02% of the value), the low *harmonic distortion* and the very low *Input Voltage Noise* of 2.2 nV/$\sqrt{Hz}$.

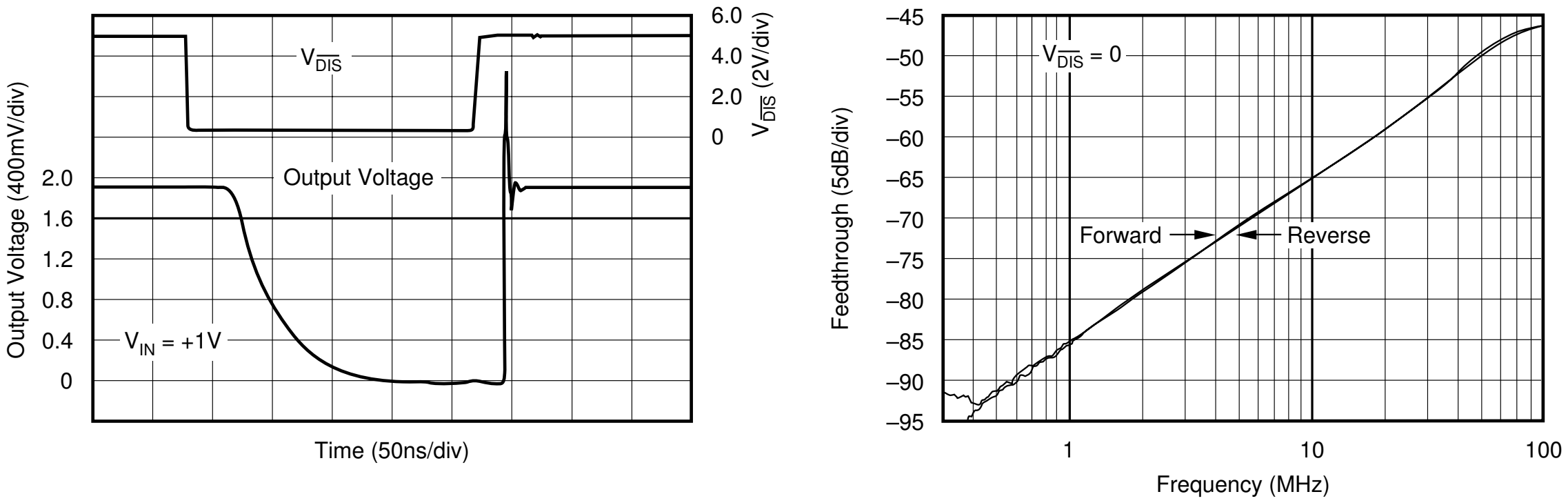

Figure 1.11: Large signal response to the $\overline{DIS}$ signal -left- and disabled feedthrough versus frequency -right- for the OPA682

This combination of features, together with the presence of a disable control pin, which can put the output into a high impedance state thus reducing drastically the feedthrough (see Figure 1.11), makes of the OPA682 an ideal choice to implement our integrated analog multiplexer and line driver. At this point a small short-term project was started called **MIDA** (*Mida Is Multiplexing Differential Amplifier*) having the aim to use a PCB CAD program (*Protel XP*) to design a suitable 2-Layers Printed Circuit Board implementing the required analog multiplexer with the OPA682 devices (made available through the Texas Instruments University Freesample Program). In short time from the simple schematic capture (see Figure 1.12 in the next page) of the required analog multiplexer circuit, implemented like a dual 5:1 MUX, we switched to the PCB design.

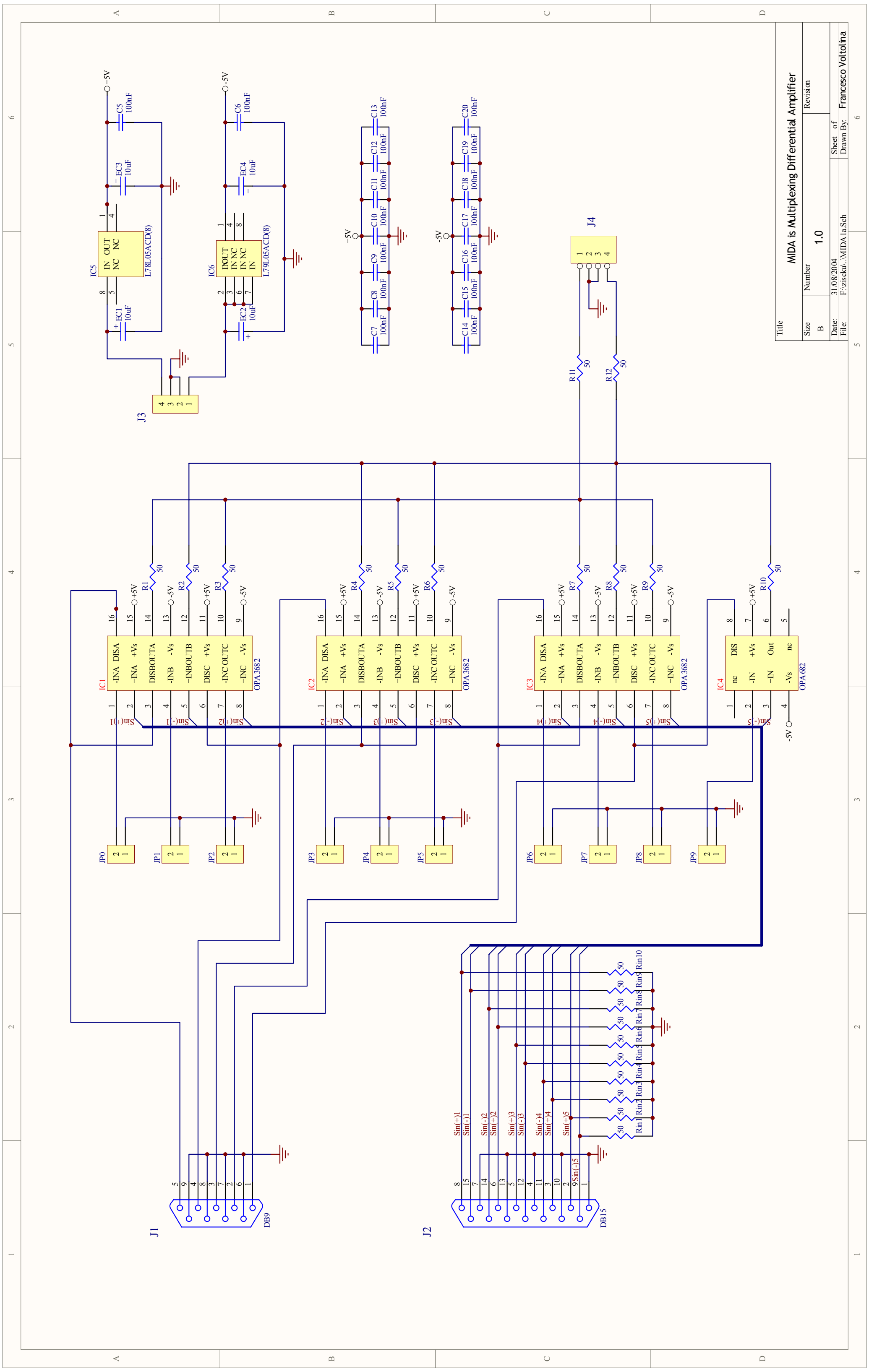

Figure 1.12: The essential schematic of the MIDA multiplexer

It involved the solution of many small problems (such as creating the footprints for the new components or converting the output files to the correct format) in addition to the main task of correctly and conveniently positioning all the components. This circuit uses two different versions of the OPA682: nine channels are implemented using the triple version OPA3682 and the remaining one uses the original OPA682, in total we obtain the required 10 channels that are used like 5 differential channels. When a satisfactory PCB layout was reached (see Fig. 1.13) we used a conversion program (*CircuitCAM*) in order to create a JOB File for the PCB milling machine (*LPKF ProtoMat 92s*) used in the laboratory to produce the prototypes.

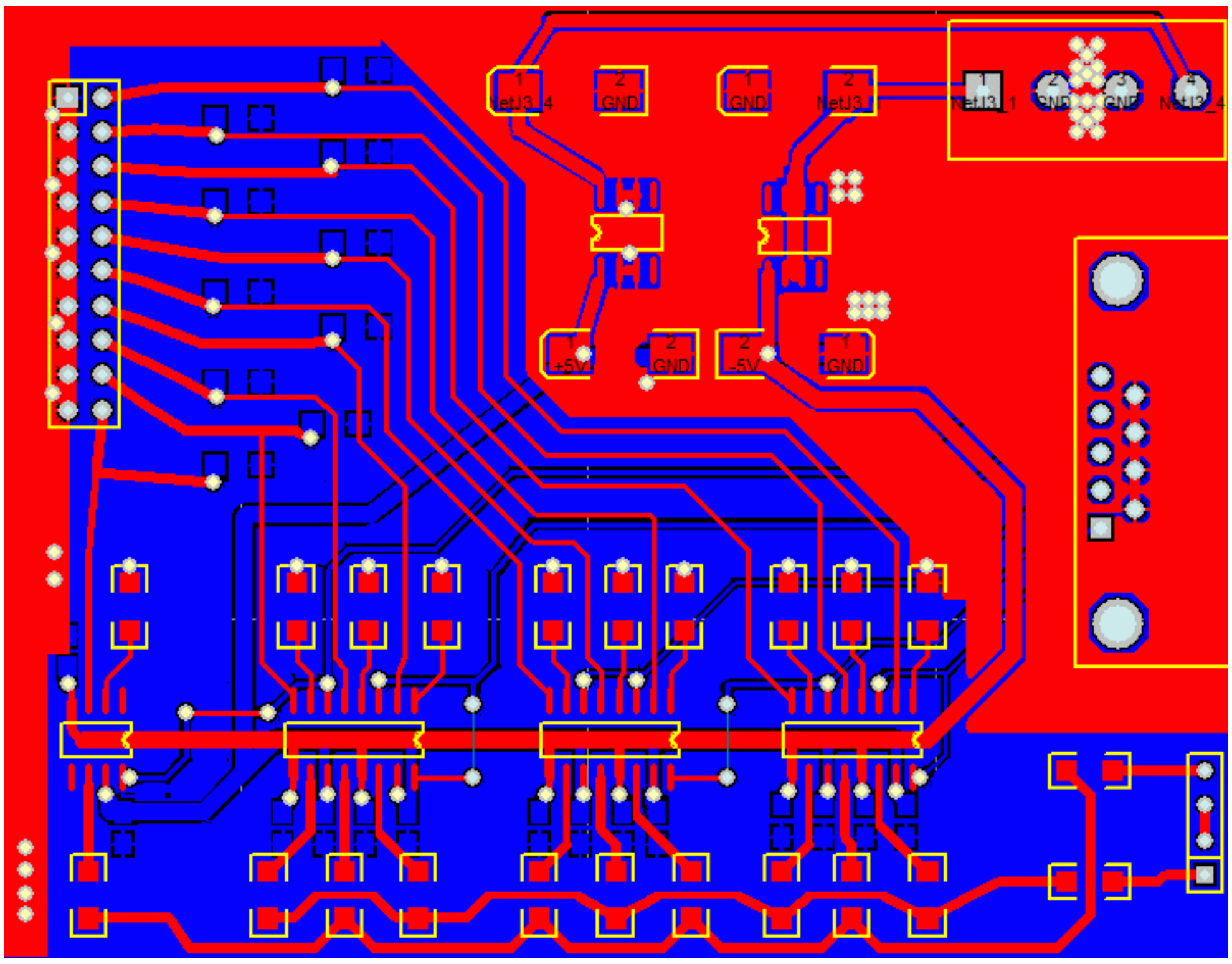

Figure 1.13: Picture of the MIDA double layer PCB in ProtelXP

After the PCB was ready the components were mounted and in short a preliminary test of the basic functionality of the circuit was made. For the successive set of measurements few new instruments were added to that already used in the previous tests with multiplexers: one Phillips Scientific NIM model 744 implementing a *reference buffer and differential driver*, one Phillips Scientific NIM model 771 implementing *four fixed Gain of 10 Amplifiers* and a LeCroy 9210 *Programmable Pulse Generator* with both 9211 and 9212 modules, giving 250 MHZ and 300 MHz (300 ps) Variable Edge Outputs. These instruments were used respectively to split a single ended signal in multiple differential ones feeding the multiplexer inputs, to amplify the output in order to see easily the noise on the oscilloscope and to check if the steepness of the input analog signals has some influence on the feedthrough, despite the constant maximum value of the waveform.

Most of the measurements were taken after appropriate shielding of the circuit in a metallic box (see Fig. 1.14) originally used for a UHF preamplifier stage in a standard TV equipment, since the initial measurements showed widely fluctuating values.

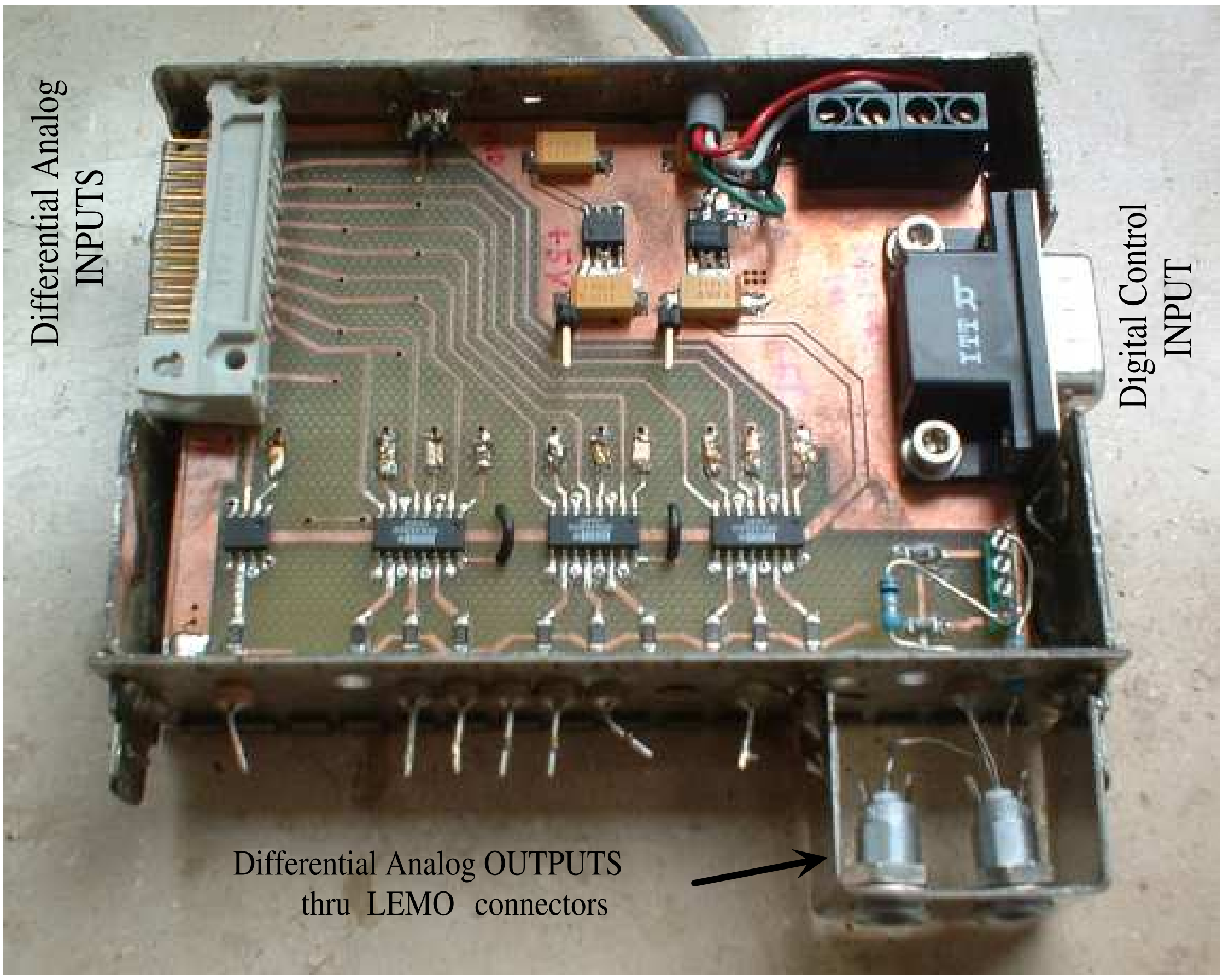

Figure 1.14: Photo of the MIDA multiplexer inside its shielding box

In the following we summarize the results with few comments for each of the main important parameters:

- **Wideband Noise** (0 to 20 MHz) at the output

  $\approx 30\ \mu V_{rms}$ with all channels disabled and constant input voltage

  $\approx 45\ \mu V_{rms}$ with only one channel enabled (steady state) and constant input voltage

  $\leq 180\ \mu V_{rms}$ with all channels enabled, a square wave (2 MHz) at all inputs and 50 kHz multiplexer rate

- **Crosstalk** always better than 80dB if the Leading and Trailing Edges of the (simulated) JAMEX output are longer than 60 ns: this constraint should be easily matched, according to the expected JAMEX output settling time of $\approx$ 100 ns
- **Offset voltage** at the output always $\leq$ 1 mV and constant characteristic of each channel (chip), hence easy to subtract with postprocessing
- **Settling Time** will never be a problem, because of the low signal frequencies involved in our design (the JAMEX output settling time is 100 ns), compared with the specifications of the OPA682 (12 ns), even when it is driving a long cable
- **Access Time** that we require lower than 4 $\mu$s and we measured separately

  **Enable Time** $\approx$ 30 ns ,very short

  **Disable Time** $\leq$ 500 ns ,comparatively long but still good

The *noise performances* we obtained after careful screening of the MIDA multiplexer are consistent with the theoretical value, calculated for a bandwidth of 20 MHz, for the *total output spot noise voltage* (obtained as in [21]) which is $\approx$ 38 $\mu$V. In other words a design close to the MIDA multiplexer (using the OPA682) is a suitable way to build the level two multiplexer for the SAXS detector, provided it is possible to use a 2 MHz clock for the JAMEX output multiplexer. Also very important would be to have a *Differential to Single Ended amplifier* with high *Common Mode Rejection Ratio* (in the order of 80 dB at 2 MHz, possible with a circuit using the AD8130 from Analog Device [22], for example) in the Custom Designed Box, in the Control Room, which is driving the DAQ system. This is needed in order to cancel all the pick-up noise collected from the long cable used for the analog signals transmission. Under these assumptions, considering that one LSB of our 14 bits acquisition system is equivalent to $\approx$ 122 $\mu$V and that the Effective Number Of Bits (ENOB) of the Digital to Analog converter is lower ($\approx$ 12.6 bits, we use the AD9240), we can expect to successfully match this resolution target after all the manipulations performed in the long signal chain of this SAXS detector design with two stages of multiplexing.

However, at this point of the design some new information, derived from the ongoing tests of the new JAMEX64I4 v2 ASIC, totally changed the situation: being considered not advisable to run the JAMEX multiplexer at the relatively high 2 MHz clock needed for this solution, a completely new idea started to take place.

This new concept will be disclosed in the following chapters since it is believed to represent the final design of this SAXS detector.

# 2

# Definition of a basic model for the SAXS detector

In the previous chapter we analyzed the negative aspects of the *Dual-Multiplexing* configuration for the SAXS detector and we ended with the not feasibility of that solution. The delays encountered in the production of the final version of the JAMEX64I4 v2 offered at this point an interesting open design option: to implement its output stage like a *single 64:1* multiplexer with single output buffer instead of *two 32:1 output multiplexers*.

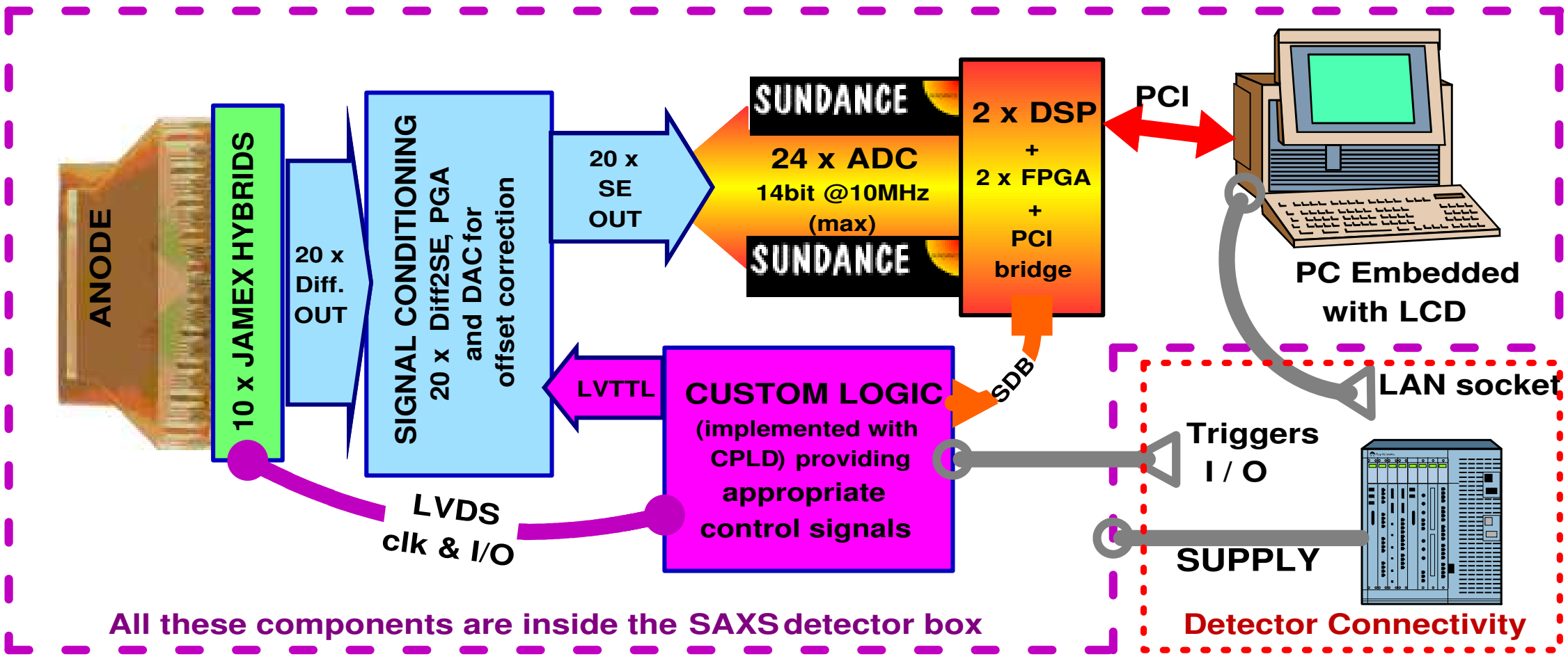

Figure 2.1: Schematic view of the new full-integrated SAXS detector design

In this way, the number of *secondary analog channels* conveying to the ADCs in the final SAXS system would be **20**: half of the number required in the alternative solution and slightly more than twice the number required for the Dual-Multiplexing configuration, making it feasible to increase the number of ADC in the system thus avoiding all the problematics related to an additional analog multiplexing stage. Other benefits of this choice would be an expected lower noise at the output of the device, consequent to the lower current drain of a JAMEX ASIC configured with a single output buffer per chip (instead of two), and the possibility of using an uniform multiplexer rate, for all the ASICs in the detector, suiting better the specifications of the Sundance acquisition system. This choice was leaving open the question whether to still implement a *splitted solution*, with differential analog transmission of the 20 analog channels, or to find a better solution involving a deeper integration of the required electronics in a single container.

## 2.1 General Detector Design

The idea to choose a *low-power PC Embedded* as the host system for the Sundance acquisition system, instead of a conventional PC, brought the design to the turning point that enabled the concept of `full-integrated SAXS detector` (see Fig. 2.1). Under standard operative conditions, in addition to the main detector box, just an external POWER Supply (common to all the SAXS detector electronics, except the Gas-Gain section that will use a separated High-Voltage device), a standard 100baseT Ethernet cable and the optional external trigger cables will be required in the Experimental Hutch.

This is a big quality step toward a detector that offers a good *ergonomy* to the users: it can be easily set in operation and positioned when needed, but also removed in short time and easily stored. The integrated PC embedded (see Fig. 2.2) will run either a WEB and an FTP server making the detector to appear like a normal PC to which it is possible to “Login” in order to set the parameter for a new measurement or to download the results.

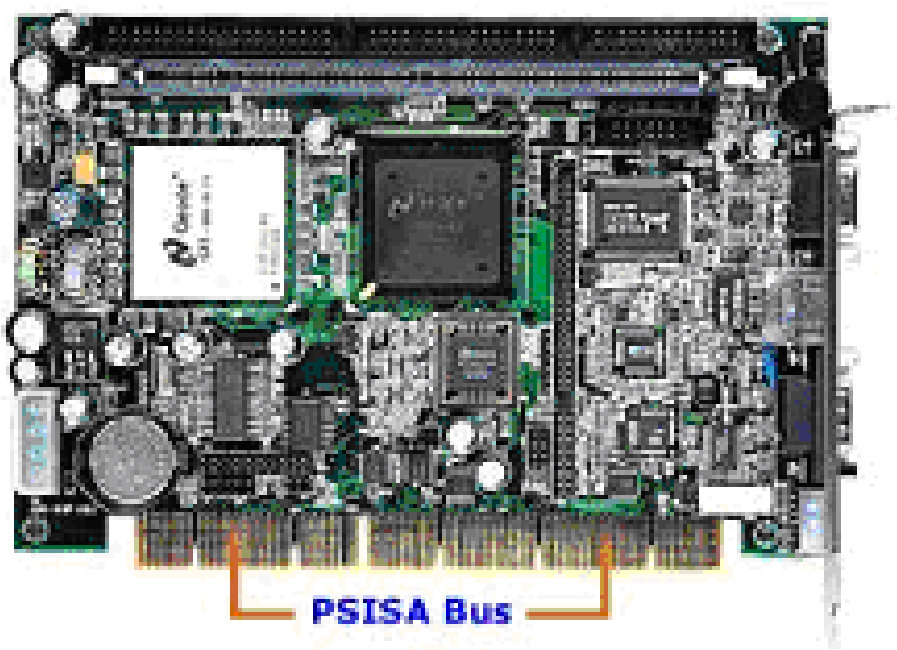

Figure 2.2: PC Embedded photo

It is possible to expect extended capability of reconfiguration, updating and troubleshooting that can be performed via remote login, even from a different country, by authorized users with Administrator privileges. Two PCI cards will be contained inside the detector and carry the standard Sundance modules providing DSP and FPGA signal processing together with 24 analog inputs channels, implemented with single ADCs featuring 14 bit resolution at a sampling rate of 10 MHz. For the proposed configuration it was necessary to modify the JAMEX multiplexer in order to provide 64:1 multiplexing and one output port per chip, like already explained: this was achieved modifying the *metalization mask* for the final production ASICs.

In this way, it is possible to connect the outputs of the JAMEX chips directly to 20 of the analog ADC inputs available while the spare 4 inputs will be used to monitor other important signals inside the detector box and for debugging purposes. Even if this design avoids the use of any additional analog multiplexer between the JAMEX Hybrids and the ADC inputs, nevertheless, since the output signals of the JAMEX64I4 v2 are balanced (differential) and the ADCs require single ended input signals, it is needed to provide such a conversion on an external interface-buffer board. The cabling between the JAMEX Hybrids and the Sundance ADC cards will take place through several connectors routed to this separate board that will also provide the required level-translation to the digital control signals (LVDS to LVTTL and vice versa) and both linear power supply and voltage references for the JAMEX ASICs. Special internal reference signals, required to test the detector electronics in absence of an X-ray source and for calibration, and additional control signals for the ASICs are provided on this board that was designated `J2S Board` (JAMEX to Sundance) and will be presented in detail in the next chapter.

The digital equipment (Embedded PC and Sundance boards) will reside inside the detector but confined in a specific shielding box, in order to prevent radiated EMI to reach critical parts like the JAMEX ASICs and the *signal conditioning electronics* on the J2S card. This configuration allows the shorter available path between JAMEX signal outputs and ADC inputs, offering clear potential improvements to this solution, regarding signal to noise and distortion performances, when compared with a differential analog signal transmission solution using long cables. A diskless solution using FlashDisks will be possibly implemented on the PC Embedded side in order to avoid either the mechanical and electrical noise of an hard disk and to give the necessary ruggedness to the detector.

## 2.2 Functional description of the SAXS detector

In the previous section we gave an overview of the proposed new structure for the SAXS detector, but a more complex characterization is needed in order to develop this idea into a concrete device. From the user point of view this detector appears like a “black box”: it can interpret some commands, transferred generally via the local area network connection, and consequentially performs some actions leading to a result, in the form of a file, which is transferred back to the user. More in detail (see Fig. 2.3) we can identify a chain of *three main blocks* linking the user to the JAMEX ASICs (that are directly responsible for the physical readout of the 1280 microstrips):

- **PC Embedded** hosting the Sundance system and implementing the USER interface
- **Sundance Acquisition System** responsible for the Real Time acquisition and processing
- **J2S Board** offering programmable analog signal conditioning and digital signal routing

These three blocks interact with each other by mean of exchanging specific commands via dedicated interfaces: the PC is communicating to the Sundance system via the PCI bus (more information about this communication will be given in Chapter 3 that also include a description of the specific commands developed for the DSP to FPGA communication) while the Sundance System controls the J2S Board via an interface and a set of command that will be defined in a successive section and we indicated in the diagram like `ZCQSI` (Zero Clock Quiet Serial Interface, indicating the aim to specify a low noise interface method using a reduced number of wires). We must mention the command and data path connecting the J2S board to the 10 JAMEX Hybrids: the characteristics of this interfacing will be explained in Chapter 4 that is totally dedicated to the J2S Board.

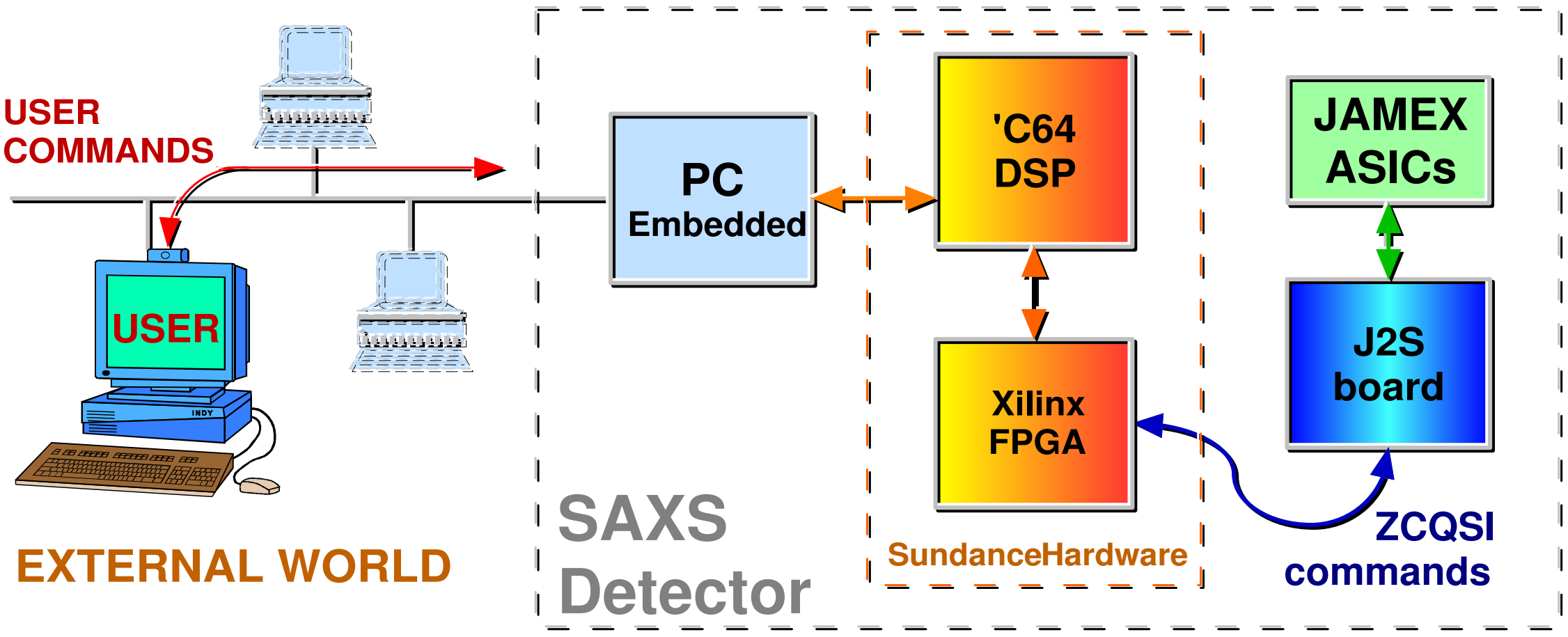

Figure 2.3: Principal command exchange paths for the SAXS detector

At this early stage of the project, the interest is concentrated on the exchange path between a generic USER and the PC Embedded: it is an urgent need to define a consistent set of `USER COMMANDS` as a base for the configuration and development (involving both software and hardware design) of all the previously mentioned main blocks. In order to familiarize with some technical terms used in the definition of this command set it is necessary to give some more information on how this SAXS detector works:

The result of the *measurement* is an `Image` (a bidimensional matrix stored in a file) for which we chose a maximum size of 1280 by 2048 words. The DSP used in the Sundance System is a 32-bit device, hence this word size was considered convenient to represent the image pixels. The number of microstrips (1280) of the detector fixes the *column size*, while the *row size* is related to the maximum number of time slots allowed in a single measurement and its index is called `TIME FRAME` number. The maximum number of TIME FRAMES (2048) is a compromise between the required *temporal resolution* in a measurement and the limited amount of RAM in which the DSP will have to store this image: $1280 \times 2048 \times 4\,(32-bit) \cong$ 10 Mbyte that can easily fit in the 32 Mbyte Memory Bank of the DSP and will not be a limit for any practical measurement task. This is possible since we decided to make variable the number of `JAMEX Cycles` (having a fixed temporal length of 128 $\mu$s) per TIME FRAME and to store the relative values in another structure (`TFCycles`) that associates univocally a 32-bit integer to any possible TIME FRAME. Hence the longest measurement will have a duration limited to $2048 \times 2^{32} \times 128\,\mu s$ $\cong$ **2141 years** (cited here just as a reference) while a measurement requiring the *highest temporal resolution* (limited by the JAMEX ASIC minimum integration time of 128$\mu$s) in *all* the TIME FRAMES must be performed on a phenomenon that is shorter than $2048 \times 128\,\mu s$ $\cong$ **262 ms**, showing that all the available choices between these extremes can suit virtually any SAXS experiment.

### 2.2.1 Real Time Acquisition and Processing

The Real Time data acquisition for a single measurement consists in getting 2048 TIME FRAMES, each consisting of *n* JAMEX CYCLES, making concurrent image processing activities, while providing intermediate results to monitor the progression, and building and storing the final image after the end of the measurement. The *main tasks* that has to be implemented in the Sundance System are:

- **JAMEX control** (generation of the main clock signals for the ASICs)
- **Image processing** (software normalization, correction and integration of the acquired data)
- **External devices control** (synchronizing all the activity with the EXTERNAL TRIGGER signals)

All the above main tasks involve specific design regarding the HOST software running on the PC Embedded and both the DSP software and the FPGA configuration of the Sundance System. A comprehensive description of these tasks will be developed in the following chapters, but we want to anticipate here some details of the required Image Processing in order to allow an easier understanding of the terms used in the following section.

The *Image Processing task* consists in building, from the Real Time acquisition of the JAMEX output signals, the matrix `Image[i,j]` (consisting of 1280 $\times$ 2048 words) in the DSP memory.

This process is described from the following equations:

$$Image[i,j] = \sum_{k=1}^{TFC(i)} \{[JAMEXOUT(k+O(i),j) - Dark(j)] \cdot Norm(j)\} \quad (2.1)$$

where

$$O(i) = \sum_{n=1}^{i} TFC(n-1)$$

| | |
|---|---|
| **JAMEXOUT(p,q)** | is the sampled value at the *p-th* JAMEX Cycle and relative to the *q-th* pixel (microstrip) |
| **TFC(i)** | is the *i-th* element of the vector TFCycles |
| **Dark(j)** | is the value defining the Dark Current subtraction for the *j-th* pixel of the detector |
| **Norm(j)** | is the value defining the Gain Normalization for the *j-th* pixel of the detector |
| **O(i)** | is a term introduced to simplify the notation, it represents the *JAMEX cycle offset* to the FIRST cycle of the *i-th* TIME FRAME |

The formula shows that the *i-th* TIME FRAME (represented as Image[i,*]) is the result computed on the DSP of the "software integration" performed adding on a per-pixel basis all the values sampled in the selected TIME FRAME. Hence, each resulting TIME FRAME is composed of 1280 JAMEX's integrated samples, a value per pixel, on whom the DSP has applied a global dark current subtraction and normalization. These DSP computations will be repeated 2048 times, for each of the *i-th* TIME FRAMEs composing the matrix Image[*,*]. The integration process can be controlled, if required in a specific experiment, with the EXTERNAL TRIGGER input signals that can determine the beginning or the end of a TIME FRAME or the transaction to the successive one; this can be useful to pause the measurement until some physical conditions (temperature or pressure for example) are reached in the sample that we want to measure. Also four EXTERNAL TRIGGER output signals are planned with the scope to synchronize the activity of any eventual externally connected device used for experimental purposes. Now we know enough about this design in order to introduce the next section describing a model for the user interaction with the detector and the relative commands.

## 2.3 Behavioural model and user commands

Having the aim to build a detector easy to familiarize with, was decided to model it in such a way that for a generic external user its operation would be no more complicated than controlling a simple state machine (see Fig. 2.4), which can exists in *three* fundamental states: **IDLE**, **READY** and **RUN**.

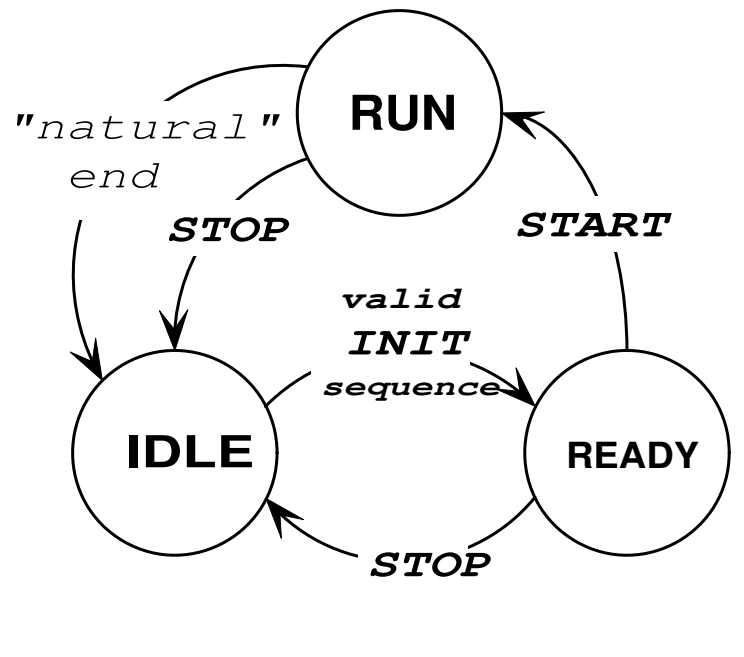

Figure 2.4: State machine

A state machine is an abstract but very effective way to describe the operation of basic up to very complex sequential devices. For our simple case a set of commands was created in order to control the transition between these states and, more importantly, the overall SAXS detector operation. We should point out again that the definition of this command set is not meant to be a rigorous and programmatic definition of the effective commands exchanged between the User and the PC Embedded, therefore it is the chance to create a common ground for all the people working on the hardware and software of this detector in order to simplify the interaction, avoiding possible misunderstanding and errors, and to achieve easily a final integration of the system. On the other hand, we will see in the following chapters that we managed to achieve the goal to have real commands not so far away from this earlier definition.

Before to give their complete synopsis and for the sake of simplicity, we will now introduce these commands showing a *typical sequence*, used to perform a complete measurement, with brief explanations aside and clearly indicating the state transitions:

- **IDLE state**: state of the detector after power-up or when a measurement has completed
- INIT PHASE: composed of many sequential sub-commands and leading, upon successful completion, to the READY state:

  `SET_INTTIME` to set the 'time structure of our measurement'

  `SET_TRIG_OUT` defines when to generate 'trigger events'

  `SET_TRIG_IN` defines if to wait for any 'trigger events'

  `SET_JAMEX_CONFIG` to load the 'configuration bitstream' for the JAMEX

  `SET_PGA_GAINS` sets a specific gain for each of the 20 PGA

  `SET_AUTO_OFFSET` to initiate an automatic offset calibration of the 20 ADCs

  `SET_COMP_OFFSET` to specify an individual offset compensation for the 1280 channels

  `SET_COMP_NORM` to specify an individual normalization for each of the 1280 channels

  `SET_REALTIME_TX` to enable an automatic real-time transmission of the measured vectors

  `SET_TESTMODE` to test the detector without X-rays via an integrated signal source
- **READY state**: the detector is waiting and the `START` command can bring it to the RUN state
- **RUN state**: the measurement and relative processing are performed. The following commands are allowed:

  `GET_IMAGE` to see the actual image

  `GET_VECTOR` to see the actual vector

`GET_ACTUAL_LOOP` to see at which point of the measurement we are

`GET_ERR_MESS` to see any eventual error message

`STOP` that acts like BREAK when in RUN state

- **IDLE state**: this state can be reached "naturally" when the programmed measurement task reached completion or via the `STOP` command that acts like an immediate BREAK, useful when, for example, the user discover to have set some wrong parameters and want to repeat the experiment without to wait.

### 2.3.1 User Commands

We will offer a *description of all the USER COMMANDS* created for this detector, listed in order of importance and observing the rule that when a prefix is present, it states the membership to one of the *four category* listed below:

1. **ACTIVITY COMMANDS** including the two commands: START and STOP that are controlling the macroscopic activity of the detector
2. **SET COMMANDS** all the commands of the INIT phase, they involve a transmission of data from the USER to the DETECTOR
3. **GET COMMANDS** all these commands involve a transmission of data from the DETECTOR to the USER
4. **WRITE COMMANDS** they was not present in the previous sequence since they exist for the only purpose of debugging eventual problems of the detector

We will try to complete the description of all the commands with some information about their importance in the framework of the detector design and their impact on the user interface.

#### STOP

This command acts as an instantaneous BREAK to the activity of the detector: no parameters need to be passed upon execution of this command. If no measurement is active (during IDLE or READY state) it can be issued as general RESET, in order to be sure the detector is set to a known state (IDLE), otherwise it terminates the actual JAMEX Cycle, forces all the JAMEXs to an IDLE condition, tells the DSP to save a pointer to the actual TIME FRAME and JAMEX Cycle and brings again the state machine to the IDLE state.

### START

The `START` command is valid only when the detector is in the READY state, otherwise it will not be executed and will set an “error flag”. When it is issued correctly, it brings the SAXS detector state machine to the RUN state.

### SET_INTTIME

Together with this command a vector of 2048 words has to be transferred to the detector. This vector is the structure (`TFCycles`) that we already encountered in the previous section and associates univocally a 32-bit integer to any possible TIME FRAME: each value represents the number of JAMEX Cycles to be integrated per every TIME FRAME in the final Image[*,*] matrix. A *null* value designates the last TIME FRAME.

E.g. [10,100,4,3,0 ...0] will define a measurement lasting for just 4 TIME FRAMES, resulting in an Image[*,*] with *1280 channels* $\times$ *4 TIME FRAMEs = 5120 elements* $\neq 0$

### SET_TRIG_OUT

Along with this command, a vector of 2048 words has to be transferred to the detector. The *n-th* element of this vector contains the information to which values has to be set the TRIGGER OUTPUT Lines before the *n-th* TIME FRAME has started and after it is finished, in this way the detector can control up to 4 external devices. The DSP will generate these control signals and the values be passed to the FPGA, which will route them to the J2S board through 4 dedicated lines. A galvanic isolation will protect the detector electronics from any problem involving the external devices.

### SET_TRIG_IN

The vector of 2048 values relative to this command contains as *n-th* element a bit’s pattern that has to be compared, before and after the TIME FRAME is finished, with the effective pattern present on the TRIGGER INPUT Lines. The result of this comparison is a Boolean value that can control, according to some rules, the beginning or the end of the n-th TIME FRAME or the transaction to the successive one. This possibility is very useful to synchronize the measurement to some physical conditions (temperature or pressure for example) evolving in the sample that we are measuring.

### SET_JAMEX_CONFIG

This command was conceived in order to reconfigure all the JAMEX ASICs in a quick way. By the word “reconfiguring” we want to intend that this command can set arbitrary custom values for the 1568 bits composing the configuration bitstream of each JAMEX (see datasheet in Appendix A). This is a very powerful command that is performing an action very close to the hardware level (since the bitstream is directly responsible for the operation of the *correlated double sampler* of the JAMEX and marginally for the GAIN of the pixel groups) and for this reason it has to be used with particular care, probably not in the routine use of the detector: since the optimal setting for most of the bits is fixed we will have them stored permanently inside the detector and only the JAMEX gains should be accessed from a generic user. For this reason, we provided another command (SET_JAMEX_GAIN) that can be safely used to select just the JAMEX pixel gains, while the rest of the bitstream is provided from a certified source like, for example, the firmware of the detector.

#### SET_JAMEX_GAIN

The operand of this command is a vector of $20 \times 32$ bits (20 words) that sets the gain of all the JAMEX. These gains have to be set at the beginning of each measurement and it is allowed to set an independent gain (16 steps) for each group of 8 pixels (channels). The user sends the vector to the PC Embedded and also the DSP reserves a memory location for it. The DSP will pass the gains to the FPGA, which will route them to the J2S board through the ZCQSI interface.

#### SET_PGA_GAINS

Issuing this command with a vector of 20 words will configure the 20 Programmable Gain Amplifiers (PGA) and buffers on the J2S Board. These PGA are used to adapt each JAMEX output signal amplitude to the respective Sundance ADC input. The user sets these PGA gains through the PC Embedded and the DSP reserves a memory location to store them. The DSP will pass the gains to the FPGA, which will route them to the J2S board through the same lines of the ZCQSI interface.

#### SET_AUTO_OFFSET

This command is provided in order to address the *offset dispersion problem* affecting both the JAMEX analog outputs and the Inputs of the Sundance ADC system that is likely to disrupt the correspondence between the analog zero level and the digital zero value after the ADC conversion. A simple algorithm will be executed on the DSP to set the proper values of the 20 DAC outputs responsible for the offset correction voltage of the 20 analog channels. This command, without any additional operand, is sufficient to perform the whole operation of global calibration.

#### SET_SPI_TEST

The operand of this command is a single word (32 bits) that either selects the target device of the successive SET_SPI_WRITE command and the signal pattern of the integrated signal source that can be used to test the detector without X-rays. This single word is sent to the PC Embedded by the user and also the DSP reserves a memory location for it. The DSP will pass the gains to the FPGA, which will route them to the J2S board through the signal lines of the ZCQSI interface.

**SET_SPI_WRITE**

The operand of this command is a single word (32 bits) that will be directly written in the serial register of the, eventually selected, SPI device (one of the 20 Offset Correction DAC that we already mentioned) causing a modification in its state. This single word is sent to the PC Embedded by the user and the DSP directly transfers these digital data to the FPGA, which will route them to the J2S board through the ZCQSI interface. No intermediate storing of these data is performed, so they are discarded as soon as they are written into the destination.

**SET_TESTMODE**

The operand of this command is a single word (32 bits). Only the Less Significative Bit (LSB) of the word has a meaning: *0 = normal mode*, *1 = test mode*. The remaining bits are reserved for future improved functionality. While *Normal Mode* is the standard settings for any real X-rays measurement with the detector, the *Test Mode* will enable the dedicated signal generator on the J2S board. This generator will produce specific input signals used to test the JAMEXs in an environment without X-rays aimed to the detector. The mode will be set using again the same signal lines of the ZCQSI interface.

**SET_COMP_OFFSET**

Along with this command a vector of 1280 words [`Dark(j)`] has to be transferred to the detector. The *n* elements of this vector contains the integer value relative to the dark current compensation for the *j-th* pixel of the detector. A value of 0 means no offset subtraction will be performed from the DSP for the relative channel.

**SET_COMP_NORM**

The operand of this command is a vector of 1280 words [`Norm(j)`] that has to be transferred to the detector. The *n* elements of this vector contains the integer value relative to the normalization factors for the *j-th* pixel of the detector. A value of 0 means a normalization factor of 1 will be used from the DSP for the relative channel (*no normalization*).

**SET_REALTIME_TX**

The operand of this command is a single word (32 bits). Only the Less Significative Bit (LSB) of the word has a meaning: *0 = NO Real Time, 1 = Real Time Enabled*. The remaining bits are reserved for future improved functionality. With Real Time Enabled the SAXS detector will automatically send, in real time during the measurement, each new calculated TIME FRAME to the Host PC Embedded and from there to the USER as a vector of 1280 words. When real time is not enabled it is responsibility of the USER to read the individual TIME FRAMES or the final Image via respectively the GET_VECTOR and GET_IMAGE command.

### GET_VECTOR

This command causes the DSP sending to the Host PC Embedded and from there to the USER the *last* calculated TIME FRAME as a vector of 1280 words.

### GET_IMAGE

This command causes the DSP sending to the Host PC Embedded and from there to the USER the *actual* `Image[*.*]` as a matrix of $1280 \times 2048 \times words\,(32\,bit) \cong 10$ Mbyte.

### GET_ACTUAL_LOOP

This command causes the DSP sending to the Host PC Embedded and from there to the USER a vector of two words that represents the number of the actual processed TIME FRAME and of the exact JAMEX Cycle within the TIME FRAME.

### GET_ERR_MESS

This command causes the DSP sending to the Host PC Embedded and from there to the USER a single word that represents the code of the actual (if any) error message.

### GET_JAMEX_CONFIG

This command was conceived in order to offer a powerful debug tool for any problem involving the JAMEX ASICs configuration process in a quick way. Upon reception of the command the DSP will send to the Host PC Embedded and from there to the USER a matrix containing the values of the 1568 bits composing the actual configuration bitstream of all the 20 JAMEX ASICs.

### WRITE_VECTOR

The operand of this command is a vector of 1280 words (a full TIME FRAME) that has to be transferred to the detector. The DSP will write them in the memory in the place of the *last* calculated TIME FRAME. This command is provided in order to offer a debug tool to solve all the problems related to data transfer, memory corruption and general failures.

**WRITE_IMAGE**

The operand of this command is a matrix of $1280 \times 2048 words$ (the full `Image[*.*]`) that has to be transferred to the detector. The DSP will write it in memory in place of the measured Image matrix. Also this command is a useful debug tool to solve problems related to data transfer, lack of bandwidth, memory corruption and general failures.

## 2.4 ZCQSI interface and protocol

In this final part we want to introduce, in an abstract way, the functionality of another important block of the SAXS detector: the *J2S Board digital logic*, with a special attention to the definition of the ZCQSI interface.

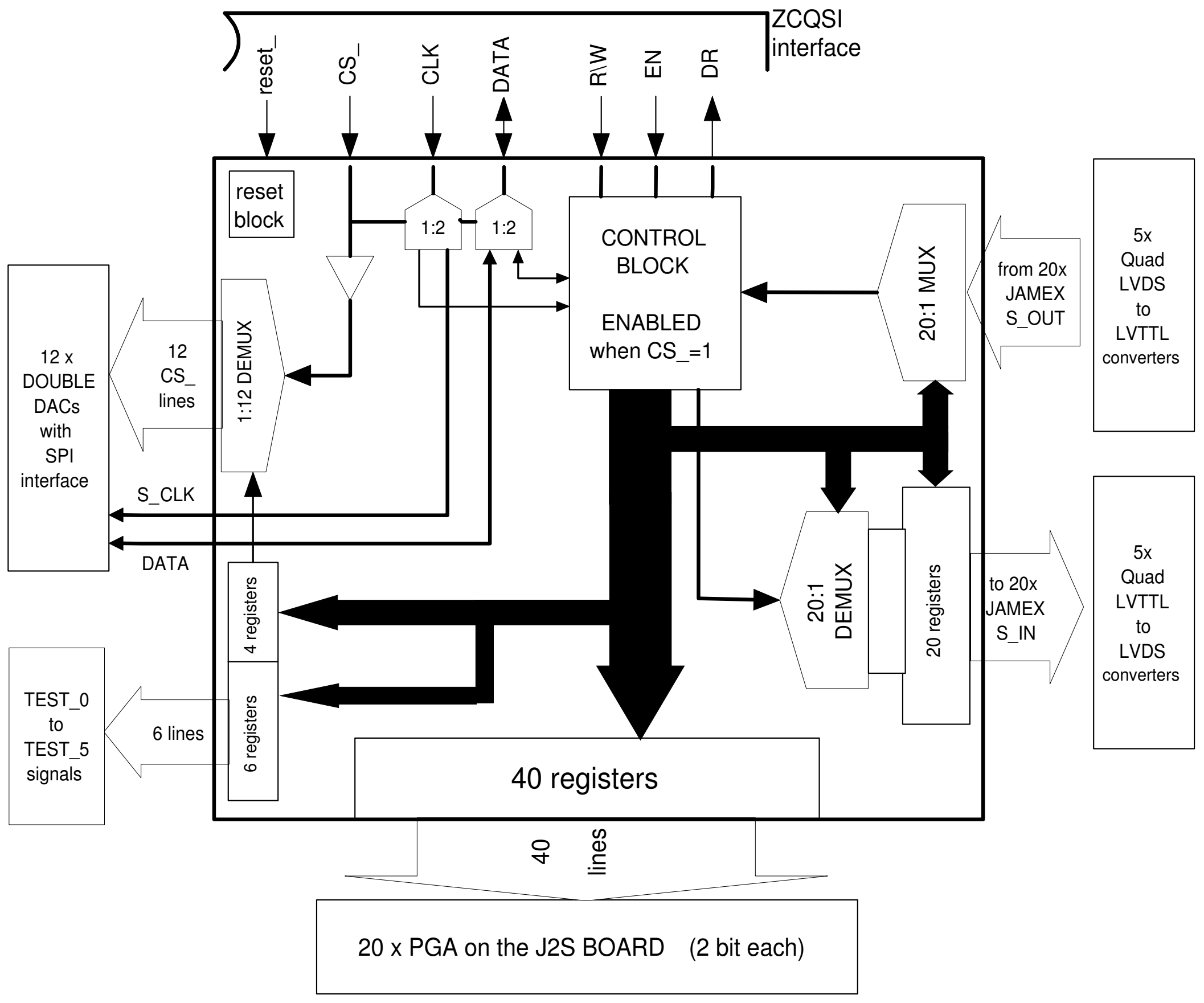

Figure 2.5: Proposed schematic view of the CPLD blocks and functions implementing the *ZCQSI interface* and all the accessory functions of the J2S Board

While its physical implementation will be presented in a later chapter, we will give here the specifications that we imposed at an early stage of the design and permitted to start a parallel work on the Sundance Acquisition System (specifically on the Xilinx FPGA Custom configuration, which was needed to implement the communication with the J2S Board through the ZCQSI interface) well before the availability of the real board. It was decided, in order to achieve a neat and easy reconfigurable design for the J2S Board digital part, to embed most of the digital logic control circuitry into a *Complex Programmable Logic Device* (*CPLD*).

Bonus of this choice is the availability of specific developer tools that accepts a wide range of possibilities to describe a design, to be implemented in a CPLD product, like *VHDL, Verilog-HDL, ABEL, State Machine and Schematic Entry*. More than this, usually some kind of Timing and Functional Simulation is already embedded in the software tools offered from many vendors: this allows understanding if a specific design is viable even before to have decided which exact CPLD device to use. In particular, when a *Post Layout simulation* feature is available, it is possible to simulate a specific design on different devices and than to decide the use and hence to buy the one that was suiting better the specific case.

Since the laboratory was already having a good experience with the Lattice Semiconductor ispLSI Family of high density and high performance CPLDs, it was decided to choose between the devices available from this vendor and consequently to use the vendor's integrated tool *ispLEVER 3.0 Starter software* in order to implement our design. Best aspect of using this family of devices is the opportunity of In-System Programming via IEEE 1149.1 (JTAG) called by Lattice the *ispJTAG Test Access Port*. A standard connector will be present on the J2S Board allowing reconfiguring of the CPLD even after the board will be ready made giving the ability to program the device again and again to easily incorporate any design modifications or upgrade.

| SIGNAL | Direction | Description |
|---|---|---|
| **1 - DATA** | *input/output* | bidirectional data signal pin |
| **2 - CLK** | *input* | asynchronous clock generated from Sundance |
| **3 - EN** | *input* | *target select*: **0**= to 40 PGA registers or 10 CONTROL registers **1**= to 20 JAMEX S_IN/OUT registers |
| **4 - R/W** | *input* | *EN=1*: **0**=Read JAMEX **1**=Write JAMEX *EN=0*: **0**=Write CONTROL **1**=Write GAIN |
| **5 - CS_** | *input* | $\triangleleft active LOW \triangleright$ when this pin is LOW, it selects the SPI interface emulation |
| **6 - DR** | *output* | data ready signal |
| **7 - RESET_** | *input* | $\triangleleft active LOW \triangleright$ |

Table 2.1: Proposed pinout and signal definitions for the ZCQSI interface

After proper configuration (see Fig. 2.5) this CPLD should operate in a very simple way: an extended serial interface counting 7 wires (ZCQSI) connects it to the Sundance Acquisition System, where timings and settings are stored in memory. In this way the user, sending the right commands to the PC Embedded, can serially program the planned 70 registers and few digital MUXes implemented in the CPLD in order to control via the many I/O pins all the S_IN and S_OUT signals for the JAMEX ASICs, the PGA settings, the OFFSET settings for each individual analog channel and the selective amplitude and enable for the TEST signals. The proposed signal definitions and pinout for the ZCQSI (Zero Clock Quiet Serial Interface) interface are presented in Table 2.1.

An important feature of this interface is that it allows a totally asynchronous operation: in order to minimize the digital noise it is possible not to feed any clock to the J2S board during the measurement sessions still allowing switching between the different modes of operation by means of changing the state of some signal lines. Clock is active only during the INIT Phase that is not noise critical, nevertheless, the CPLD must keep fixed all the values programmed into the registers until then: a pure combinatorial design is compulsory for this device. In short, the CPLD is going to act like a smart SERDES (serializer-deserializer) with output latch capability that allows writing and reading of data in an asynchronous way via a serial line, it also performs some digital signals multiplexing in order to use the same ZCQSI serial lines to program many different SPI devices and to READ or WRITE all the 20 JAMEX ASICs.

Setting in the proper way the signal lines of the ZCQSI interface it is possible to make the CPLD of the J2S Board to perform the following commands:

- `Write_PGA_GAIN` (this command requires 40 bits of data to be completed)
- `Write_CONTROL` (this command requires 10 bits of data to be completed)
- `Write_SPI` (this command allows a direct control of the selected SPI device)
- `Write_JAMEX` (this command requires 20 bits of data to be completed)
- `Read_JAMEX` (this command returns 20 bits of data upon completion)

We require that as soon as a new bit is written into a register of the CPLD the corresponding pin changes its value. There are 40 registers in the CPLD whose status directly controls the 20 PGAs: they are written via the `Write_PGA_GAIN` command and it is not possible to read their status. They are called *GAIN Registers* and organized in such a way that groups of 2 bits are responsible for the gain setting of each PGA according to the scheme: ||***00*** *= -4dB*||***10*** *= 2dB*||***01*** *= 8dB*||***11*** *= 14dB*||.
For each of the 20 PGAs the relative bits are stored sequentially in the registers with the LSB first. Proper operation and programming should be guaranteed via careful programming timings for the ZCQSI interface.

There are 10 more registers that are accessed via the `Write_CONTROL` command: the Test signals enable status and SPI selected target device are stored here. This register is organized in this way: 6 bits to individually enable the Test signals and 4 bits to select which SPI Device to program via the successive `Write_SPI` command. We currently need to program 12 different SPI devices: storing a decimal value of **0** in the 4 bits register means we select the first device and we can select up to the last device storing the value **11**. But since we can represent up to 16 different values using 4 bits we stated that a value of 12 to 15 do not select any device allowing a Write Protect state for the SPI devices. These registers are called *CONTROL Registers*, they are "write-only" and the LSB is stored first there too.

The last 20 registers are called JAMEX Registers and directly connected (via LVTTL to LVDS conversion) to the S_IN input signals of the JAMEX ASICs. These are write only registers and they are accessed via the `Write_JAMEX` command.

A `SAFE PROTECT` state for the ZCQSI interface is defined like the status in which the individual signal lines should be set when no command and thus no operation is required: EN_=1 , R/W=0, CS_=0 with the Write Protect state already programmed in the Control Register. This configuration can guarantee a stable state of all the signals during a measurement is performed.

### 2.4.1 ZCQSI Commands

We will present now the already mentioned commands one by one together with an *operational description* of the signals required on the ZCQSI lines in order to perform each of them. At the same time, we will establish a *relation with the User Command* to whom they could relate and the description of each of them will be completed with some *graphical representations* of the correct timings for all the signals involved.

#### `Write_PGA_GAIN`

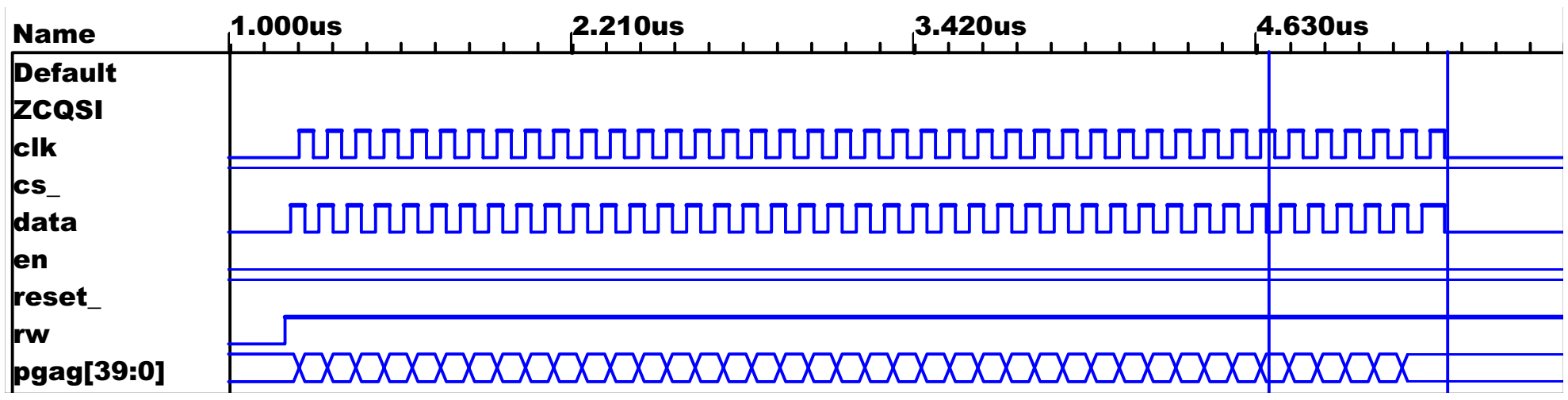

Figure 2.6: Timings relative to the command `Write_PGA_GAIN`

This command is clearly related to the User command SET_PGA_GAINS of the previous section: in the process of passing the gains to the FPGA, which routes them to the J2S board through the ZCQSI interface, the DSP will extract the needed 40 bits from the original vector of 20 words passed by the user. The timing of the signals given in Figure 2.6 can be operatively described in this way: prior to execute this command the **CS(5)** pin has to be set HIGH from the Sundance system and has to preserve this state until the command is executed completely. The first valid data bit should be presented to the **DATA(1)** pin, which the CPLD has configured as input immediately after the Sundance system has set the **EN(3)** to LOW and **R/W(4)** to HIGH. Now the Sundance system can rise HIGH the **CLK(2)** pin of the ZCQSI and, while keeping EN and R/W in the already fixed states, it needs to complete *40 clock periods* (the clock frequency is 10MHz) in order to correctly load the registers with the remaining 39 data bits sequentially inputted at pin DATA(1). As soon as the last bit is written the CPLD should toggle HIGH the **DR(6)** pin in order to notify the Sundance DSP of the correct end of the task. This signal is set LOW upon initiation of any ZCQSI command.

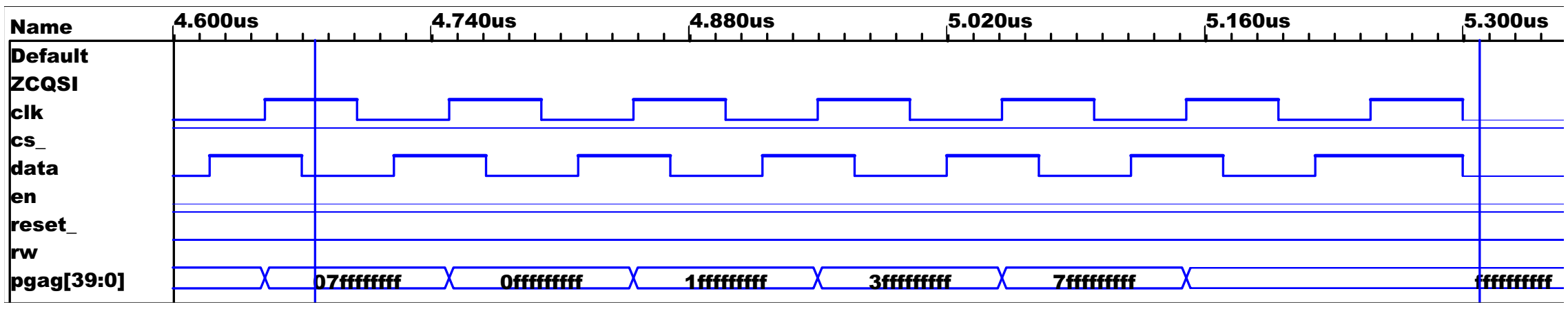

Figure 2.7: Magnified timings relative to the final part of the `Write_PGA_GAIN` command

At this point the CLK(2) pin can be set stable LOW until a new command request will be performed and the other pins can be set to the SAFE PROTECT configuration.

Observing carefully Figure 2.7 we can notice the progressive state change of the 40 PGA lines as soon as a new HIGH bit is presented at the pin DATA(1). This simulation feature all the bits loaded to the DATA(1) as HIGH state resulting in a progressive increment of the registers value.

### `Write_CONTROL`

This command is related to the User command SET_SPI_TEST: in the process of passing the word to the FPGA, which routes it to the J2S board through the ZCQSI interface, the DSP will extract the needed 10 bits from the original word passed by the user. In Figure 2.8 we can see the timings of the involved signals, which we can operatively describe in this way: prior to execute this command the **CS(5)** pin has to be set HIGH from the Sundance system and has to preserve this state until the command is executed completely.

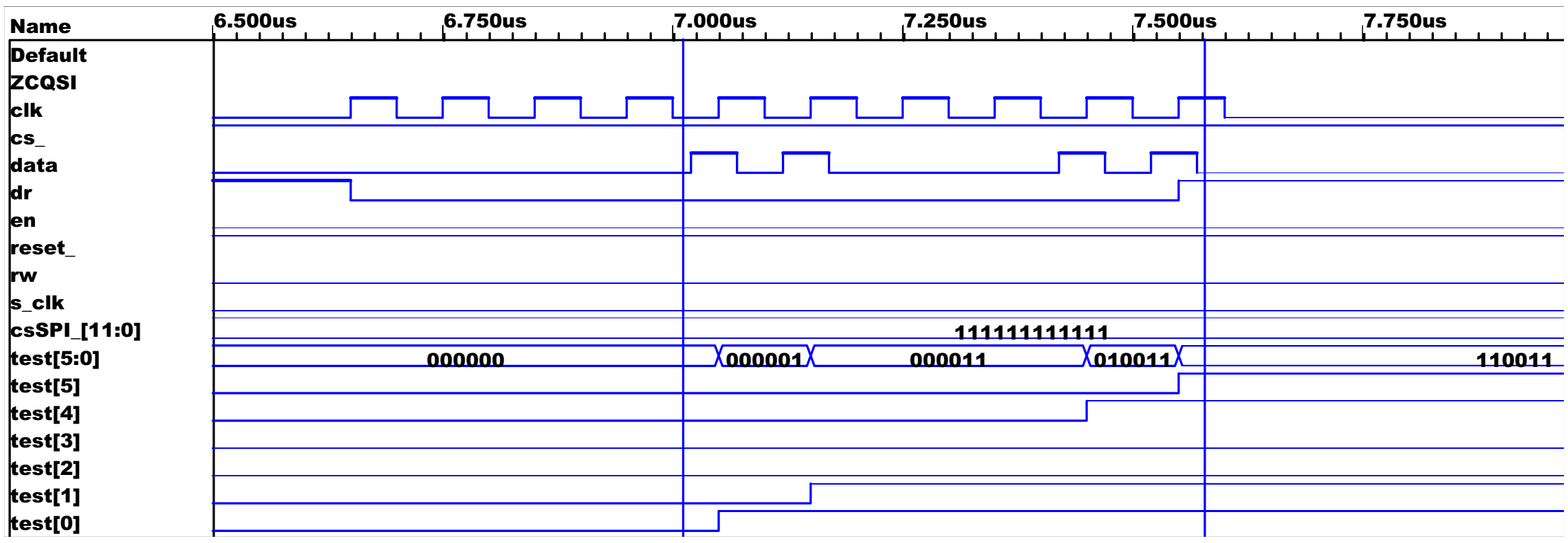

Figure 2.8: Timings relative to the command `Write_CONTROL`

The first valid data bit should be presented to the **DATA(1)** pin, which the CPLD has configured as input immediately after the Sundance system has set both the **EN(3)** and **R/W(4)** to LOW. Now the Sundance system can rise HIGH the **CLK(2)** pin of the ZCQSI and, while keeping EN and R/W in the already fixed states, it needs to complete *10 clock periods* (the clock frequency is 10MHz) in order to correctly load the registers with the remaining 9 data bits sequentially inputted at pin DATA(1). As soon as the last bit is written the CPLD should toggle HIGH the **DR(6)** pin in order to notify the Sundance DSP of the correct end of the task. This signal is set LOW upon initiation of any ZCQSI command. At this point the CLK(2) pin can be set stable LOW until a new command request will be performed and the other pins can be set to the SAFE PROTECT configuration.

In the specific case presented in Figure 2.8 we can notice how the binary pattern **110011** is sequentially loaded in the 6 bits of the CONTROL Register dedicated to the TEST Signals ENABLE. We can also see how the individual 6 test lines are changing their state according to the pattern upon configuration. In this Figure the 4 bits of the SPI Control are loaded with all zeros, enabling eventually the first SPI device conditionally to the state of the CS(5) line that does not manifest any activity thus giving no additional information. A more useful graph is given in Figure 2.9 where it is shown both the CONTROL Register programming and some activity on the SPI lines.

### `Write_SPI`

This command is the hardware level interpretation of the SET_SPI_WRITE user command: it is needed to program the selected `SPI` device (SPI stands for Serial Peripheral Interface and is a low-cost communication option for short distances which is used to communicate with the 12 dual DACs on the J2S Board) using some of the ZCQSI lines as they would be directly connected to the device itself, hence requiring the correct timings for the specific component to be observed. At this stage, we are just interested in having the relevant signals correctly routed with minimum delay when this command is performed.

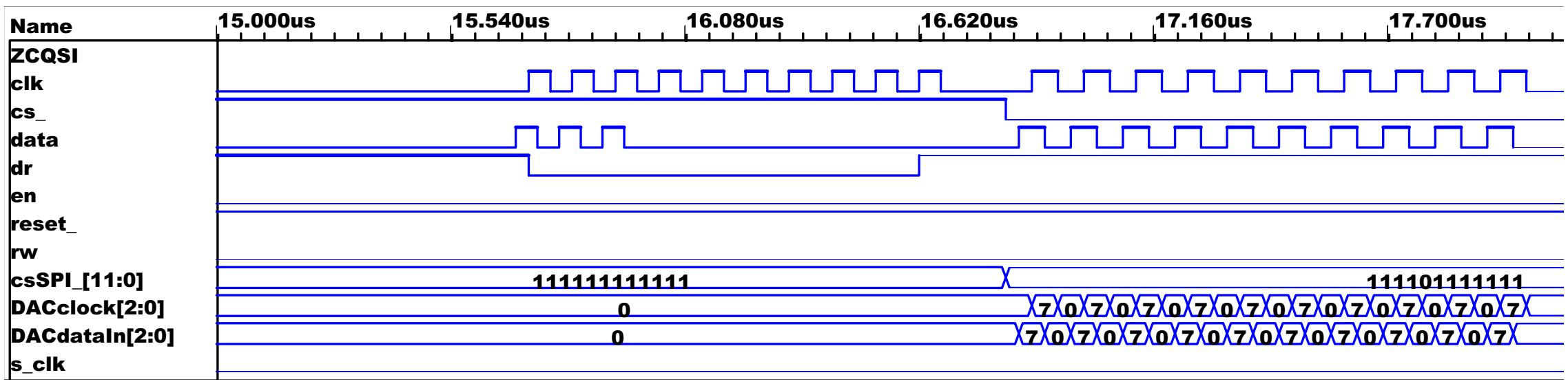

Figure 2.9: Timings relative to the command `Write_SPI`

The involved signals are:
*||**CLK(2)** routed to `DACclock`||**DATA(1)** routed to `DACdataIN`||**CS(5)** to the `CS`||* and their correct routing can be observed in Figure 2.9. According to the SPI protocol CS(5) is active LOW and in our CPLD design, it has priority over the EN(3) and R/W(4) pins that can be any state.

### `Write_JAMEX`

This command is the analogous hardware level command for the user command SET_JAMEX_CONFIG that is used to reconfigure all the JAMEX ASICs in a quick way.

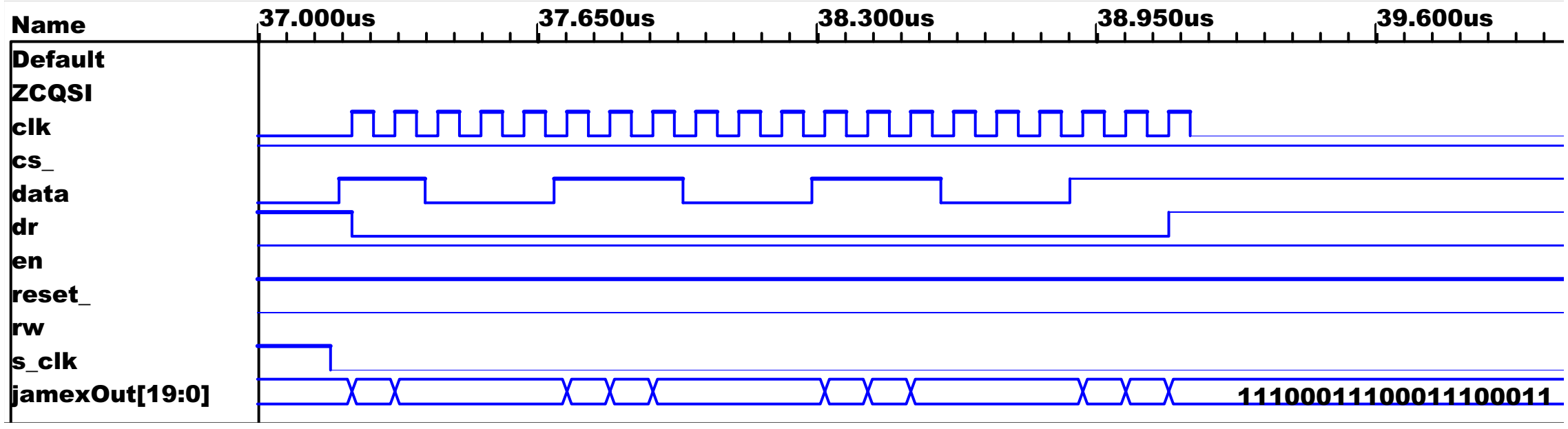

Figure 2.10: Timings relative to one cycle of the `Write_JAMEX` command

The main difference between the User command and this ZCQSI command is that the first one is setting the chosen custom values for the 1568 bits composing the configuration bitstream of each of the 20 JAMEX ASICs at once, while the latter has to be issued sequentially 1568 times in order to perform the equivalent task because it is changing each time just one of the 1568 bits on all the 20 JAMEX.

This command can be performed only if both `S_MODE1` and `S_MODE2` are HIGH (setting the JAMEX ASIC in Programming Mode, see: Appendix B) and the Sundance system has set HIGH **CS(5)** that has to preserve this state until the command is executed completely. The operational description of how to perform this command (to be repeated 1568 times) is like this:

> The first valid data bit should be presented to the **DATA(1)** pin, that the CPLD has configured as input immediately after the Sundance system has set the **EN(3)** to HIGH and **R/W(4)** to HIGH. Now the Sundance system can rise HIGH the **CLK(2)** pin of the ZCQSI and in addition, while keeping EN and R/W in the already fixed states, it needs to complete *20 clock periods* (the clock frequency is 10MHz) in order to correctly load the registers with the remaining 19 data bits sequentially inputted at pin DATA(1). As soon as the last bit is written the CPLD should toggle HIGH the **DR(6)** pin in order to notify the Sundance DSP of the correct end of the task. This signal is set LOW upon initiation of any ZCQSI command. At this point, the CLK(2) pin can be set stable LOW until a new command request will be performed and the other pins can be set to the SAFE PROTECT configuration.

In normal conditions, 1568 of these cycles (see Figure 2.10) has to be performed in order to correctly sequentially programming all the JAMEX ASICs: in fact each of these cycles has to be completed before a successive period of the `S_CLK` (the JAMEX primary clock) signal can be generated. In the same figure, we can check the consistency between the bitstream presented in input through the ZCQSI DATA(1) line and the 20 signals, represented by the vector `JAMEXOUT[19:0]`, that are connected to the 20 `S_IN` lines of the JAMEX ASICs.

### Read_JAMEX

This command is the hardware level equivalent command for the user command GET_JAMEX_CONFIG that is used to read the whole configuration bitstream of all the JAMEX ASICs in a quick way.

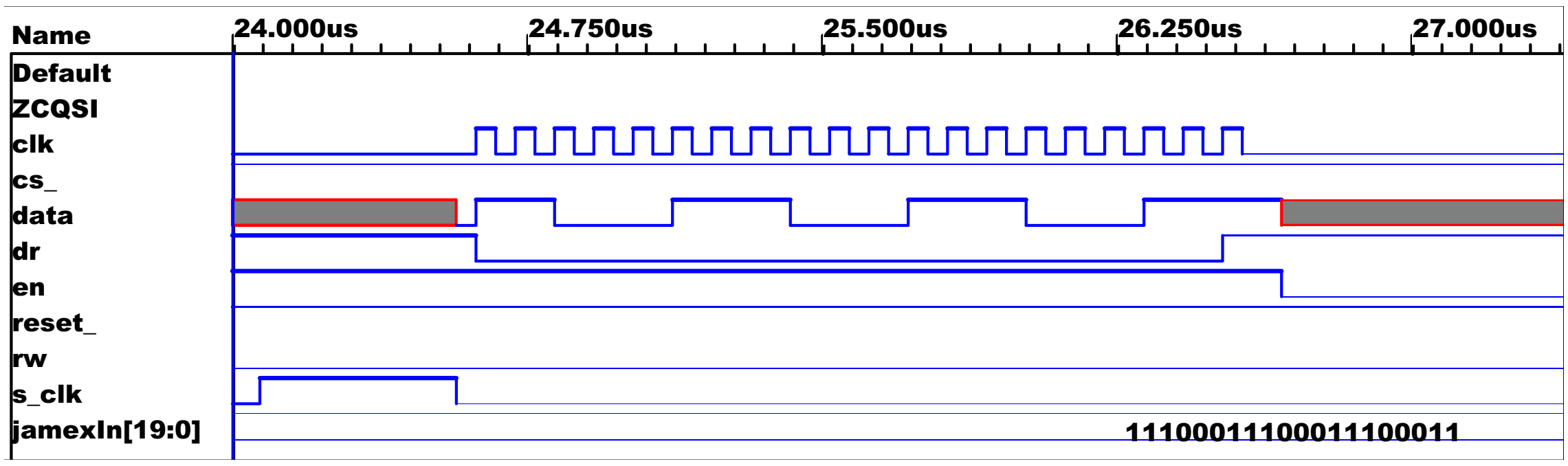

Figure 2.11: Timings relative to one cycle of the `Read_JAMEX` command

Like for the Write_JAMEX command, the main difference between the analogous user command and this ZCQSI command is that the first one is setting the chosen custom values for the 1568 bits of each JAMEX ASICs at once, while the latter has to be issued sequentially 1568 times in order to perform the equivalent task.

This command can be performed only if both `S_MODE1` and `S_MODE2` are HIGH (setting the JAMEX ASIC in Programming Mode, see: Appendix B) and the Sundance system has set HIGH **CS(5)** that has to preserve this state until the command is executed completely. The operational description of how to perform this command (to be repeated 1568 times) is like this:

> The first valid data bit will be presented by the J2S Board on the ZCQSI's **DATA(1)** pin, that the CPLD has configured as output immediately after the Sundance system has set the **EN(3)** to HIGH and **R/W(4)** to LOW. Now the Sundance system can rise HIGH the **CLK(2)** pin of the ZCQSI and, while keeping EN and R/W in the already fixed states, it needs to complete *20 clock periods* (the clock frequency is 10MHz) in order to output sequentially at pin DATA(1) the remaining 19 data bits. As soon as the last bit is presented the CPLD should toggle HIGH the **DR(6)** pin in order to notify the Sundance DSP of the correct end of the task. This signal is set LOW upon initiation of any ZCQSI command. At this point the CLK(2) pin can be set stable LOW until a new command request will be performed and the other pins can be set to the SAFE PROTECT configuration.

In normal conditions, 1568 of these cycles (see Figure 2.11) has to be performed to correctly read the programmed configuration of all the JAMEX ASICs and each of these cycles has to be completed before a successive period of the `S_CLK` (the JAMEX primary clock) signal can be generated. In the same figure we can check the consistency between the 20 signals, represented by the vector `JAMEXIN[19:0]`, that are connected to the 20 `S_OUT` lines of the JAMEX ASICs and the bitstream presented at the output through the ZCQSI DATA(1) line.

#### Interleaved `Write/Read_JAMEX`

This is not a separate command for the ZCQSI interface but the result of the observation that on a practical point of view it is expected to perform only Interleaved Write/Read_JAMEX commands, repeated 1568 times, in order to reprogram all the chips while testing the previous programming. This is required since the JAMEX ASICs implement a `FIFO` (First In First Out) register to store the bitstream, hence the ASIC cannot output its bitstream without requiring a new reconfiguration, thus this modality is the only useful way to perform its programming. Operatively we can describe in this way the sequence required to program all the JAMEX ASICs, while reading their previous bitstream:

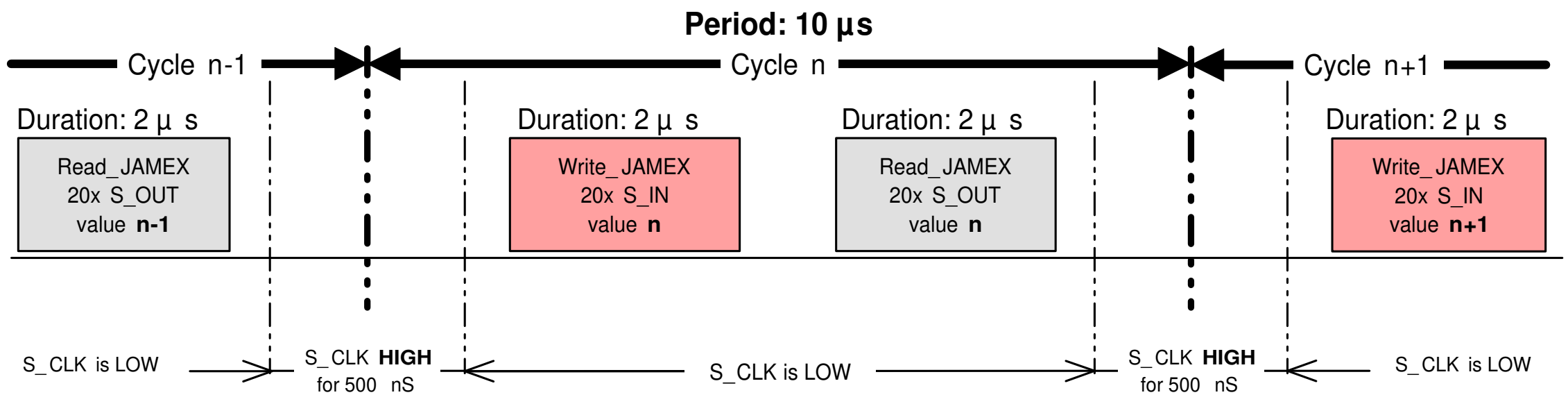

Figure 2.12: Timings for the `Interleaved Write/Read_JAMEX` sequence

This sequence (see Figure 2.12) can be performed only if both `S_MODE1` and `S_MODE2` are HIGH (setting the JAMEX ASIC in Programming Mode, see Appendix B) and the Sundance system has set HIGH **CS(5)**.

- while `S_CLK` is **LOW**, we can use the command `Write_JAMEX` to store in each JAMEX ASIC register the first of the 1568 bits relative to the new configuration
- then we can use the command `Read_JAMEX` to read back the first of the 1568 bits relative to the old bitsream for the 20 JAMEX ASICs
- now `S_CLK` is pulsed **HIGH** for a duration of 500 ns while we are performing no operations and the pulse repetition is 100 KHz (Period T = 10 $\mu$s)

We must repeat this loop 1567 times more in order to complete the programming of all the ASICs.

Each cycle begins with a S_CLK HIGH pulse and before the next pulse, during the S_CLK LOW phase, we must perform the read and write through ZCQSI of one bit from all the 20 JAMEX. The expected total time for the write and verify of all the ASICs will be $1568 \times 10\,\mu s$ = 15.68 ms.

In the following chapter we will describe the Sundance Acquisition system, which is going to be interfaced to the J2S board. This system will implement the communication via the ZCQSI interface and also will be responsible for the generation of all the previously presented timings.

# 3

# Sundance Acquisition System

A key idea in the SAXS detector design is the decision to build the acquisition electronics around a *commercially available system* in order to shorten the development time required for the whole device. The Daresbury Laboratory in England, one of the partner of this project, already had some experience in designing a Detector System that took advantage of commercially available hardware: for that purpose they focused on the flexible digital acquisition (DAQ) and processing cards offered by the company Sundance [23]. Since the range of ready available solutions offered from this company was fitting our requirements and in order to preserve the valuable experience gained at the Daresbury Laboratory, it was decided that Sundance will provide the hardware and software tools for this project. Moreover, considering the not indifferent complexity of the required acquisition system, it was considered a wise choice to involve the same people of Sundance in developing the “JAMEX application” that is the name chosen to designate the skeleton of the software and hardware for the SAXS detector.

This chapter is organized in few main sections: in the first will be introduced the special Sundance hardware components used in the project; in the second will be given an overview of the software development tools required to develop the JAMEX application and in the third will be presented a preliminary version of the JAMEX application.

## 3.1 Sundance Hardware

The Sundance acquisition system for the SAXS detector is built using some standard components belonging to the wide range of Modular Systems cards and Boards available from the company, which are designed to meet the life cycle challenge posed by the actual high-end acquisition and Digital Signal Processing systems. The available components feature a modular and scalable DSP/FPGA multiprocessing architecture consisting of DSP, ADC, DAC and FPGA modules. It is a matter of the designer to select a carrier board among the many standard available (*PCI, compactPCI, VME, VXI or standalone*) and to populate the required carrier board with appropriate modules.

The main strength of this solution, apart from the high reconfigurability (see Fig. 3.1), is in the module to module communication’s speed that can reach 400 Mbytes/s since all data transfers are done through fast proprietary buses.

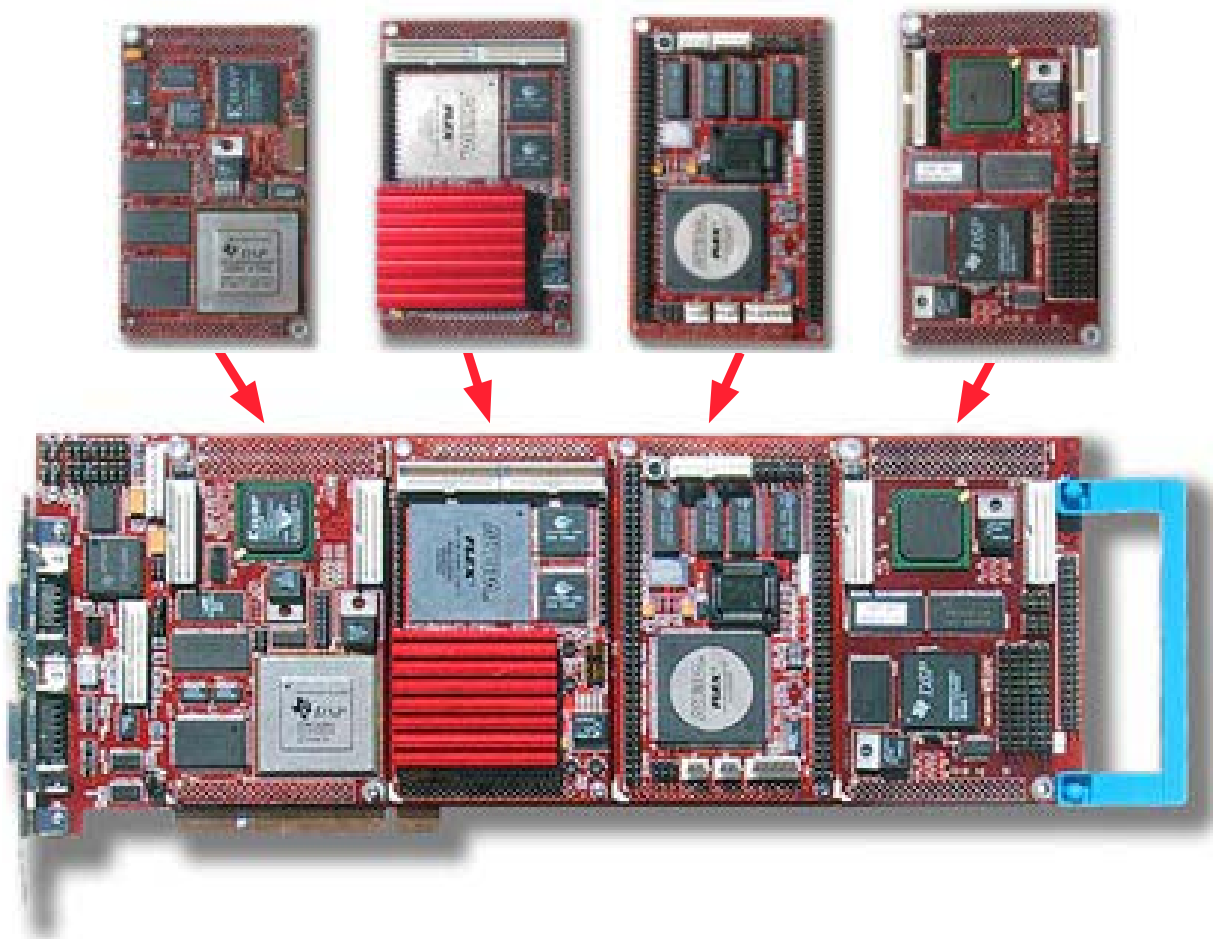

Figure 3.1: Modularity of the Sundance components

This is not a secondary aspect in our detector design since it involves several high bandwidth paths: assuming a run with integration time slots of 128 $\mu s$ with all 1280 channels stored like 32-bit words, it results in an aggregated data flux of **40 Mbyte/s**. We already introduced some of the requirements for the acquisition system: the ADC subsystem will include a single PCI bus board that can carry up to 4 standard Sundance modules (called *TIM*) but in our configuration only 3 modules, providing 8 ADC channels each, will be used for a total of 24 single-ended analog channels. The two additional modules composing the DSP/FPGA subsystem, featuring both a 320C64x DSP chip and one Virtex FPGA each, will reside on a second PCI carrier board.

There are several different ways to *exchange data* between Sundance hardware modules and other peripheral devices: the maximum data rate required by the specific application will usually determine the choice of which to use. Standard TIMs allow data exchange by providing a number of different sets of signals, most of which are normally present on processor (DSP) TIMs and optionally may also be available on other types of TIM like the ADC modules. These signals led to the following data exchange mechanisms:

1. **Communication ports** (Comports): they can connect two TIMs together to provide a bi-directional path for exchanging streams of 32-bit values. This is the basic communication method for all TIMs and is usually used to load programs into processor TIMs. Comports can sustain a data rate of about 20 Mbyte/s.

2. **Sundance Digital Link** (SDL): it is similar to a comport but faster and not compatible with comports, hence it is used in exceptional cases.

3. **Sundance Digital Bus** (SDB): is a high-speed 16-bit communications device and these signals are available through connectors on the TIMs themselves. The DSP accesses this device 32 bits at a time and they are presented at the connector as pairs of 16-bit accesses.

4. **Global Bus**: its signals are presented on the Global Bus connector at the back of the TIM. This connects to the carrier board (only the first TIM slot has a global bus connector) where the signals are passed on to the host interface or used to access the board's resources

We will cover in the successive sections all the system components, with relative characteristics and explanatory figures, of the Sundance hardware used in the detector.

### 3.1.1 The SMT310Q Carrier Board

The SMT310Q (ref. [24]) is a full-length PCI board that can carry and provide access to a maximum of four TIM (*Texas Instruments Module*) format (ref. [25]) processor modules. This board is required since we cannot use a TIM on its own, but we need to hold it physically, provide power and to have some means of connecting it to the outside world. This is done using this carrier board that will also communicate through the PCI bus with the host PC Embedded to which it will be connected.

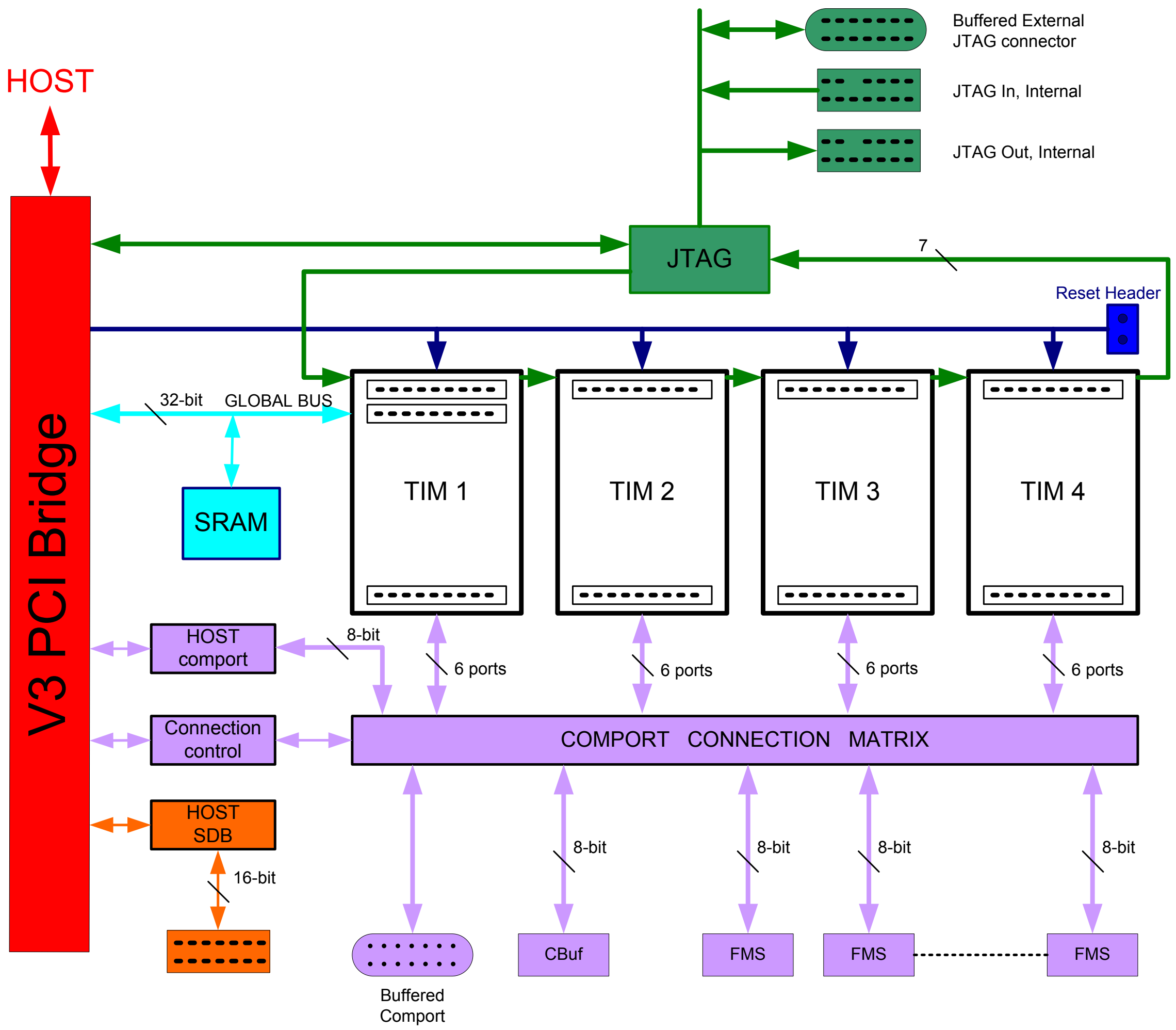

Figure 3.2: Block diagram representing the SMT310Q Carrier Board

The SMT310Q can support up to four TI320C6x DSP modules (see Fig. 3.2) and these can exchange data using either comports and SDBs. It also provides a software-configurable *routing matrix* to allow certain comport connectivity of different modules without the need of external cables. An on-board SDB interface to the PCI bridge allows the card to be interfaced to any SDB standard modules directly. The first TIM on the carrier is known as the *Master Module* and an enhanced PCI interface allows data packets to be exchanged between this module and the PCI bus at burst speeds approaching a maximum of 100 Mbyte/s. The DSP has access to the **V363EPC** (implementing the PCI bridge) internal registers to control DMAs, mailbox events and interrupts. Moreover, a shared SRAM of 1 Mbyte is mapped onto the Global Bus and can be accessed as a global resource by the Host system across the PCI bridge or by the Master Module; an integrated FLASH ROM offers 512 Kbytes of memory that suits any special configuration need.

An on-board *JTAG controller* allows the systems to be debugged using Code Composer Studio. This JTAG controller also has buffered outputs and inputs that can be accessed using connectors on the carrier's back panel: in this way some off-board devices (like the J2S board's CPLD) can be integrated in the same JTAG chain. Headers are provided for the RESET_IN and RESET_OUT signals to allow multiple SMT310Q carriers to be connected together and synchronized.

The SMT310Q takes the 3.3 V and 5 V power from the PC's power supply. The following *current consumption* figures are specified during a quiescent period: a current of 440 mA is drawn from the 3.3 V supply and 100 mA from the 5 V supply. The board size, important for the future integration in the SAXS detector box, is 312 mm $\times$ 120 mm. A sample of the board and several TIM modules is shown in Figure 3.1.

### 3.1.2 The SMT361 DSP/FPGA Processor Module

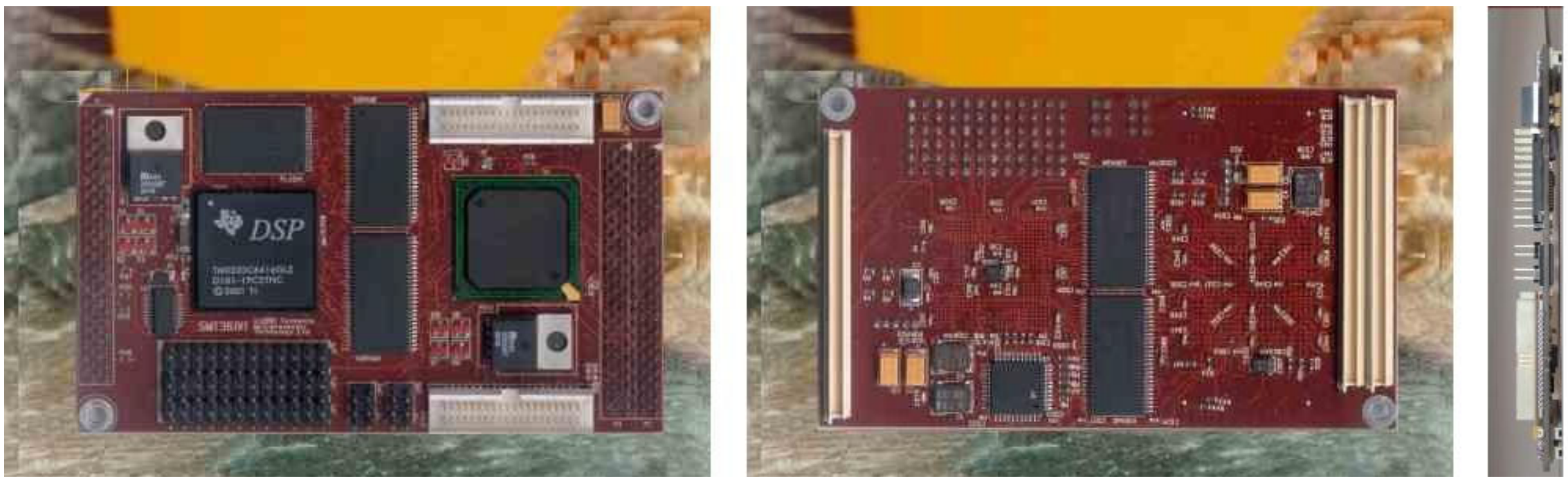

Figure 3.3: Photos of the SMT361 DSP/FPGA Processor Module

The SMT361 module (ref. [26]) is based on the **TMS320C6416** Fixed Point Digital Signal Processors (DSPs) from Texas Instruments, offering a very high processing power (peaking at 4800 Mips) that has been implemented with a scalable solution using Xilinx

Field Programmable Gate Arrays (FPGAs). The TIM modular design was created in the early 90's by Texas Instruments and all Modules have since been backward compatible: the latest module is no exception, but whereas the first series of Modules had a 40 MHz DSP and 1 Mbyte of memory, the SMT361 (see Fig 3.3) has a 600 MHz DSP with 32 Mbyte of synchronous DRAM (SDRAM) memory and, for added performance, a 1 million gate **Xilinx Virtex-II** FPGA.

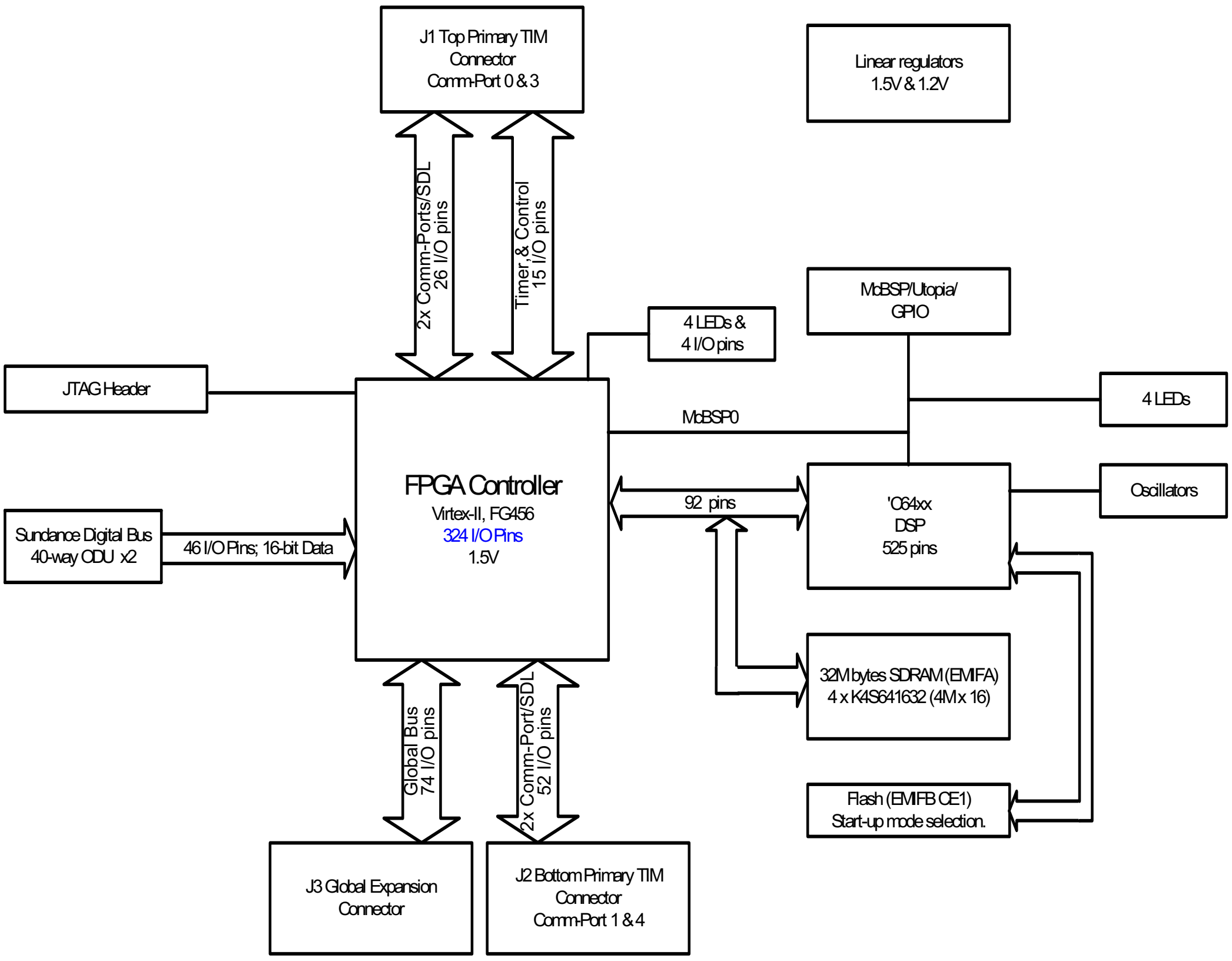

Figure 3.4: Schematic view of the elementary blocks composing the SMT361 Module

The SMT361 is supported by the `Code Composer Studio` environment for single DSP applications and, since the SMT361 was designed to support MultiDSP applications (like the SAXS detector, which is using 2 DSPs), also a ported version of the `3L Diamond RealTimeOS` is available. This software suite with integrated operative system provides effortless parallelization of sequential code and hassle-free control of on-chip resources like DMA channels, Host PC interaction and peripheral interface. The Virtex-II FPGAs is used (see Fig 3.4) to manage global bus accesses and to implement four communication ports and two Sundance Digital Buses. Each one of the two SDB interfaces present on the SMT361 module can provide more than 200 Mbytes/s of data transfer rate from the ADC Modules SMT356 that we are going to use in our detector.

This fast data transfer mechanism can be used for both I/O (e.g. reading data from the ADCs) and InterDSP communication, which can be used for `pipeline processing` of data in applications like Radar, Medical Imaging, Image Processing or similar data processing DSP-intensive applications of which the SAXS detector is again a concrete example.

The Xilinx FPGA is configured from the data held in 2 Mbyte of *Flash memory* and sets up the communication ports, the global bus and the Sundance Digital Buses. This step must have been completed before that data can be sent to the module from any external sources such as the host or other TIMs: this accounts for a short delay, of around 1 second, between the release of the board reset and the successful FPGA configuration. The same Flash memory also contain the boot code for the C64xx DSP and some free space available to the user: in our case most of this space will be used to store the configuration bitstream for the ADC modules and for the JAMEX ASIC standard configuration.

### 3.1.3 The SMT356 ADC Module

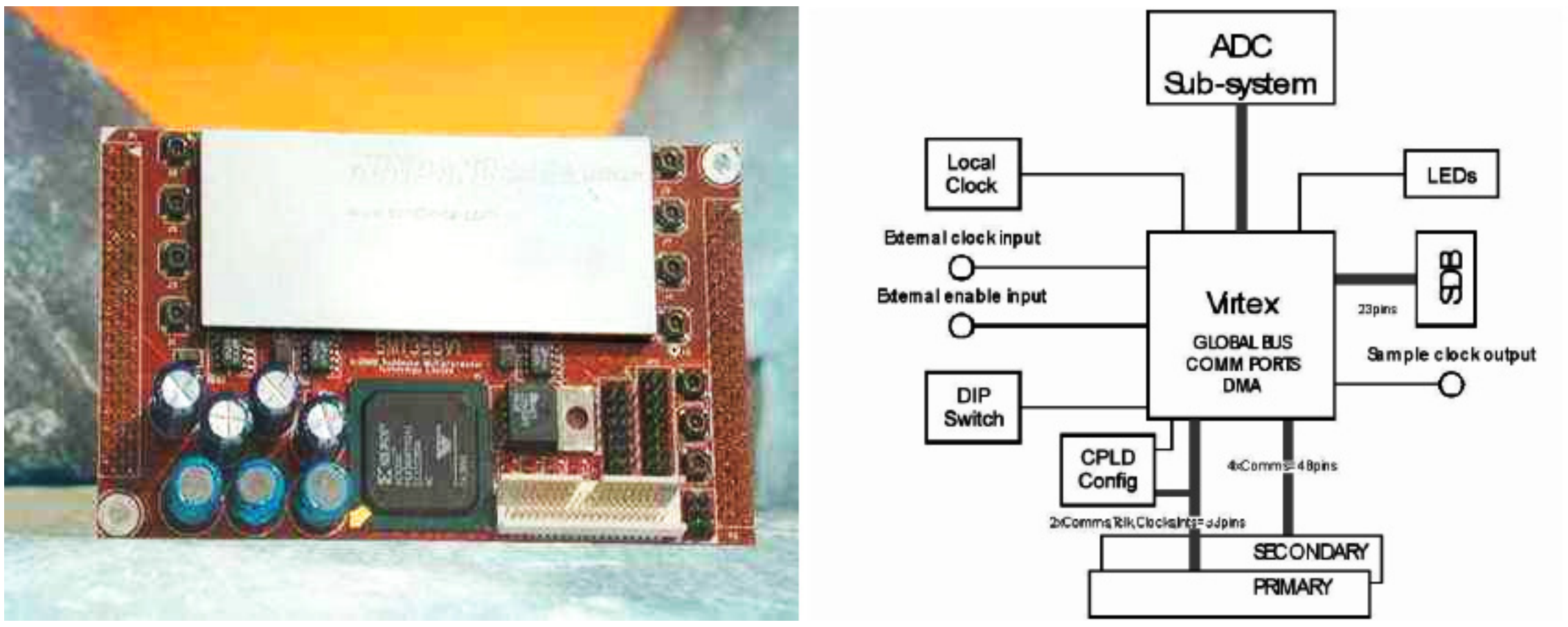

Figure 3.5: Photo and general architecture of the SMT356 ADC Module

The SMT356 is an ADC converters Module (ref. [27]) offering 8 channel of 14-bit simultaneous sampling ADC in the comparatively small footprint (see Figure 3.5) of a standard single-sized TIM module. The main component that controls the operation of the module is a Xilinx **XCV300 Virtex** FPGA that directly interfaces to the eight Analog Devices **AD9240** converters. These CMOS ADCs provide an overall system performance showing a Spurious Free Dynamic Range (*SFDR*) $\geq$ 70 dB for an Effective Number of Bits (*ENOB*) of 12 (minimum). The flexible FPGA-based design allows the implementation of `SoftDSP algorithms` like over-sample, decimate or more complex option in dependence of the user configuration that, since this device is volatile, requires to be reconfigured every time the module is powered on. The configuration data (bitstream) must be presented through a comport by a processor module like the SMT361.

The sampled data is output with high bandwidth via a single 16-bit SDB connector and this interface implementation on the module is as an output only. Once the sample clock is enabled, all ADCs are sampled at the same time at a frequency selected by a Jumper Bank or a software divider. The 14 bit samples occupy the higher 14 bits of the SDB word, while the two lower LSBs are undefined: the sampled data is presented, one channel at a time, to the most significant 14 data bits of the SDB connector. These data will be transmitted on the SDB as a packet with all the sample data buffered within a 256 word FIFO. The maximum sampling rate per ADC is 10 MHz, which indicates that the maximum SDB word rate would be 160 MHz (with the channel number and sample number feature enabled). The SDB specification does not allow this data rate, so a compromise must be taken. Either the extra sample information should be disabled, the number of channels reduced, or the sample rate reduced. For the JAMEX application the additional features are not needed, since we will use an *hardware synchronization* to keep track of the sample number. Hence, the SDB word rate will be always lower than 80 MHz with all channel enabled: this figure is well beyond the SDB specification at 100 MHz.

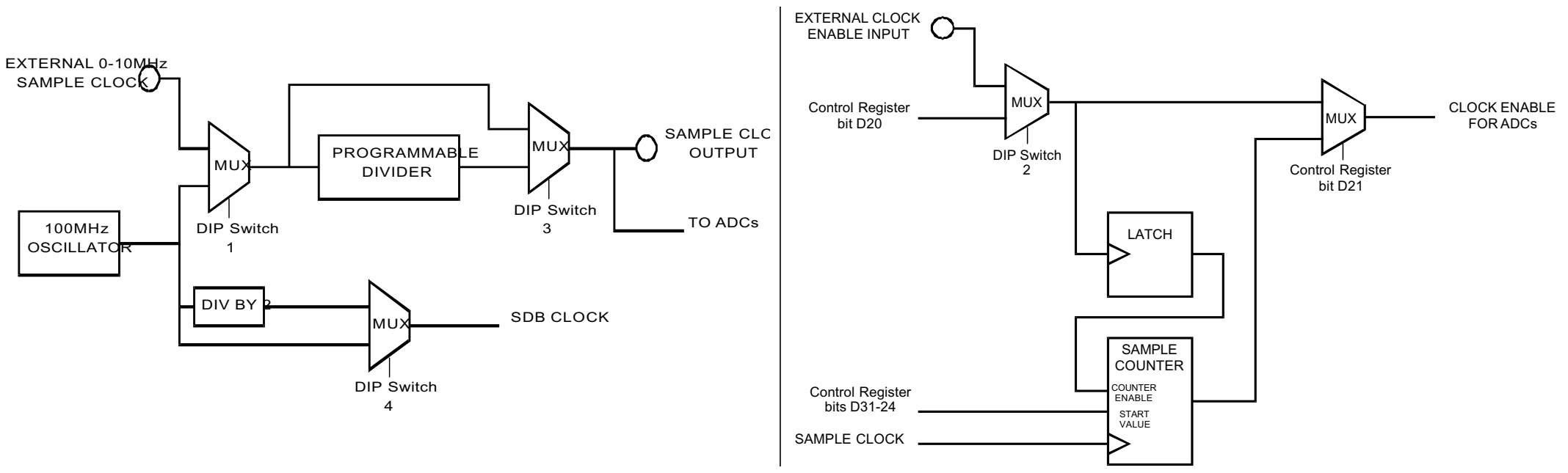

Figure 3.6: Diagrams showing the the *sample clock* and *clock enable* routing in the SMT356

The *sample rate* of the ADC converters (see Figure 3.6) is derived from one of two sources: either from an external clock input or via the on-module 100 MHz reference. The maximum external clock frequency is 10 MHz: we will use an externally generated, TTL compatible, 4 MHz clock signal that will be passed through a programmable divider inside the SMT 356 module and then both output to a connector and routed to the ADC converters. An 8-bit divider is provided which allows the ADC clock to be provided from a divided clock or can be bypassed in order to get a division of 1.

The *clock enable* (ADC sample enable) is controlled from either the external enable input or from an enable register. The *enable register* gates the ADC clock via software, allowing the selection of a *latched counter mode* where a user programmable value, stored in another register, determines the number of samples automatically acquired in *internal clock enable* mode. The range for this value is from 1 to 256, making this functionality very powerful for the implementation of an automatic asynchronous sampling of the JAMEX output values at an ADC module level.

## 3.2 Sundance Software

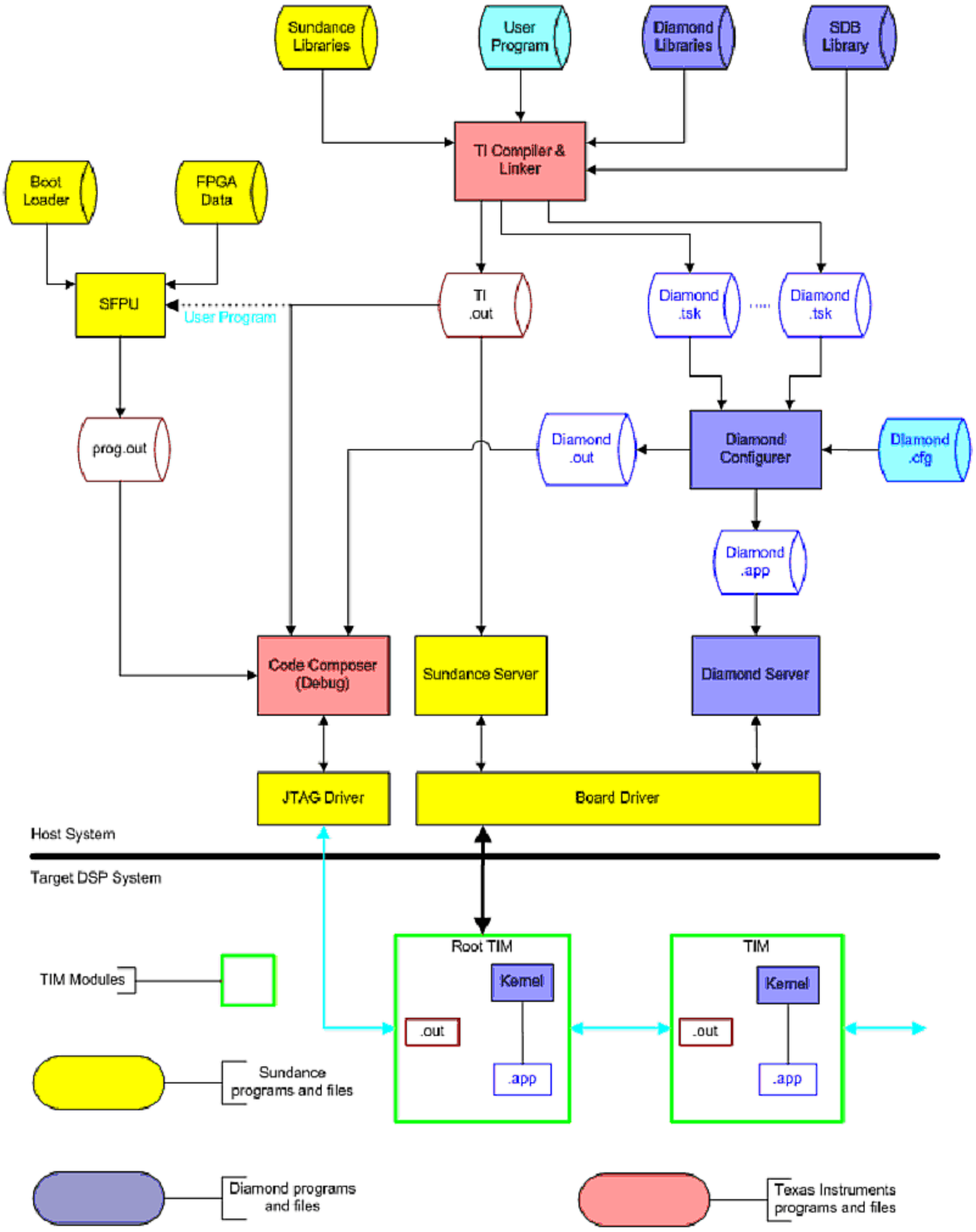

Figure 3.7: Reciprocal relations between the software needed to develop a Sundance application

Creating an application to be run on the Sundance acquisition system involves two distinct components: the *algorithms* running in the DSPs and a *control program* running on the host PC.

The control program is responsible for loading the DSP and for providing the user interface to the application. To build the applications, Sundance recommends to use the *3L Diamond* real-time operating system since this efficiently supports the features of the Sundance hardware and handles transparently the variations between different processor TIMs. The main advantage of the Diamond RTOS is that it allows easy and efficient creation of applications that run on multiple TIMs: a MultiProcessing application running concurrently on two DSPs, like the JAMEX application, can be easily developed and tested in this environment.

Most of the software provided by Sundance to support the JAMEX application assumes that the user has access to various tools for the `C6000 DSP architecture` from Texas Instruments: particularly the C compiler, the assembler and the linker. A few software components also assume the presence of TIs Code Composer debugger that nevertheless is a useful option to have.

In the next sections will be presented all the tools (see Figure 3.7) needed to develop and debug the JAMEX application: they will be ordered according to the optimal sequence of installation in a suitable Personal Computer running Windows 2000 or XP.

### 3.2.1 Code Composer Studio

Code Composer Studio from Texas Instruments is an integrated environment for debugging DSP applications. It communicates with Sundance TIMs using `JTAG`, thus leaving the host comport free for the application and allowing concurrent debug. Although it is possible to use Sundance products without having CCS, its presence is required for debugging efficiently the DSP programs and managing the flash ROM using the SMT6001 product that will be presented later. It is planned to use this tool until a stable final version of the software for the detector will be available, possibly, at this point, this environment will be “un-installed” in order to free the HD space in the host PC and make it available for the measurements.

The Code Composer Studio (CCS) IDE extends the basic code generation tools with a set of debugging and real-time analysis capabilities. The CCS IDE supports all phases of the development cycle: *design, code build, debug and analyse*. The fully integrated development environment includes real-time analysis capabilities, easy to use debugger, C/C++ Compiler, Assembler, linker, editor, visual project manager, simulators, XDS560 and XDS510 JTAG emulation drivers and DSP/BIOS support. Familiar tools and interfaces allow users to get started in short time and easily add complex functionality to the application thanks to sophisticated productivity tools.

It is possible to create applications for the Sundance system with CCS: in this case it is needed to write a file (*.cmd*) that gives the memory layout of each TIM in the system and explicitly allocates the sections of the program to the available memory. The "Processor TIM support package" SMT6400 gives some examples of how to do this: for each processor TIM (the SMT361) in the system it is required to compile the relevant *.c* and *.h* files into *.obj* object files and then link these into a *.out* executable file. In this case the source files will need to be changed if to move from using one type of TIM to another (e.g. to use a faster DSP).

When using CCS, it is necessary to explicitly load each *.out* application on one DSP at a time and, if the application requires the DSP processors to be started in a particular order, it is needed to arrange to do that explicitly each time we run the system. This loading is done via the slow solution using the JTAG.

This way is much more difficult and error-prone than the equivalent Diamond procedure, thus the SAXS detector will use the *3L Diamond* real-time operating system to load the applications.

The specific configuration required for CCS in order to support the Sundance hardware will be discussed in the next section that is dedicated to the Sundance drivers.

### 3.2.2 SMT6012 JTAG device driver for CCS

This product gives the device drivers that are needed by Code Composer Studio and the SMT6001 to communicate with the TIMs via a JTAG interface that is provided on the Sundance carrier board (SMT310q) by the JTAG controller chip. This chip enables the various processor elements on the carrier board to transfer data between each other and it is possible to use this interface to load and run programs, as we already saw, but it is generally more efficient to use the SMT6300 device driver. For this reason JTAG is often left to its intended application: to support debugging.

CCS needs to be given several parameter before it can use correctly this driver: they include the order and type of processors in the JTAG chain and the type and I/O address of the carrier board. Recently Sundance created a *Wizard* to perform this configuration automatically. In our experience it is better to perform the configuration stage manually in order to have a better control on the environment and to track easily eventual mistakes.

### 3.2.3 SMT6300 basic driver software

The SMT6300 package installs a number of *host side tools* and the documentation for the carrier boards that are useful to check the basic functionality of the Sundance boards and components: the main tools are the `BoardInfo` application and the `E2P` application.

The *BoardInfo* application gives information about the boards installed in the system, can reset a whole board or individually the TIMs it holds, configures the on-board comport connections and also can run a simple confidence test. This utility can also be called directly from the Diamond server or any other host program and does not use JTAG, so it is independent from CCS and very useful in case of problems.

The *E2P* Application allows the user to reprogram the EEPROM on the Sundance carrier boards in the rare cases when this is necessary.

### 3.2.4 SMT6025 Windows Server

The SMT6025 is a software development kit (ref. [28]) that is used to develop host software to interact with a DSP application: it is a collection of software modules that provide an interface to Sundance PCI carrier boards (see Figure 3.8 -left-) and also a high-speed PCI communication through the global bus. It allows the user to control these boards from the host as well as to exchange data between the carrier board and the host. The SMT6025 is ideal for users that wish to develop their own code to interface with Sundance hardware, like will be required for the SAXS detector.

The simplest and most general I/O mechanism that can be used to communicate between the host and the Root DSP is the *host comport*. It is a sequential, bi-directional link that gives typical transfer speeds of up to 2 Mbyte/s that is also used for loading programs into the DSP.

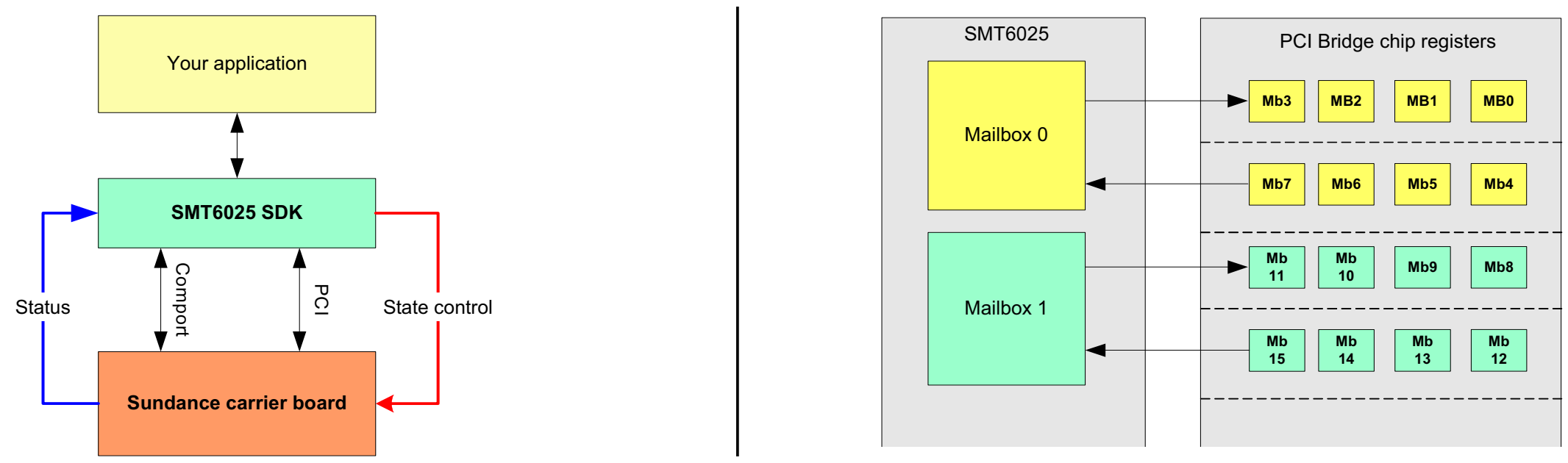

Figure 3.8: Interfaces to Sundance hardware and Mailboxes in the SMT6025

Each DSP TIM in the system will load a *bootloader* from its FLASH ROM once the reset has completed. This bootloader performs various housekeeping operations to initialize the TIM and then waits until data arrives on any of it's comports. The first comport to become active is selected and the data it provides are loaded into the DSP and executed. The host comport only gives access to the Root TIM of a DSP board. It is needed to load any other TIMs in the system indirectly via the root with explicit code. This is done automatically when loading a 3L Diamond (presented in the next section) application in the form of a single *.app* file.

Another method of communication is the *mailboxes* provided by the PCI bridge chip that allows the host and the Root DSP to signal each other with information used for synchronization. These mailboxes are intended as a signaling mechanism and not as a way of passing large amounts of data: the V3 bridge chip provides *16* $\times$ *8-bit* mailboxes and they are combined to form two independent 32-bit, bi-directional mailboxes (see Figure 3.8 -right-).

There is another powerful software mechanism (built using the mailboxes) for the host and root DSP to communicate that is called the *High-speed channel* (HSC) and is the preferred strategy (see Figure 3.9 -left-) to exchange large blocks of data over the PCI bus while achieving higher speeds than can be achieved using the host comport. The SMT6025 provides 8 High Speed Channels to the user and any communication across an HSC is always initiated by the DSP. The DSP sends a control word, along with any associated data, to the host's channel handler, which performs some resulting action and replies with a reply word and associated data: each of the 8 channels has an associated handler on the host.

There is a standard handler associated with each channel by default, but the DSP can replace this with a user defined one because an handler is just plain software that runs on the host and performs actions on behalf of the DSP (see Figure 3.9 -right-). Handlers take the form of user-defined *Dynamically Loaded Libraries* (DLLs) that are used to extend the functionality of the host code. Handlers are most easily used in conjunction with 3L Diamond but it is possible to write the handler code without using Diamond, although this would be very complicated. The DSP can instruct the host, sending a command word, to load a handler and then communicates directly with it (e.g if a DSP program wishes to write data to a file on the host system, a specific handler can be loaded by the DSP to perform the file access).

When the handler receives the DSP's command word, it performs some action and this action may or may not involve the host side software. Some of the commands sent to the host have associated data that need to be passed to the host and similarly, the reply from the host may also include associated data: this is the crucial point in moving large block of data with this method.

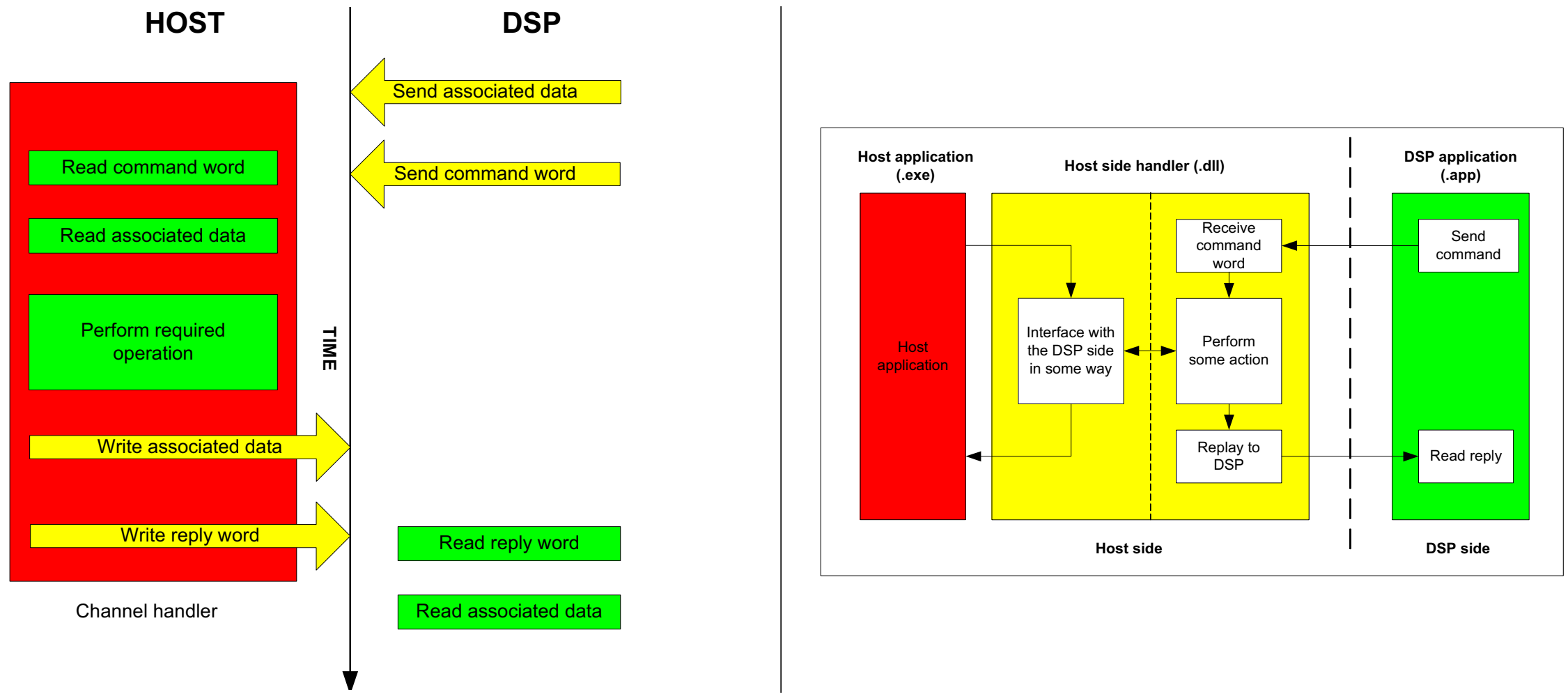

Figure 3.9: Standard HSC transaction and operation of a handler for the SMT6025

These data are transmitted using, like an intermediate buffer, either the carrier board's SRAM or the host memory.

When using the *carrier board SRAM*, each HSC has an associated region of 64 KByte in this SRAM, called the *argument area*, to be used for transmitting associated data. The DSP can read and write to this memory using the global bus, hence at very high speed, and at the same time the carrier board's SRAM is directly accessible in the host address space. The SMT6025 provides utility functions that allow the host to access the argument area for each channel and similarly, on the DSP side, 3L Diamond provides utility functions with which the argument area can be accessed as well.

If using the *host memory* as a buffer, the host can lock down a region of its memory and can make this available to the DSP over the PCI bus. The host memory needs to be locked down in order the DSP can gain a safe access to this memory area. The host places Information necessary for the DSP to access this memory in a section of the carrier board's SRAM reserved for this purpose and known as a *memory descriptor list* (MDL). The host operating system will not page to disk any locked down memory, and will ensure that the memory remains at a fixed physical location.

Another functionality of the SMT6025 is to give access to the PCI bridge chip register, although should not be needed any direct access to the PCI registers for most systems, it is also to consider that writing incorrect values to these registers will almost certainly crash the host with consequent data loss.

It is useful to indicate the *performance figures* for various transfer types obtained with the SMT6025: using the HSC to write or read to contiguous memory shows transfer rates in the order of 50 Mbyte/s compared to the 2 Mbyte/s allowed by the host comport. This higher speed is very important in our application that needs to transfer a huge amount of data between Host and DSP in short amount of time in order to avoid any kind of loss of data.

### 3.2.5 3L Diamond Multiprocessor DSP RTOS

The 3L-Diamond system (ref. [29]) provides a convenient way of developing efficient applications that make use of all the features of Sundance TIMs. Since Diamond was designed for multiprocessor systems operating in Real Time, it includes everything is needed for programs running on several TIMs and to reduce the time needed to create a high-performance implementation of an application.

An important feature of the Diamond software suite is the integrated RTOS based on a *small microkernel* providing support for multiprocessor activity, including multi-tasking, multi-threading, inter-processor communication and management of interrupt handling, events, semaphores and timers. This microkernel implements a pre-emptive, priority-based real-time scheduling, with context switch times around 0.5 $\mu$s, and provides access to host services ('C' standard I/O and Windows GUI) for all DSPs in the system. It uses an *automatic loader* to handle the issue of loading multiple processors and starting them in a synchronized manner. An *optimized interfaces* to the Sundance peripherals, such as comports, SDBs and global bus is provided: simply declaring these features automatically initializes the necessary support code that has been developed to maximize performance. Also the *placement of software components* on the available processors is allowed in a straightforward and reconfigurable way because a single file is holding the complete multiprocessor application, removing the possibility of accidentally getting the wrong component on any of the processors.

Moreover, it features an automatic management of differences between TIMs, such as the location of external memory and the addresses of peripheral devices, preventing accidental errors while giving maximum *freedom in allocating memory*, either fully automatically, semi-automatically, or completely under user control. Diamond uses the standard TI Code Composer Studio compiler, assembler, and linker to generate the individual object files called TASKS.

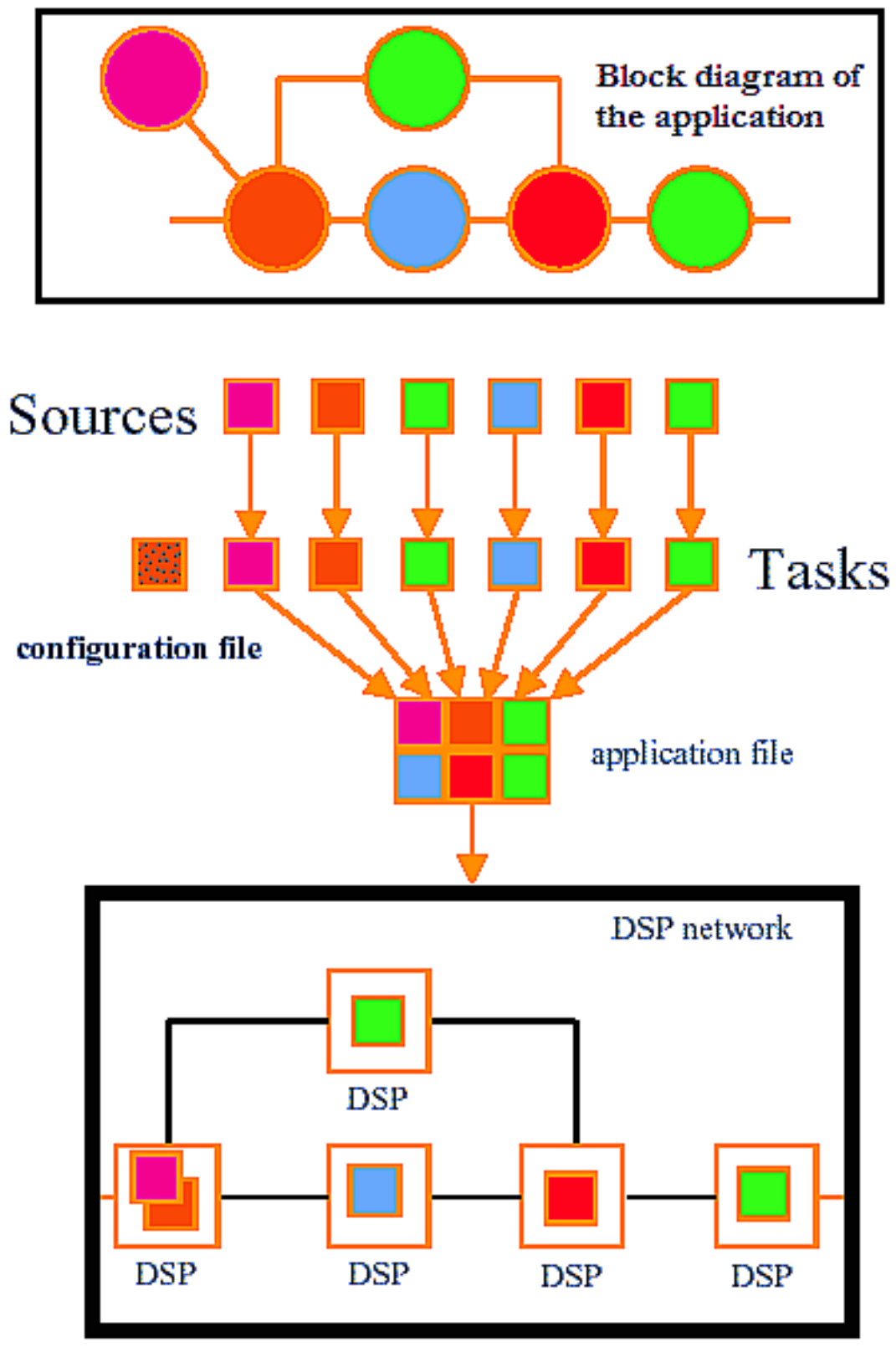

Figure 3.10: 3L Diamond operation

A tool called the configurer combines and optimizes these binary files together into a single executable application file as it is schematically shown in Figure 3.10. The configurer is driven by a text file that describes the target network, the tasks and their interconnections, which is provided by the user. The key idea of loading DSP applications with Diamond is that it takes care of this instead of the user because all the information needed to load the relevant code and data to the correct processors in the system are built into the already generated *.app* application file: this means that, automatically, each processor will be loaded with only the code needed for that processor.

To load a Diamond application the user simply resets the system and send the whole *.app* file through the host comport, operation that is usually done by the Diamond Windows Server. If the physical connections in the system are changed, it is just needed to simply reflect the changes in the *.cfg* configuration file and to build a new application file with the Diamond configurer.

A small minus is that Diamond *.app* files are proprietary and do not use the *Common Object File Format* (COFF) from Texas Instruments because this format does not provide any of the facilities required for multiprocessor operation.

Most of the accessory operations are performed by the *Diamond Server* that is a host program that controls the execution of any Diamond application in the DSP network. It includes the following features: automatic application loading, implementation of *standard 'C' streams* (`stdin`, `stdout` and `stderr`) and *'C' file I/O* (`fprintf`, `fopen`, `fscanf`,...).

Another option, very useful in order to test simple applications or for debugging purposes, is the ability to run *stand-alone applications*: when a normal Diamond application executes an I/O call (e.g. printf), some code from the Diamond run-time library communicates with the host server over the host comport. This communication uses a protocol agreed by both the server and the run-time library. Since Diamond allows to specify that an application does not communicate with the host server, such a stand-alone application does not need to include code on the DSPs to manage the communication protocol and the host comport bandwidth will be left free for the application's use. This option is generally used for embedded applications where there will be no host to communicate with, or for applications that will communicate exclusively with user-written code on the host.

### 3.2.6 SMT6001 flash programming utility

This is a package that allows changes, corrections and custom configurations to the contents of the FLASH ROMs on the processor TIMs (like the SMT361): these are delivered with standard contents, adequate for generic applications, but they can be tuned to achieve faster operation, improved stability or additional functionality.

It will be used in the SAXS detector, during the test phase, in order to choose between different firmware configurations and to test their relative efficiency. For example, it is very useful in order to write additional custom data inside the FLASH ROM of the SMT361 TIM: the JAMEX configuration bitstreams and the configuration data for the SMT356 ADC board can be conveniently stored in the FLASH ROM. In this way the startup speed of the JAMEX application will increase of an order of magnitude and a more stable system will result.

## 3.3 The JAMEX application

Considering the complexity, evident from the previous sections, of the software and hardware components required to build the acquisition system for the SAXS detector, it is understandable the choice to involve Sundance directly in developing the “JAMEX application”. Since the final application software would require not only developing of the code for the Host PC and the DSPs, but also the required *firmware* to be loaded in the Virtex FPGA that should implement the ZCQSI interface. A deep knowledge of the Sundance product’s hardware and software is required in order to avoid the *hardware damages* that will certainly occur as a result of a wrong configuration bitstream loaded in the main FPGA of the DSP modules.

### 3.3.1 Firmware implementation of the ZCQSI interface

The Firmware for a Sundance module is just binary data that sets programmable logic (the Virtex FPGA) into a state where it implements a particular hardware design. It starts as a textual description of an electronic circuit which is processed to produce a binary file that can be loaded into a FPGA to realize the design.

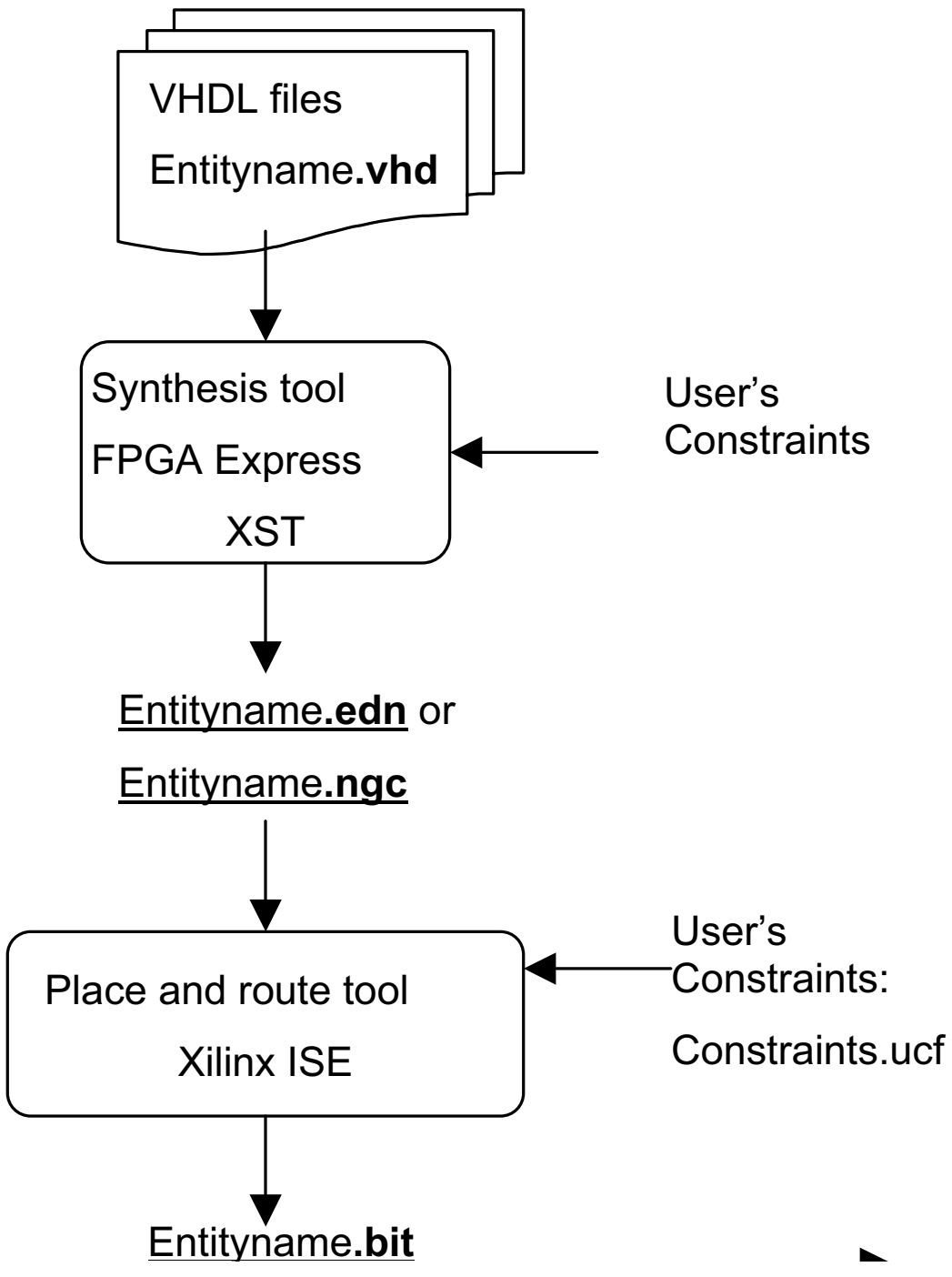

Figure 3.11: From the VHDL description of the configuration to the final *bitstream file*

Sundance uses *modularity* in its firmware designs to improve reusability and maintainability. This approach creates a number of *resource blocks* that are combined to make a complete design. These resource blocks are used to implement both the glue logic required on the module in order to interconnect the DSP with memory and all the Sundance peripherals: comports, SDBs and eventual custom defined interfaces. Different processor TIMs may include a diverse numbers of resources: for example, some processor TIMs contain six comports and two SDBs, while others may contain only four comports. The basic comport firmware block is shared among these modules and the source code for all of them is provided from Sundance, subject to a non-disclosure agreement: they are described in VHDL (VHSIC Hardware Description Language) and compose a library of basic firmware blocks that can be used for a custom design.

When creating firmware for a DSP processor module, like the SMT361, the DSP is interfaced to the basic firmware blocks by additional blocks: for example some FIFOs for the comports and SDBs. From the DSP's point of view, the FPGA is mapped into the memory space of the DSP via the External Memory Interface (EMIF) bus, operating at 100 MHz. The DSP accesses the registers implemented in the FPGA by reading and writing this EMIF. This is the way for the firmware to connect the user program running on the DSP to the outside world: the ZCQSI interface itself is implemented in this way.

The SMT361 DSP module has several connectors routed to TIM-specific input and output pins on the FPGA: the default firmware implemented on the TIM connects the signals of the firmware blocks to the appropriate I/O pins of the FPGA, which are in turn connected to the physical connectors. It is possible to modify this standard firmware and thereby change the connection of signals on the external connectors: in our case one of the SDB connectors will change its original functionality in order to accommodate the ZCQSI signals and the additional clocks and control signals required to run the JAMEX ASICs.

The implementation of a generic firmware (see Figure 3.11) consists of two two stages: *synthesis* and *place and route*. During the synthesis stage, the VHDL sources files are processed by the `Xilinx Synthesis Technology tools` to produce a net-list file. The latest versions of any Sundance firmware assume the use of Xilinx Synthesis Technology (XST) while older versions of the firmware were generated using `FPGA Express` by Synopsis.

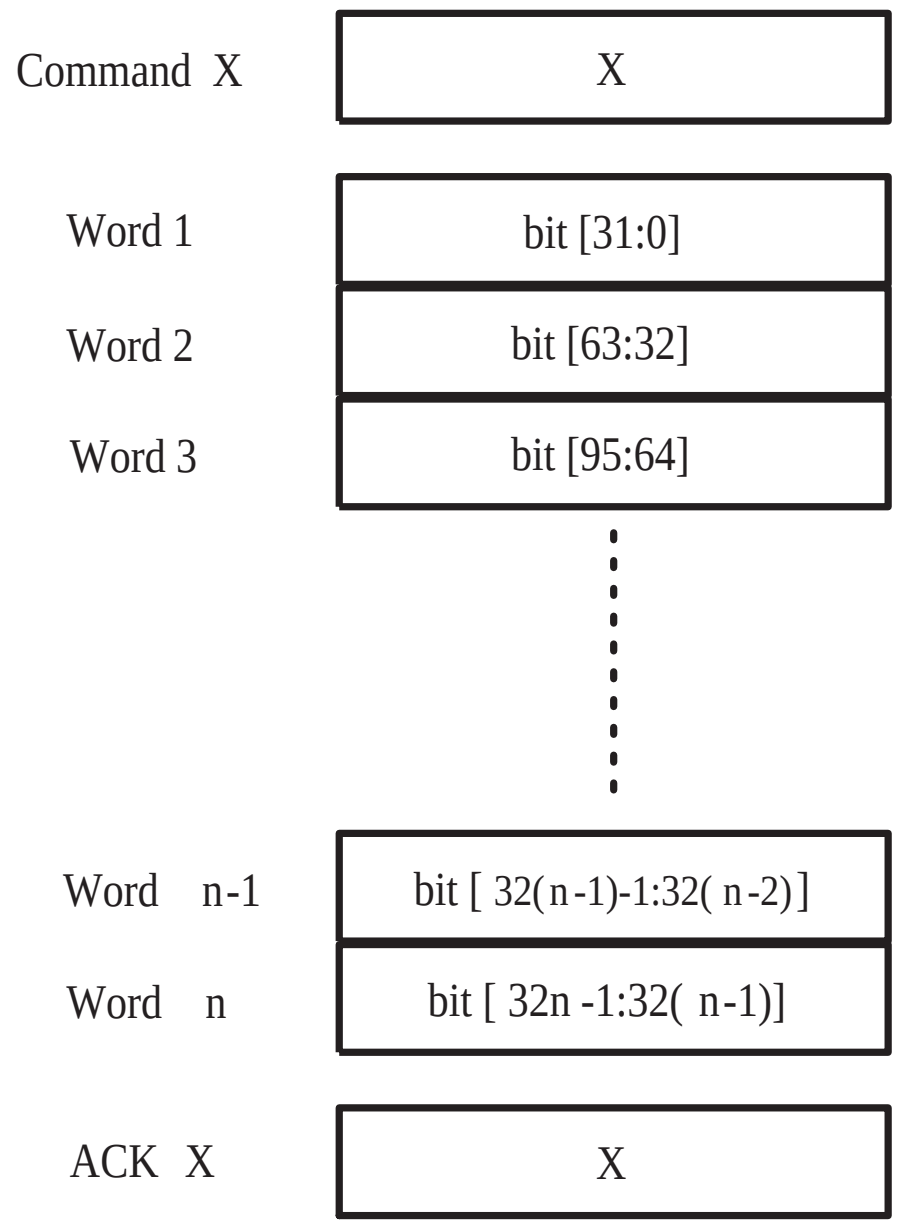

Figure 3.12: Command style

For the place and route phase Sundance uses the Xilinx ISE place and route tools to generate a bitstream for a specific family of FPGAs. These tools take the net-list produced by the synthesis tool and apply a constraint file describing timing constraints and pin locations. The custom Firmware required for the SMT361 Master in the SAXS detector was developed in VHDL and implemented with the tools provided with XILINX ISE 5.1.03i: the syntesis tool is XST. The VHDL source code files were grouped in different directories (ref. [30]) to distinguish the files related to the standard SMT361 capabilities from that related to the new capabilities for J2S control.

The basical idea for the `Communication Protocol` between FPGA and DSP is to divide each new functional mode, implemented with this custom firmware, in some tasks. Each task becomes a `command` that the DSP gives to the FPGA through the *SDB J2S FIFO*. Each command is constituted by a fixed number of 32-bit words like in Figure 3.12. The *first word* identifies the command and is called "header" that may be followed, except in some cases, by a variable number of words like summarized in Table 3.1. The *controller* implemented by the new firmware begins the task only when all data needed by the command are present in the SDB J2S FIFO. When the executed task ends, the controller sends back to the DSP an acknowledge word (ACK).

| Command Code | Command Name | Data Sent | Data Read |
|---|---|---|---|
| 0x10000000 | J2S ZCQSI PGA_GAINS_CONFIG. | 20 words | ACK |
| 0x11000000 | J2S ZCQSI WRITE_CONTROL | 1 word | ACK |
| 0x12000000 | J2S ZCQSI SPI_CONFIGURATION | 1 word | ACK |
| 0x13000000 | RESERVED_ CONFIGURATION | * | * |
| 0x14000000 | Write_JAMEX | 1 word | ACK |
| 0x15000000 | Read_JAMEX | / | 1 word |
| 0x20000000 | WRITE TRIGGER_OUT_PHASE_0 | 1 word | ACK |
| 0x21000000 | WRITE TRIGGER_OUT_PHASE_1 | 1 word | ACK |
| 0x22000000 | WRITE TRIGGER_OUT_PHASE_2 | 1 word | ACK |
| 0x23000000 | READ TRIGGER_IN | / | 1 word |

Table 3.1: List of the Commands implemented in the custom Firmware

This is performed writing the header, correspondent to the last command, in the SDB J2S FIFO: in other words, the DSP has to poll the SDB J2S FIFO until it reads back the header of the last command. We can notice from Table 3.1 how the commands characterized from a command code of **'0x1'** closely resemble the commands given in Chapter 2 for the definition of the ZCQSI interface. In this way we can be pleased to have reached the goal of having a uniform tassonomy for the equivalent commands at all the levels in the detector structure. Some new commands are here present, with command codes starting for **'0x2'**, to control another additional functionality of the new Firmware: to provide the basic clock and control signals that will be routed through the J2S board to the JAMEX ASICs.

These signals that will be provided by some pins of the *SDB connector B* of the SMT361 Master and both buffered and converted to LVDS on the J2S board are:

- **ADC_CLK** [output]: a 4 MHz clock signal that accommodates the constraints given in the *RUNNING MODE Timing* datasheet (Appendix B)

- **ADC_EN** [output]: it is the TTL ADC clock that must meet the Sundance SMT356 specification

- **MUX_CLK** [output]: is a 1 MHz clock operating in RUNNING MODE only

- **S_CLK** [output]: according to the JAMEX datasheet definition for the RUNNING MODE it is a 1 MHz clock synchronous to MUX_CLK; in PROGRAMMING MODE it is a HIGH level pulse of 500 ns with repetition period of 10 $\mu s$ (100 KHz)

- **S_MODE1 and S_MODE2** [output]: are according to JAMEX datasheet and Timings (Appendix A and B)
- **MUX_IN** [output]: is shown in Figures 3.13 and 3.14
- **MUX_OUT** [input]: is provided as an optional line

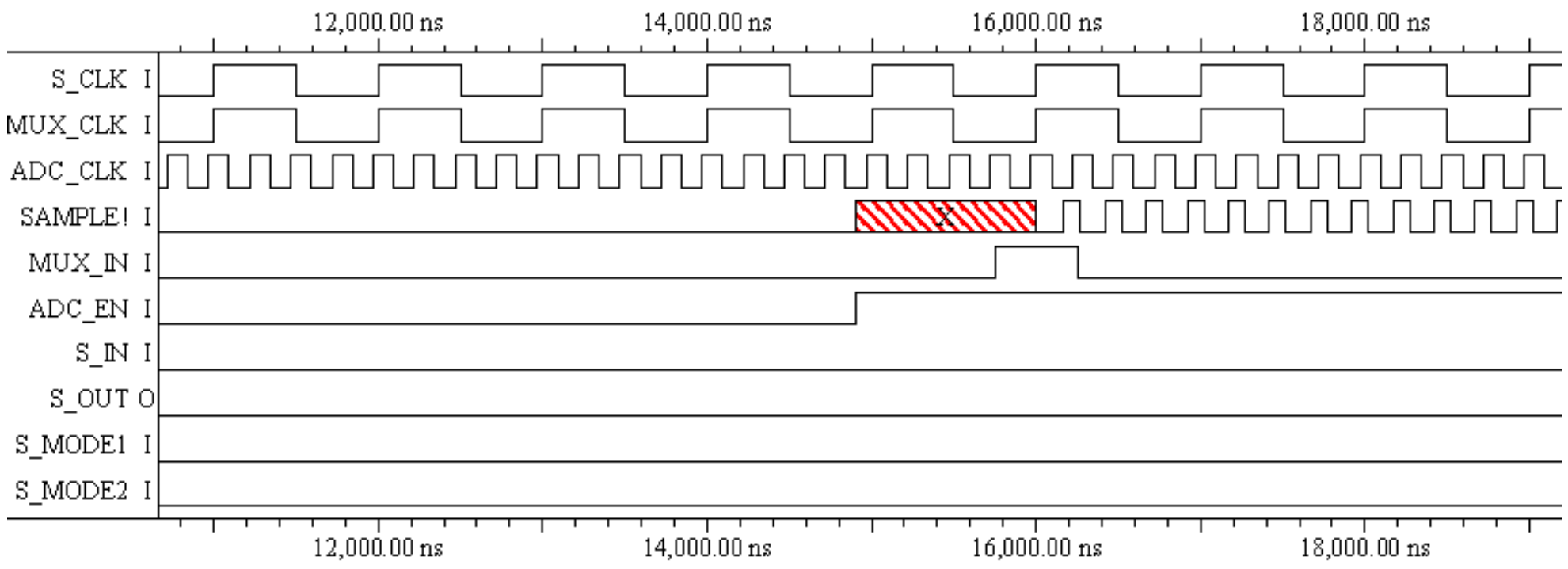

Figure 3.13: Timing relative to the initial part of a RUNNING MODE cycle

In these two Figures we can see the optimal timing for all the signals we have just defined, as required to sample the JAMEX outputs while producing the lowest interference to the ongoing measurement.

Specifically, during the RUNNING MODE all the clock signals are cycling with a period of 128 $\mu s$ (the period of 1 JAMEX Cycle) and the two relevant phases are: the beginning of valid output values given in Figure 3.13 (originated from the signal MUX_IN pulsed HIGH) and the end of one cycle of the JAMEX output multiplexer, which is occurring 64 $\mu s$ later, like depicted in Figure 3.14.

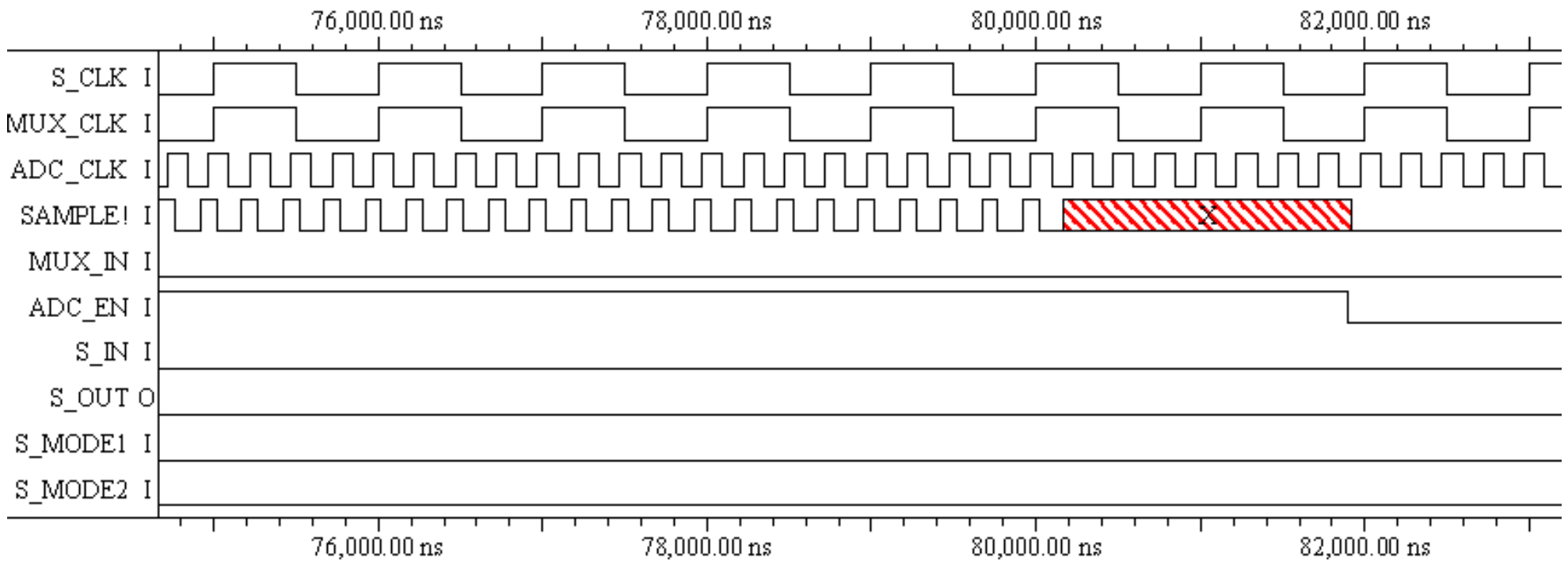

Figure 3.14: Timing relative to the final part of a RUNNING MODE cycle

The purpose of the commands with code of **’0x2’** is exactly to generate these timings correctly and to synchronize the overall activity of the SAXS detector with the I/O signals eventually provided from the external devices connected to the detector by means of the *TRIGGER_IN* and *TRIGGER_OUT* Lines. We introduced this capability in Chapter 2 when we was describing the SET_TRIG_OUT and SET_TRIG_IN commands.

| | Pin | Signal | Signal | Pin |
|---|---|---|---|---|
| CLK | 1 | CLK | GND | 2 |
| CS_ | 3 | D0 | GND | 4 |
| DATA | 5 | D1 | GND | 6 |
| EN | 7 | D2 | GND | 8 |
| R/ W | 9 | D3 | GND | 10 |
| DR | 11 | D4 | GND | 12 |
| RESET_ | 13 | D5 | GND | 14 |
| S_ CLK | 15 | D6 | GND | 16 |
| MUX _CLK | 17 | D7 | GND | 18 |
| S_ MODE1 | 19 | D8 | GND | 20 |

| | Pin | Signal | Signal | Pin | |
|---|---|---|---|---|---|
| ADC _CLK | 21 | D9 | GND | 22 | |
| S_ MODE2 | 23 | D10 | GND | 24 | |
| ADC _EN | 25 | D11 | GND | 26 | |
| MUX _IN | 27 | D12 | GND | 28 | |
| MUX _OUT | 29 | D13 | GND | 30 | |
| OUT 0 | 31 | D14 | GND | 32 | |
| OUT 1 | 33 | D15 | GND | 34 | |
| OUT 2 | 35 | UD0 | DIR | 36 | OUT 3 |
| IN 0 | 37 | WEN | REQ | 38 | IN 1 |
| IN 2 | 39 | UD1 | ACK | 40 | IN3 |

Figure 3.15: New signal definitions for the SDB - B connector

In order to control the 4 output and 4 input *TRIGGER experimental lines* we need to reserve some signals on the SDB - B and provide them the correct LVTTL interfacing and galvanic isolation. As shown in Figure 3.15, the pins 31 - 33 - 35 - 36 of the original SDB - B connector are set as output pins to provide the *TRIGGER_OUT* Lines. In the same way the pins 37 - 38 - 39 - 40 will act like LVTTL complaint standard input pins to provide the *TRIGGER_IN* Lines. Further conversion of these signals will be performed on an external *Optocouplers Board*.

### 3.3.2 Software for Host PC and dualDSP

The final JAMEX Application Software was developed to run on a full system composed of 2 $\times$ SMT361 + 3 $\times$ SMT356 and therefore including the integration of the final Firmware application functions into the final Software application (ref. [31]). This section will offer a starting point to understand how the available components fit together to build the complex processing systems.

A graphical explanation of the connections required between the two carrier boards and several modules, composing the final complete system, are given in Figure 3.16. The Host software is developed using Microsoft Visual Studio while the DSP software is developed using 3L Diamond and Code Composer Studio.

The two workspaces are called respectively Host_M6.dsw and DSP_M6.dsw.

The *Host_M6.dsw* is the workspace that implements the basics of the User interface and the drivers to communicate with the DSP embedded system.

The *DSP_M6.dsw* is the workspace that implements the final DSP application over the full MultiDSP system.

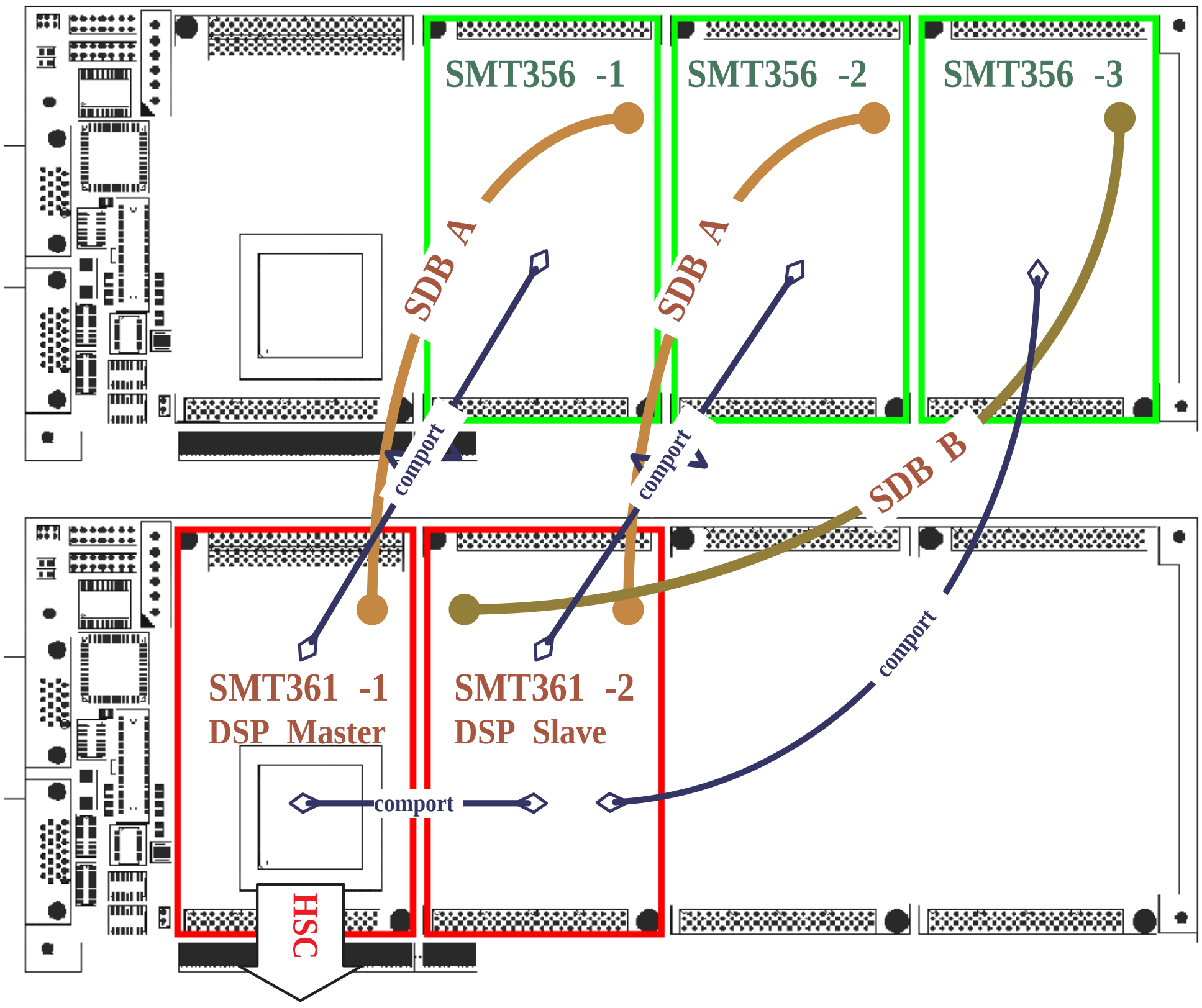

Figure 3.16: Layout for the carrier boards and modules integration

### The HOST_M6 workspace

The Host workspace builds a *Win32 Console Application* and it is composed of the following files: *Host.cpp*, *StdAfx.cpp*, *Host.h*, *StdAfx.h* and *Smtdrv.h*.

The `HOST.cpp` source file is the main source file of the workspace and implements the user interface of the project. It is composed of the following main parts:

- `Main()`: it provides the required PCI interface to the carrier board at a driver level and gets information about the boards found in the system; this code is transparent to the user and it is handled by the calls to the *Smtdrv.dll*
- `HostControl()`: this function implements the basic user interface and build a system to communicate with the DSP board. It consists of the following main parts:

  - Memory allocation and initialization on host and check of the DSP memory allocation

  - DSP Board reset

  - DSP application loader

- User Interface

The *User Interface* implements the following commands: `EXIT` (exits the application), `INIT`, `START`, `GET_JAMEX_CONFIG` (get the JAMEX Configuration data), `GET_IMAGE` (get the JAMEX Image from the DSP boards' RAM: the first portion of image is from the SMT356_1), `WRITE_VECTOR` (write a TIME_FRAME vector to the DSP master that will be used for testing purposes) and `WRITE_IMAGE` (write a JAMEX Image to the DSP master, will be used for testing).

During the **INIT** phase the Host reads the following I/O files and transfers the contents into memory arrays on the DSP board: *SetJAMEXGain.out, SetOpGains.out, SetIntTime.out, SetExpTrigOut.out, SetExpTrigIn.out, SetJAMEXConfig.out, SetOffset.out* and *SetNorm.out*. It also executes the following Application Firmware commands: *J2S ZCQSI PGA_GAINS_CONFIG., JAMEX CONFIGURATION* (via Read_JAMEX and Write_JAMEX), *WRITE_CONTROL* and *SPI_CONFIGURATION*. Then it will CONFIGURE the SM356s' FPGAs of the system: it will read and process the content of the file *v-top.mcs* and then it will transfer the relevant configuration bitstream to the SMT361s which will configure the SMT356s that are in charge with.

The **START** command brings the system into *RUN state* and performs the programmed acquisition.

It will report the real-time results of the JAMEX application and in particular of the following actions:

- `LOOP TIME_FRAMES`

  READ_TRIGGER_IN

  WRITE_TRIGGER_OUT_PHASE0

  LOOP JAMEX_CYCLES (loops WRITE_TRIGGER_OUT_P1 and READ_TRIGGER_IN)

  WRITE_TRIGGER_OUT_P2

  Real-Time TIME_FRAME

  ACTUAL TIME_FRAME

The file `STDAFX.cpp` includes only the *StdAfx.h* header file, which is the standard Win32 system file. `SMTDRV.h` is the header file of the *Smtdrv.dll* and so it includes all the Sundance board windows driver calls.

#### The DSP_M6 workspace

The DSP code is organized in six tasks working in parallel, synchronized and running over two DSPs.

The **DSP Master** runs the following tasks:

- `SMT361M` JAMEX application (handling 8 JAMEX output channels) + Host interface on Master DSP.

- `Int512M` Integration of 512 samples per JAMEX cycle on the Master DSP (it supports real-time data acquisition from the SMT356_1).

Whereas the **DSP Slave** runs the following tasks:

- `SMT361S` JAMEX application on Slave DSP (handling 8 JAMEX output channels).
- `Int512S` Integration of 512 samples per JAMEX cycle on the Slave DSP (it supports real-time data acquisition from the SMT356_2).
- `SMT361S` JAMEX application on Slave DSP (handling 4 JAMEX output channels)
- `Int256S` Integration of 256 samples per JAMEX cycle on the Slave DSP (it supports real-time data acquisition from the SMT356_3).

The whole MultiDSP application is a complex multi-tasking application, synchronized using a total of 13 *virtual channels* and *6 physical channels* (comports), running on both DSPs. The high speed SDBs links are used for the data acquisition from the ADC modules. The DSP code architecture is *modular* and *symmetric* and implements 3 JAMEX sub-applications according to the three different SMT356 ADC modules.

Essentially, we can summarize the whole *JAMEX application* in 3 complex parts:

1. *Sub-JAMEX application 1* runs over the sub-system composed of the SMT356_1 module and the SMT361 Master module and is based on the SMT361M and Int512M tasks.
2. *Sub-JAMEX application 2* runs over the sub-system composed of the SMT356_2 module and the SMT361 Slave module and is based on the SMT361S and Int512S tasks.
3. *Sub-JAMEX application 3* runs over the sub-system composed of the SMT356_3 module and the SMT361 Slave module and is based on the SMT361Sbis and Int512Sbis tasks.

# 4

# Design issues in the J2S board family

The name given to this set of Printed Circuit Boards is the acronym of *JAMEX to Sundance* because it is designed in order to allow interaction and connection of two different entities: one consists of the *JAMEX ASICs*, mounted in couples on each SAXS Hybrid, which are integrated circuits capable of directly sensing the charges on the microstrips of the SAXS detector's anode and the other is represented by the Sundance ADC acquisition cards and digital processing electronics.

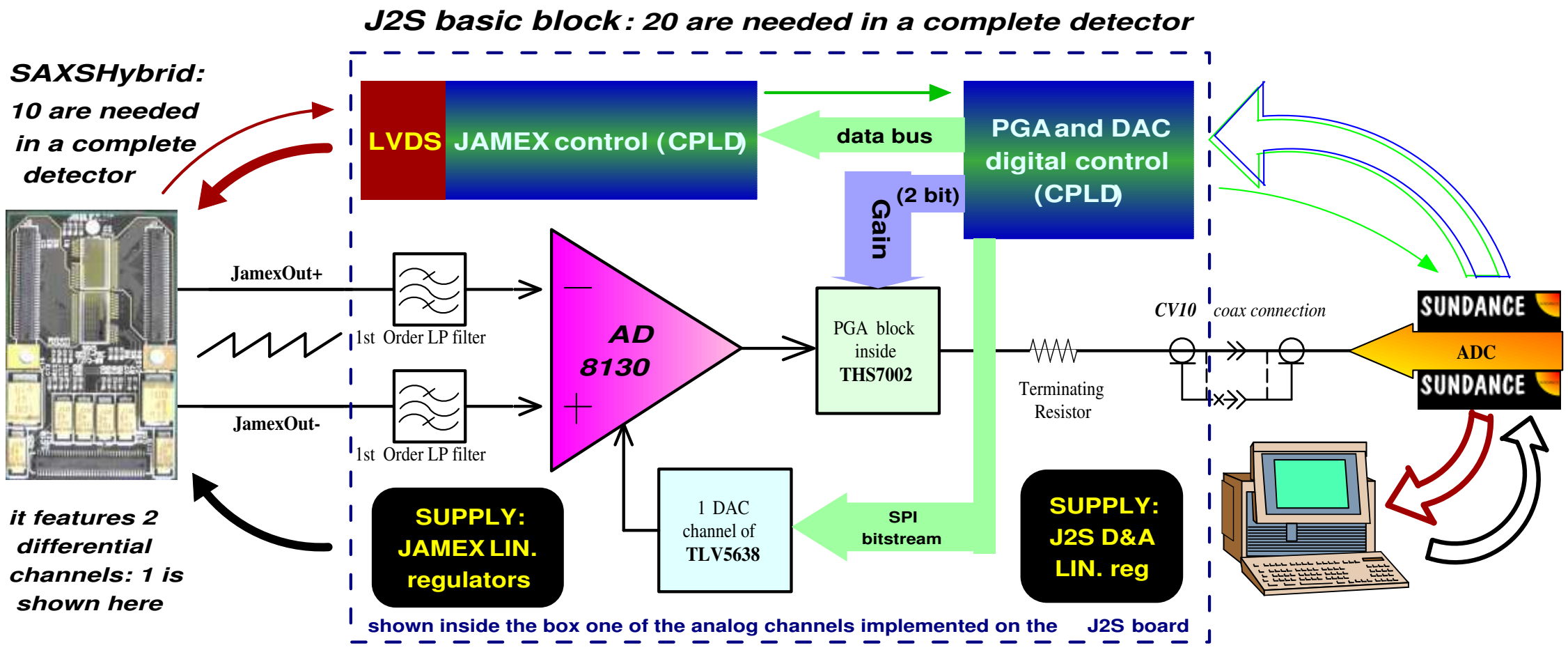

Figure 4.1: Exemplification of the functionalities offered by the J2S board family

## 4.1 J2S boards overview

The Sundance acquisition system was not custom designed in order to meet the JAMEX ASICs' requirements but it is near "off the shelf" general purpose electronics. Hence, it is natural to expect that some signal interfacing and conditioning will be required (see Figure 4.1) to interface these two realities. In order to achieve good noise performances and high supply efficiency, the JAMEX ASIC implements a pretty *tricky analog output* *stage*: while being a *differential* one indeed, its signals vary between **0** and **1.5 V** on the positive output and between **0** and **-1.5 V** on the negative one. Hence, the JAMEX has a *null common mode voltage* at the outputs that avoids any current flow at the terminating resistors when no charge is sensed. In this way, it is possible to minimize both the ASIC quiescent power dissipation and any induced noise on the supply rails when sensing low charges, hence increasing the total dynamic range.

But on the Sundance side we have the SMT356 ADC module that provides 8 `single ended` bipolar inputs accepting **-1** to **1 V**, for the full scale, and are terminated with 150$\Omega$ resistors. The J2S board must perform the required differential to single ended conversion, a convenient offset shifting to exploit the full resolution of the ADC converters and possibly some additional signal conditioning: a first order low pass filtering can reduce the overall noise input to the ADCs, a PGA (programmable gain amplifier) is useful to increase the flexibility and a built in signal clamping function helps to avoid ADC corrupted samples caused by input overvoltage. Another important function of the J2S board is to provide all the digital and analog signals needed to operate the 20 JAMEX ASICs, used in the detector, while embedding also some functionality to allow some basic testing of the detector operation without the exploitation of an X-rays source.

At a preliminary stage, it was supposed viable to integrate all the required interfacing electronics on a single board called J2S board, but with the passing of time and consequent progressive clarification of the final configuration of the detector, this original idea was superseded by a *distributed configuration*. During a parallel study on the custom box, dedicated to contain the whole detector, many constraints were imposed on the PCB design for the J2S board:

- avoid a PCB footprint bigger than the Sundance carrier boards, in order to preserve the compactness of the overall detector design

- accommodate some coaxial high voltage cables that are mandatory passing through the volume, of the detector box, dedicated to the J2S board

- create a convenient flux of air and optimized PCB surface to allow a reliable fan cooling of either the SAXS Hybrids and the J2S board analog electronics

The last point proofed to be not a minor concern in the overall design, since during the preliminary tests on the JAMEX ASICs it appeared that the noise of its input stage is strongly related to the operating temperature. Since the first stage is the most critical for the overall noise performance of the system, it is mandatory to provide proper cooling of the 20 JAMEX (dissipating in total a maximum of *10 W*, a substantial value for an embedded application) while avoiding any unnecessary source of additional power dissipation, thus encouraging the use of *low-power components* in the design of the J2S board.

The decision to split the complex functionality of the original J2S board in many PCBs required some additional considerations in order to achieve a final configuration with a high degree of optimality. Essentially, this configuration includes a PCB with the dimensions of a TIM-40 single Sundance module, called **J2S.SR**: it will be hosted on one Sundance carrier board and will be mainly responsible of buffering some clock signals for the ADCs, converting to LVDS the primary clocks for the JAMEX ASICs and offering some connectors aimed to an easier signal debug. Another similar PCB will contain the optocouplers and isolated DC/DC converters necessary to offer the required *galvanic isolation* to the experimental trigger inputs and outputs: it will be placed in the other free place on the Sundance carrier board and will be called **J2S.OPTO**.

A third and bigger PCB will embed most of the functionalities of the original J2S board and will be produced in two versions called respectively **J2S.MINI** and **J2S.FULL**: the first one implements all the J2S digital control functionalities required by a full 1280 channels detector. On the other hand, it supports only 256 analog channels and its purpose is to proof the validity of the PCB design before to produce the final full version. This will be a 6-layers PCB implementing all the required analog channels and will lack of the debug connectors present in the J2S.MINI, in order to save space.

The last PCB in this series is the first that chronologically was produced: it is called **J2S.PIG**, meaning it is aimed to be mounted “piggyback” on the SAXS Hybrids. For this reason, and assuming a final production of four SAXS detectors, this PCB was realized in many exemplars (*60 pieces*) from the professional company TVR, which accepted the standard *Gerber files* describing the PCB that were created using the circuit design program *Protel XP*.

## 4.2 J2S.PIG

The SAXS Hybrid board (ref. [32]) is a small (it measures 32$\times$50 mm.) 6-layers PCB, laminated on aluminum for dimensional stability and protective reasons, which mounts a couple of JAMEX ASICs via direct gluing and wire bonding. It features three very fine pitch proprietary connectors (produced from Fujitsu, with 80 pins and 0.5 mm pitch): two of them are connecting it to the detector’s anode microstrips while the last one gives access to all the JAMEX signals.

In the original idea it was expected to produce the J2S board in a *specialized multi-layer process* incorporating an *additional kapton layer* that would have offered some flexible extensions to the main PCB on which will be mounted the required connectors to interface with the SAXS Hybrids.

The complexity involved in producing such a PCB, the desire to have the JAMEX signals accessible from a less fragile connector than the Fujitsu one and also the need to offer some additional mechanical support to the SAXS Hybrids via the third connector inspired the design of the J2S.PIG PCB.

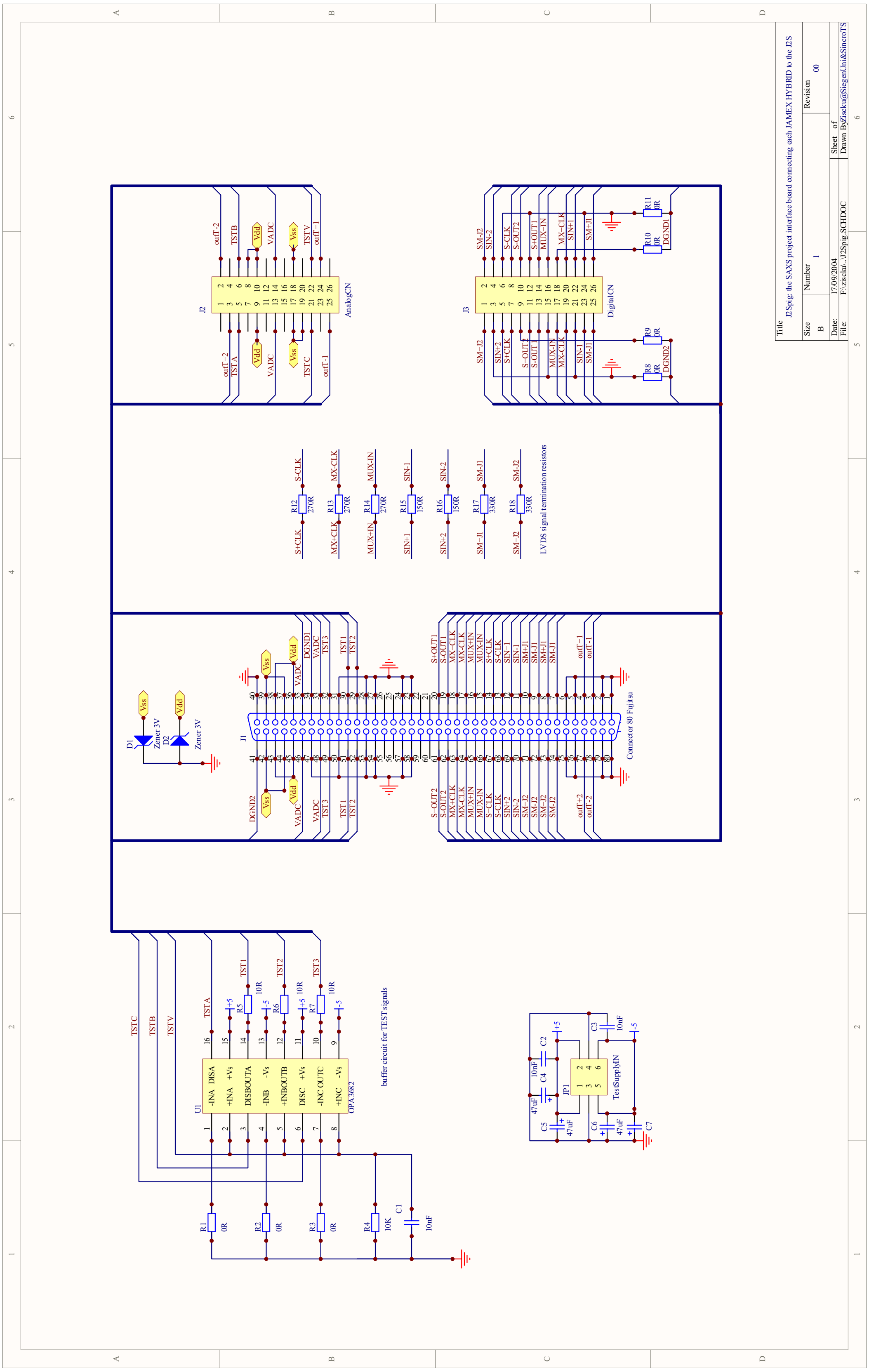

J2Spig: the SAXS project interface board connecting each JAMEX HYBRID to the J2S
AnalogCN
DigitalCN
Connector 80 Fujitsu
LVDS signal termination resistors
buffer circuit for TEST signals
OPA3682
TestSupplyIN

It addresses all these problematics and both incorporates the final buffering of the detector's internal test generator and terminates with resistors the LVDS control signals as close as possible to the JAMEX ASICs, since this feature is not provided on the SAXS Hybrids.

As it is possible to see from the schematic, this is not a complex design. Nevertheless, many imposed constraints made it very difficult to finalize and to produce. Since this small board was going to be used piggyback to the SAXS Hybrids, which are stacked side to side on the anode with only 1 mm gap between each other, it needs to be no wider than the 32 mm of the Hybrid and no longer than 70 mm, in order to avoid any mechanical stress to be applied to the edge connector.

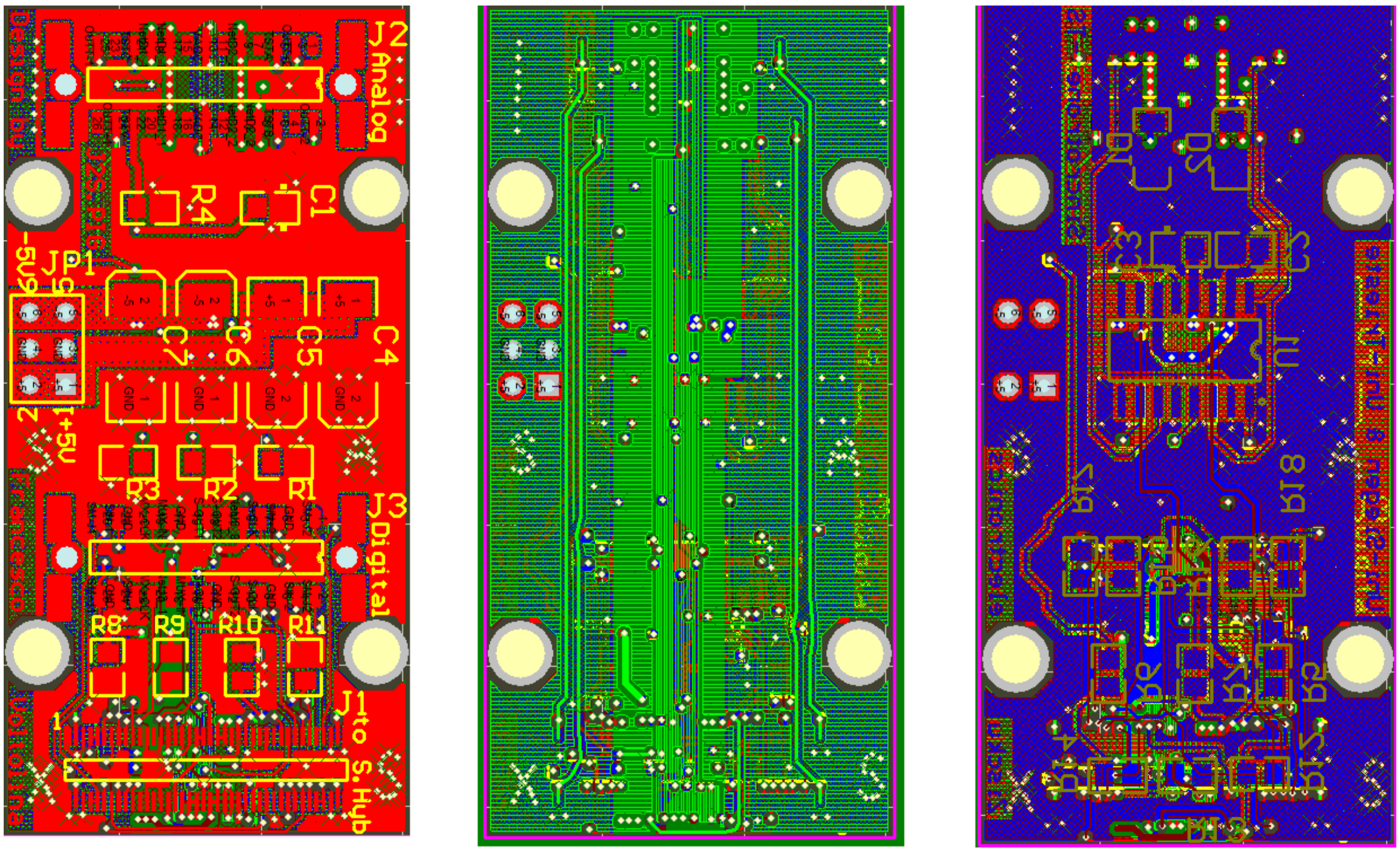

Figure 4.2: The J2S.PIG pcb layers as seen in Protel XP: top layer and ground plane -left-, analog and reference layers -center-, digital and bottom layers -right-

After a careful selection of the components it has been decided to produce the final PCB as a *6-layers board* in order to have enough freedom to route conveniently apart, and to separate with ground planes, the digital signals from all the analog supplies, reference voltages and output signals (see Figure 4.2). The PCB was produced with 2 mm thick FR4 material in a final dimension of 32×60 mm and an *electrochemical deposition of gold* over the copper pads was performed in order to facilitate the successive soldering of the very fine pitch components. The minimum track and gap of the PCB is 0.1 mm and most of the vias are characterized by a hole diameter of 0.2 mm: only in this way it was possible to shrink the PCB dimensions while routing all the required signals out of the small 80-pins Fusjitsu connector (see the rightmost connector in Figure 4.3 -left-).

The J2S.PIG board routes and separates the signals available at the SAXS Hybrid 80 pin connector into two 26 pin, 0.8 mm pitch, ERNI connectors: one is dedicated to the *analog signals*, while the other to the *digital connectivity* (see the Schematic for the effective pinouts). This separation is convenient in order to minimize the interference between digital and analog signals when they will be transferred on the approximately 15 cm of flat cable connecting every J2S.PIG to the J2S.MINI or J2S.FULL board. Convenient low-voltage SMD zener diodes are provided on the PCB to smooth any eventual voltage spike or misregulation artifact on the JAMEX ASICs supply lines: they offer a good protection since their conductance rises very fast when the supply voltage exceeds 2.7 V.

The only *integrated circuit* present on the PCB is the **OPA3682** (ref. [20]) from Texas Instruments (see U1 in Figure 4.3 -right-). It is the same device that has been already presented in the First Chapter and used to implement the MIDA multiplexer. A similar configuration is used here in order to provide a selective routing of the TEST signals reference voltage, produced on the J2S.MINI or J2S.FULL and buffered with this triple fixed-gain amplifier from TI, in order to easily drive the three, 50Ω terminated, TEST inputs on the SAXS Hybrid.

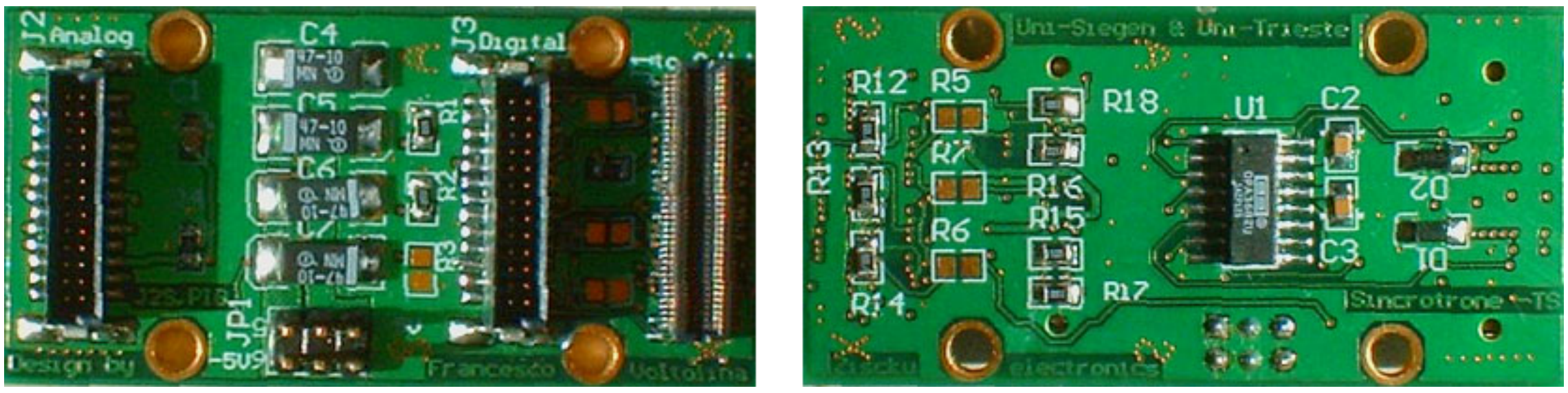

Figure 4.3: The final PCB of the **J2S.PIG**, with mounted components, is reproduced in this two pictures in its *real dimensions*: the top view is on the left, while on the right is the bottom one

The special design of the detector box in conjunction with the component placement on the J2S.PIG allows a direct cooling of this OPA3682 integrated circuit through a copper “heat pipe”. All the heat, dissipated by the 10 devices used in a complete detector, will be conveniently removed from inside the detector box by the heat pipe, increasing the overall efficiency of the primary fan cooling system.

## 4.3 J2S.MINI

The J2S.MINI has been designed and built to *evaluate in a short time* the performance achievable from the components selected to build the final full-version called J2S.FULL. In order not to incur in the considerable expenses related to the fixed costs of a small PCB production from a professional company without to have a proof of the validity of the chosen design, it was decided to produce this test PCB as a *2-layers board* of small dimension in order to try to find out any possible critic problem of the design before to produce the final one.

A preliminary evaluation of the components required in this design was performed after having built a rough prototype of the basic *analog J2S board channel* (see the symbolic representation in Figure 4.1) on an experimental PCB board with embedded ground plane. This circuit used *two key-components*: the **AD8130**, a 270 MHz Differential Receiver Amplifier (ref. [22]) from Analog Devices, and the **THS7002**, a programmable gain amplifier (ref. [33]) from Texas Instruments.

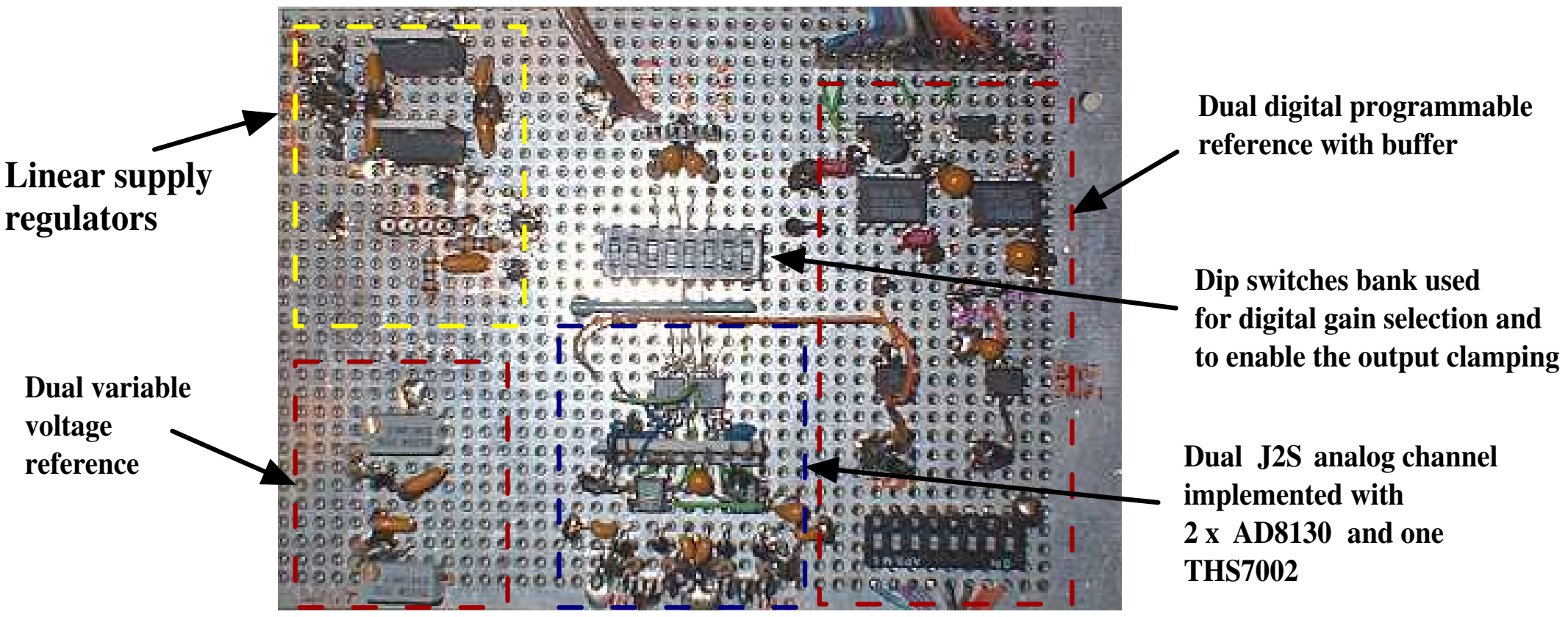

Figure 4.4: First prototype of the J2S board analog channel, built on an experimental PCB with the help of a microscope to help soldering of the very fine pitch TSOP components

The JAMEX differential outputs are fed through a *first order RC low pass filter* to the high impedance inputs of the AD8130 that is designed as receiver for the transmission of high-speed signals over twisted-pair cables: it has a high Common Mode Rejection Ratio even at high frequency and here it is used for converting the JAMEX differential signals to a single-ended one. The high CMRR of this device, approaching 70 dB at 10 MHz, allows the use of *unshielded flat cables* to transfer the JAMEX output signals without fear of corruption by external noise sources or crosstalk.

A definitive advantage of this device over a conventional operational amplifier is the ability to change the polarity of the gain merely by swapping the differential inputs. This trick is used to implement the offset subtraction without any additional component and requiring just a positive reference voltage (this characteristic will proof to be very useful in the J2S.MINI, where an unipolar positive DAC output is used to implement the adaptive offset subtraction feature without any additional component).

The single ended output signal is directly connected to one channel of the THS7002 high-speed programmable gain amplifier: each channel on this device consists of a separate low noise input preamplifier and a programmable gain amplifier (PGA). The *preamplifier* is a voltage feedback type offering a low 1.7 nV/$\sqrt{Hz}$ voltage noise with 100 MHz bandwidth, which was not used in the prototype (it will be used to implement, on the J2S.MINI, all the accessory buffering stages since the input and output pin are fully accessible), whereas the AD8130 output signal is directly connected to the 3-bit digitally controlled PGA that provides a **-22 dB** to **20 dB** attenuation / gain range with **6 dB** step resolution. In addition, the PGA provides separate high and low *output clamp protections* that are useful to prevent the output signal from swinging outside the input range of the ADCs present on the SMT356 modules. The PGA provides a wide bandwidth of 70 MHz, which remains relatively constant over the entire gain / attenuation range, and a low input noise that suits very well our requirements.

Both these devices allows a low voltage operation with a dual $\pm 5$ V supply: this permits to achieve a low power dissipation that is perfectly suited to our embedded device, which employs a conservative voltage swinging amplitude of the analog signals in order to minimize distortion. Despite the prototype board was featuring a quite unusual "dead bug" mounting (see Figure 4.4) for the usually critical high-bandwidth operational amplifiers, it manifested no tendency to oscillatory behaviour and gave very encouraging results in the preliminary measurements, indicating a good suitability to the required noise and dynamic characteristics.

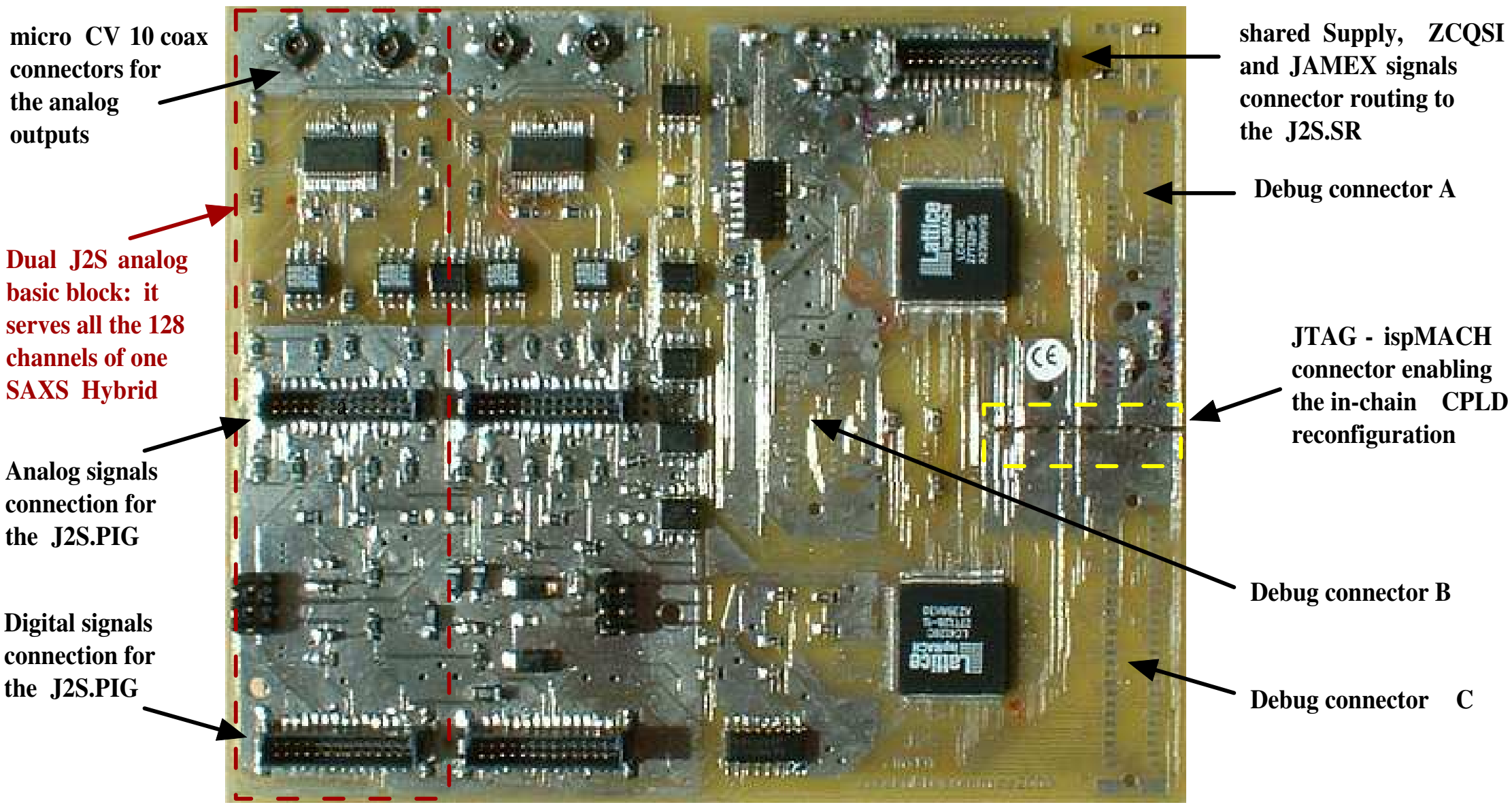

Figure 4.5: Picture of the J2S.MINI PCB *front side*

This fact rose the interest in building a more refined board, the J2S.MINI (see Figure 4.5 or 4.6 for the final picture with mounted components or Appendix C for the schematic and PCB), offering many additional functionality. We wanted to include not only an evolution of the circuit, which has been already tested in the prototype, used for the signal conditioning and conversion of the JAMEX analog output signals, but also to integrate the additional logic to provide the 20 JAMEX ASICSs with all the necessary signals for a correct operation (in connection with the Sundance system) during both the *Programming* and *Operating* mode.

This would require the implementation of a full featured *ZCQSI interface* and command set (explained in the previous Chapters) that necessary requires a convenient CPLD to be chosen in order to realize this design. It was decided to use a CPLD from `LATTICE Semiconductor` and in particular the **ispMACH 4128C** (ref. [34]) that is available in a very compact 128 pin TQFP Package with a fine pitch of 0.4 mm. It features a very *low power dissipation* combined with *low propagation delay* and fully configurable inputs and outputs banks. The relatively limited number of register and I/O pins suggested the use of 2 of these devices in order to implement the complex combinatorial logic, while allowing some more headroom for any modification, required in the future, that could be easily accommodated through the *In System Programming* characteristic of these devices.

From an electrical point of view the signals required to control the JAMEX ASICs are needed to be according to the *LVDS* standard: in particular the input pins are true LVDS (they can be terminated with 100 or 50 Ω) while its outputs are *pseudo LVDS*, since the common mode voltage is correctly set around 1.25 V, but they manifest a full ±1.25 V swing and they don't accept any resistive termination.

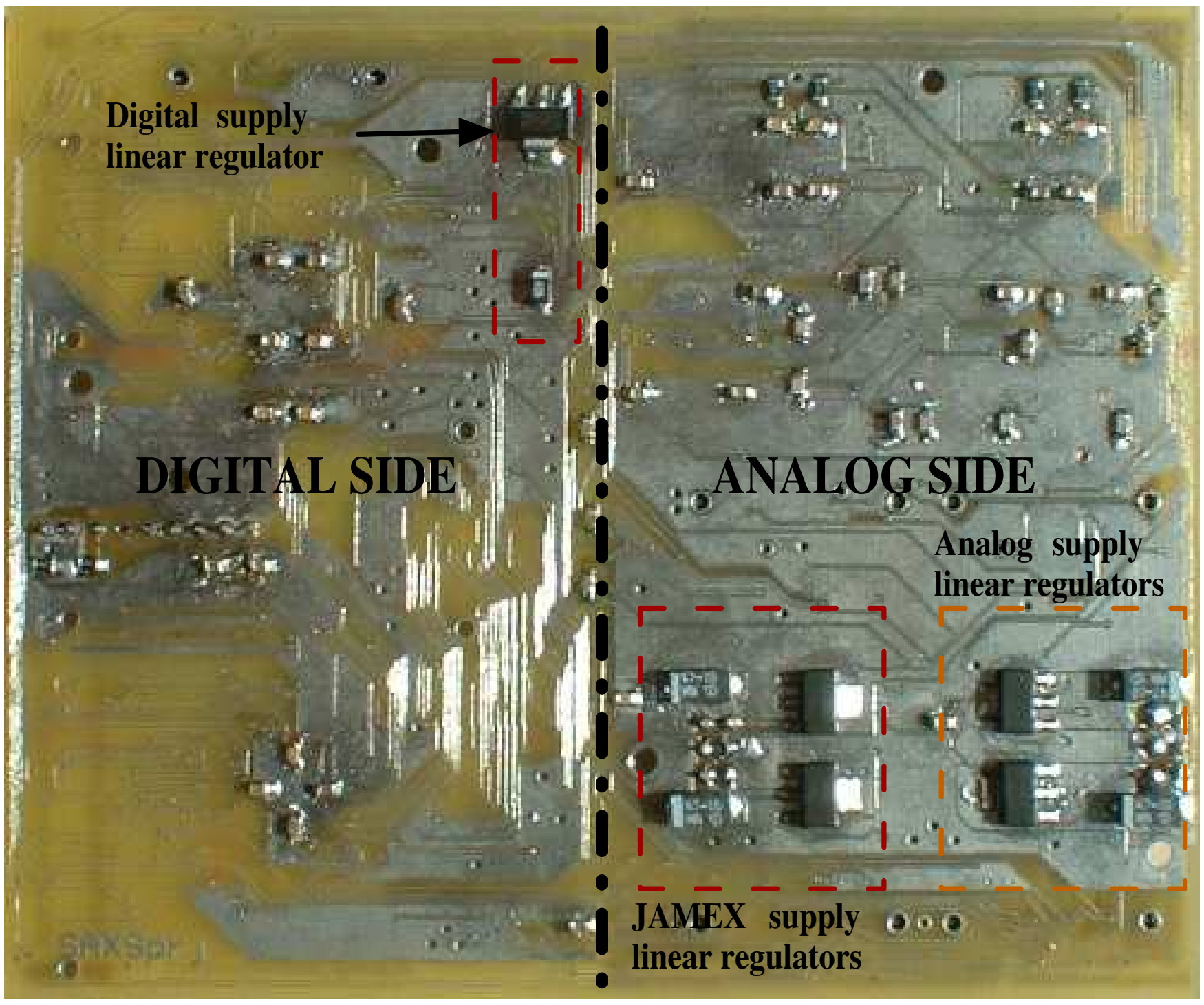

Figure 4.6: Picture of the J2S.MINI PCB *bottom side* in real scale

The chosen MACH CPLD does not embed any LVDS capability, but it is possible to fulfil those requirements using standard electronics components that perform the driver and receiver functions: the parts used in this design are the **SN65LVDM31** (ref. [35]) and **SN65LVDS32** (ref. [36]), respectively a High-Speed Differential LINE DRIVER and a RECEIVER from Texas Instruments. They contain either 4 receivers or 4 drivers per chip that are used to translate the LVDS signals into conventional LVTTL signals and vice versa: in this way it is possible to control the JAMEX ASICs via any standard LVTTL digital logic device. From the schematic (see Appendix C) it is possible to see that the J2S.MINI board allows also a direct connection of the JAMEX pseudo LVDS outputs to the MACH device in order to evaluate the possibility of omitting these components in the final design.

Not only the linear regulators for all the required analog and digital voltages are present on this board, but also some DACs, the low power SPI device TLV5638 from Texas Instruments (ref. [37]), are employed in order to create the required voltage references for an optimal operation of both the JAMEX ASICs and the integrated signal generator. This last feature allows the injection of a *programmable amount of charge*, via the three test lines present on each JAMEX ASIC that are capacitively coupled to their inputs pads, simulating the operation in presence of X-rays.

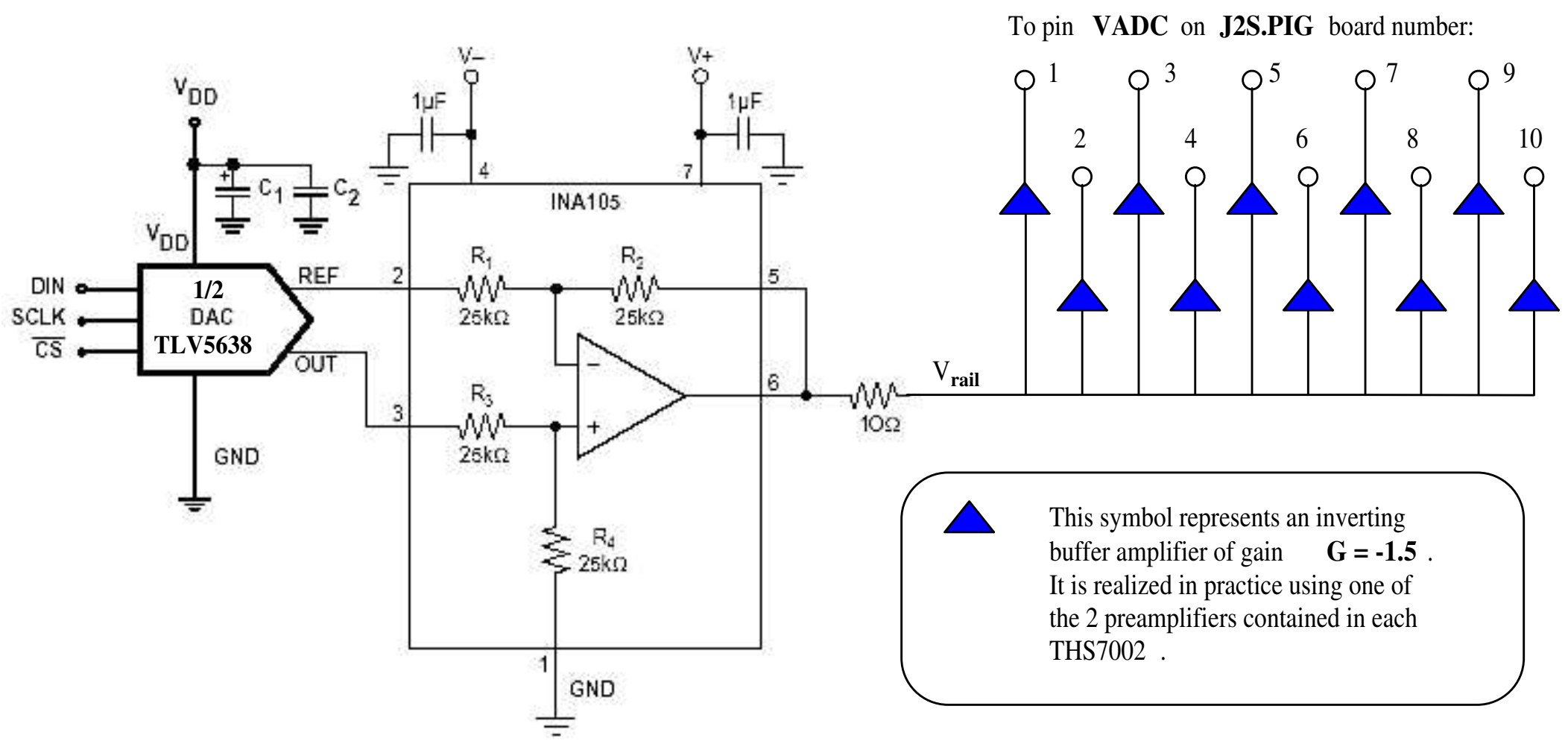

Figure 4.7: The simple circuital configuration used to generate the required *variable reference voltage* for the JAMEX input stage negative supply: it feature a range of ±3 V

The arbitrary amplitude step signal generation and control circuitry is implemented on the J2S.MINI and it uses the OPA3682 buffer amplifiers hosted on the J2S.PIG PCBs as a final routing and signal buffering device. A similar circuit (see Figure 4.7) is used to provide the variable reference voltage for all the JAMEX ASICs: in this case, since a low maximum current of 1 mA is required for each ASIC, the buffer functionality will be implemented not with the OPA3682, but using the independent preamplifier sections contained inside the THS7002 dual channel integrated PGA circuits. Particularity of this stage is the ability to provide both the negative and positive voltages needed to assure a *proper operation* and a *stable power down condition* to the JAMEX ASIC.

This PCB was designed, since the creation of the schematic, with *Protel XP* and its final realization was authored by `PCB-Pool`. This company is based in Germany and produces very high quality PCBs from standard Gerber files at very competitive prices since they use a proprietary method of combining the individual customers layouts in panels of bigger dimensions. In this way the fixed costs are shared and it is possible to produce in short time single prototypes at low cost, without to sacrifice the quality and allowing the very small feature sizes and vias required in today's high-density designs.

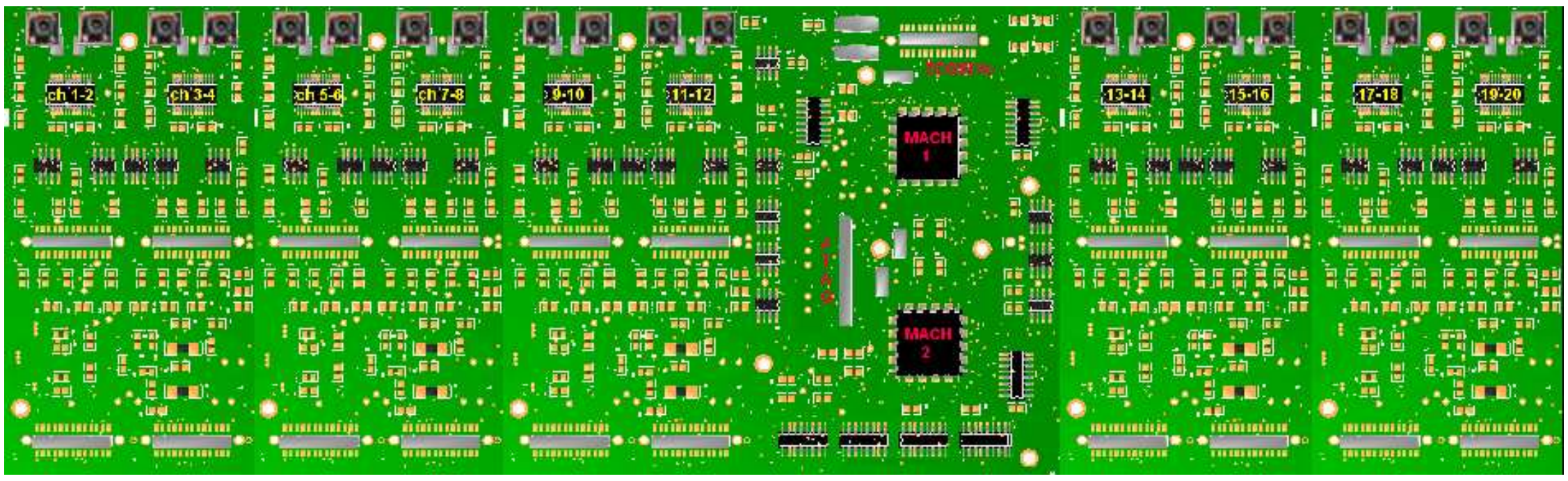

Figure 4.8: A pictorial view of the final J2S.FULL PCB generated with Protel XP

With the help of the electronic group at the University of Siegen, using a specialized *rework station*, both the MACHs and the THS7002 devices were optimally soldered to the J2S.MINI PCB, while the rest of the components was placed and soldered by hand. After some quick routine test and a safe power up of the board, the In System Programming of the two MACH CPLDs, connected in daisy-chain, was *completed successfully* via a PC Parallel JTAG "dongle", used in conjunction with ispVM System, the LATTICE's proprietary software programmer.

This design showed with time an impressive reliability and convenience, thus it will be implemented with minor changes in the final J2S.FULL board (see Figure 4.8). While the digital section will require just few refinements, a particular care will be taken in routing the analog signals on the final board, possibly using the same strategy as in the J2S.PIG small PCBs, in order to minimize noise and crosstalk.

## 4.4 J2S.SR

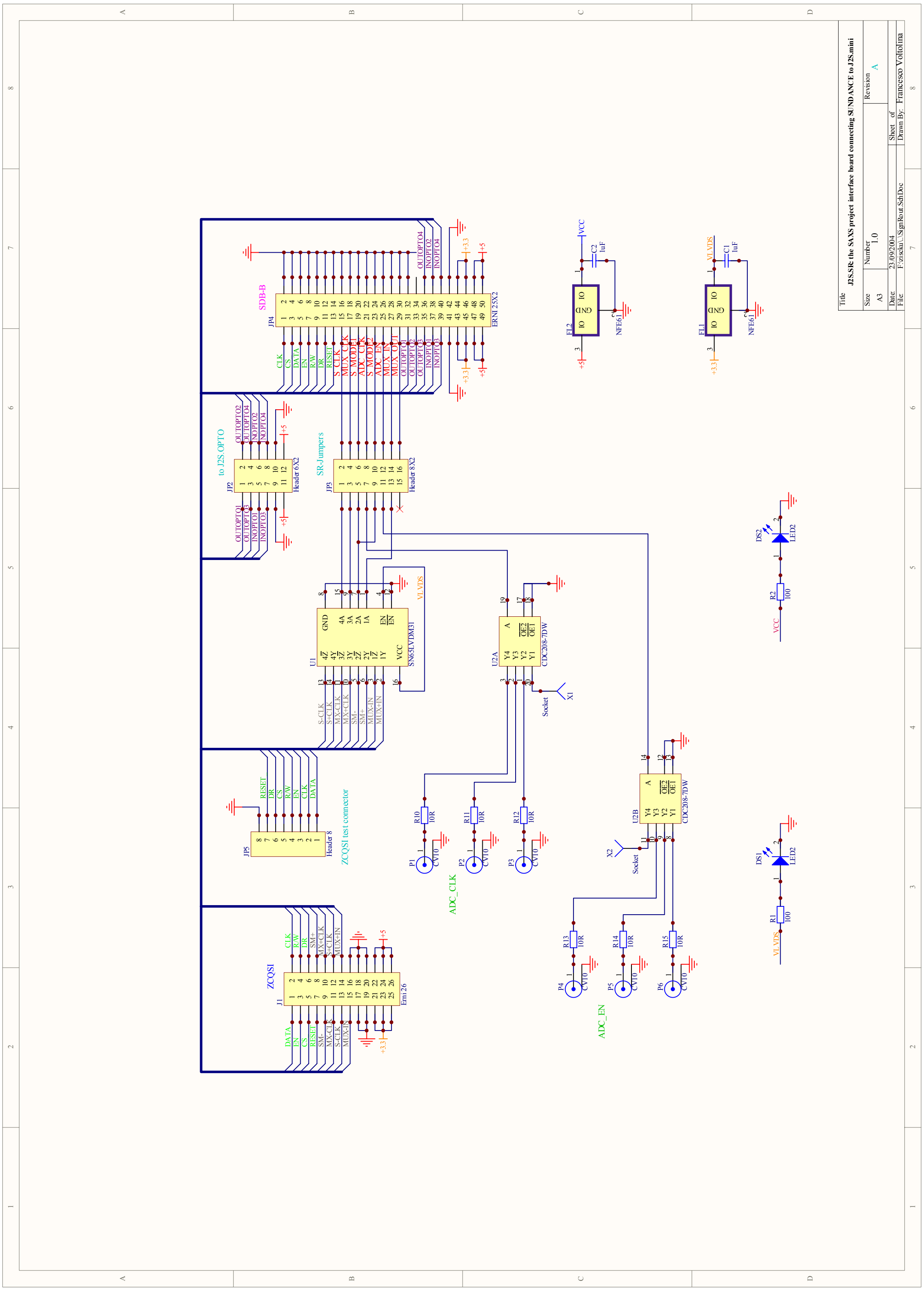

J2S.SR: the SAXS project interface board connecting SUNDANCE to J2S.mini
Revision A
Size A3
Number 1.0
Date 23/09/2004
Drawn By: Francesco Voltolina

The main concern in the whole SAXS detector design is noise: in specific, the lowest is the *electrical noise* produced inside the detector box, the highest performance, in term of available dynamic range at a fixed gain, can be expected when using the detector in a real measurement.

A weak point of the *old design*, featuring a single J2S board, was the following: the LVTTL signals relative to the ZCQSI interface and primary clocks for the JAMEX ASICs were transferred via flat cable from the Sundance acquisition system to the J2S board. Thanks to the constraints imposed on the ZCQSI interface, not to have any signal switching during an active measurement, that signals were not a concern, but the primary clocks of the JAMEX, necessary active during a measurement and traveling on the flat cable in proximity of the analog channels, were a potential risk.

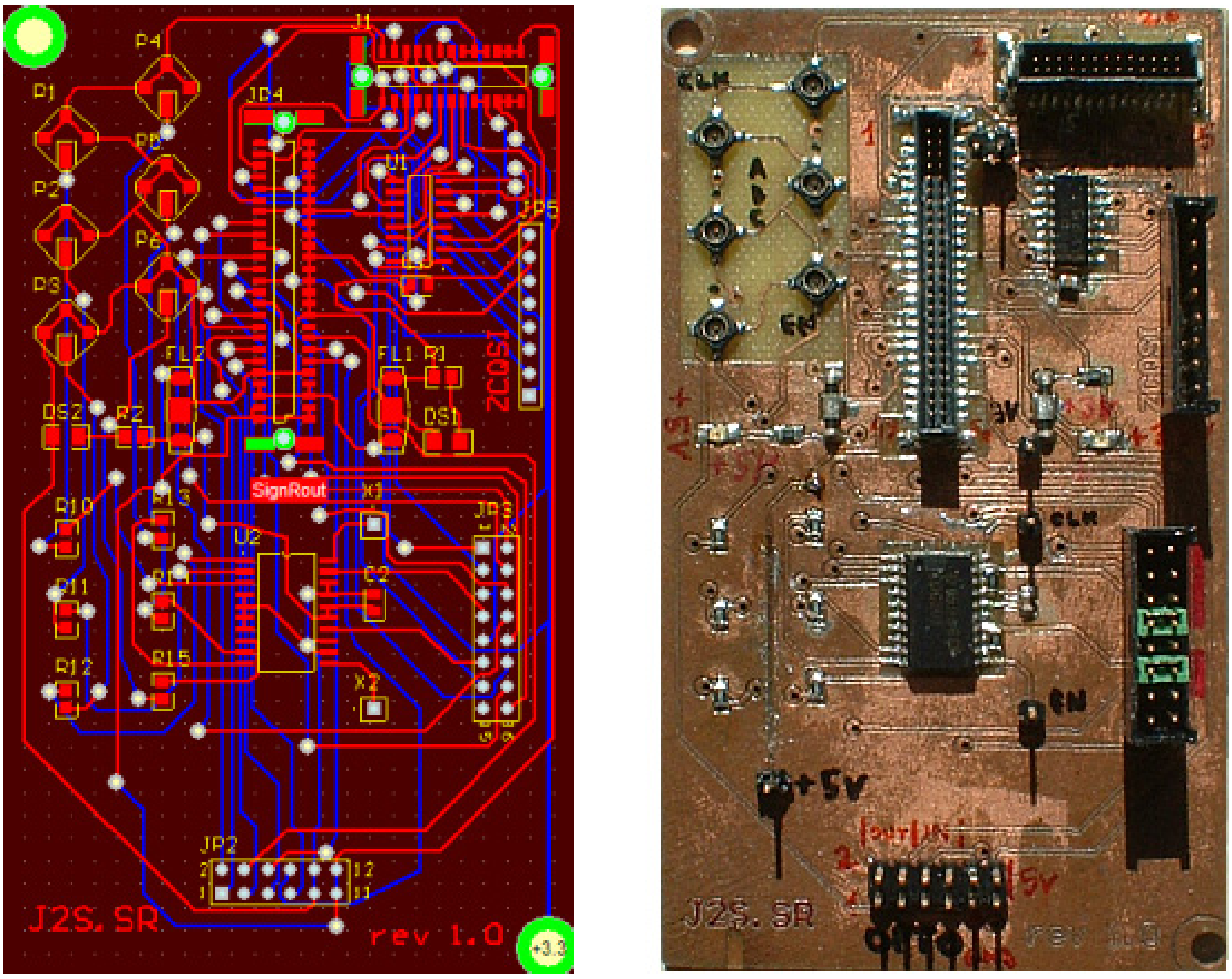

Figure 4.9: Picture, in real scale, relative to the J2S.SR PCB as seen in Protel XP -left- and a photo of the same board with already mounted components -right-

The solution to this problem came out from the J2S.SR design: this board is hosted on a free TIM-40 module site of the Sundance acquisition system and translates the potentially interfering LVTTL clock signals to LVDS, before to be routed on the flatcable. In this way any interference radiated from the flat cable connecting the J2S.SR to the J2S.MINI board will be minimized. The convenient placement of this board on the Sundance acquisition system suggested the idea to implement on it also the signal buffering functionality needed to create and route the three copies of the `ADC_CLK` and `ADC_EN` signals required for the ADC modules.

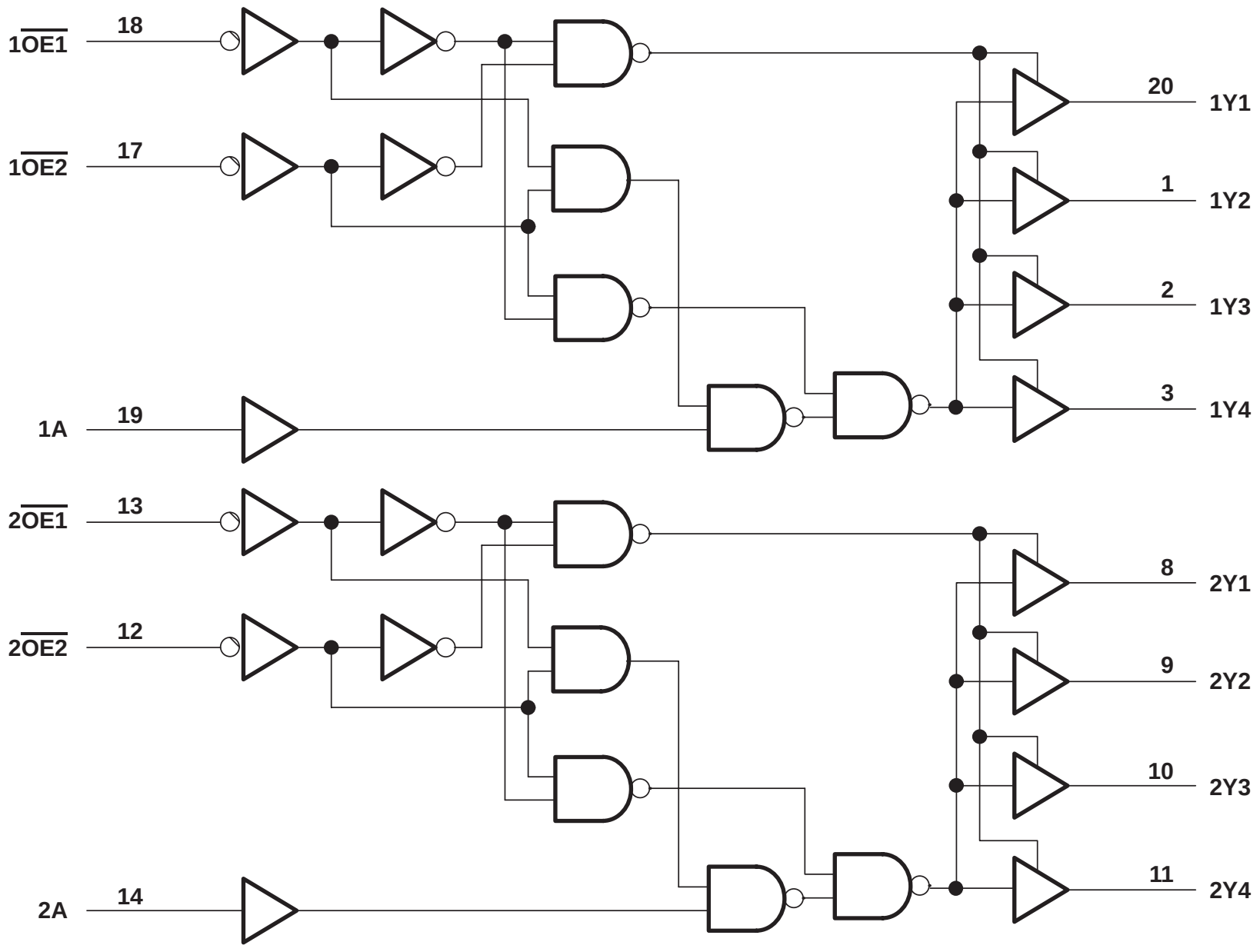

Figure 4.10: Logic diagram relative to the CDC208 CLOCK DRIVER

This functionality is realized by the **CDC208** CLOCK DRIVER with 3-STATE OUTPUTS from Texas Instruments (ref. [38]), which contains dual clock driver circuits that fanout one input signal to four outputs. It features minimum skew on the short propagation delay of 8 ns and is especially suited for clock distribution (see Figure 4.10). The J2S.SR board mounts the same micro-coaxial connectors (CV10 from JAE) present on the Sundance ADC modules and the relative connections are actuated via custom produced CV10 to CV10 cables.

Some LEDs to indicate the presence of the digital supply rails and many connectors to access all the signals available on the modified SDB connector are also present on this board. In particular, it was conceived the possibility to control all the J2S boards and JAMEX ASICs via another external DSP board (the *EVM56002* from Motorola) using these connectors. This possibility resulted very important to debug the entire system and will be presented in the next chapter.

# 5

# SAXS detector prototype integration and measurements

Since the beginning, the design phase of this SAXS detector was characterized by a close interaction between the specialized electronics and the environment to be created for it. In many cases the electronics dictated the overall shape of the housing for the detector, as in the case of the fixed dimensions of the Sundance acquisition system, but in general, and this is especially true if to consider the J2S board family, also the electronic design had been deeply influenced by the mechanical constraints.

This is especially evident if to consider the anode, already finalized well before the electronic design had started, whose integrated mechanical support for the SAXS Hybrids suggested the idea to split the original J2S board in a family of PCBs that included the J2S.PIG devices aimed to support the free sides of the hybrids.

The successful realization of the housing design of this SAXS detector has been made possible by the support of the mechanical design office at Elettra and specifically the person of Claudio Fava. He translated, using the 3D modeling CAD program *ProEngineer*, a huge amount of information, contained in many free hand sketches, into the final production drawings.

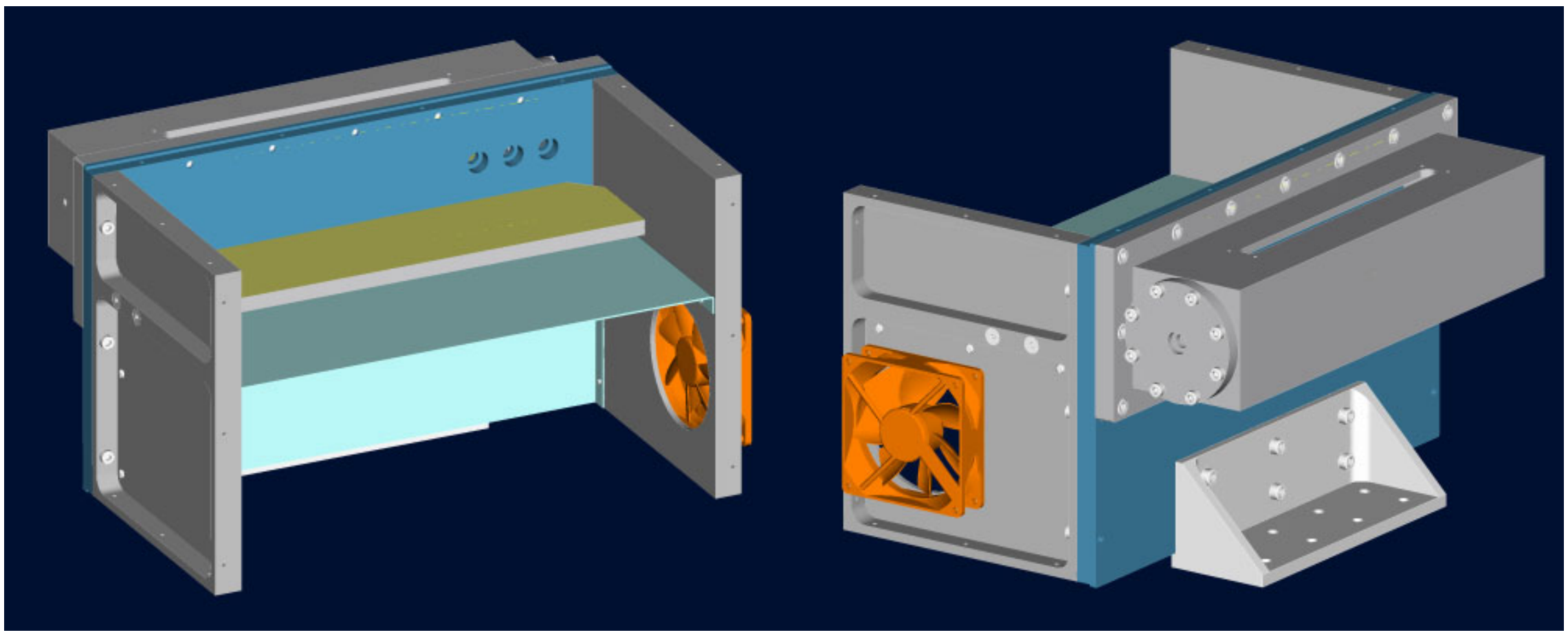

Figure 5.1: Two photorealistic views of the SAXS detector generated with ProEngineer

## 5.1 Integrating the different components

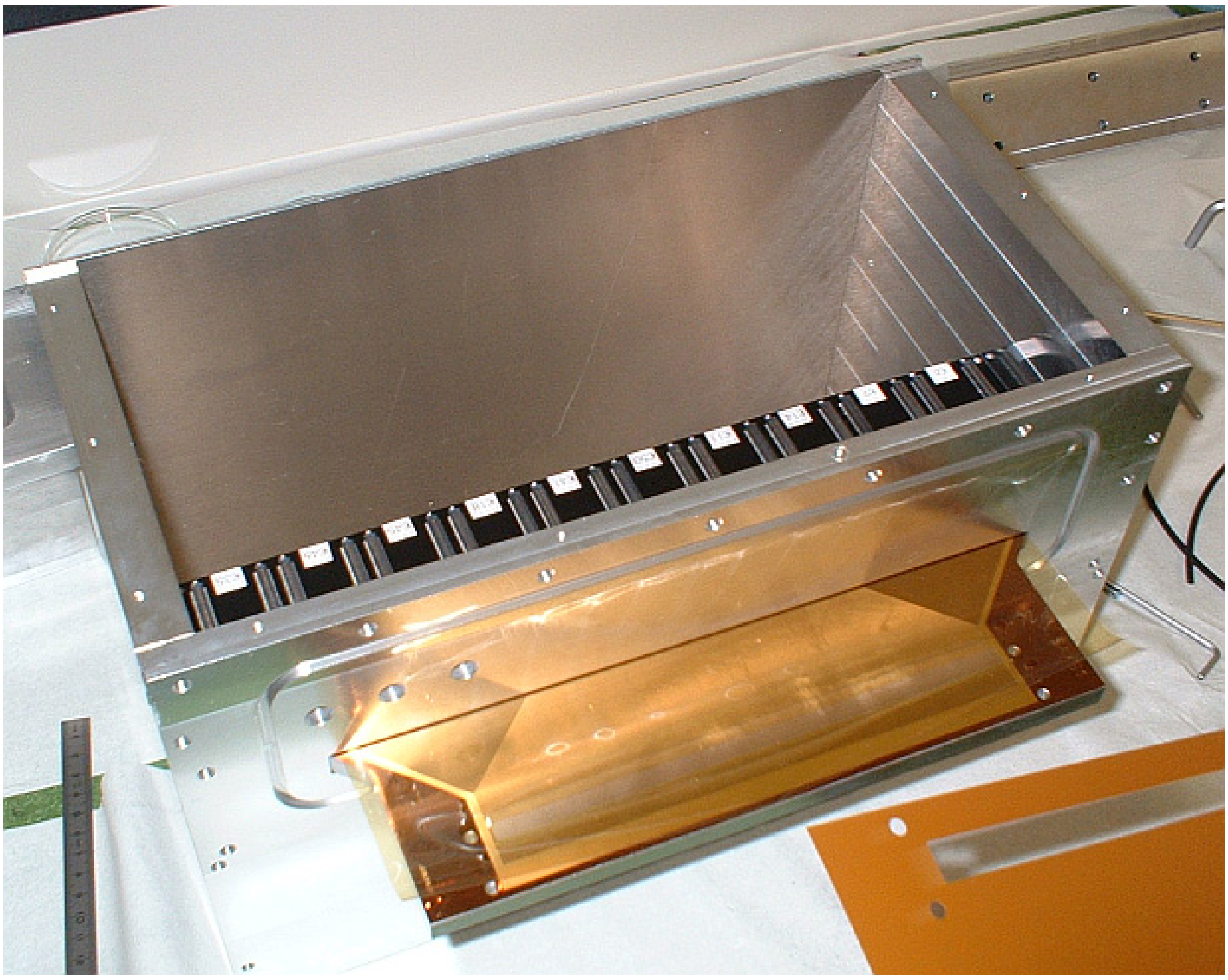

Figure 5.2: Picture of the SAXS detector housing and glued anode, missing the pressure vessel

The SAXS detector housing was produced in the mechanical workshop at the University of Siegen, using a numerically controlled milling machine. The anode was precisely glued in *clean room*, in order to avoid any contamination that could adversely affect the whole detector operation (see Figure 5.2).

Some preliminary tests involved mounting the PC Embedded, its accessory components and the complete Sundance acquisition system in the half box of the detector that is dedicated to host the digital electronics (see Figure 5.3).

An external modified Personal Computer power supply was used, and is still in use, in order to run the system, while the final version of the SAXS detector will feature a custom designed power module with integrated microcontroller. This compact, external power supply will provide either the digital voltages for the acquisition system and the, linearly regulated, analog voltages dedicated to the J2S boards and JAMEX ASICs.

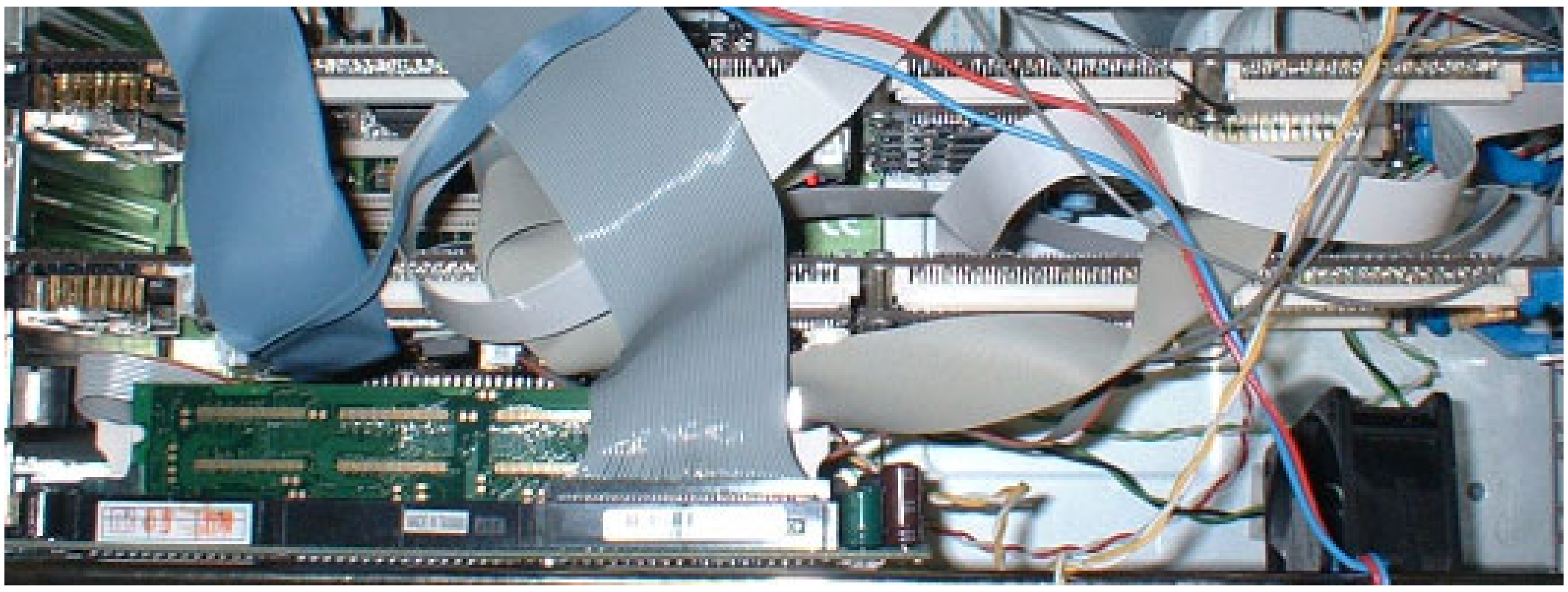

Figure 5.3: Top view of the acquisition system, mounted inside the tight box of the detector

A correct power sequencing is needed to avoid *latch-up* problems to the JAMEX: while in the actual detector this is provided manually, in the final supply the sequencing will be decided from the integrated microcontroller that will be also responsible for detecting any abnormal condition on the supply rails.

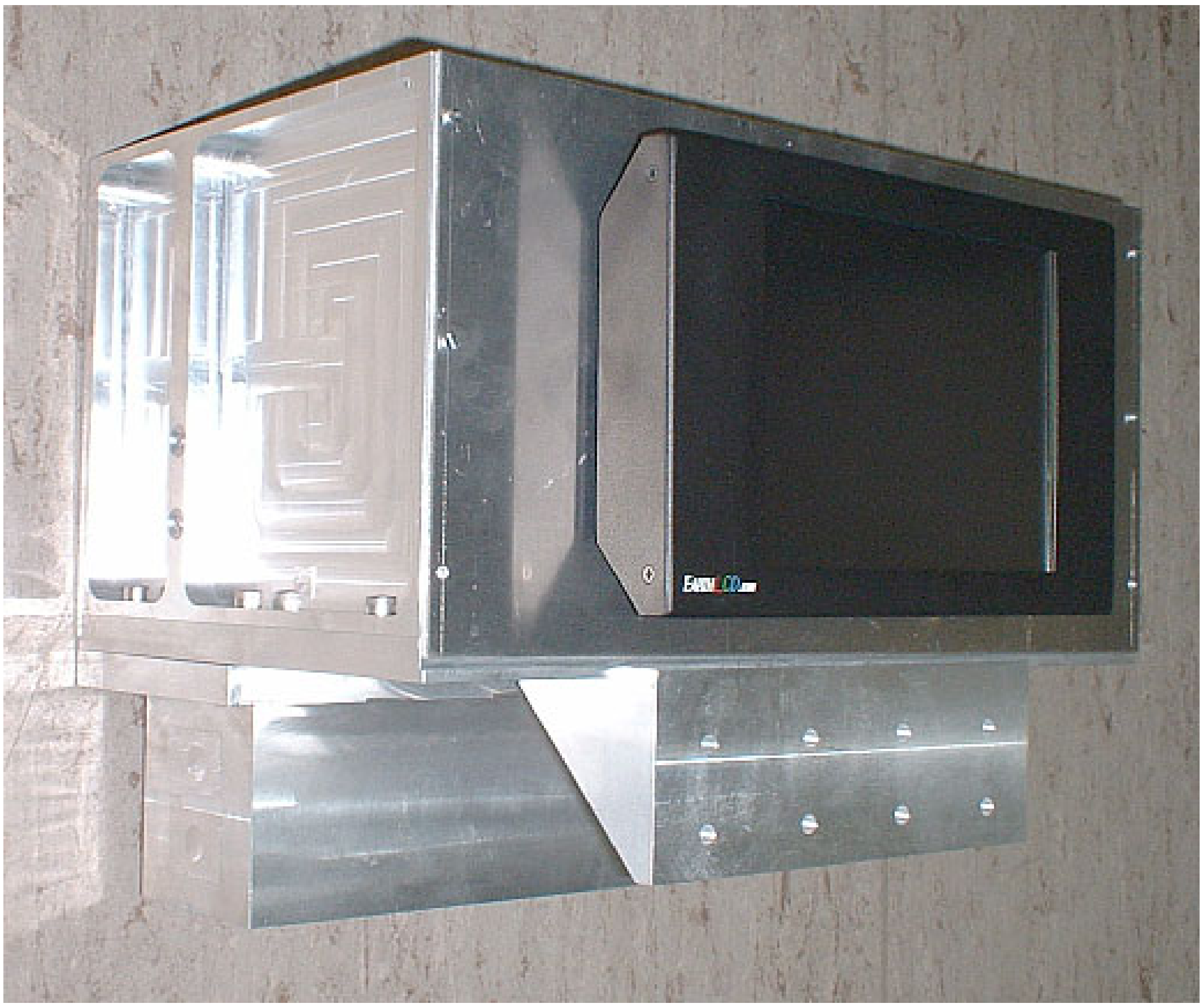

Figure 5.4: Side view of the aluminum detector housing with integrated industrial LCD monitor

An external industrial LCD panel with touch screen (see Figure 5.4) is provided to give the possibility of accessing the detector, during the measurement setup phase, to an eventual user located in the *experimental hutch* of the beamline. Otherwise the screen can be switched off in order to minimize any interference.

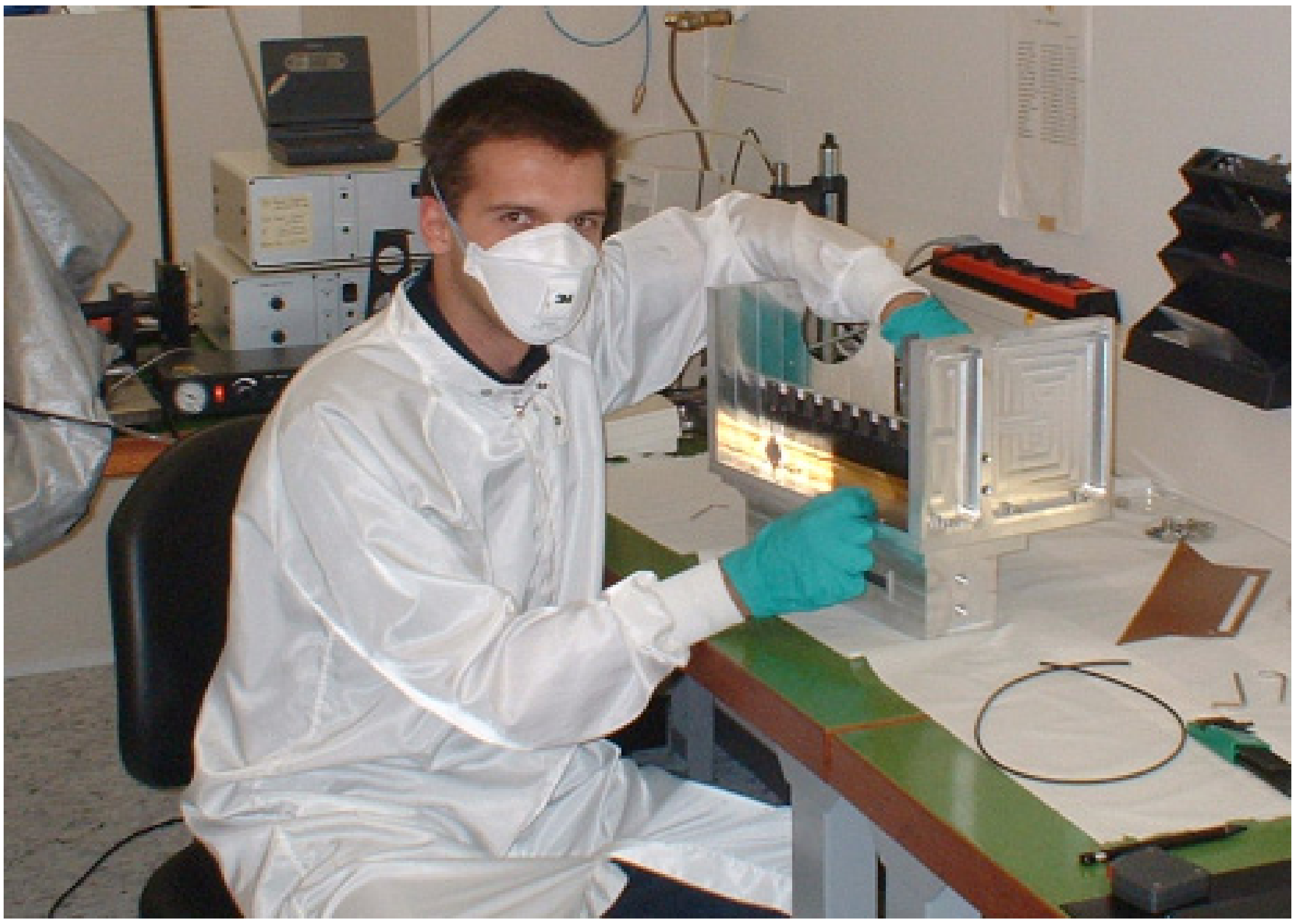

Figure 5.5: Picture showing a preliminary mounting of the SAXS Hybrids on the anode, performed in the *clean room* of the FEC Group at DESY, Hamburg

The acquisition system and relative PC Embedded were fitting correctly the SAXS detector housing but they was not conveniently accessible: before to proceed further with the measurement on the detector, it was decided to move these components to a standard PC case. In this way, it resulted easier to mount the 10 SAXS Hybrids on the detector's anode (see Figure 5.5) and to perform some noise measurements on the JAMEX ASICs. Before to give this results, in the following section will be presented an application, created for the Sundance acquisition system, used to visually evaluate the noise performance of each analog channel.

## 5.2 Custom software used for the measurements: Zscope

Since the original JAMEX application was not offering any direct possibility to evaluate the progressing of a measurement nor the integrity of the acquired data, it has been decided to create a far simpler application, with a real time graphical output, to address these needs.

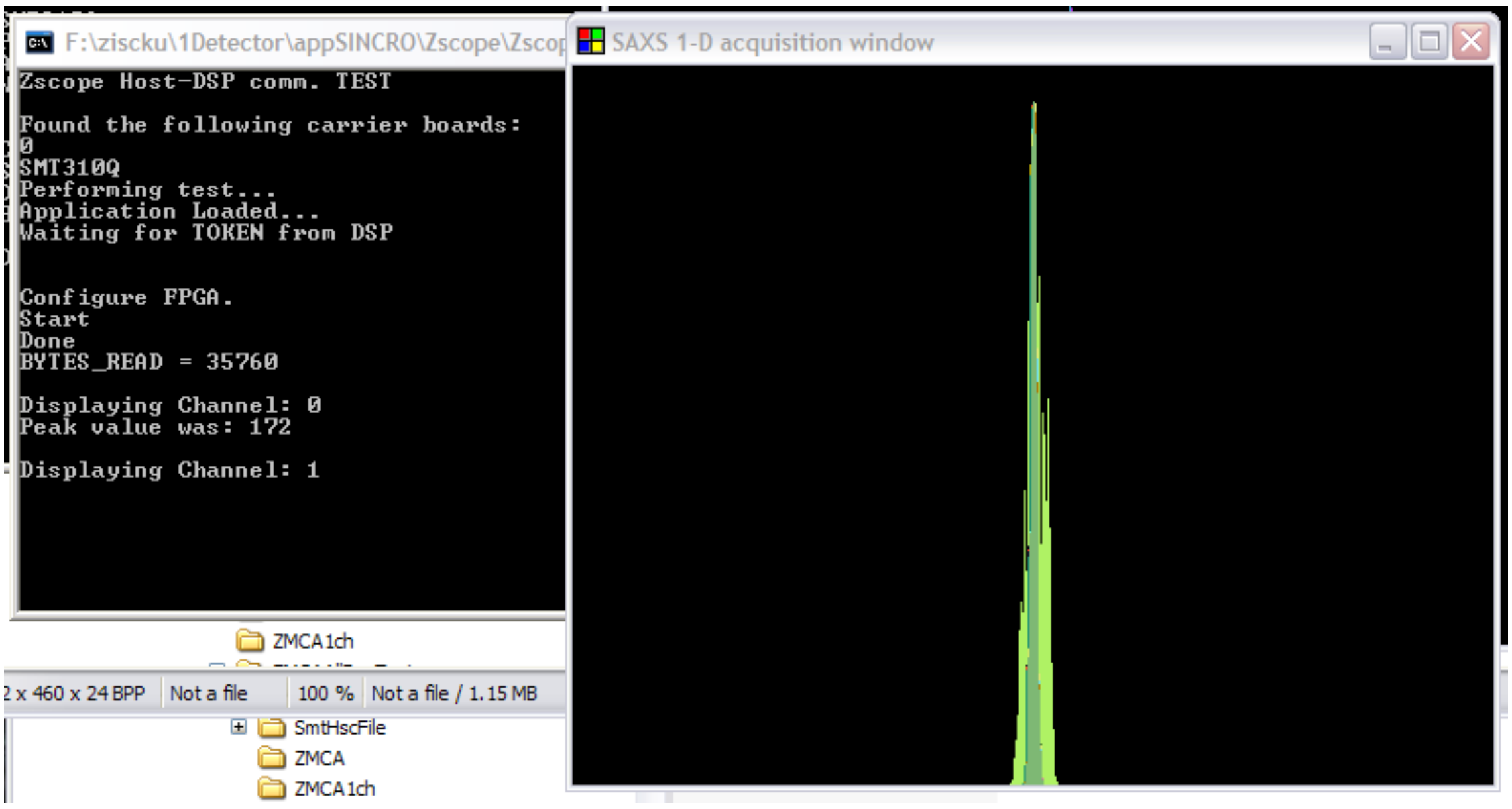

Figure 5.6: Screen capture relative to the execution of the Zscope application

The Host software is developed using Microsoft Visual Studio while the DSP software is developed using 3L Diamond and Code Composer Studio, the main files are available in Appendix D.

The two workspaces are called respectively ZsCPgraph.dsw and ZMCA1ch.dsw.

The `ZsCPgraph.dsw` is the workspace that implements the graphical presentation of the data, based on the OpenPTC library, and the drivers to communicate with the DSP embedded system.

The `ZMCA1ch.dsw` is the workspace that implements the DSP application responsible for the real time acquisition of one ADC channel at a time, selected sequentially.

The Zscope application acts like a basic Multi Channel Analyzer (MCA): the actual sampled value increments the correspondent bin counter while the graphical result is continuously updated many times per second and normalized to full height.

Different options are available to represent the results on the graphical window: graphs can be magnified, automatically centered around the main peak and also a *visive persistence* feature is realized that helps the comparison between different channels (like in Figure 5.6 where it is possible to see, represented in two different colors, the histograms showing the different open-input noise of two consecutive channels).

This program resulted very useful to indicate noise related problems in the system and to identify the noisiest channel of each ADC module. It was also modified in many ways in order to accommodate variable sample rates, the possibility to save the acquired data in a RAW format or to perform specific processing on the acquired data.

## 5.3 The configuration used for the noise measurements

A first preliminary measurement of the JAMEX output noise was performed in Hamburg with the original test equipment of Werner Buttler. The detector board channels have been grouped in three for test purposes and excited by connecting test input 1, 2 and 3 sequentially to a pulse generator. By overlaying all data streams in one plot, like in Figure 5.8, all 64 channels can be seen and compared. Furthermore, the offset signal stream has been monitored and added to the plot, indicated with a black line.

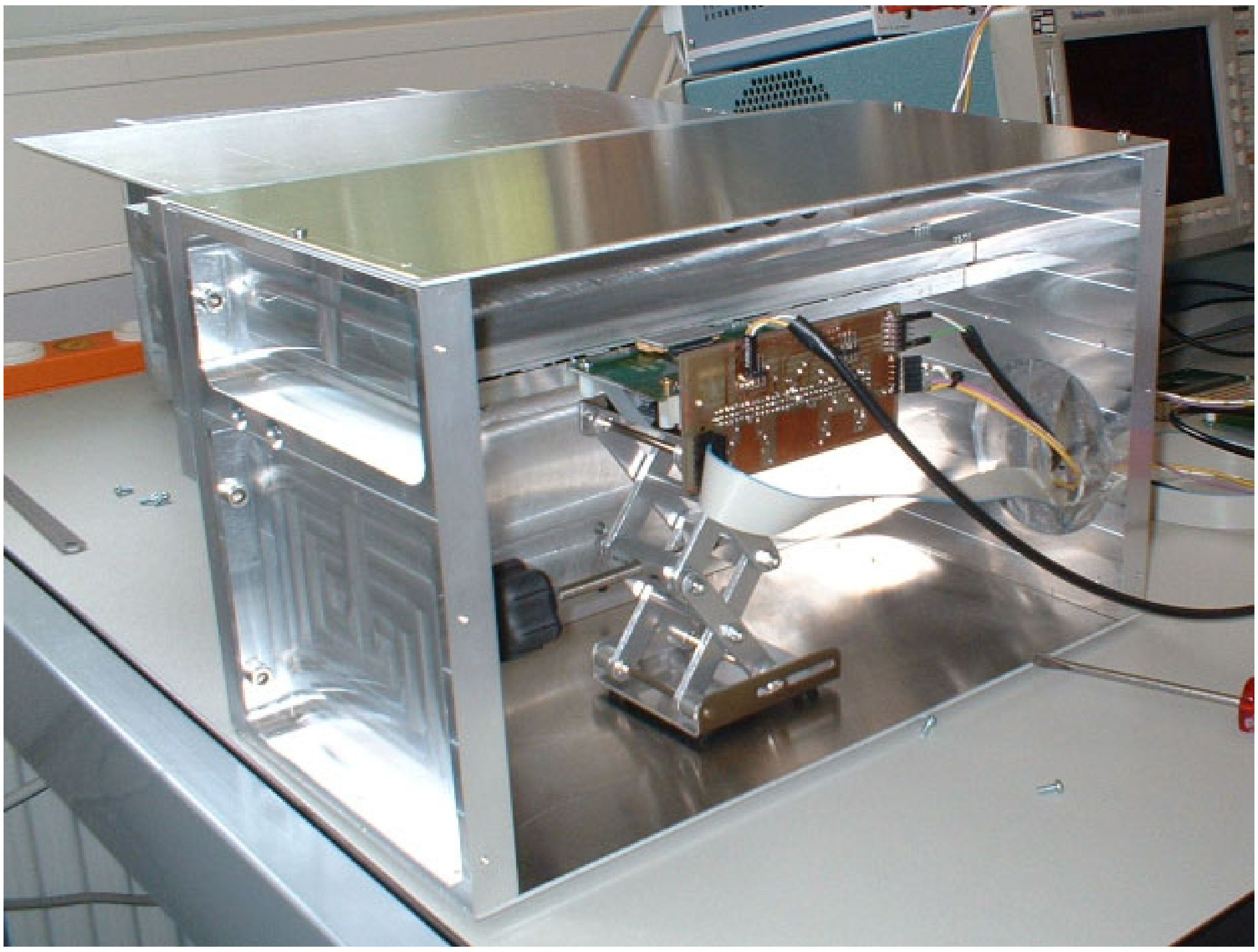

Figure 5.7: Picture showing one of the SAXS Hybrids, interfaced with the control signal generator and a digital storage oscilloscope, ready for the noise measurement

The graph shows the data stream from the negative and positive output (raw data) relative to the first JAMEX of the Hybrid. Further plots were extracted from the offset compensated differential output signals. To achieve this, the mean value over all channels were calculated separately for test 1, 2, 3 and the offset.

Analysing the recorded data stream with a digital oscilloscope shows the noise measurements as histograms. Channel by channel were considered 18 sample values (e.g. 18×64 points for the offset noise): due to the fact that sequences of 4 bursts were taken, the given sigma corresponding to *rms noise* has to be multiplicated by two.

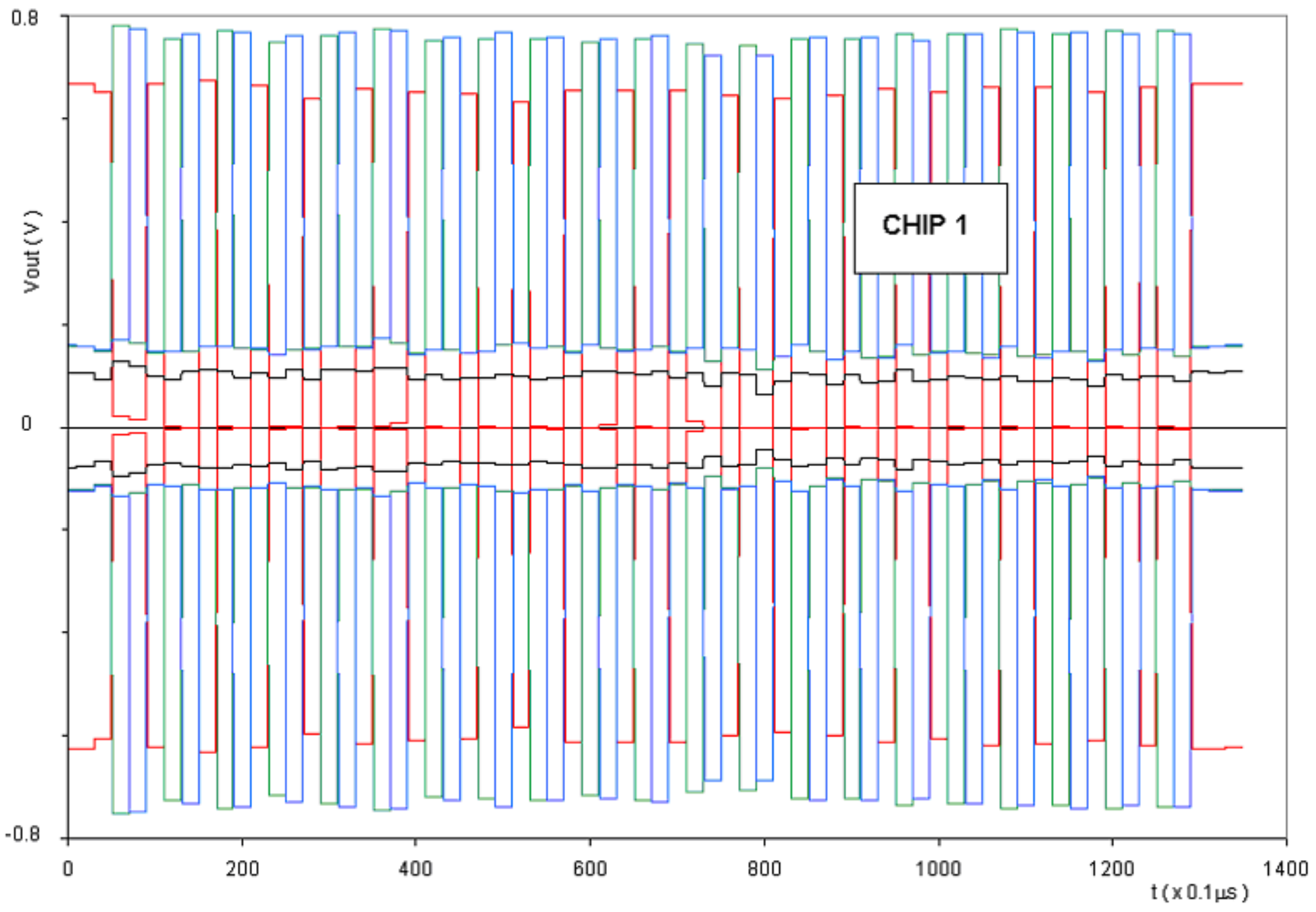

Figure 5.8: Plots relative to the differential output signals of the JAMEX ASIC

Comparing the offset noise with the signal noise (see Figure 5.9) it can be seen that clocking increases the noise by 10%. The amount of the positive output dominates the noise of a 60% more than the negative output.

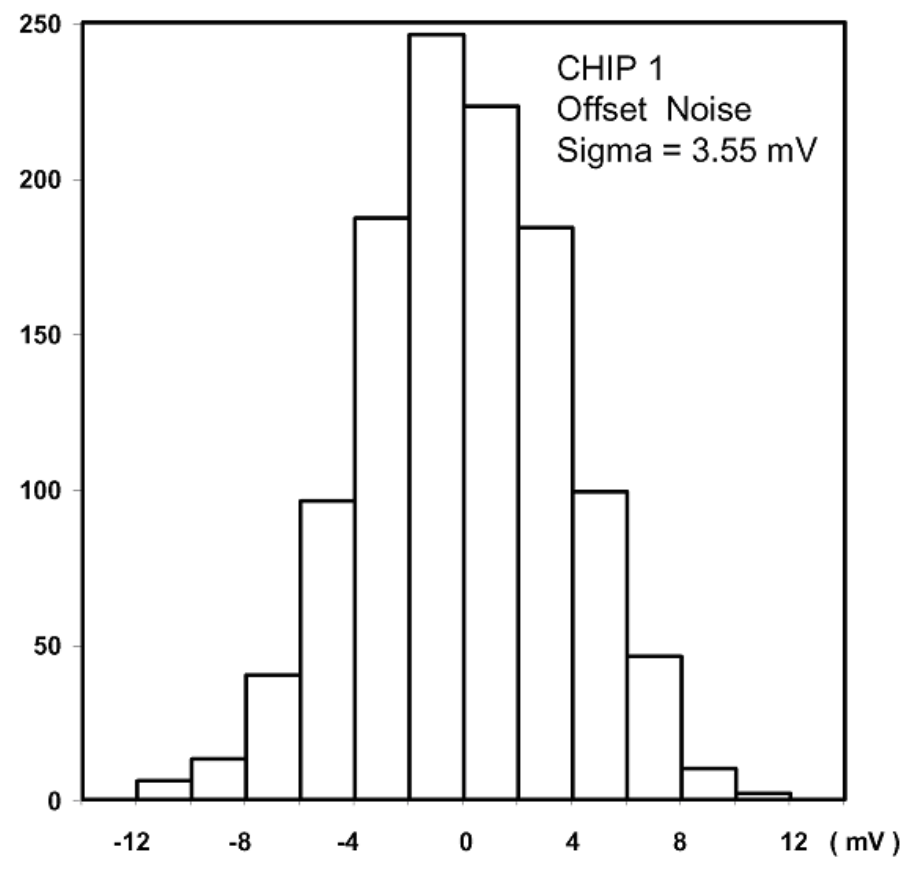

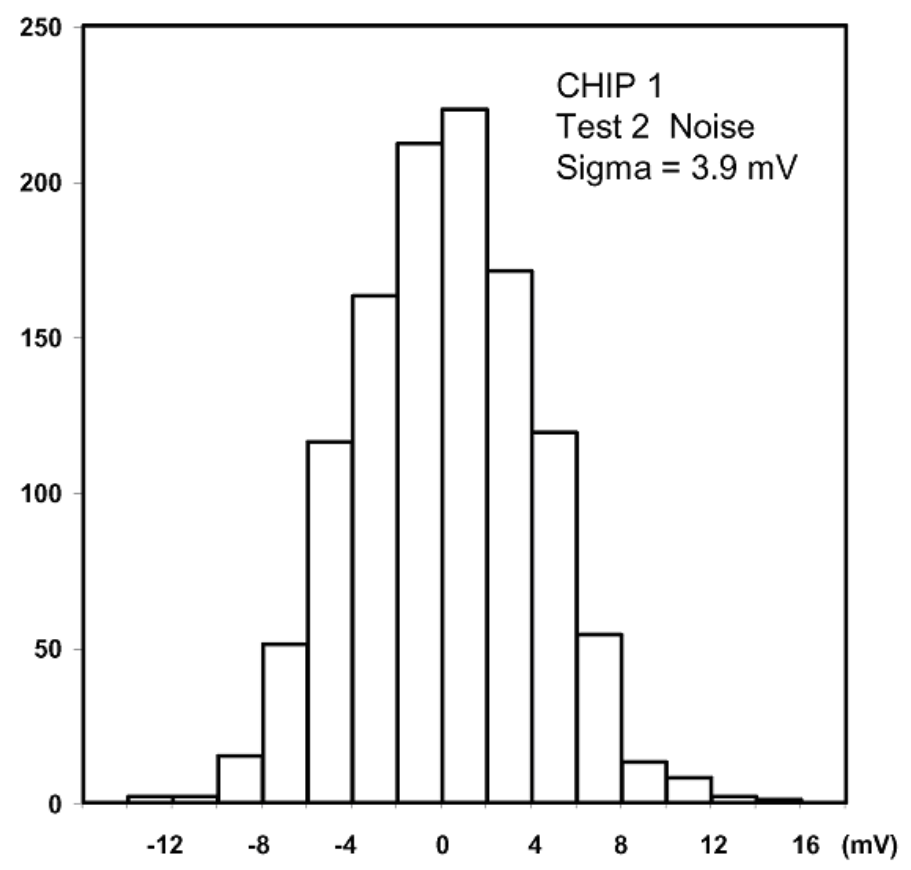

Figure 5.9: Histograms relative to the *offset noise* -left- and *signal noise* -right-

Similar results regarding the offset noise were achieved at the University of Siegen, with a different setup (see Figure 5.10) employing the Sundance acquisition system and a modified version of the Zscope application as a replacement for the digital oscilloscope.

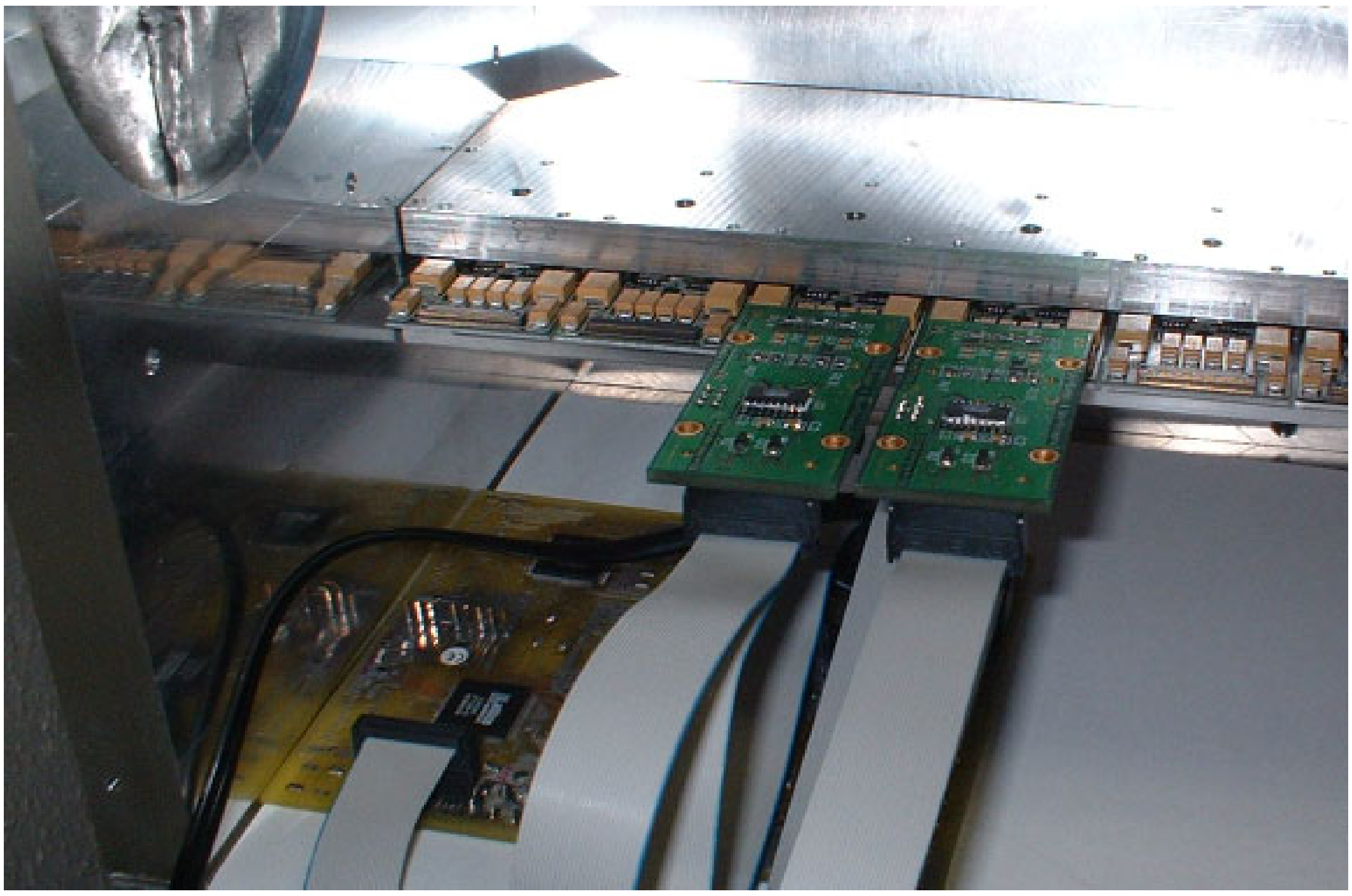

Figure 5.10: Picture of the setup used to measure the JAMEX offset noise via J2S boards connected to the Sundance acquisition system -not shown-

In the following we summarize the results of the preliminary noise and offset measurements on the Sundance acquisition system connected to the J2S board family and in particular to the J2S.MINI board:

- **Noise** (DC to the Nyquist Frequency)
  - $\approx$ $200\,\mu V_{rms}$ for the ADC only, open input
  - $\approx$ $250\,\mu V_{rms}$ with the ADC connected to the J2S.MINI (gain 0)
  - $\approx$ $310\,\mu V_{rms}$ with the ADC connected to the J2S.MINI (gain 1)
  - $\approx$ $430\,\mu V_{rms}$ with the ADC connected to the J2S.MINI (gain 2)
  - $\approx$ $580\,\mu V_{rms}$ with the ADC connected to the J2S.MINI (gain 3)
- **Offset dispersion** for the ADCs is high and around 15 mV; since it is a constant characteristic of each ADC channel, it is easy to eliminate using the Offset Correction DACs of the J2S.MINI Board

This measurements are the result of an average taken over many measurements and several, different, analog channels. Recalling that the J2S.MINI is a 2-layers PCB, which was not routed specifically to optimize the noise, these are quite good results, especially when compared to the JAMEX output noise.

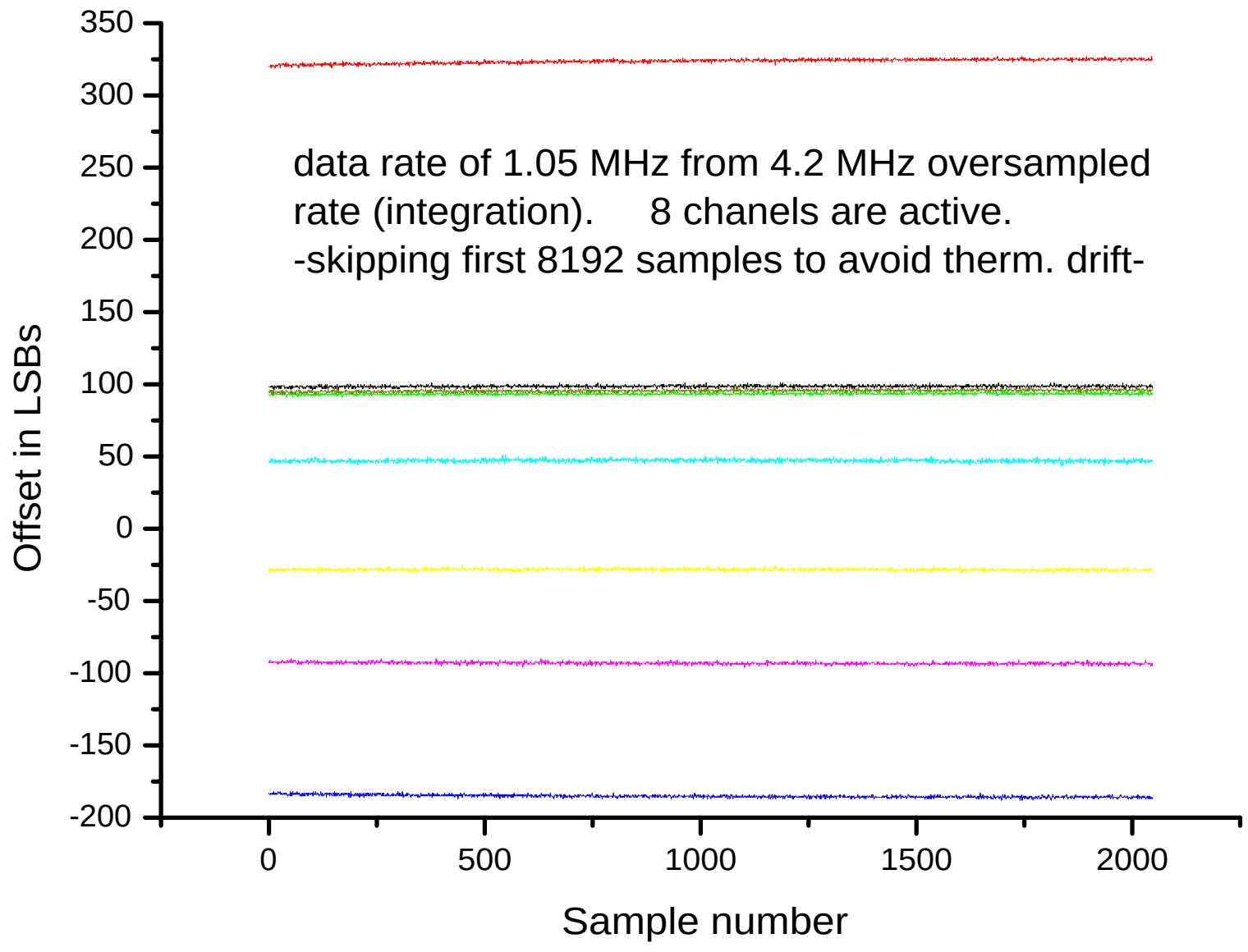

Figure 5.11: Offset variation and noise of the 8 ADC channels of the SMT356

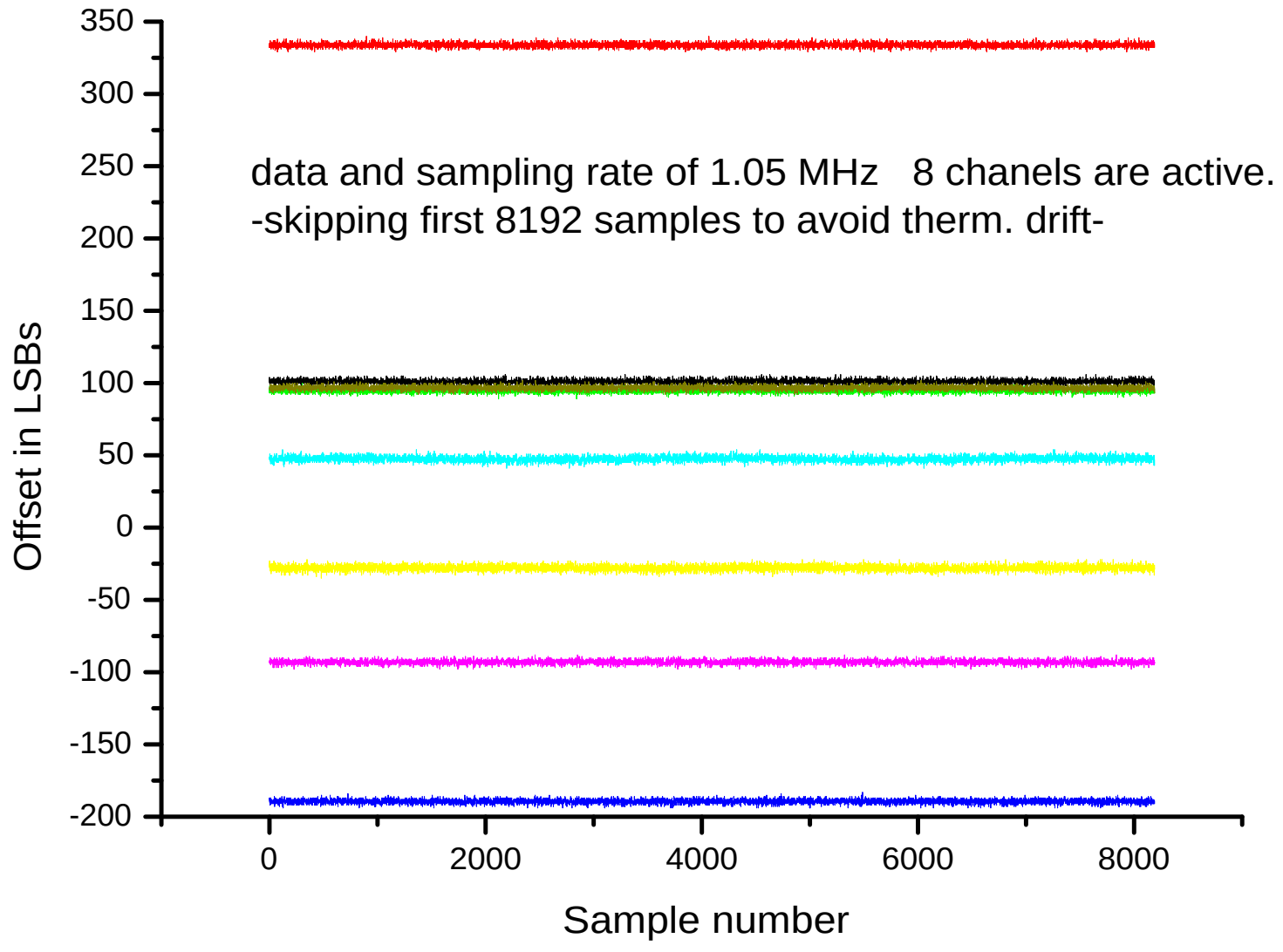

Figure 5.12: Offset variation and noise of the 8 ADC channels of the SMT356

If considering that a simple wall supply was used to provide the dual voltage required for the J2S.MINI board operation, we can realize the robustness of this design.

In Figure 5.11 it is shown the convenience to use a *four times oversampling + software integration* scheme, instead of choosing the lower sampling frequency, when running the JAMEX Application. The 4×oversampling effectively lowers the noise of the ADC to $\approx\ 100\ \mu V_{rms}$, open input: this is noticeable even comparing the two graphs.

## 5.4 Evaluation of the J2S digital logic

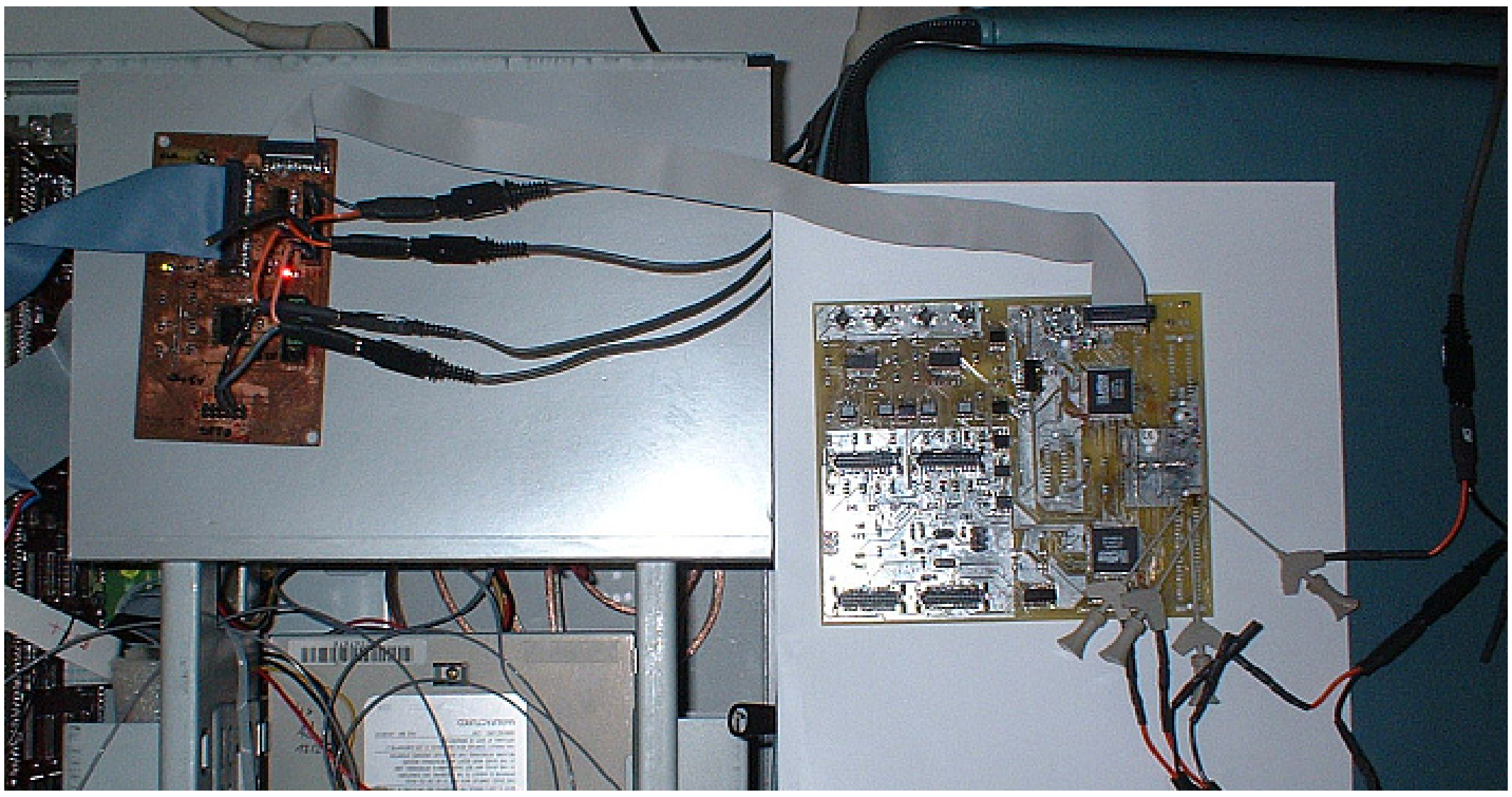

Figure 5.13: Use of the Tektronix TLS216, 16-channels DSO, to test the entire system

In order to evaluate the correct operation of both the Sundance firmware and the J2S.MINI CPLD programming, it was decided to perform some tests on the Sundance acquisition system running the JAMEX Application. A digital oscilloscope with 16 channels (the Tektronix TLS216) was used to record the waveforms relative to the main timings and patterns of the command executions of interest.

Some timings, showing the correct operation of both the CPLD on the J2S.MINI and the firmware of the Sundance system, are shown in Figure 5.14 and 5.15: the correct execution of the command Write_PGA_GAIN is achieved for any possible pattern of values sent through the ZCQSI interface, of which some examples are given in the four graphs.

Also the discover of some problems (see Figure 5.16) in generating stable timings, relative to the execution of a correct Write/Read_JAMEX cycle, was highlighted thanks to this storage oscilloscope. Some modification to the VHDL that describes the new functionalities, integrated in the Sundance FPGA, will be required to guarantee a correct programming of the JAMEX ASICs and, thus, the full operation of the SAXS detector.

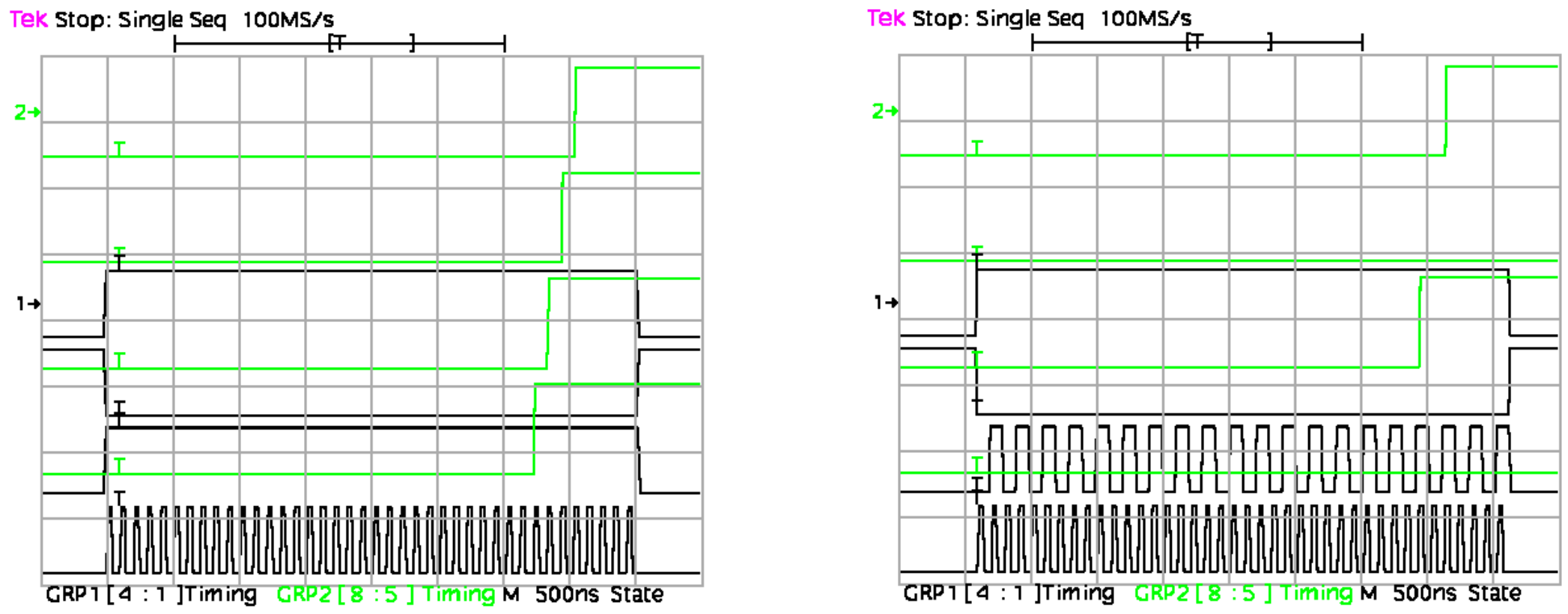

Figure 5.14: Measured timings upon execution of the Write_PGA_GAIN command with a uniform selection of *gain 3* -left- and *gain 2* -right- for all the 20 channels

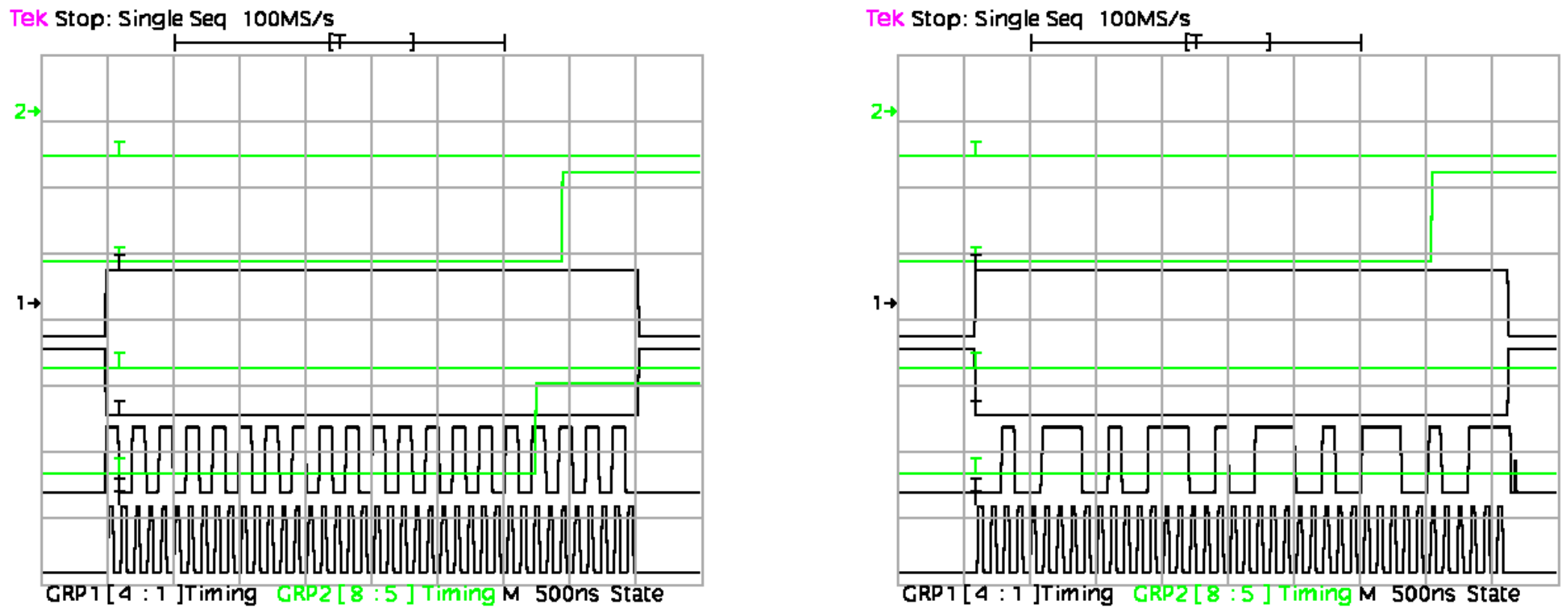

Figure 5.15: Measured timings upon execution of the Write_PGA_GAIN command with a uniform selection of *gain 1* -left- and *gain 0 and 1* -right-

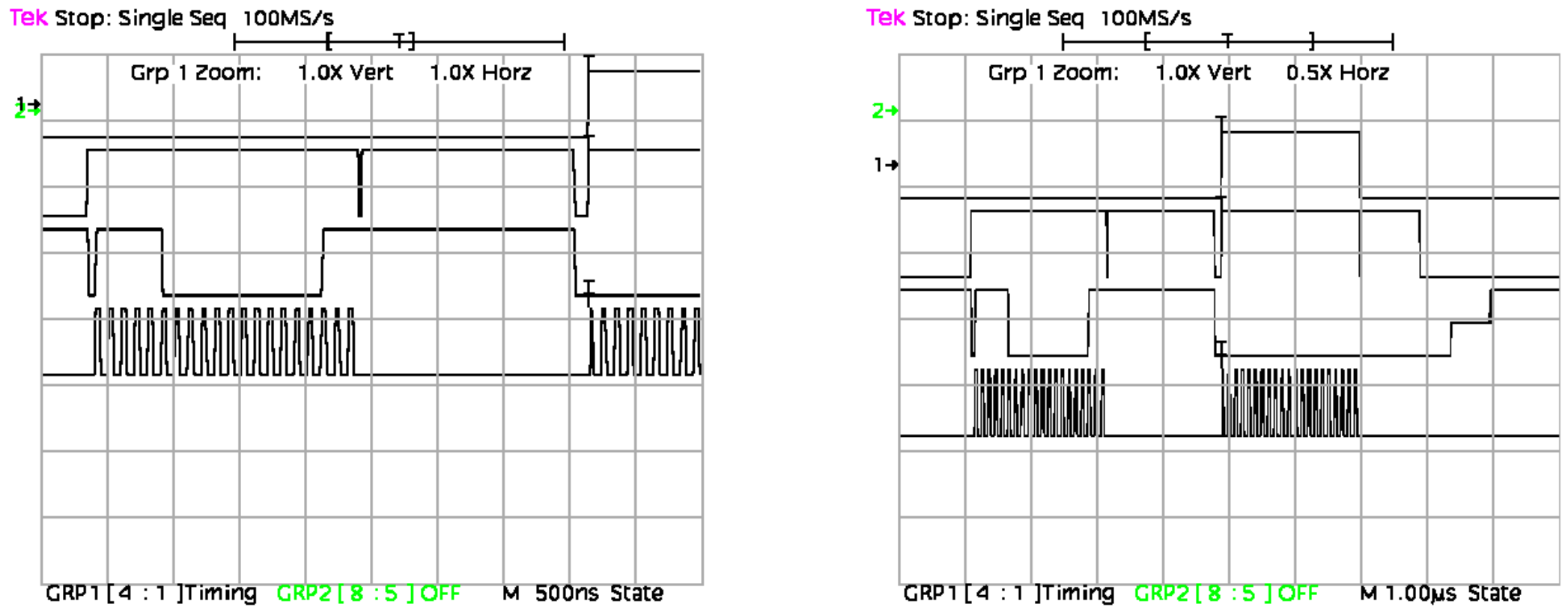

Figure 5.16: Measured timings relative to the execution of a Write/Read_JAMEX cycle

# Conclusions

The aim of this thesis work was to realize the final version of the one dimensional SAXS detector planned in the framework of an European Project devoted to improve the facilities for small angle X-ray scattering within Europe.

A big effort was put in order to reach the final target in a short time, but it was soon realized that, because of the complexity of the designed system, this priority would not be easily resolved as initially expected. Nevertheless, a lot of experience was gained thanks to the stimulating experimental work that took place either in the Instrumentation and Detector Laboratory belonging to ELETTRA or in the University of Siegen.

The main task, designing the circuit boards dedicated to interface the Sundance acquisition system to the JAMEX ASICs, reached a good level of completion with the production of the definitive version of the J2S.PIG boards and of some well specified prototypes, called the J2S.MINI and J2S.SR boards, that can be developed in a definitive design with marginal changes.

On the side of the firmware and software required for the Sundance acquisition system, the question is still open, and it is expected that a longer experimentation and some modifications are still needed to suit the final detector specifications.

It is expected that, upon realization of this design, a really evolutionary SAXS detector will result, thus fulfilling the original idea to rise the scientific potential of the existing beamlines.

# APPENDIX

Contents:

- APPENDIX A : JAMEX64I4 v2 Datasheet
- APPENDIX B : JAMEX64I4 v2 Timings
- APPENDIX C : J2S.MINI Board Schematic and PCB
- APPENDIX D : Program sources relative to the Zscope application

# JAMEX64 I4 – V2

## monolithic readout system for gas detectors

**The readout system JAMEX64 I4 has identical 64 readout channels and includes the following components:**

- 64 low noise charge amplifier
- digital selectable gain - 16 steps - common for 8 channels
- 64 SC-noise filter using 8 times correlated double sampling
- 64 sample and hold amplifier
- two outputs with multiplexer for serial readout
- differential 50Ω output buffer
- programmable digital shift register for timing control

**Revision: 2.1**

More information: Werner Buttler
Ueberruhrstr. 476, D-45277 Essen
Tel. 49 201 8585973, Fax. 49 201 8585974
e-mail: werner.buttler@t-online.de

# JAMEX64 I4 – V2

| | | |
|---|---|---|
| **General informations → preliminary !!!** | | |
| **Maximum voltages** | | |
| $V_{DD}$ | | + 2.5V |
| $V_{SS}$ | | – 2.5V |
| all inputs | | max. $V_{DD}$, min. $V_{SS}$ |
| **Power supply** | | |
| $V_{DD}$ | | + 2.5V |
| $I_{DD}$ | at 300K | --.-- mA |
| $V_{SS}$ | | – 2.5V |
| $I_{SS}$ | at 300K | --.-- mA |
| $V_{DGND}$ | | 0.0V |
| $V_{REF_B}$ | | – 2,5V (var.: 0 → -2.5V) |
| $I_{REF_B}$ | | --.-- mA |
| $V_{REF_S}$ | | – 2,5V (var.: 0 → -2.5V) |
| $I_{REF_S}$ | at 300K | --.-- mA |
| $V_{REF_V}$ | | – 2,5V (var.: 0 → -2.5V) |
| $I_{REF_V}$ | at 300K | --.-- mA |
| **Digital inputs** | | |
| control levels | LVDS - input | CML: 1.2V - ±200mV |
| | pseudo LVDS - output | CML: 1.25V - ±1250mV |
| | | $C_L$ < 20pF |
| | | no resistive load |
| **Output buffer** | | |
| load | | 2x |
| | | $R_L$ = 50 Ω |

# JAMEX64 I4 – V2

| **System amplification** | | |
|---|---|---|
| integration capacitance | | |
| $C_f$ (fix) | | 0,4 pF |
| $C_{fv1}$ select by GAIN1 | | 2,0 pF |
| $C_{fv2}$ select by GAIN2 | | 5,0 pF |
| $C_{fv3}$ select by GAIN3 | | 10,0 pF |
| $C_{fv4}$ select by GAIN4 | | 30,0 pF |
| | | |
| **system performance** | | |
| noise at the output | | |
| output stability | | |
| output TK | | |
| | | |
| **Timing parameters** | | |
| reset pulse with | R1, R2 | min. 1 µs |
| setup time before switching of the CDS-filter | | min. 1 µs |
| CDS-filter pulse with | S_SEL | min. 200 ns |
| S&H pulse with | S&H | min. 1 µs |
| timing clock pulse (run or load) | S_CLK | min. 125 ns |
| readout time of the multiplexer | MUX_CLK | min. 500 ns |
| the S_CLK and the MUX_CLK should have an fixed (integer) relation | | |

# JAMEX64 I4 – V2

## Block diagram

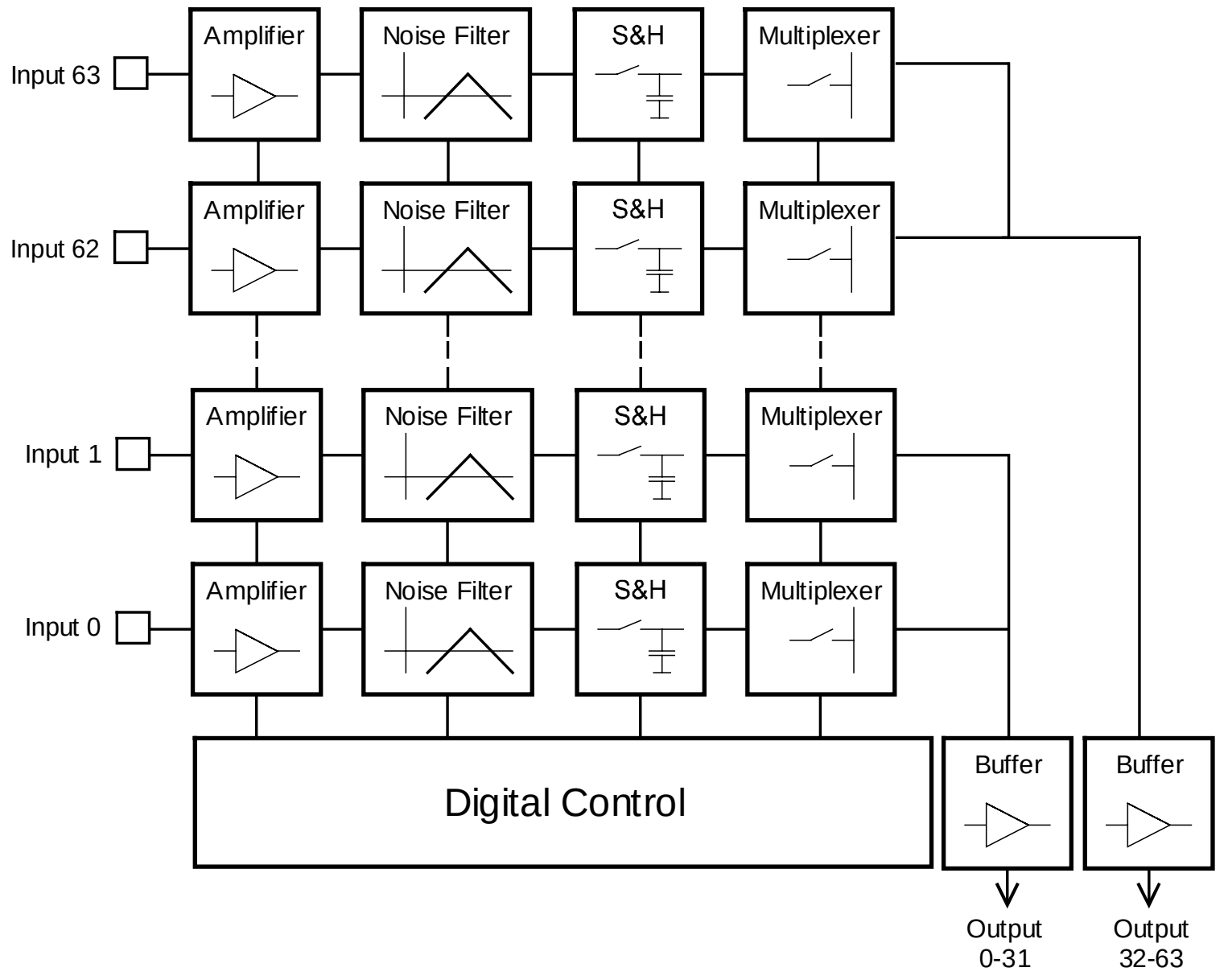

## Circuit diagram of JFET-CMOS amplifier

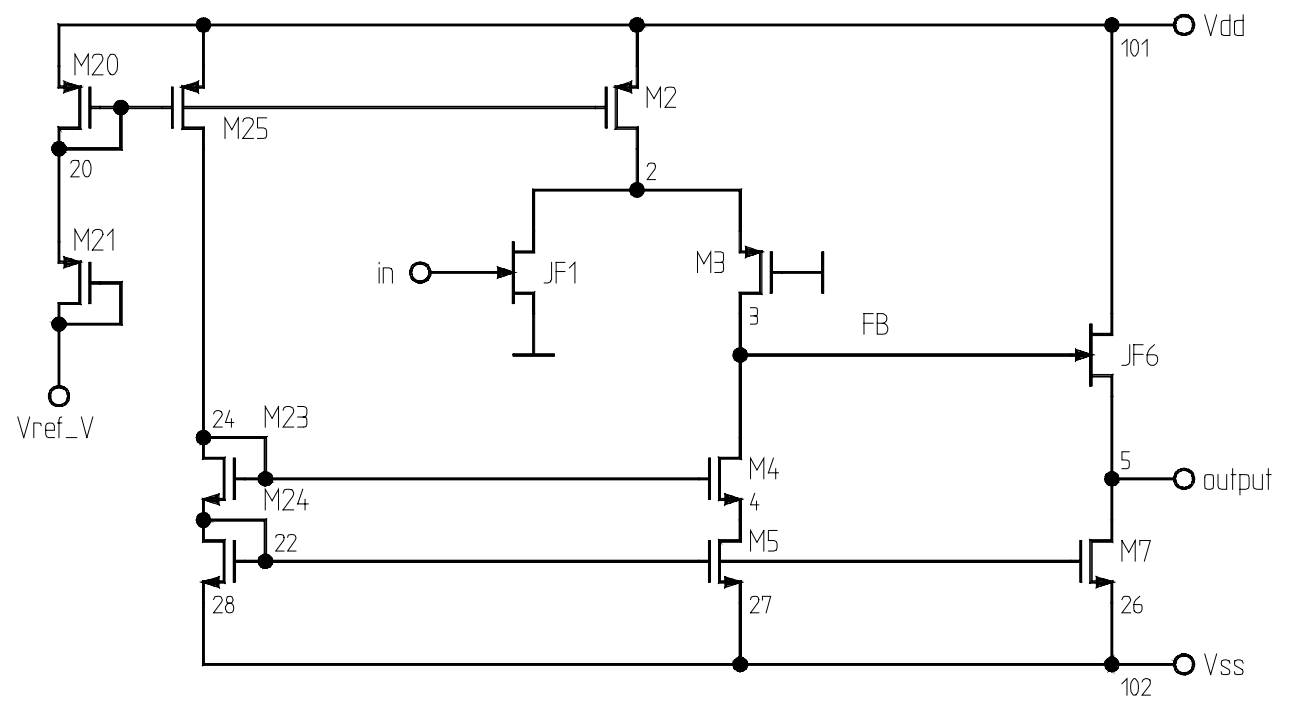

# JAMEX64 I4 – V2

**Scheme of a readout channel**

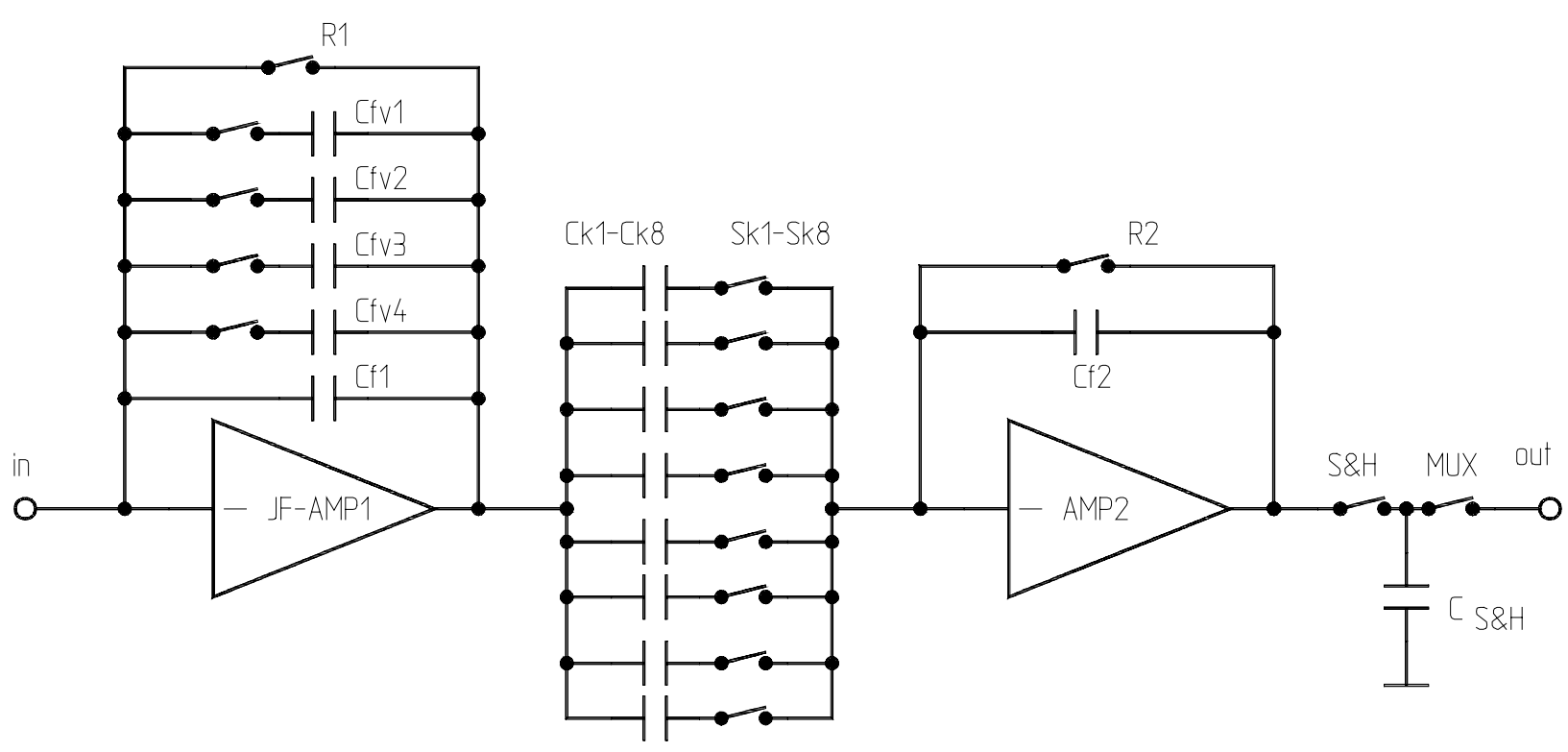

**Timing of the switches**

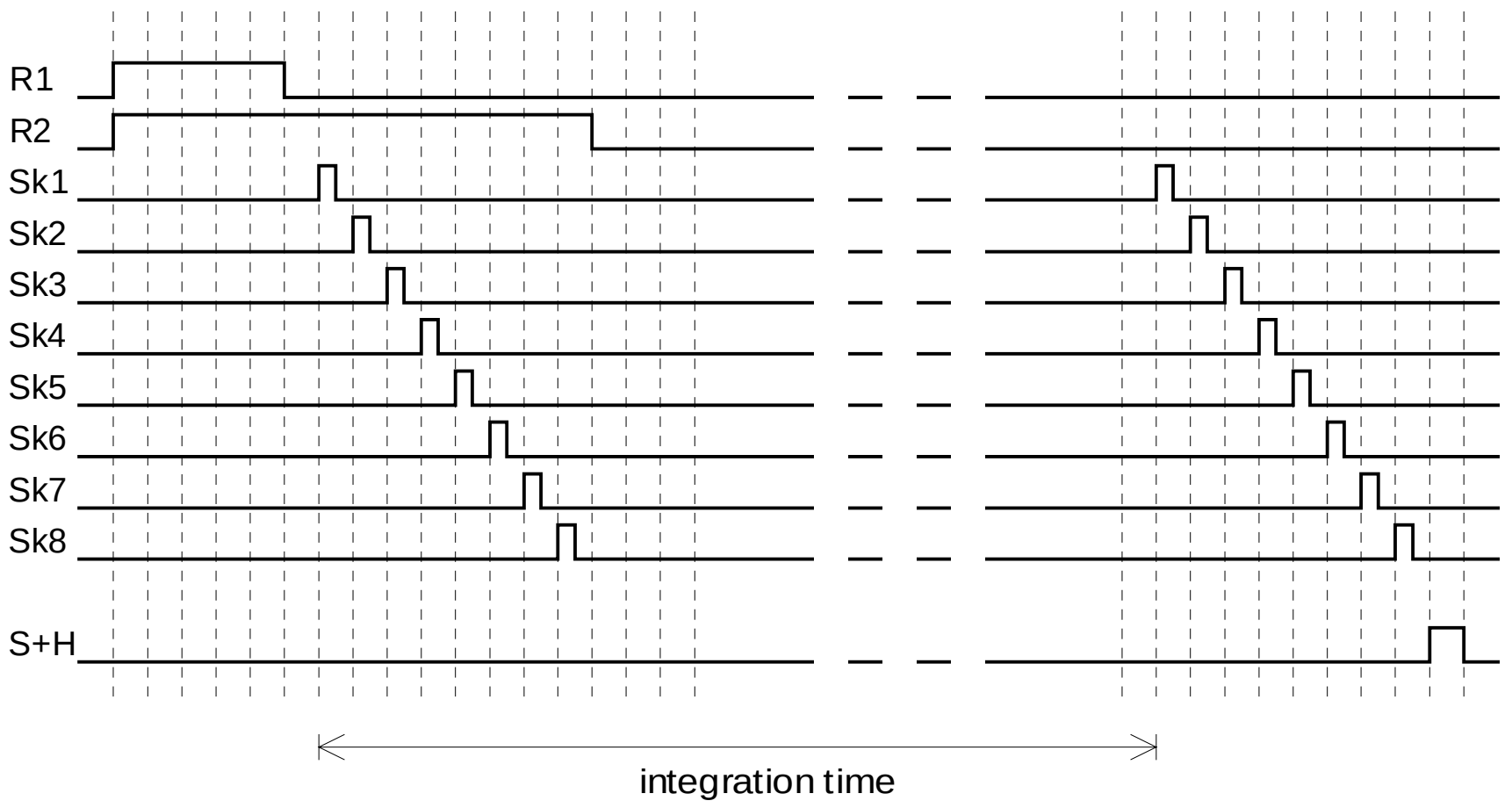

# JAMEX64 I4 – V2

## Digital timing and control register

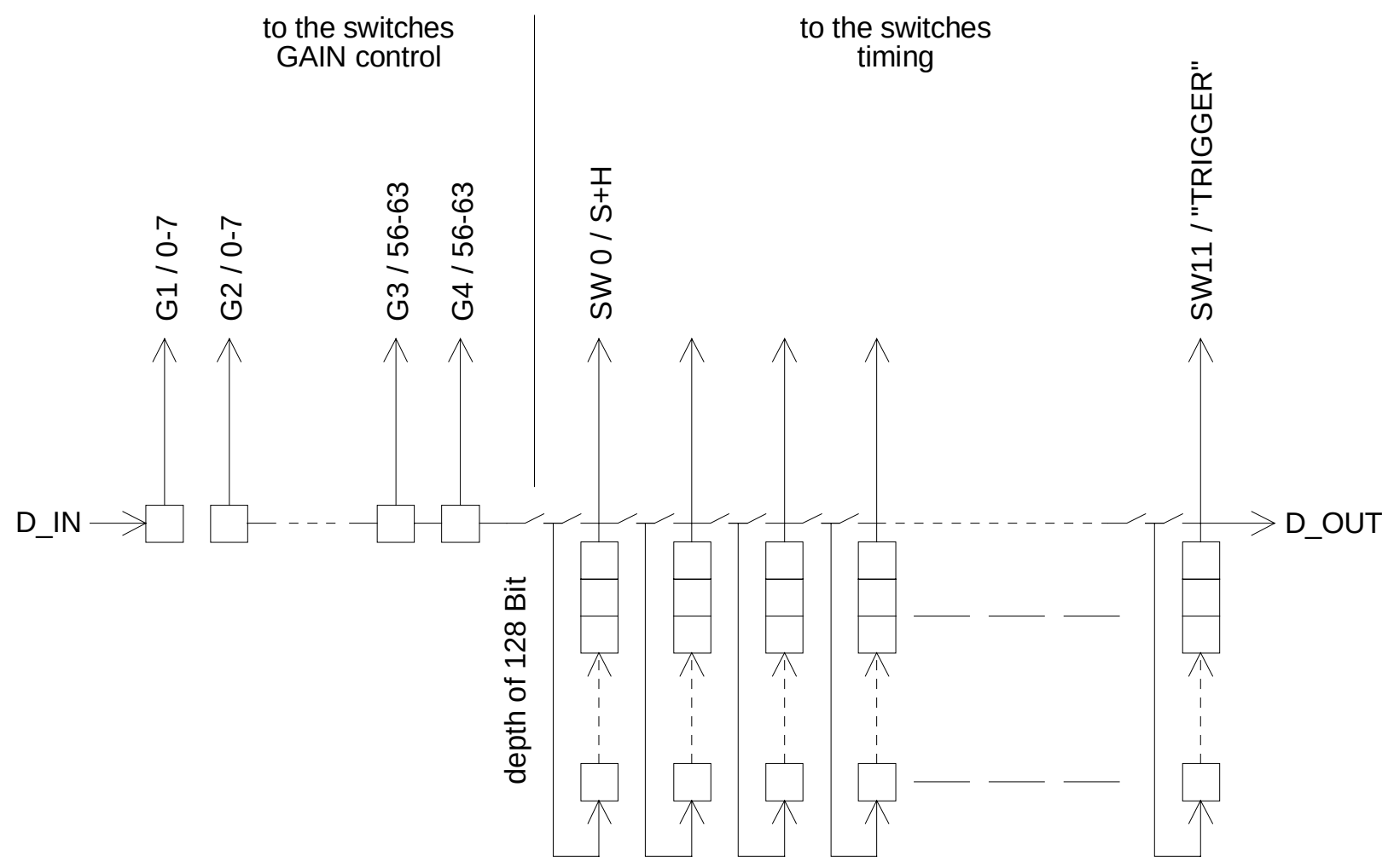

## Timing of the shift registers

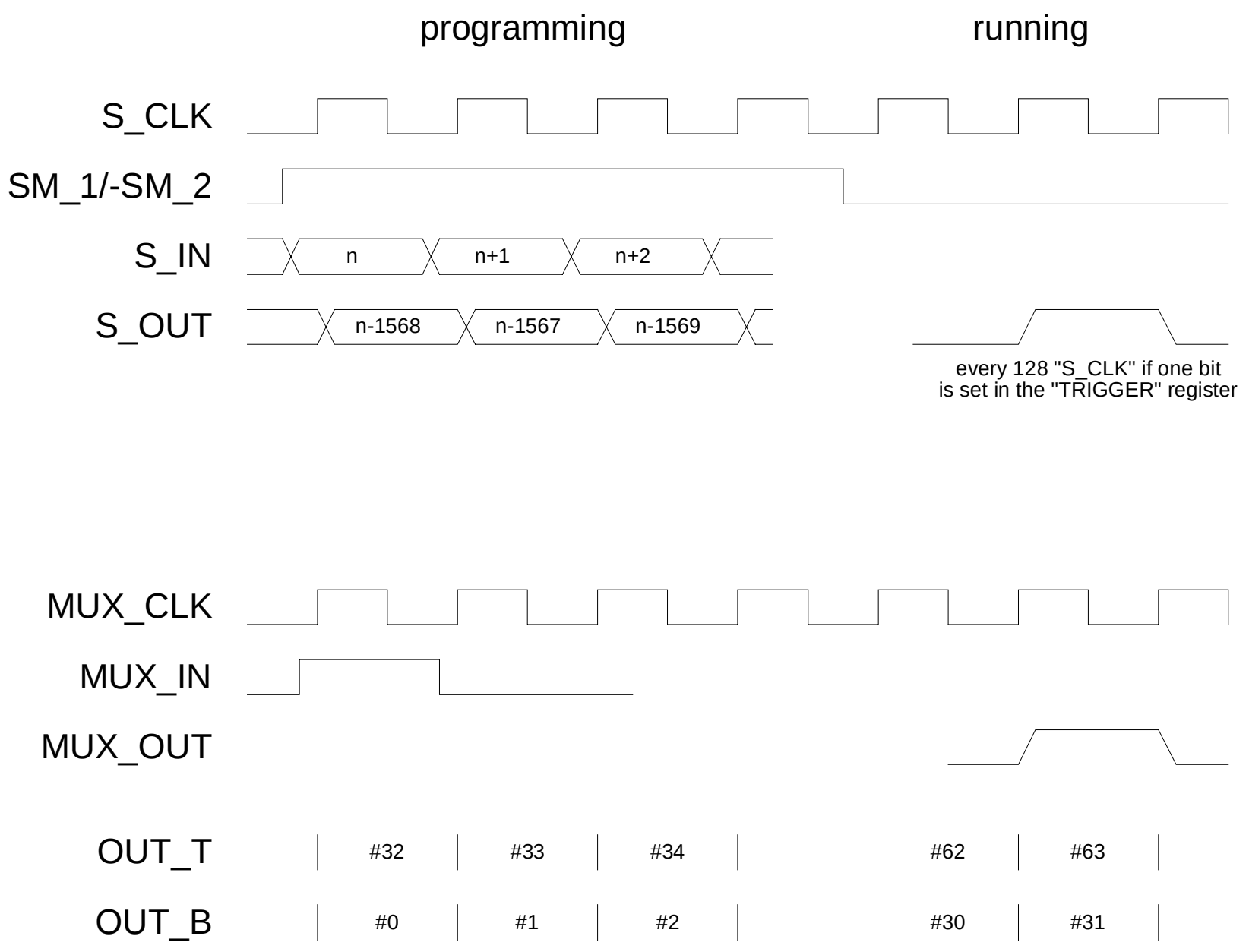

# JAMEX64 I4 – V2

## Data bits in the timing and control register

| | | | | | function | |
|---|---|---|---|---|---|---|
| fist | bit # | 0 | - | 127 | TRIGGER | |
| | bit # | 128 | - | 255 | R1 | |
| | bit # | 256 | - | 383 | SK8 | |
| | bit # | 384 | - | 511 | SK7 | |
| | bit # | 512 | - | 639 | SK6 | |
| | bit # | 640 | - | 767 | SK5 | |
| | bit # | 768 | - | 895 | SK4 | |
| | bit # | 896 | - | 1023 | SK3 | |
| | bit # | 1024 | - | 1151 | SK2 | |
| | bit # | 1152 | - | 1279 | SK1 | |
| | bit # | 1280 | - | 1407 | R2 | |
| | bit # | 1408 | - | 1535 | S+H | |
| | bit # | 1536 | | | GAIN4 | channel 56-63 |
| | bit # | 1537 | | | GAIN3 | |
| | bit # | 1538 | | | GAIN2 | |
| | bit # | 1539 | | | GAIN1 | |
| | bit # | 1540 | | | GAIN4 | channel 48-55 |
| | bit # | 1541 | | | GAIN3 | |
| | bit # | 1542 | | | GAIN2 | |
| | bit # | 1543 | | | GAIN1 | |
| | bit # | 1544 | | | GAIN4 | channel 40-47 |
| | bit # | 1545 | | | GAIN3 | |
| | bit # | 1546 | | | GAIN2 | |
| | bit # | 1547 | | | GAIN1 | |
| | bit # | 1548 | | | GAIN4 | channel 32-39 |
| | bit # | 1549 | | | GAIN3 | |
| | bit # | 1550 | | | GAIN2 | |
| | bit # | 1551 | | | GAIN1 | |
| | bit # | 1552 | | | GAIN4 | channel 24-31 |
| | bit # | 1553 | | | GAIN3 | |
| | bit # | 1554 | | | GAIN2 | |
| | bit # | 1555 | | | GAIN1 | |
| | bit # | 1556 | | | GAIN4 | channel 16-25 |
| | bit # | 1557 | | | GAIN3 | |
| | bit # | 1558 | | | GAIN2 | |
| | bit # | 1559 | | | GAIN1 | |
| | bit # | 1560 | | | GAIN4 | channel 8-15 |
| | bit # | 1561 | | | GAIN3 | |
| | bit # | 1562 | | | GAIN2 | |
| | bit # | 1563 | | | GAIN1 | |
| | bit # | 1564 | | | GAIN4 | channel 0-7 |
| | bit # | 1565 | | | GAIN3 | |
| | bit # | 1566 | | | GAIN2 | |
| last | bit # | 1567 | | | GAIN1 | |

# JAMEX64 I4 – V2

## PAD description

| PAD-Nr. | Name | Type | Function |
|---|---|---|---|
| 1 (top) | $V_{DD}$ | supply | positive supply voltage |
| 2 | GND | supply | ground |
| 3 | + OUT_T | + output - analog | channel 32 - 63 |
| 4 | – OUT_T | – output - analog | channel 32 - 63 |
| 5 | DGND | supply | digital ground |
| 6 | + SM_2 | + input - digital | CML: 1.2V – +200mV / –200mV |
| 7 | – SM_2 | – input - digital | CML: 1.2V – –200mV / +200mV |
| 8 | + SM_1 | + input - digital | CML: 1.2V – +200mV / –200mV |
| 9 | – SM_1 | – input - digital | CML: 1.2V – –200mV / +200mV |
| 10 | + S_IN | + input - digital | CML: 1.2V – +200mV / –200mV |
| 11 | – S_IN | – input - digital | CML: 1.2V – –200mV / +200mV |
| 12 | + S_CLK | + input - digital | CML: 1.2V – +200mV / –200mV |
| 13 | – S_CLK | – input - digital | CML: 1.2V – –200mV / +200mV |
| 14 | + MUX_IN | + input - digital | CML: 1.2V – +200mV / –200mV |
| 15 | – MUX_IN | – input - digital | CML: 1.2V – –200mV / +200mV |
| 16 | + MUX_CLK | + input - digital | CML: 1.2V – +200mV / –200mV |
| 17 | – MUX_CLK | – input - digital | CML: 1.2V – –200mV / +200mV |
| | | | |
| 18 | + S_OUT | + output - digital | +2.5V / 0V – CL < 20pF |
| 19 | – S_OUT | – output - digital | 0V / +2.5V – CL < 20pF |
| 20 | + MUX_OUT | + output - digital | +2.5V / 0V – CL < 20pF |
| 21 | – MUX_OUT | – output - digital | 0V / +2.5V – CL < 20pF |
| 22 | $V_{REF_B}$ | input - analog | bias current of the S&H, ouput buffer |
| 23 | $V_{REF_S}$ | input - analog | bias current of the 2. amplifier |
| 24 | $V_{REF_V}$ | input - analog | bias current of the 1. amplifier |
| 25 | + OUT_B | + output - analog | channel 0 - 31 |
| 26 | – OUT_B | – output - analog | channel 0 - 31 |
| 27 | $V_{SS}$ | supply | negative supply voltage |
| 28 | TEST 3 | input - analog | test input 3 for channel 2, 5, 8, ... 50 Ω to GND |
| 29 | TEST 2 | input - analog | test input 2 for channel 1, 4, 7, ... 50 Ω to GND |
| 30(bottom) | TEST 1 | input - analog | test input 1 for channel 0, 3, 6, ... 50 Ω to GND |

**JAMEX64 I4 – V2**

**Measurement Setup**

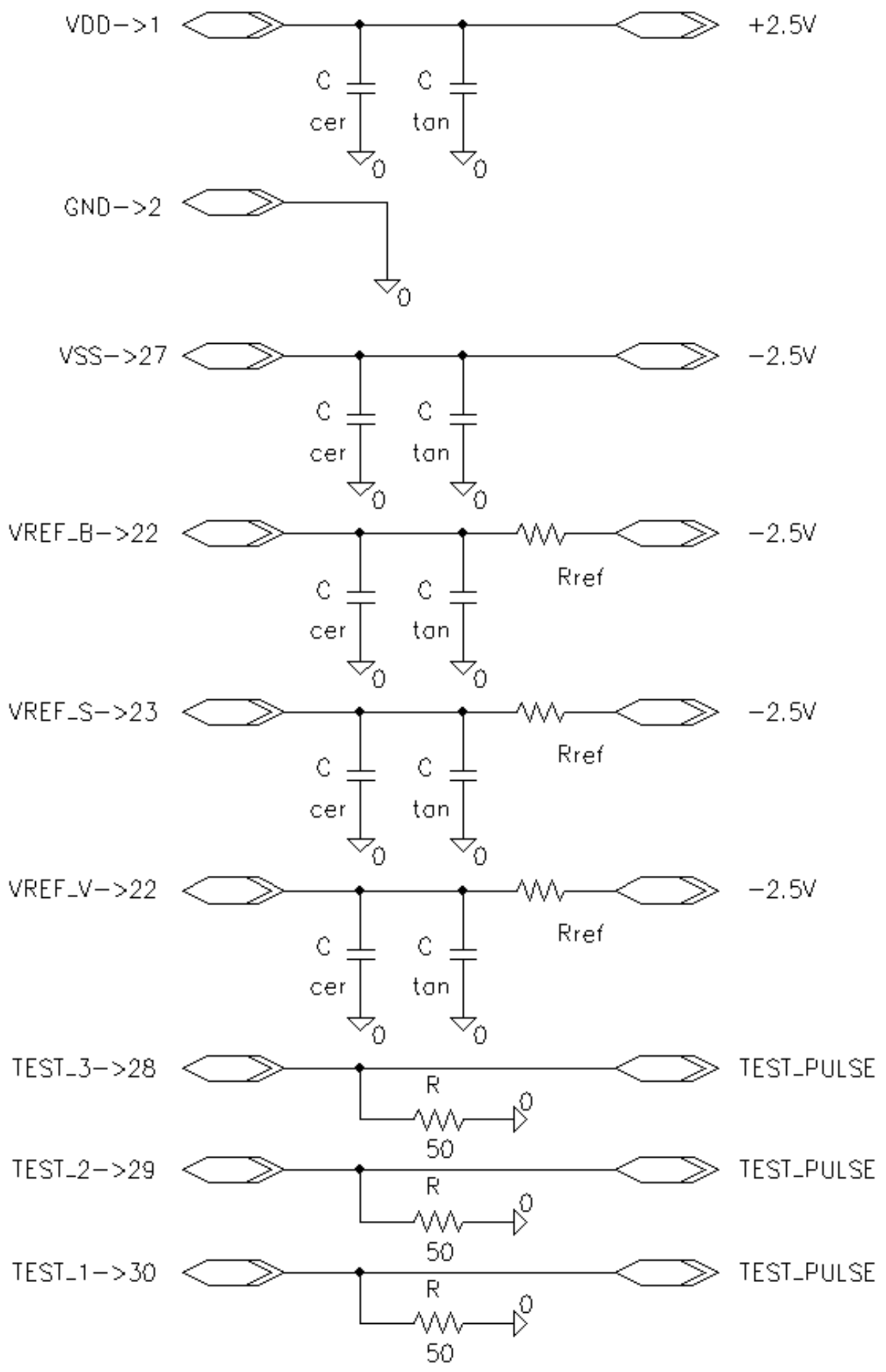

# JAMEX64 I4 – V2

## Chip layout

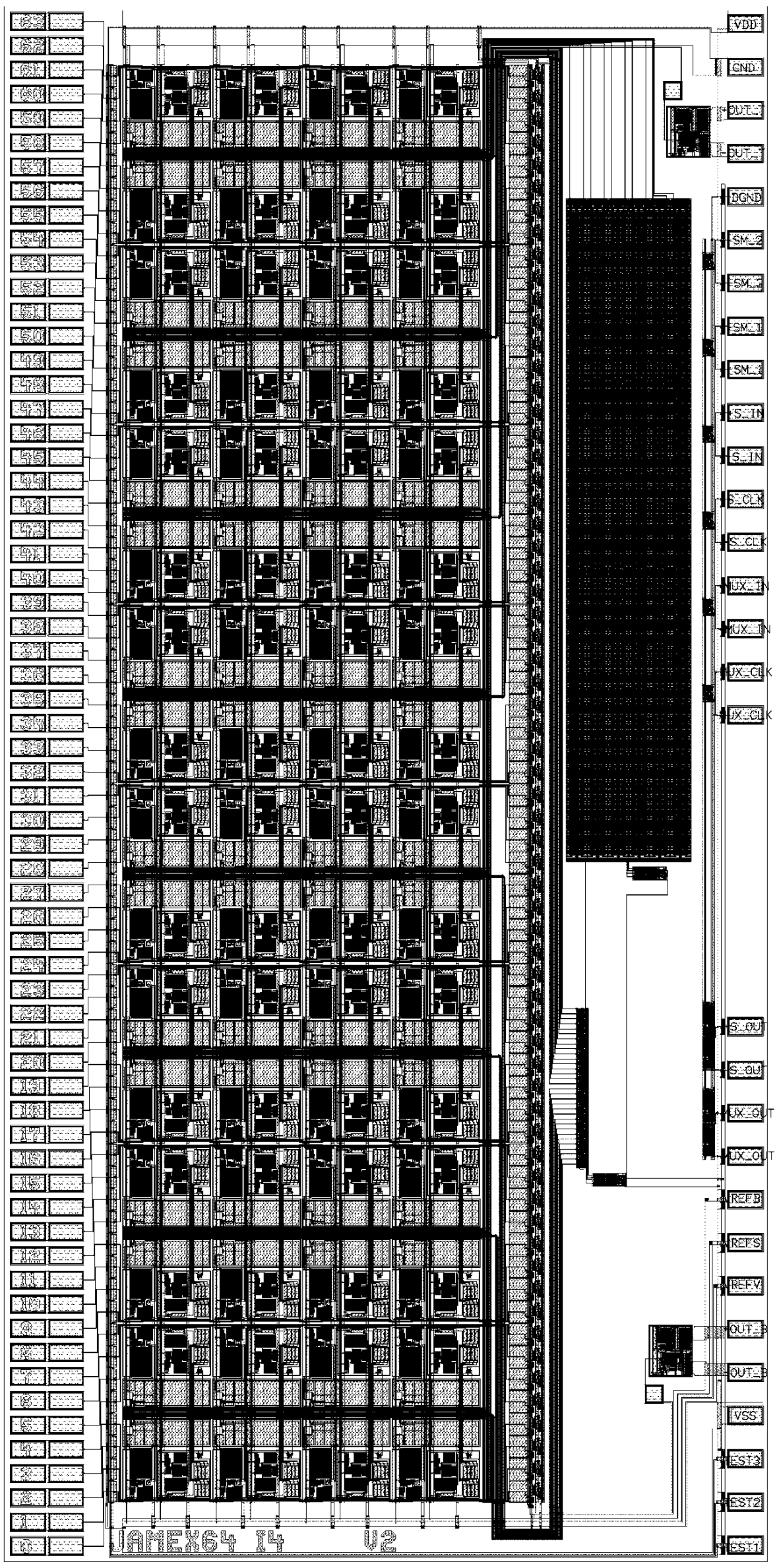

| JAMEX64 I4 – V2 |
|---|

| **Die size** |
|---|
| Total: X: 4640µm Y: 9220µm |

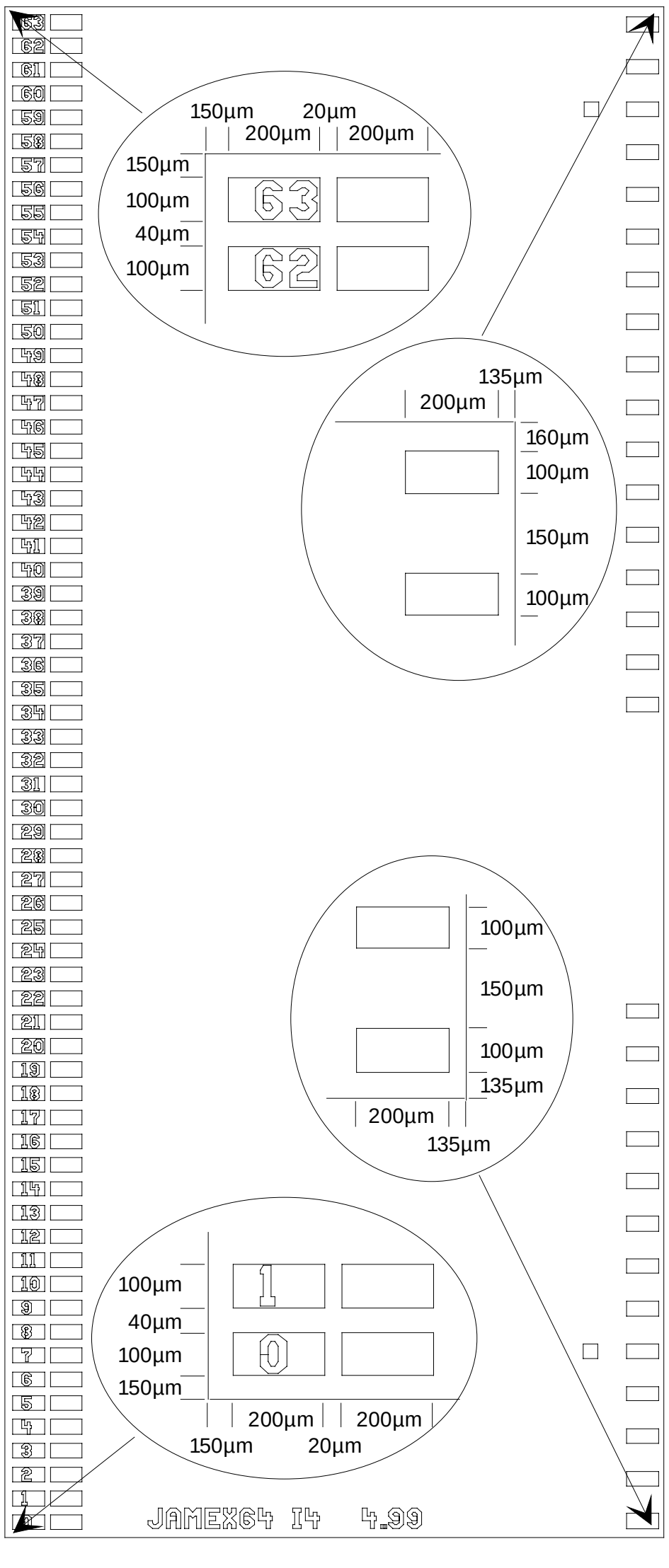

**different modes of operation**

- **STOP MODE**

  - no operation
  - all signals on “LOW“

- **PROGRAMMING MODE**

  - writing the data pattern into the JAMEX
  - 20 individual pattern into the 20 JAMEX’s
  - pattern length 1568 bits each

- **RUN MODE**

  - operating the JAMEX
  - 128 clocks for one operation cycle
  - 64 data conversions in one cycle to read out the JAMEX

## JAMEX64 I4 - OPERATION MODES

*Ingenieurbüro Buttler*
*45276 Essen*

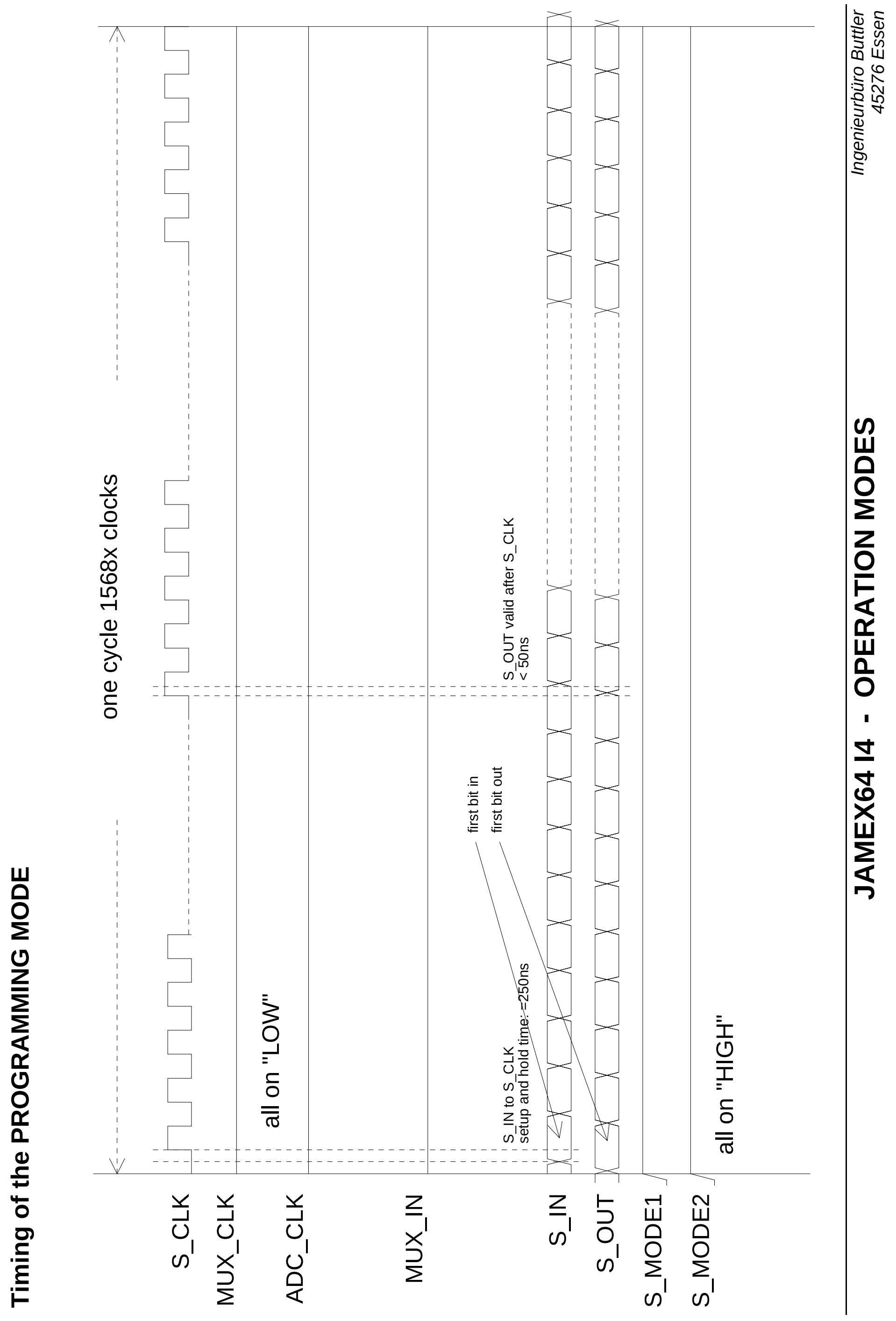
Timing of the PROGRAMMING MODE
one cycle 1568x clocks
S_CLK
MUX_CLK
ADC_CLK
MUX_IN
S_IN
S_OUT
S_MODE1
S_MODE2
all on "LOW"
all on "HIGH"
S_IN to S_CLK
setup and hold time: =250ns
first bit in
first bit out
S_OUT valid after S_CLK
< 50ns
JAMEX64 I4 - OPERATION MODES
Ingenieurbüro Buttler
45276 Essen

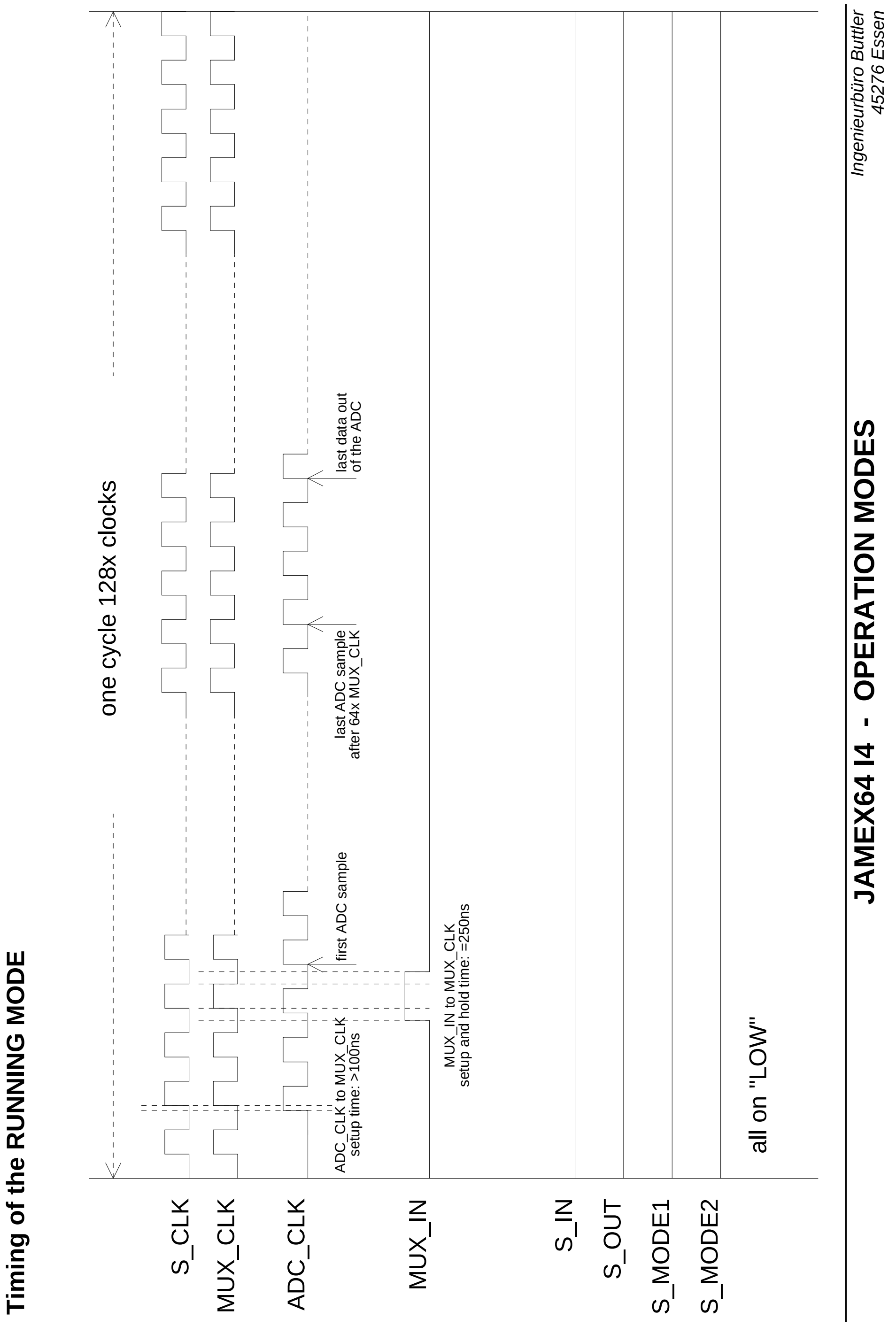

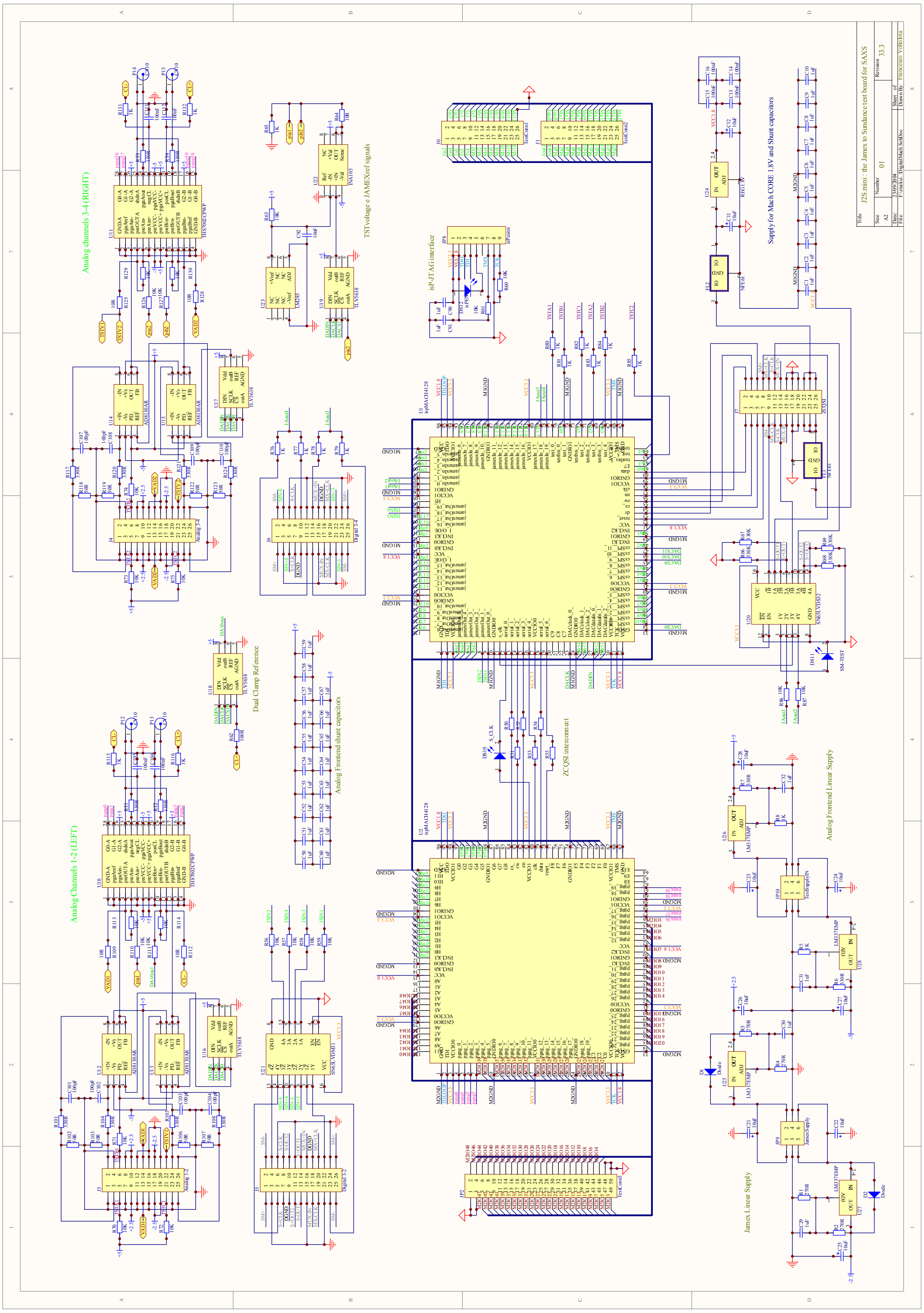

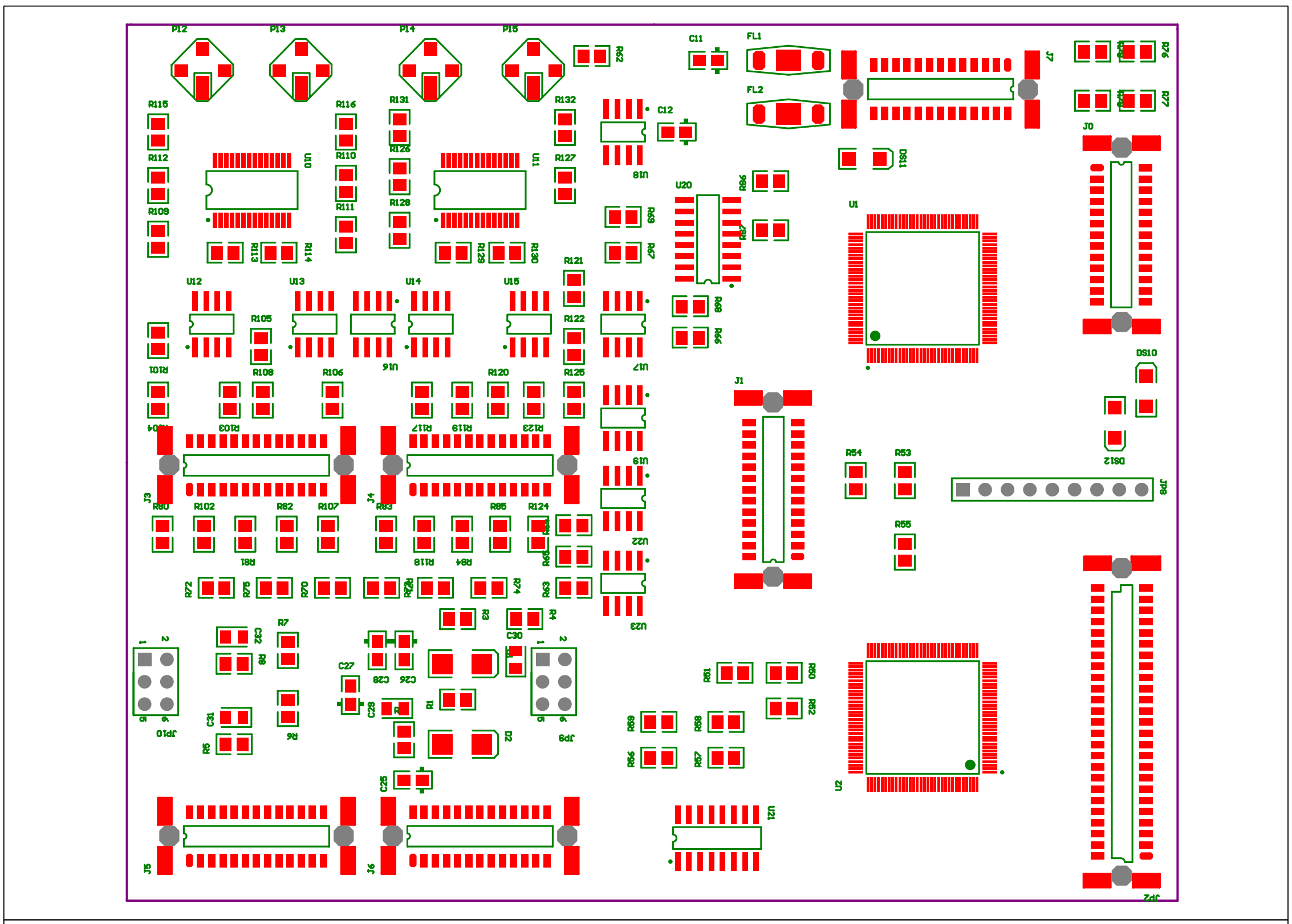

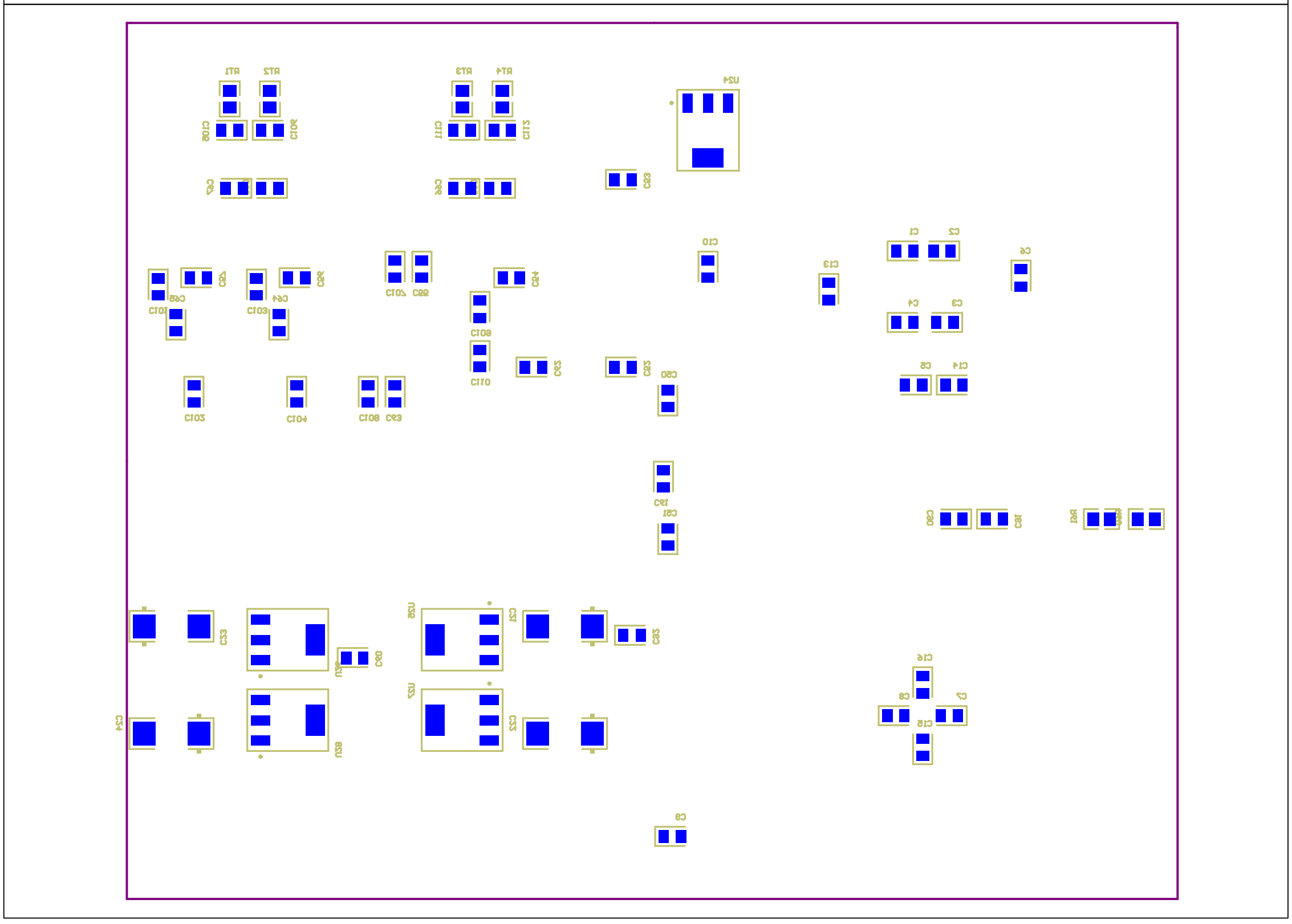

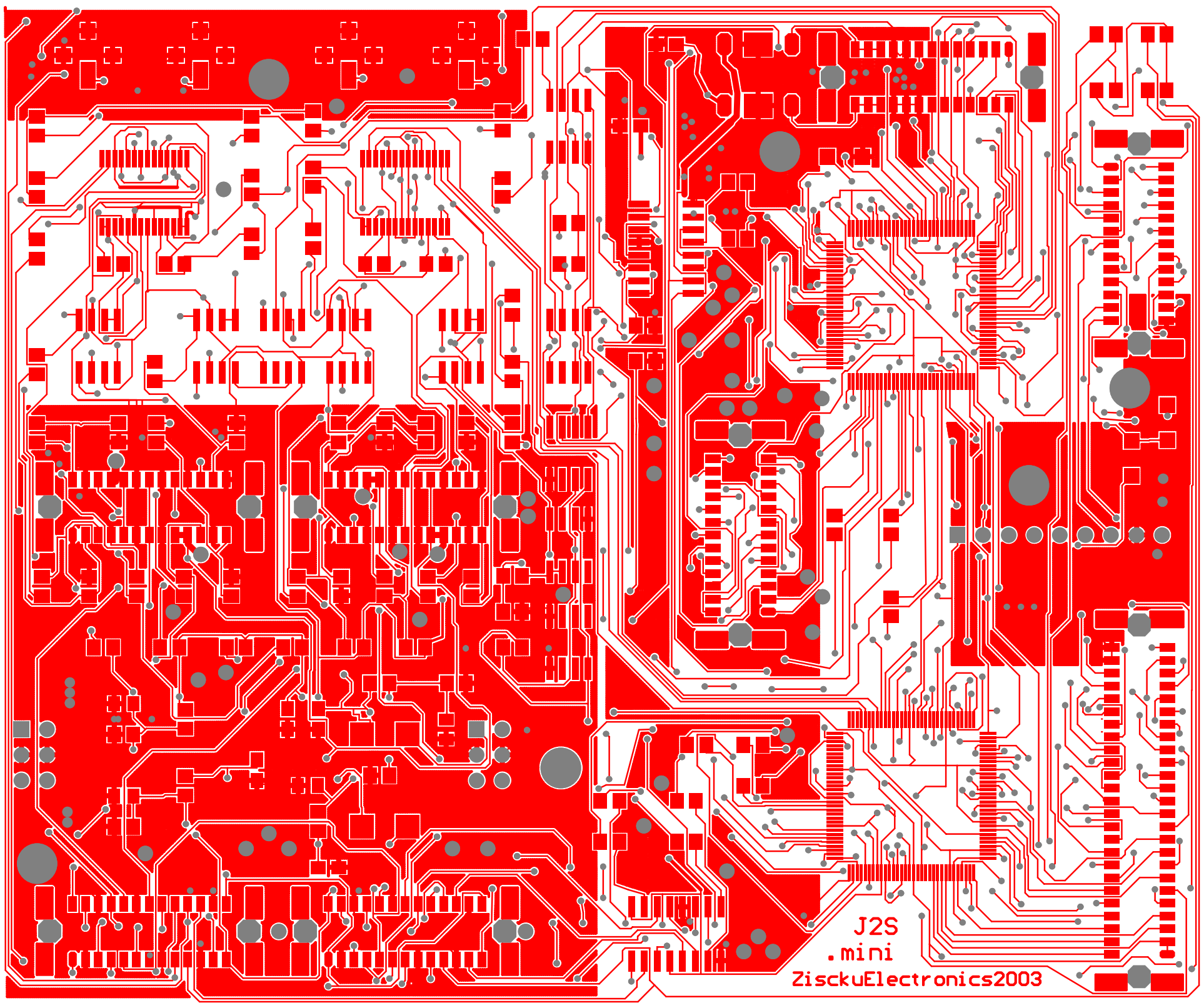
J2S
.mini
ZisckuElectronics2003

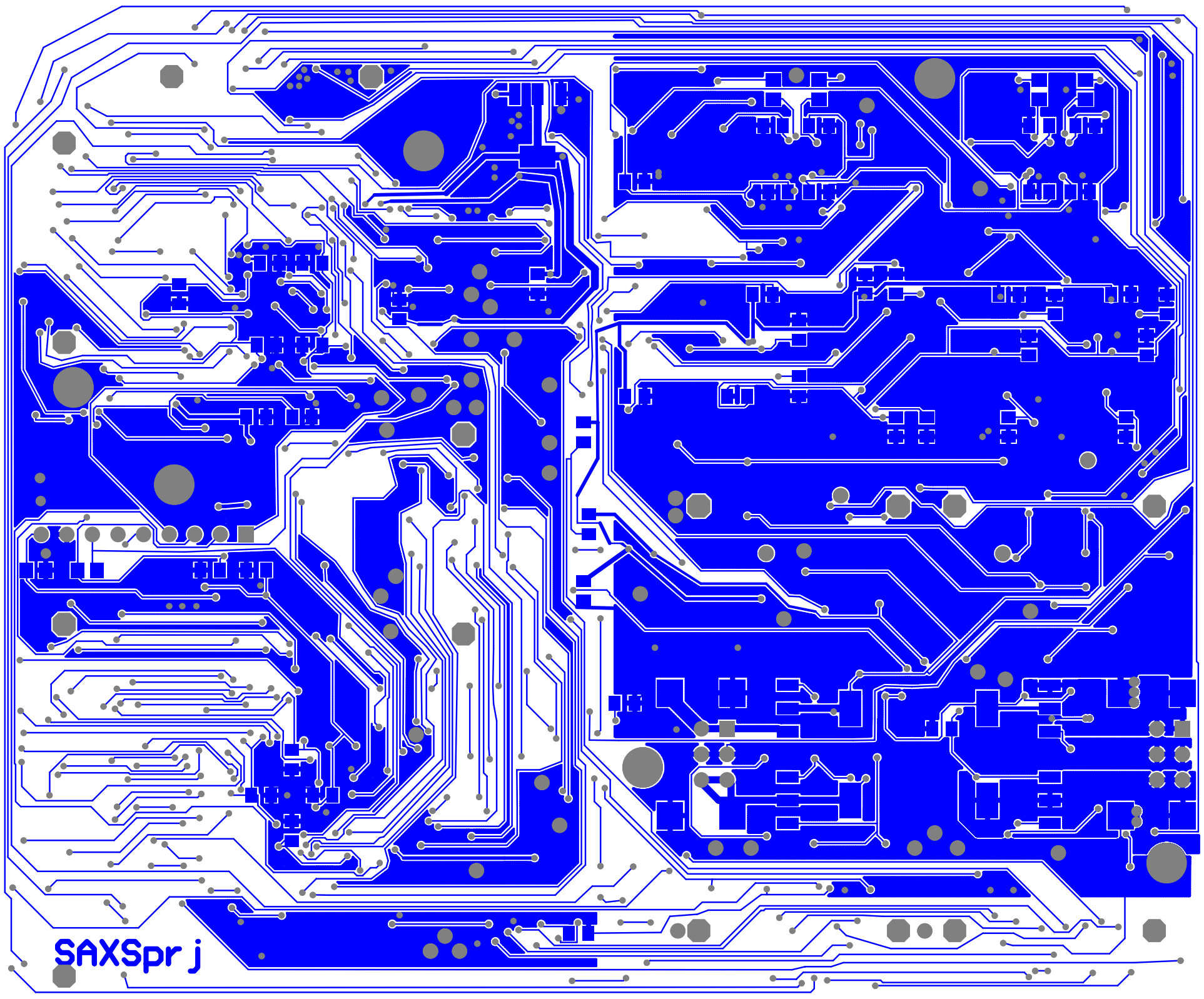
SAXSprj

```
/*//////////////////////////////////////////////////////////////////
 /////////////////////////////////////////////////////////////////
  Ziscku Electronics inc.

  www.zisckuelectronics.org

  ZscopeGraphics May 2004
/////////////////////////////////////////////////////////////////


  In order to test the comport, we need to have a program running
  on the DSP.  In this sample, a Diamond .app is downloaded to the
  DSP.  The app reads values from the comport, multiplies them by
  two and writes them back to the comport.
  Before the .app file is downloaded, the TIMs are reset to make
  sure that the bootloader is running and the TIM is ready to receive
  the .app file.

  On the PC side, we write values to the comport, read them back and
  test to see if the value read back is twice the value written.

  NOTE: It is important that you specify the correct type of TIM
  you are using in the .cfg file for the DSP project.  Use the nmake
  utility to rebuild the DSP side if you needed to change the TIM type.
//////////////////////////////////////////////////////////////////*/



#include "stdafx.h"
#include <iostream.h>
#include <conio.h>
#include <stdio.h>

#include "grafik.h"
#include "smtdrv.h"                 /* Include header file for SmtDrv.dll */
#pragma comment(lib, "smtdrv.lib")  /* Link to the library for the SmtDrv.dll */

void PerformTest(IFHw * pBoard);

//////////////////////////////////////////////////////////////////
unsigned char fgetcc(FILE *fp)
{
    unsigned char ch;
    ch=fgetc(fp);
    return(ch);
}

//////////////////////////////////////////////////////////////////
unsigned char hex2dec(unsigned char hexnum)
{
    if(hexnum<0x3a) return(hexnum-0x30);
    else return(hexnum-0x37);
}

//////////////////////////////////////////////////////////////////

FILE *fp;

/* main */

int not_end=1, byte_count, address, record_type, bytes_read=0;
int iiii, i, load_word, chan;
int load_address=0, record_decoded;
int iadc[16385];
unsigned char ch;


int main(int argc, char* argv[])

{

  cout << "Zscope Host-DSP comm. TEST" << endl << endl;

  SMTRet ret = SmtOpen(); /* Open the library */

  if (ret!=SMT_OK) {
```

```
    cout << "Could not open SmtDrv library.  The error was (" <<
         SmtGetError(ret) << ")" << endl;
    return 0;
  }

  int nFound = SmtGetBoardCount();
  if (nFound==0) {
    cout << "No Sundance carrier boards found" << endl;
    return 0;
  }

  SMTBI bi;
  cout << "Found the following carrier boards:" << endl;
  for (int n=0; n<nFound; n++) {
    SmtGetBoardInfo( n, bi );
    cout << n << endl;
    cout << bi.cszType << endl;

    IFHw * pBoard = SmtOpenBoard(n); /* Obtain an interface to the carrier board */

    if (pBoard)
      PerformTest(pBoard);
    else
      cout << "Could not open the board" << endl;
  }

  return 0;
}


/* END OF MAIN!!!!!  ----------------------------------------   */



/* PerformTest */
void PerformTest( IFHw * pB )
{
  int nWrite, nRead, tokenRead, iadcRead, MaxRead, Peak;
  try {
    cout << "Performing test..." << endl;

    pB->ResetTIMs();
    pB->BinaryLoad( "ZsCpDSP.app" );
    cout << "Application Loaded..." << endl;


    for (nWrite=0; nWrite<1025; nWrite++) {
      pB->CpWrite( &nWrite, sizeof(int) );
      pB->CpRead( &nRead, sizeof(int) );
     // cout << nRead << endl;
      //printf(".");
      if (nRead != nWrite*2) {
        cout << "Data corruption.  Test failed" << endl;
        return;
      }
    }

    //cout << nRead << endl;
    cout << "Waiting for TOKEN from DSP" << endl;

    pB->CpRead( &tokenRead, sizeof(int) );
    //  cout << tokenRead << endl;

      if (tokenRead != 0) {
        cout << "Corrupted TOKEN from DSP" << endl;
        return;
      }


    /**************************************************************/
    /**** START - SMT356 CONFIGURATION                         ***/
    /**************************************************************/
                if (!( fp = fopen("v-top.mcs","rb") )) {
                    printf("\n\nCould not open v-top.mcs file.\n");
                    exit(0);
                }
```

2

```
printf("\n\nConfigure FPGA. \nStart\n");

    not_end=1;
    bytes_read=0;
    load_address=0;
    iiii=0;

while (not_end) {

    //cout << endl << "step" << endl;
    //cout << iiii << endl;
    iiii++;

    ch=fgetcc(fp);                              // read start character ':'

    byte_count =hex2dec(fgetcc(fp))<<4;
    byte_count+=hex2dec(fgetcc(fp));            // read byte count

    address =hex2dec(fgetcc(fp))<<12;
    address+=hex2dec(fgetcc(fp))<<8;
    address+=hex2dec(fgetcc(fp))<<4;
    address+=hex2dec(fgetcc(fp));               // read address

    record_type =hex2dec(fgetcc(fp))<<4;
    record_type+=hex2dec(fgetcc(fp));           // read record type
    record_decoded=0;

    //cout << endl << record_type << endl;

    pB->CpWrite( &record_type, sizeof(int) );

    if (record_type==1) {
        not_end=0;                             // end of file
        record_decoded=1;
    }

    if (record_type==2) {                      // extended address
        record_decoded=1;
        load_address =hex2dec(fgetcc(fp))<<12;
        load_address+=hex2dec(fgetcc(fp))<<8;
        load_address+=hex2dec(fgetcc(fp))<<4;
        load_address+=hex2dec(fgetcc(fp));      // read load address

        load_address<<=4;
                           // printf("Address:%08X\n",load_address);
    }

    if (record_type==4) {                      // extended address
        record_decoded=1;
        load_address =hex2dec(fgetcc(fp))<<12;
        load_address+=hex2dec(fgetcc(fp))<<8;
        load_address+=hex2dec(fgetcc(fp))<<4;
        load_address+=hex2dec(fgetcc(fp));      // read load address

        load_address<<=16;
                          //  printf("Address:%08X\n",load_address);
    }

    address+=load_address;                     // add address offset

    if (record_type==0) {                      // data record

        record_decoded=1;

        pB->CpWrite( &byte_count, sizeof(int) );

        for(i=0;i<(byte_count/4);i++) {
    //  for(ii=0;ii<(byte_count/4);ii++) {

            load_word =hex2dec(fgetcc(fp))<<4;
            load_word+=hex2dec(fgetcc(fp))<<0;

            load_word+=hex2dec(fgetcc(fp))<<12;
            load_word+=hex2dec(fgetcc(fp))<<8;
```

```
                    load_word+=hex2dec(fgetcc(fp))<<20;
                    load_word+=hex2dec(fgetcc(fp))<<16;

                    load_word+=hex2dec(fgetcc(fp))<<28;
                    load_word+=hex2dec(fgetcc(fp))<<24;

                    bytes_read+=4;
                                        //    printf("%X\n",load_word);
                    //printf(".");
                    pB->CpWrite( &load_word, sizeof(int) );
                }
            }

            if (record_decoded==0) {
                printf("\n\nERROR - invalid record type : %X\n\n",record_type);
                while(1);
            }

            ch=fgetcc(fp);                                // read checksum
            ch=fgetcc(fp);
            ch=fgetcc(fp);                                // read cr-lf
            ch=fgetcc(fp);

        } //End while(not_end)

        fclose( fp );

        printf("Done\nBYTES_READ = %X\n",bytes_read);


        /*****************************************************************/
        /**** END - SMT356 CONFIGURATION                              ***/
        /*****************************************************************/


    //cout << "FPGA config OK....." << endl;



    for(chan=0; chan<8; chan++)


    {

    cout << endl << "Displaying Channel: " << chan << endl;
   // cout << chan << endl;
    //cout << "Waiting for TOKEN2 from DSP" << endl;

    pB->CpRead( &tokenRead, sizeof(int) );
        //cout << tokenRead << endl;

      if (tokenRead != 0) {
        cout << "Corrupted TOKEN from DSP" << endl;
        return;
      }



//  for(i=0;i!=16000;i++);



    ITUConsole c; // a console

    c.open("SAXS 1-D acquisition window", // console title
             512,
             400,
        false);              // console fullscreen ?

    //ITUwindow_width,   // console width
    //  ITUwindow_height,  // console height

    // while the user does not press any key...
```

4

```
while(!c.keypressed())

{
    c.lock(); // lock memory

    MaxRead=0;
    Peak=0;
    nWrite=1;


    pB->CpWrite( &nWrite, sizeof(int) );

    for (int i=0; i<16384; i++) {

        pB->CpRead( &iadcRead, sizeof(int) );
        iadc[i]=iadcRead;

        if (MaxRead<iadcRead) {
            MaxRead=iadcRead;
            Peak=i; }

    }

    //c.clear(); // clear screen

    int colour = rand()%256; // colour is 0-255
    int rred   = rand()%256,
        rgreen = rand()%256,
        rblue  = rand()%256;

    // put 1000 pixel at random places

    for (i=0; i<512; i++) {

    // put one Bar on screen

    int startx = i;
    int starty = 0;
    int rwidth = 1;
    int rheight= ((iadc[Peak+i-256]*380)/MaxRead);

    for (int y=starty; y<starty+rheight; y++)
        for (int x=startx; x<startx+rwidth; x++)
            c.putpixel(x,y,rred,rgreen,rblue);

        }


        c.update(); // unlock and copy to window
}


    nWrite=0;

    pB->CpWrite( &nWrite, sizeof(int) );


    c.close(); // close and release window

    cout << "Peak value was: " << (Peak-8192) << endl;

        }


cout << "Waiting for TOKEN3 from DSP" << endl;

pB->CpRead( &tokenRead, sizeof(int) );
    cout << tokenRead << endl;

  if (tokenRead != 0) {
    cout << "Corrupted TOKEN from DSP" << endl;
    return;
```

5

```
        }

      cout << "All OK! Board Test passed" << endl;

    }

    catch (SMTExc &e) {
      cout << "An exception occured. (" << e.GetError() << ")" << endl;
    }
  }
```

6

```
/*********************************************************************************
** File Name:              ZMCA1ch.C
**
** Description:     Multi Channel Analyzer (1ADC chann.) on Sundance (Cont. Sampling)
**                       16384 channels , 32 bit accumulators , 1 ADC channel
**                 NO threshold.
**                  Client-Server Application structure, RT Graphics Window !!
**
** Author:          Francesco Voltolina
**
**
** Data:                April 2004
**
**              ZisckuElectronics inc.
**
*********************************************************************************/

#include <stdio.h>
#include <stdlib.h>
#include <link.h>
#include <SMT_SDB.h>
#include <timer.h>

#define USE_SDB 0           // SDB A
#define K 1024


main()

{

    int w=0, i, load_word, byte_count, not_end=1, record_type, HOST_LINK=3 ;

    unsigned samples, leds, channels, loops, divider, active_channels, ctrl_word;

    int j, l, n, value1, value2;

    int *buf = memalign(128, 128*K*sizeof(int));

    int *adc1 = memalign(128, 32*K*sizeof(int));

    SMT_SDB *Sdb = SMT_SDB_Claim(USE_SDB);


  while (w!=2048) {
    link_in_word( &w, HOST_LINK  );
    w = w * 2;
    link_out_word( w, HOST_LINK  );
  }

    link_out_word( 0, HOST_LINK  ); // Token READY send ################


//   Config FPGA Routine Start ****************************************************


while (not_end) {

                link_in_word( &record_type, HOST_LINK );

                if (record_type==1) {
                    not_end=0;                 // end of file
                }

                if (record_type==0) {

                    link_in_word( &byte_count, HOST_LINK );

                    for(i=0; i<(byte_count/4); i++) {

                        link_in_word( &load_word, HOST_LINK );  //load_word

                        link_out_word(load_word, 0);
                    }
```

```
                } //if (record_type==0)

            } //End while (not_end)

//   Config FPGA Routine END   ***************************************************

//
//                    SDB e co. initialization
//



    if (!buf) {
        // printf("Cannot allocate memory for buffer\n");
            exit(1);
        }


        if (!Sdb) {
        // printf("Cannot claim SDB\n");
        exit(1);
        }



  loops=256;
  samples=256;
  channels=0x01;
  active_channels=1;
  divider=05;


  for (l=0; l<8; l++) {

    not_end=1;
       channels=0x01<<l;
     for (n=0; n<128*K; n++) buf[n] = 0;
     for (n=0; n<32*K; n++) adc1[n] = 0;


//
//                    start of scope loop
//



        leds=0;

        SMT_SDB_Control(Sdb, SDB_CLRIF|SDB_CLK);

        ctrl_word=(samples-1<<24)+0x00F00000+(leds<<16)+(channels<<8)+(divider-1); // 03 is orig
ctrW
        //printf("%08X\n",ctrl_word);
        link_out_word(ctrl_word,0);

        //printf("After ctrl_word written to adc!\n\n");


        SMT_SDB_CacheArea(Sdb, 128*K*4, buf, 0);

        SMT_SDB_Read (Sdb, 2*samples*active_channels*loops, buf);   // throw (2*samples*active_ch
annels*loops) bytes


        link_out_word( 0, HOST_LINK  ); // Token2 send !!!       ###################


    while (not_end) {

        leds=10;
```

2

```
        for (n=0; n<32*K; n++) adc1[n] = 0;

        SMT_SDB_Read (Sdb, 2*samples*active_channels*loops, buf);   // Read 65536 samples

        for (j = 0; j < samples*active_channels*loops/2; j++) {

        //value1 = (buf[j]&0x0000FFFF);
        //value2 = ((buf[j]>>16)&0x0000FFFF);

        value1 = (*(buf + (j))&0x0000FFFF);
        value2 = ((*(buf + (j))>>16)&0x0000FFFF);

                value1 >>= 2;
                value2 >>= 2;

                adc1[value1]++;     // DO istogramming !!
            adc1[value2]++;

        }

        link_in_word( &w, HOST_LINK  ); // Token !not_end! !!! ##############
        not_end=w;

            for (i = 0; i < 16384; i++) {

                link_out_word( adc1[i] , HOST_LINK  );  // send Histogram on Cp.

            }



            }   // ******** end of WHILE not_end


        leds=4;
        ctrl_word=(samples-1<<24)+0x00100000+(leds<<16)+(channels<<8)+(divider-1);
        //printf("%08X\n",ctrl_word);
        link_out_word(ctrl_word,0);


        }

    link_out_word( 0, HOST_LINK  );  // TOKEN3 #########################

}
```

ZsCpDSP.cfg

```
! **********************************************************
! ZsCpDSP       configuration file for ZsCpDSP
! Ziscku electronics inc.
! NOTE: You need to specify the type of TIM you are using.
! See the 3L Diamond documentation for more details.
! **********************************************************


processor root smt361 SDBRX=0
!processor node smt361

!wire ? root[1] node[4]     ! This assumes
                            ! EITHER:   a physical comport cable from
                            !           T1C1-O (TIM1) to T2C4-I (TIM2)
                            ! OR:       the quick switches on the motherboard
(SMT310Q)
                            !           have been set appropriately.
                            !           From the server, use:
Board/Properties/Configuration

! Task declarations indicating channel I/O ports and memory requirements

task ZsCpDSP  data=2M
!task ZsCpDSP    ins=1 outs=1 data=30k

! Set up the connections between the tasks.
!connect ? ZsCpDSP[0] ZsCpDSP[0]
!connect ? ZsCpDSP[0] ZsCpDSP[0]

! Assign software tasks to physical processors
place ZsCpDSP root
!place ZsCpDSP  node
```

Page 1